\documentclass[letterpaper,12pt,twopage]{report}
\usepackage{cite}
\usepackage{longtable}
\usepackage{tabularx}
\usepackage[margin=1in]{geometry}
\usepackage{booktabs}
\usepackage{capt-of}
\usepackage[utf8]{inputenc}
\usepackage[spanish.lcroman]{babel}
\usepackage{graphicx}
\usepackage{rotating, graphicx}
\usepackage{anysize}
\usepackage{float}
\usepackage{fancyhdr}
\usepackage{mathtools}
\usepackage{amsmath, amsthm, amssymb}
\usepackage{placeins}
\usepackage{float}
\usepackage{amsmath, amssymb, graphics, setspace}
\usepackage{graphicx}
\newcommand{\mathsym}[1]{{}}
\newcommand{\unicode}[1]{{}}
\usepackage{textcomp}
\usepackage{eurosym}
\usepackage{amsfonts}
\usepackage{array}
\usepackage{amsthm}
\usepackage{changes}
\usepackage{lipsum}
\definechangesauthor[name={Per cusse}, color=orange]{per}
\usepackage{bm}

\usepackage[colorlinks=true,
            linkcolor=blue,
           urlcolor=blue,
           citecolor=blue]{hyperref}
\usepackage{amsmath, amsfonts, amssymb}
\usepackage{bm, array}
\usepackage{mdframed}
\usepackage{grffile}
\usepackage{amssymb}
\usepackage{caption}
\usepackage{ulem}
\usepackage[T1]{fontenc}
\usepackage[utf8]{inputenc}
\usepackage{subfigure}

\providecommand{\U}[1]{\protect\rule{.1in}{.1in}}
\newcommand{\be}{\begin{equation}}
\newcommand{\ee}{\end{equation}}

\newcommand{\mincir}{\raise
-3.truept\hbox{\rlap{\hbox{$\sim$}}\raise4.truept\hbox{$<$}\ }}
\newcommand{\magcir}{\raise
-3.truept\hbox{\rlap{\hbox{$\sim$}}\raise4.truept\hbox{$>$}\ }}

\def\e{\mathbf{e}}

\def\udot{\dot{u}}
\def\ex{e_1{}^1}
\def\ey{e_2{}^2}
\def\ez{e_3{}^3}

\def\y{\vartheta}
\def\z{\varphi}

\usepackage{amsfonts}
\usepackage{amsthm}
\usepackage{amsmath}
\usepackage{rotating}

\usepackage{pdflscape}
\RequirePackage{graphicx}
\RequirePackage{mathptmx}      
\RequirePackage{flushend}
\usepackage{textcomp}
\usepackage{eurosym}
\usepackage{widetext}
\usepackage{array}
\usepackage{bm}
\usepackage{graphicx}
\usepackage{bm, array}
\usepackage[T1]{fontenc}
\def\be{\begin{equation}}
\def\ee{\end{equation}}
\def\beq{\begin{eqnarray}}
\def\eeq{\end{eqnarray}}

\newcommand{\ben}{\begin{eqnarray}}
\newcommand{\een}{\end{eqnarray}}

\usepackage{color}
\usepackage{enumerate}
\usepackage{soul}
\def\bea{\begin{eqnarray}}
\def\eea{\end{eqnarray}}

\def\be{\begin{equation}}
\def\ee{\end{equation}}
\def\e{\mathbf{e}}

\def\udot{\dot{u}}

\def\ex{e_1{}^1}
\def\ey{e_2{}^2}
\def\ez{e_3{}^3}

\def\y{\vartheta}
\def\z{\varphi}

\newtheorem{theorem}{Teorema}[section]

\theoremstyle{definition}
\newtheorem{definition}{Definición}[section]

\theoremstyle{remark}
\newtheorem{remark}{Nota}[section]

\marginsize{4cm}{2.5cm}{4cm}{2.5cm}
\graphicspath{ {figuras/} }

\title{Análisis cualitativo de modelos Einstein-æther con fluido perfecto y campo escalar} 
\author{Alfredo David Millano Mejías} 

\makeatletter 
\let\newtitle\@title
\let\newauthor\@author
\makeatother

\AtBeginDocument{}
\AtBeginDocument{}
\AtBeginDocument{}
\begin{document}
\pagenumbering{gobble}

\thispagestyle{empty}
\begin{center}
    \includegraphics[scale=0.5]{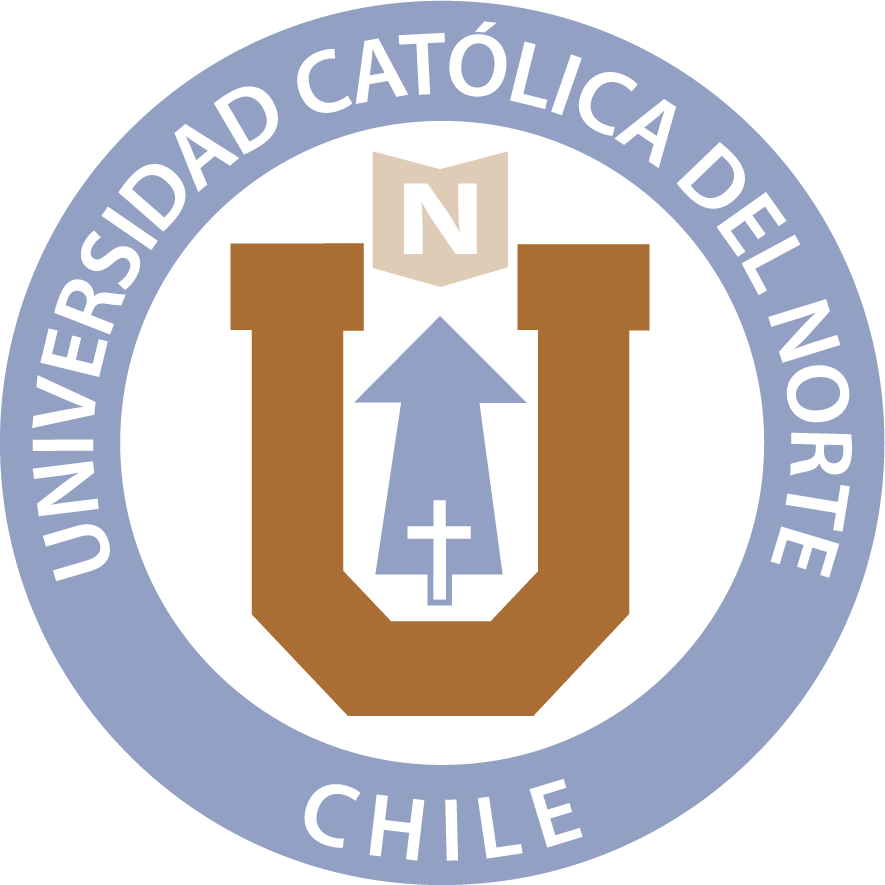}\\
    \vspace{1cm}
    \begin{large}
        \textbf{UNIVERSIDAD CATÓLICA DEL NORTE}\\
        \vspace{0.3cm}
        FACULTAD DE CIENCIAS\\
        \vspace{0.3cm}
        Departamento de Matemáticas\\
        \vspace{1cm}
    \end{large}
    \begin{large}
        \textbf{\MakeUppercase{\newtitle}}\\
        \vspace{1.5cm}
    \end{large}
    \begin{large}
        Tesis para optar al grado de Magíster en Ciencias Mención Matemática\\
        \vspace{1cm}
        \textbf{\MakeUppercase{\newauthor}}\\
        \vspace{1cm}
        Profesor guía: Doctor Genly Leon Torres.\\ 
        \vspace{0.5cm}
        Antofagasta, Chile.\\
        Junio, 2020 
    \end{large}
\end{center}
\newpage
\thispagestyle{empty}
\begin{flushright}
\textit{Esta tesis está dedicada a Isabel Montenegro, Sic Parvis Magna.}
\end{flushright} 

\chapter*{Agradecimientos}
\noindent Agradezo principalmente a la Agencia Nacional de Investigación y Desarrollo por financiar esta tesis a través del Proyecto FONDECYT Iniciaci\'on grant no. 11180126. Agradezo también a la Dra. Elva Ortega, Directora del Programa, por sus consejos y atención y a la planta académica de la UCN por haberme formado durante el Magíster. Finalmente, agradezco a mi tutor de tesis el Dr. Genly Leon por su apoyo incondicional durante la realización de este trabajo.  

\pagenumbering{roman}
\cfoot{\thepage}
\renewcommand{\headrulewidth}{1pt}
\renewcommand{\footrulewidth}{1pt}
\tableofcontents
\addcontentsline{toc}{chapter}{\textbf{Índice de figuras}} 
\listoffigures 
\addcontentsline{toc}{chapter}{\textbf{Índice de tablas}} 
\listoftables 
\include{resumen} 
\pagenumbering{arabic}

\addcontentsline{toc}{chapter}{\textbf{Introducción.}} 
\chapter*{Introducción}
\cfoot{\thepage}
\renewcommand{\headrulewidth}{1pt}
\renewcommand{\footrulewidth}{1pt}

\noindent

En la Física moderna, se distinguen tres áreas de estudio, que son relevantes en el contexto de la presente investigación, a saber:
\begin{itemize}
    \item Astronomía: es la ciencia que se ocupa del estudio de los cuerpos celestes del universo, incluidos los planetas y sus satélites, los cometas y meteoroides, las estrellas y la materia interestelar, los sistemas de materia oscura, gas y polvo llamados galaxias y los cúmulos de galaxias; por lo que estudia sus movimientos y los fenómenos ligados a ellos. La Astronomía también abarca el estudio de la formación y el desarrollo del Universo en su conjunto mediante la Cosmología, y se relaciona con la Física mediante la Astrofísica, la química mediante la Astroquímica y la biología con la Astrobiología. 
    \item Astrofísica: trata sobre el desarrollo y estudio de la Física aplicada a la Astronomía. Estudia las estrellas, los planetas, las galaxias, los agujeros negros y demás objetos astronómicos como cuerpos de la Física, incluyendo su composición, estructura y evolución. La Astrofísica emplea la Física para explicar las propiedades y fenómenos de los cuerpos estelares a través de sus leyes, fórmulas y magnitudes. Se dedica al estudio de la Física de procesos fuera del entorno de la tierra (pero incluyendo algunos como procesos en la magnetosfera, auroras, rayos cósmicos, etc.). Trata sobre la modelación, observación, simulación de procesos como la transferencia radiativa en cuerpos celestes, evolución de estrellas, galaxias, de cúmulos de galaxias, abundancias químicas en nubes, planetas, estrellas etc..
    \item Cosmología: en rasgos generales estudia la historia del universo desde su nacimiento. Hay numerosos campos de estudio de esta rama de la Astronomía. Varias investigaciones conforman la Cosmología actual, con sus postulados, hipótesis e incógnitas. La Cosmología física comprende el estudio del origen, la evolución y el destino del Universo utilizando los modelos de la Física. La Cosmología física se desarrolló como Ciencia durante la primera mitad del siglo XX como consecuencia de diversos acontecimientos y descubrimientos encadenados durante dicho período. Se centra en el conjunto del universo, como por ejemplo en el Big Bang o las agrupaciones de las galaxias y sus movimientos; la modelización de la aceleración de la expansión de universo a través del estudio de la energía oscura (y alternativas) y su interrelación (e interacción) con la materia oscura. 
\end{itemize}

La Cosmología se puede decir que es una rama de la Física que informalmente se enfrenta a preguntas de las cuales no se puede dar una respuesta sencilla, según Carl Sagan tales preguntas solo fueron tratadas en mitos y en ámbitos religiosos.
Formalmente, la Cosmología estudia el origen y evolución del universo, las propiedades del mismo desde un enfoque a "`gran escala"'. Los origenes de la Cosmología se pueden rastrear hasta culturas antiguas como la Babilónica y Egipcia aunque con un enfoque dirigido hacia la espiritualidad, fe y mistificación. El entendimiento actual de la Cosmología es el conjunto de ideas científicas que estudian el Cosmos. En el siglo XVI, Nicolás Copérnico impulsó el cambio del sistema geocéntrico al heliocéntrico esto ayudo a alcanzar grandes resultados científicos, entre estos esta el estudio de Isaac Newton sobre la interacción de las fuerzas gravitacionales durante el siglo XVII y muchos otros que desembocaron en el siglo XVIII o mejor conocido como siglo de la ilustración. Debido a esto, se puede interpretar como uno de los primeros resultados de la Cosmología moderna el hecho de dejar atrás una mentalidad medieval y comprender donde estamos en el universo. Los experimentos y las matemáticas formaron parte fundamental en el estudio de los fenomenos que describen el comportamiento del universo, esto dio paso al ahora llamado Método Científico. En el albor del siglo XX, Albert Einstein propuso la unificación del espacio y el tiempo, lo llamó "`Espacio-tiempo"' en su famosa Teoría General de la Relatividad, la cual sería el marco para estudiar el universo desde el enfoque físico-matemático. La visión de Einstein del universo era la de uno estático, homogéneo e isótropo, lo que significa que en el contexto cosmológico, o sea, a grandes escalas es igual en todas las direcciones y en todos los puntos.  Una pregunta a principios de Siglo XX era si la Vía Láctea contenía todo el universo. El famoso Edwin Hubble en sus estudios realizados desde $1919$ hasta $1924$, calculó que la galaxia Andrómeda y la galaxia del Triángulo estaban muy lejos para ser parte de nuestra galaxia y que por lo tanto eran tres galaxias diferentes. Los hallazgos de Hubble cambiaron fundamentalmente la visión científica del Universo y abrió el camino futuros astrónomos y cosmólogos. En $1920$ Willem de Sitter formula su modelo estático vacío de materia. Pero en $1921$ Alexander Friedmann presentaba la primera solución matemática de las ecuaciones de Einstein, con un universo en expansión. Claramente esto presentaba un choque de visiones y modelos de universo, esto hizo que incrementara la tensión. Esto dio como resultados la Ley de Hubble, el efecto Doppler y la radiación de fondo de microondas. Este último fue de crucial importancia para establecer al modelo del Big Bang como un elemento adicional al modelo estándar del universo.

Sin embargo, en las últimas dos décadas ha habido un cambio de paradigma con relación a la composición del universo, dado que un universo dominado por materia bariónica, con un remanente de radiación no da cuenta (a través de la ecuación de Raychaudhuri) del fenómeno de  la «expansión acelerada del universo». 
La «expansión acelerada del universo» o «universo en expansión acelerada» son términos con los que se designa el hecho descubierto en el año 1998 de que el universo se expande a una velocidad cada vez mayor. Este hecho fue un descubrimiento inesperado, ya que hasta ese descubrimiento se pensaba que, si bien el universo ciertamente estaba en expansión, esta, lo hacía a un ritmo que iba decreciendo por efecto de la atracción gravitatoria mutua entre galaxias distantes (aunque lentamente debido a la densidad de materia-energía baja presente en el universo, $\Omega_m$, que se estima en ser igual a $0.29$)

A finales de los años 1990 unas observaciones de supernovas tipo A (clase Ia) arrojaron el resultado inesperado de que la expansión del universo ha ido acelerándose desde hace unos 5000 millones de años. Estas observaciones parecen más firmes a la luz de nuevos datos.

Puesto que la energía causante de la aceleración del espacio-tiempo no ha podido ser observada en forma directa, se ha dado en llamarla energía oscura. Dos candidatos teóricos que podrían hacer las veces de esta energía son una constante cosmológica no igual a cero (que pudo haber causado la inflación cósmica) y una energía repulsiva más general llamada quintaesencia. De todas maneras una expansión acelerada no entra en contradicción frontal con la formulación original de la Teoría General de la Relatividad, que ya ocasionó en su tiempo una polémica entre Albert Einstein, quien en un tiempo introdujo la constante cosmológica en su ecuación de campo retirándola después, y varios científicos: Alexander Friedman, Georges Lemaître, Howard Percy Robertson y Arthur Geoffrey Walker, quienes probaron que existían soluciones estables no estacionarias sin el término proporcional a la constante cosmológica.

Algunos hitos cruciales de la Cosmología en los últimos 20 años son:

\begin{itemize}
	\item El descubrimiento de la expansión aceledara del universo en 1998, por S. Perlmutter B. P. Schmidt y A. G. Riess, ganadores del Premio Nobel de Física en $2011$. 
	\item El descubrimiento del Bosón de Higgs, que fue una pieza clave en el modelo de Física de partículas estandar. Peter Higgs recibió el Nobel de Física en $2018$ por su teoría formulada en $1964$.
	\item El descubrimiento de las ondas gravitacionales por el LIGO (Laser Interferometer
Gravitational Wave Observatory) en el año $2016$ por los físicos R. Wiess,
B. Barish y K. Thorne.
\item En el año $2019$ se otorga el Premio Nobel de Física a J. Peebles, M. Mayor y D.
Queloz por descubrimientos en Cosmología física, en especial el descubrimiento
de un exoplaneta que orbita una estrella tipo solar.
\end{itemize}

En esta tesis, se estudiará un modelo con fluido perfecto en el marco de los modelos temporales autosimilares esféricamente simétricos, generalizando los resultados obtenidos por Uggla, Nilsson y Goliath en \cite{Goliath:1998mx} para el caso de Relatividad General. Los modelos con fluido perfecto se usan en la teoría de la Relatividad General (GR) para modelar distribuciones idealizadas de la materia, como el interior de una estrella o un universo isotrópico  \cite{Nilsson:1995ah,Goliath:1998mx}. En el último caso, la ecuación de estado del fluido perfecto puede usarse en las ecuaciones de Friedmann – Lemaître – Robertson – Walker (FLRW) para describir la evolución del universo. También es de interés estudiar un modelo cosmológico con campo escalar con un potencial adecuado bajo el mismo marco de los modelos temporales autosimilares esféricamente simétricos.
En particular, nos interesa la caracterización del espacio de las soluciones de las ecuaciones diferenciales que resultan de considerar fluidos de materia de tipo campo escalar o fluido perfecto en la teoría Einstein-æther, así como determinar criterios de estabilidad para las soluciones de estado estacionario del sistema. 

La teoría Einstein-æther consiste en la Relatividad General acoplada a un campo vectorial de tipo tiempo unitario, llamado el ``æther''. En esta teoría efectiva, la invarianza de Lorentz es violada, pero la localidad y la covarianza son preservadas en presencia del campo vectorial. Para la formulación matemática de los modelos se usa el formalismo $1+3$ que permite escribir las ecuaciones de campo para métricas nohomogéneas esféricamente simétricas como un sistema de ecuaciones diferenciales parciales en dos variables. En particular, usando la formulación diagonal homotética, dichas ecuaciones diferenciales parciales pueden ser escritas como ecuaciones diferenciales ordinarias más restricciones algebraicas, usando el hecho que la métrica se adapta a la simetría homotética. Las ecuaciones resultantes son muy similares a las de los modelos con  hipersuperficies espaciales homogéneas \cite{Goliath:1998mw}. Esto permite el estudio cualitativo de las soluciones usando técnicas de la teoría local de los sistemas dinámicos.  Los resultados análiticos serán verificados mediante integración numérica.

\section*{Estado del arte}

Con el descubrimiento de la aceleración del universo en 1998 usando supernovas de tipo Ia \cite{Riess:1998cb}, el concepto de la Energía Oscura fue introducido en el modelo cosmológico éstandar para dar cuenta del $68\%$ de la energía del universo \cite{Ade:2013sjv}.
Mediciones de anisotropías de la radiación cósmica de microondas (CMB) (radiación de fondo de microondas) obtenidas de experimentos como los realizados por los satélites WMAP \cite{Bennett:2012zja} y Planck \cite{Ade:2013zuv}, han brindado un fuerte apoyo para modelos cosmológicos con energía oscura (y específicamente con constante cosmológica, $\Lambda$).
Sin embargo, existen algunas diferencias entre las mediciones locales de la tasa de expansión del Hubble a partir de las supernovas Ia \cite{Riess:2011yx} y de otros datos cosmológicos \cite{Ade:2014xna}.

Es difícil de concebir una constante cosmológica $\Lambda$ cuya densidad de energía deba ser ajustada (Problema de ajuste--fino,  o sea, se deben ajustar parámetros de un modelo de manera precisa para  adaptarse a ciertas observaciones) en
$\sim 55$ ordenes de magnitud para dar cuenta de la aceleración actual del universo \cite{Martin:2012bt}. Varios intentos de explicar la aceleración cósmica en el contexto de la Relatividad General sin energía oscura, por ejemplo usando la reacción inversa de inhomogeneidades en dinámica cósmica  \cite{Buchert:1999er} o postulando que vivimos cerca del centro de un gran vacío en un universo dominado por polvo (en Cosmología se maneja todo a gran escala, el sistema solar, las galaxias se tratan como \textit{partículas de polvo}) \cite{Alnes:2005rw},  han sido poco convincentes. Por tanto, una alternativa es abandonar Relatividad General y modificar la acción de Einstein-Hilbert. 

Una teoría muy interesante es la llamada teoría  Einstein-æther 
\cite{Gasperini:1986ym, Kostelecky:1989jp, Jacobson:2000xp,  Lim:2004js, Kanno:2006ty, Zlosnik:2006zu,Jacobson:2008aj, Donnelly:2010cr,  Carruthers:2010ii, Blas:2014aca, Latta:2016jix, Coley:2015qqa}, que consiste en Relatividad General acoplada a un campo vectorial temporal unitario llamado æther. En esta teoría efectiva, se viola la invarianza de Lorentz  pero el sistema dinámico preserva la localidad y la covarianza en presencia del  æther \cite{Jacobson:2000xp, Kanno:2006ty, Zlosnik:2006zu,Donnelly:2010cr, Carruthers:2010ii}. El impacto de estas teoría en el CMB  ha sido explorado en  \cite{Kanno:2006ty, Zlosnik:2006zu,Carruthers:2010ii, Blas:2014aca}. En particular los espectros primordiales de perturbaciones generados por la inflación en presencia de un campo vectorial temporal que viola la invarianza de Lorentz, han sido calculados y se encontró que la amplitud de los espectros cambia, lo que en general conduce a  una violación de la relación de consistencia inflacionaria
\cite{Lim:2004js}.
Con exactitud, la teoría 
Einstein-æther es una teoría de campo efectiva en la cual la acción de Einstein--Hilbert action se modifica mediante la introducción de un campo vectorial dinámico unitario, $u^{a}$, el æther, el cual está acoplado covariantemente, a nivel lagrangiano, hasta derivadas de segundo orden de la métrica del espacio--tiempo, $g_{ab}$, excluyendo derivadas totales. La unitaridad es impuesta introduciendo un multiplicador de Lagrange que se agrega a la acción. Esta teoría tiene características muy interesantes para los matemáticos, los físicos y los cosmólogos: a) Viola la invarianza de Lorentz, pero preserva localidad  y covarianza; b) Tiene algunas huellas en el escenario inflacionario; c) Satisface las condiciones para la estabilidad linealizada, la energía positiva y satisface las restricciones a los parámetros post-newtonianos;   d) Para valores genéricos de las constantes de acoplamiento, el  æther y el elemento de línea isotropizan en el futuro (aunque para ángulos de inclinación grandes o derivadas respecto de los ángulos de inclinación grandes, hay un comportamiento desbocado en el que las anisotropías aumentan con el tiempo, y pueden aparecer algunas singularidades); y e) 
cada solución de Einstein æther hipersuperficie-ortogonal es una solución de Ho\v{r}ava (la mayoría de las soluciones estudiadas), etc., ver, por ejemplo,  \cite{Gasperini:1986ym,Kostelecky:1989jp,Jacobson:2000xp,Carroll:2004ai,Eling:2004dk,Lim:2004js,Kanno:2006ty,Zlosnik:2006zu,Garfinkle:2007bk,Jacobson:2010mx,Donnelly:2010cr,Carruthers:2010ii,Garfinkle:2011iw,Barrow:2012qy,Sandin:2012gq,Alhulaimi:2013sha,Jacobson:2013xta,Blas:2014aca,Coley:2015qqa,Latta:2016jix,Alhulaimi:2017ocb,VanDenHoogen:2018anx,Coley:2019tyx,Leon:2019jnu,Roumeliotis:2018ook,Roumeliotis:2019tvu,Paliathanasis:2020bgs,Paliathanasis:2019pcl}. 

La teoría de Einstein-æther tiene aplicaciones en varios contextos anisotrópicos y no homogéneos, por ejemplo, en \cite{Coley:2015qqa, Latta:2016jix, Alhulaimi:2017ocb, VanDenHoogen:2018anx}. En \cite{Coley:2015qqa} se implementó el formalismo de marco ortonormal $1 + 3$, adoptando el gauge comóvil para el æther, para obtener ecuaciones de evolución en variables normalizadas, adecuadas para cálculos numéricos y el análisis de espacio de fase, estudiándose modelos de Kantowski-Sachs espacialmente homogéneos. En \cite{Alhulaimi:2017ocb} el campo escalar interactúa tanto con el escalar de  expansión del campo æther como con el escalar de cizalla a través del potencial, estudiándose la estabilidad frente a la curvatura espacial y las perturbaciones anisotrópicas. El atractor de la teoría es la fase expansiva del vacío de Sitter. En \cite{Coley:2015qqa,Coley:2019tyx,Leon:2019jnu} se estudiaron métricas estáticas para un fluido perfecto no inclinado con ecuaciones de estado lineales y politrópicas, y con un campo escalar con potenciales exponenciales o monomiales. Otras soluciones han sido estudiadas, como son: vacío Bianchi Tipo V en \cite{Roumeliotis:2019tvu}; métricas Friedmann - Lema\^{\i}tre - Robertson - Walker  (FLRW) en \cite{Paliathanasis:2019pcl, Roumeliotis:2018ook}; una métrica Bianchi Tipo III, localmente rotacionalmente simétrica (LRS) en \cite{Roumeliotis:2018ook}; Cosmología de campo escalar modificada con interacciones entre el campo escalar y el æther \cite{Paliathanasis:2020bgs} basado en teoría Einstein-- æther \cite{Kanno:2006ty,Donnelly:2010cr}. 
Se hizo hincapié en la cuestión de la existencia de soluciones de las ecuaciones reducidas, la clasificación de las singularidades y el análisis de estabilidad. Estas teorías son diferentes de las teorías tenso--escalar, y son similares a modelos de la creación de partículas, viscosidad aparente, teorías de vacío variable, o teorías de partículas de materia oscura con masa variable \cite{Pan:2016jli,Li:2009mf,Pan:2016bug,Basilakos:2009wi,Basilakos:2009ms,Pan:2017ios,Oikonomou:2016pnq,Leon:2009dt}. En \cite{Paliathanasis:2020axi} se estudió la existencia de varias soluciones exactas y analíticas de una teoría de Einstein-æther que incorpora un campo escalar acoplado no mínimamente al æther a través de un acoplamiento efectivo, $B(\phi)$, definido en términos de los parámetros del æther \cite{Kanno:2006ty}. Se selecciona la función de acoplamiento $ B \left(\phi\right) = 6 B_0 \phi ^ 2 $, lo que implica que las variables dinámicas de las ecuaciones de campo evolucionan en un espacio plano bidimensional de signatura Lorentziana, que es máximamente simétrico, encontrándose cinco familias de potenciales de campo escalar de la forma $ V_{A}\left (\phi \right) = V_{0} \phi ^{p} + V_{1} \phi ^ {r} $, donde $ p, r $ son constantes específicas, que conducen a  sistemas integrables según Liouville, y que admiten leyes de conservación cuadráticas en el momento.

Siguiendo una estrategia análoga en \cite{Paliathanasis:2020pax} se determinaron soluciones analíticas y exactas de las ecuaciones de campo gravitacional para el modelo Einstein-æther con campo escalar con un espacio de fondo Bianchi I con interacciones no lineales del campo escalar con el campo æther.  El lagrangiano puntual de las ecuaciones de campo depende de tres funciones desconocidas. Se derivaron leyes de conservación para las ecuaciones de campo para formas específicas de las funciones desconocidas de tal manera que las ecuaciones de campo son integrables según Liouville. Además, la evolución de las anisotropías se estudia determinando los puntos de equilibrio y analizando su estabilidad.

En \cite{Coley:2019tyx} se investigaron las ecuaciones de campo en la teoría de Einstein-æther para espacios espaciales simétricos esféricamente estáticos y una fuente de fluido perfecto y, posteriormente, con la adición de un campo escalar (con un potencial de auto-interacción exponencial). Se introdujeron variables dinámicas más apropiadas que facilitaron el estudio de los puntos de equilibrio del sistema dinámico resultante y, además, se discutió la dinámica en el infinito. Las propiedades cualitativas de las soluciones fueron de particular interés, así como su comportamiento asintótico y si admiten singularidades. Se presentaron varias soluciones nuevas.
Continuando con esta línea, en \cite{Leon:2019jnu} se investigó la existencia de soluciones analíticas para las ecuaciones de campo en la teoría de Einstein-æther para un espacio-tiempo estático esféricamente simétrico,  proporcionándose un análisis detallado del sistema dinámico correspondiente.

Recientemente en \cite{Barrow:2012qy,Sandin:2012gq,Alhulaimi:2013sha} se estudiaron modelos isotrópicos en teoría Einstein-aether con un potencial exponencial. En el artículo \cite{Coley:2015qqa} se estudiaron  modelos esféricamente
simétricos en teoría Einstein-aether con una fuente de materia de fluido perfecto.

En esta tesis, se estudiarán las métricas conformalmente estáticas (constituyendo los llamados modelos esféricamente simétricos auto-similares tipo temporal) con fluidos perfectos y con un campo escalar no homogéneo con potencial exponencial, que  satisface la simetría homotética, mediante el uso de técnicas de la teoría cualitativa de los sistemas dinámicos. 

En el sentido matemático, se puede usar el formalismo $1 + 1 + 2$ \cite{Clarkson:2002jz, Clarkson:2007yp} o el formalismo  $1 + 3$   \cite{Coley:2015qqa,wainwrightellis1997} para escribir las ecuaciones de campo como un sistema bien definido de ecuaciones  en derivadas parciales (PDE) de primer orden en dos variables. Luego, mediante el uso de simetrías adicionales, se pueden transformar estas ecuaciones a sistemas de ecuaciones diferenciales ordinarias.

Es importante mencionar que encontrar soluciones exactas a los problemas que aparecen en esta tesis es una tarea difícil. Primero se debe elegir las coordenadas adecuadas para simplificar los calculos. Además se sabe que el sistema completo (Un conjunto de 10 ecuaciones diferenciales muy complicadas de segundo orden acopladas y no lineales, que en el contexto de gravedad modificada pueden ser de cuarto orden) es generalmente no integrable, incluso para modelos más simples con materia realista, por ejemplo soluciones de estrellas de neutrones o para soluciones en el vacío en el contexto de gravedad modificada. Luego,  se deben usar técnicas que no involucran resolver las ecuaciones de campo. Esto es útil en el sentido de evitar el problema de la elección de coordenadas, la aparición de singularidades, etc.
En efecto, la descomposición semi-tétrada \cite{Ganguly:2014qia} permite escribir las ecuaciones de campo como un sistema autonomo de cantidades definidas covariantemente \cite{Clarkson:2002jz, Clarkson:2007yp}.  Esto permite eliminar todas las singularidades que puedan aparecer por una mala definición de coordenadas. Por otro lado este sistema autónomo puede ser simplificado incorporando simetrías de Killing del espacio-tiempo.

Para el análisis de nuestro modelo, es mejor considerar la descomposición covariante $ 1 + 3 $ \cite{Coley:2015qqa,wainwrightellis1997} en lugar del enfoque $ 1 + 1 + 2 $ \cite{Clarkson:2002jz, Clarkson:2007yp}, y seguir la misma estrategia que en \cite{Goliath:1998mx}. Los modelos  autosimilares tipo tiempo esféricamente simétricos se caracterizan por un grupo de simetría homotético de 4 dimensiones $H_4$ que actúa múltiplemente transitivamente sobre superficies de tiempo tridimensionales. El elemento de línea, escrito en forma diagonal, donde una de las coordenadas está adaptada a la simetría homotética, tiene la forma \cite{Bogoyavlensky}: 
\begin{align*}
& d\tilde{s}^2 = e^{2 t} ds^2= e^{2 t}\Big[- b_1^{-2}(x)  dt^2 + dx^2   +b_2^{-2}(x) (d\y^2 + \sin^2 (\y)  d\z^2)\Big].
\end{align*} 
Es bien conocido que para que los campos escalares no homogéneos $\phi(t,x)$, con auto--interacción  $V(\phi(t,x))$, respeten la simetría homotética de la métrica conformalmente estática,  tienen que ser de la forma \cite{Coley:2002je}:
\begin{align*}
& \phi(t,x)=\psi (x)-\lambda t, \quad  V(\phi(t,x))= e^{-2 t} U(\psi(x)), \quad  U(\psi)=U_0 e^{-\frac{2 \psi}{\lambda}},    
\end{align*}
donde por convenio se asume $\lambda>0$, tal que si $\psi>0$, $U\rightarrow 0$ cuando $\lambda \rightarrow 0$.

Luego, el \textbf{objeto de estudio} de esta tesis, son los modelos temporales autosimilares esféricamente simétricos con fluido perfecto y campo escalar con potencial de interacción que satisface la simetría homotética antes dicha.

Después de hacer una revisión extensa de la literatura relevante al área de estudio, y de la elaboración del marco téorico, se formulan las siguientes 
\textbf{preguntas de investigación}:
\begin{itemize}
	\item ¿Es posible utilizar herramientas de sistemas dinámicos para describir cualitativamente modelos cosmológicos/astrofísicos con fluido perfecto y campo escalar para métricas temporales autosimilares esféricamente simétricas?
	\item ¿Es posible generalizar los resultados de Uggla, Nilsson y Goliath obtenidos en \cite{Goliath:1998mx} para  Relatividad General a partir de los resultados del análisis cualitativo de nuestros modelos?
\end{itemize}
Para responderlas nos trazamos como \textbf{objetivo general}:
 \begin{itemize}
     \item  Analizar cualitativamente sistemas de ecuaciones diferenciales ordinarias en el marco de los modelos temporales con métricas autosimilares esféricamente simétricas con fluido perfecto y campo escalar en  teoría Einstein-æther.
     \end{itemize}
A su vez, hemos trazado tres \textbf{objetivos específicos}:
\begin{itemize}
	\item Analizar cualitativamente modelos cosmológicos/ astrofísicos con fluido perfecto, dentro del marco de los modelos temporales autosimilares esféricamente simétricos, mediante el uso de técnicas de sistemas dinámicos.
	\item Analizar cualitativamente modelos cosmológicos/ astrofísicos con campo escalar, dentro del marco de los modelos temporales autosimilares esféricamente simétricos, mediante el uso de técnicas de sistemas dinámicos.
	\item Generalizar los resultados obtenidos en \cite{Goliath:1998mx} por Uggla, Nilsson y Goliath para modelos de Relatividad General.
\end{itemize}

\textbf{Hipótesis de investigación}\newline 
Se formula como hipótesis de investigación que es posible obtener información relevante -desde el punto de vista cualitativo- de sistemas autonomos de ecuaciones diferenciales, que provienen del contexto cosmológico, mediante el uso de herramientas cualitativas de sistemas dinámicos. En particular el uso de variables normalizadas, y la aplicación de resultados relacionados con la estabilidad de puntos de equilibrio.

\textbf{Novedad Científica}\newline
Durante el proceso de investigación fueron obtenidos los siguientes resultados novedosos:
\begin{enumerate}
    \item[\checkmark] Se recuperan los resultados de Uggla, Nilson y Goliath para relatividad general.
    \item[\checkmark] Se obtienen condiciones de estabilidad para puntos de equilibrio de modelos con fluido perfecto.
    \item[\checkmark] Se obtienen condiciones de estabilidad para puntos de equilibrio de modelos con campo escalar.
\end{enumerate}

La tesis tiene la siguiente 
\textbf{estructura}. 
En el capítulo \ref{ch_1} se hace una revisión de resultados de la teoría cualitativa de los sistemas dinámicos. Los resultados para sistemas lineales se presentan en la sección \ref{seccion1.1}. En la sección \ref{seccion1.2} se enuncia el Teorema de Hartman Grobman que se aplica para estudiar la estabilidad los puntos de equilibrio hiperbólicos. En la sección \ref{seccion1.3} se enuncia el Teorema de la Variedad Estable e Inestable. En la sección \ref{Strogatz} se discute un ejemplo de flujo en una dimensión.

En el capítulo \ref{ch_2} se discute el marco teórico.  La sección \ref{aetheory} se dedica a la teoría de la gravedad de Einstein-æther, que contiene a la teoría de Relatividad General en un caso límite. En la sección \ref{section2.2}  se discute el formalismo $1+3$, y en la sección \ref{homotetica} se introduce la formulación diagonal homotética que  tiene la ventaja de que las ecuaciones diferenciales parciales que describen el modelo (provenientes del formalismo $1+3$) se reducen a ecuaciones diferenciales ordinarias - dado que la métrica se adapta a la simetría homotética- las cuales son muy similares a las de los modelos con hipersuperficies espaciales homogéneas \cite{Goliath:1998mw}, por lo que se puede usar las técnicas de la teoría cualitativa de los sistemas dinámicos que se aplican exitosamente para el análisis de dichas teorías.

En el capítulo \ref{ch_3} se estudia el modelo con fluido perfecto. En la sección \ref{model3} se usa la formulación diagonal homotética  para escribir las ecuaciones de campo como un sistema de ecuaciones diferenciales ordinarias más restriciones. En la sección  \ref{thetaperfectfluid} se normalizan las ecuaciones para obtener un sistema en variables adimensionales y se hace un análisis cualitativo de los puntos del sistema resultante. En la sección \ref{SL}, se estudia la estabilidad de la superficie de no extendibilidad de las soluciones, la cual contiene una curva singular que esta relacionada con posibles cambios de causalidad del modelo. Se analiza su estabilidad haciendo un rescalamiento no monótono de la variable independiente que permite analizar localmente la estabilidad de la curva singular mediante el estudio de las perturbaciones lineales, obteniéndose analíticamente y gráficamente las condiciones de estabilidad de los puntos sobre dicha línea. En la sección \ref{tiltfluidoperfecto} se estudian los conjuntos invariantes ($v=\pm 1$) correspondientes a los casos de inclinación extrema.  En la sección \ref{fluidosinpresion} se estudia un fluido perfecto sin presión  (con índice barotrópico $\gamma=1$). En la sección \ref{general} se estudia el caso general $v\neq 0$. En la sección \ref{vcero} se estudia el conjunto invariante $A=v=0.$ En la sección \ref{DiscusionC1} se  resumen los resultados obtenidos en este capítulo.

En el capítulo \ref{ch_4} se estudia un modelo con campo escalar analizando cualitativamente varios sistemas de ecuaciones diferenciales ordinarias que modelan distintas situaciones de interés físico. En la sección  \ref{SECT:4.1} se formulan los modelos temporales autosimilares esféricamente simétricos con campo escalar. 
En la sección \ref{Sect:4.2} se discute la $\theta$-normalizacion de las ecuaciones. Dada la dificultad computacional de la obtención de los puntos de equilibrio del sistema general \eqref{reducedsystSF} analíticamente, en diferentes subsecciones se estudian algunos casos particulares de interés físico. En particular en la sección \ref{gamma=2/3} se discute el caso especial $\Omega_t=v=0$ y $\gamma=2/3$, correspondiente a
un fluido cosmológico en forma de gas ideal con ecuación de estado $p_m =(\gamma-1)\mu_m$, con $\gamma=2/3$ que mimetiza un espacio-tiempo FLRW  con curvatura no nula. 
En la sección \ref{Isotrop} se estudia el  conjunto invariante $\Sigma=0$ (no confundir con los modelos isotrópicos). En la sección \ref{tilt2} se estudian los conjuntos invariantes $v=\pm 1$ que corresponden a inclinación extrema. Finalmente, en la sección \ref{Section4.7} se estudia el conjunto invariante $A=v=0$. En todos los casos se procede de la misma manera que en el capítulo anterior, presentando condiciones de estabilidad de los puntos de equilibrio.  Los resultados parciales se discuten en la sección \ref{SECT:4-3} y en la sección 
\ref{progreso} se comenta brevemente sobre trabajo en progreso. El capítulo \ref{ch_5} está dedicado a un resumen final y a conclusiones.

\lhead{Capítulo \ref{ch_1}}
\rhead{Teoría de sistemas dinámicos}
\cfoot{\thepage}
\renewcommand{\headrulewidth}{1pt}
\renewcommand{\footrulewidth}{1pt}

\chapter{Teoría de sistemas dinámicos}\label{ch_1}

 En este capítulo se resumen técnicas de la teoría cualitativa de sistemas dinámicos \cite{Smale,Wig,Stro,Coley:2003mj,wainwrightellis1997}, las cuales se utilizarán para analizar la estabilidad de las soluciones de los modelos bajo estudio en esta tesis. El capítulo esta dividido en cuatro secciones. La primera sección se dedica al estudio de los sistemas lineales autónomos. La segunda sección se dedica al estudio de estabilidad los puntos de equilibrio hiperbólicos y al Teorema de Hartman Grobman, que es una de las herramientas principales que se emplearán en esta tesis. La tercera sección está dedicada a discutir algunos resultados de la Teoría de la Variedad Estable e Inestable. Finalmente, la cuarta sección se enfoca en la manera en la cual se analizan los flujos en una dimensión. 

\section{Sistemas lineales autónomos}
\label{seccion1.1}
En esta tesis se estudian sistemas de EDO's en espacios de dimensión finita de la forma 
\begin{equation}\label{sistema no lineal}
\dot{x}=f(x),
\end{equation}
donde $\dot{x}\equiv \frac{dx}{dt}$, $x \equiv (x_1,...,x_n)\in \mathbb{R}^n$ y $f:\mathbb{R}^n \rightarrow \mathbb{R}^n$ es al menos de clase $C^1$ en $\mathbb{R}^n$, llamados \emph{sistemas autónomos}, es decir, no hay dependencia explicita de $t$. En muchos problemas aparecen sistemas del tipo 
\begin{equation}
\dot{x}=f(x,t),
\end{equation}
llamados \emph{sistemas no autónomos}. En general en  \eqref{sistema no lineal} la función $f$ es no lineal. Los  \emph{sistemas lineales} puede escribirse como 
\begin{equation}\label{sistema lineal}
\dot{x}=Ax,
\end{equation}
con $A\in M_{n\times n}(\mathbb{R})$ 

Se llamará a $x$ vector de estado del sistema y a $\mathbb{R}^n$ espacio de fase, $f$ puede ser vista como un \emph{campo vectorial} ya que $f(x)=(f_1(x),...,f_n(x))$ se puede interpretar como un vector en $x$.

\begin{definition}{\textbf{Solución de un sistema de EDO's:}}
Una función $\phi:\mathbb{R}\rightarrow \mathbb{R}^n$ es una solución de \eqref{sistema no lineal} si 

\begin{equation}\label{Solucion}
\dot{\phi}(t)=f(\phi(t))
\end{equation}
se cumple para todo $t\in \mathbb{R}$.

\end{definition}

\begin{remark} No se espera encontrar soluciones exactas de \eqref{sistema no lineal} cuando $n\geq 2$ por tanto, se ocuparán métodos cualitativos y numéricos para deducir el comportamiento asintótico del sistema.
\end{remark}

\begin{definition}{\textbf{Punto de equilibrio:}}\label{pto eq} Un punto $x_0\in \mathbb{R}^n$ que cumpla $f(x_0)=0$ se llama punto de equilibrio de \eqref{sistema no lineal}.
\end{definition}

\begin{remark} Los puntos de equilibrio también son llamados puntos \emph{singulares} o \emph{fijos}.
\end{remark}

\begin{definition}{\textbf{Punto de equilibrio hiperbólico y no hiperbólico:}}\label{pto eq hyp} Sea $x_0$ como en la definición \ref{pto eq}. Este se llama hiperbólico si  todas las partes reales de los autovalores de  $Df(x_0)$  son distintas de cero. En caso contrario $x_0$ se llama no hiperbólico.
\end{definition}

\begin{definition}{\textbf{Matriz de linealización:}} 
La matriz derivada de $f:\mathbb{R}^n \rightarrow \mathbb{R}^n$ definida por 
\begin{equation}\label{mtriz linealizacion}
Df(x_0)=\left(\frac{\partial f_i}{\partial x_j}\right)_{x=x_0}, i,j=1,...,n
\end{equation}
\newline se llama matriz de linealización del sistema \eqref{sistema no lineal} donde $f_i$ son las componentes de $f$.
\end{definition}

\begin{definition}{\textbf{Flujo de un sistema:}}\label{flujo} Sea $x(t)=\phi_a(t)$ una solución de \eqref{sistema no lineal} con condición inicial $x(0)=a$, se define el flujo $\{g_a^t\}$ en función de de la solución por $g_a^t=\phi_a(t)$. 
\end{definition}

\begin{definition}{\textbf{La órbita:}} La órbita en $a$ se define por
\begin{equation}\label{orbita}
\gamma(a)=\{x\in \mathbb{R}^n|x=g_a^t, \forall t\in \mathbb{R}\}.
\end{equation}
\end{definition}

\begin{remark}
Las órbitas se clasifican en tres tipos
\begin{enumerate}
  \item \textbf{Órbitas puntuales:} Corresponden a puntos singulares en los cuales $g^t_a=a$ para todo $t\in \mathbb{R}$, entonces $\gamma (a)=\{a\}$.
  \item \textbf{Órbitas periódicas:} describen un sistema que evoluciona periodicamente en el tiempo, esto ocurre si existe un tiempo $T>0$ tal que $g^T_a=a$.
  \item \textbf{Órbitas no periódicas:} Esto ocurre si $g^t_a\neq a$  para todo $t\neq 0$.
\end{enumerate}
\end{remark}

\begin{definition}{\textbf{Matriz exponencial:}} Sean $A\in M_{n\times n}(\mathbb{R})$ y $t\in \mathbb{R}$, se define 
\begin{equation}\label{power matrix}
e^{At}=I+A t+\frac{1}{2!} A^2t^2+\cdots=\sum_{k=0}^\infty \frac{A^k t^k}{k!}
\end{equation}
\end{definition}
\begin{remark} $I$ es la matriz identidad de tamaño $n$, $A^2=AA$ además, para $t=1$ se tiene la definición clasica de la matriz exponencial $e^A$. Esta serie matricial converge si las $n^2$ series infinitas correspondientes a las entradas convergen en $\mathbb{R}$, esta serie siempre converge para toda matriz  real $A$ \cite{Smale}.
\end{remark}
Dada una matriz no singular $P$ tal que $B=P^{-1}AP$ se dice que $B$ y $A$ son similares y se puede tomar $P$ de tal forma que $A$ se reduzca a su forma de canónica de Jordan. En particular para cualquier $A\in M_{2\times 2}(\mathbb{R})$ existe  una matriz no singular $P$ tal que 
\begin{equation}
J=P^{-1}AP, 
\end{equation}
y $J$ es una de las siguientes
\begin{equation}\label{Jordan forms}
\begin{pmatrix}
\lambda_1&0\\
0        &\lambda_2\\
\end{pmatrix},
\begin{pmatrix}
\lambda&1\\
0        &\lambda\\
\end{pmatrix},
\begin{pmatrix}
\alpha&\beta\\
-\beta&\alpha\\
\end{pmatrix}.
\end{equation}

\begin{remark} 
Si $B=P^{-1}AP$ entonces $e^B=P^{-1}e^AP$ y se puede calcular $e^A$ para cualquier matriz.
\end{remark}
El siguiente resultado es una herramienta importante en el estudio de sistemas lineales.
\begin{theorem}{\textbf{Teorema fundamental para sistemas autónomos lineales:}}
Sea $A\in M_{n\times n}(\mathbb{R})$. Entonces para un $x_0\in \mathbb{R}^n$ dado, el problema con valor inicial 

\begin{equation}\label{Teo fndmtl lineales}
\begin{split}
\dot{x} &=  Ax \\ 
x(0) &=  x_0
\end{split}
\end{equation}
tiene una única solución dada por 

\begin{equation}
x(t)=e^{At}x_0
\end{equation}
\end{theorem}

\begin{definition}{\textbf{Equivalencia lineal:}}
Dados dos sistemas de EDO's 
\begin{equation}\label{2 sistemas lineales}
\dot{x}=Ax,\quad \dot{y}=By
\end{equation}
con $A,B$ matrices, se dice que los sistemas \eqref{2 sistemas lineales} son equivalentes si y solo si existe una matriz no singular $P$ y una constante $k>0$ tal que
\begin{equation}
e^{At}=P^{-1}e^{B k t}P
\end{equation}
para todo $t\in \mathbb{R}$.
\end{definition}

Considerando las tres formas canónicas de Jordan \eqref{Jordan forms} se presentarán los siguientes casos según las direcciones propias de $A$.\\

\textbf{Dos direcciones propias:} Existe una matriz $P$ tal que $J=P^{-1}AP$, con 
\begin{equation}
J=\begin{pmatrix}
\lambda_1&0\\
0        &\lambda_2\\
\end{pmatrix}
\end{equation}
luego el sistema \eqref{sistema lineal} es equivalente a 
\begin{equation}\label{sistema linealizado}
\dot{y}=Jy
\end{equation}
y el flujo es 
\begin{equation}
e^{Jt}=\begin{pmatrix}
e^{\lambda_1 t}&0\\
0        &e^{\lambda_2 t}\\
\end{pmatrix},
\end{equation}
los autovectores son $e_1=(1,0)^T,e_2=(0,1)^T$. Por el teorema \ref{Teo fndmtl lineales} la solución es $y(t)=e^{Jt}b$ con $b\in \mathbb{R}^2$. Es decir,

\begin{equation}
\begin{split}
y_1(t)=e^{\lambda_1 t}b_1\\
y_2(t)=e^{\lambda_2 t}b_2
\end{split}.
\end{equation}
Para calcular las ecuaciones que definen las órbitas de \eqref{sistema linealizado} se escribe
\begin{equation}
\ln\left[\left(\frac{y_1}{b_1}\right)^{\frac{1}{\lambda_1}}\right]=t=\ln\left[\left(\frac{y_2}{b_2}\right)^{\frac{1}{\lambda_2}}\right].
\end{equation}
Luego, 
\begin{equation}
\left(\frac{y_1}{b_1}\right)^{\frac{1}{\lambda_1}}=\left(\frac{y_2}{b_2}\right)^{\frac{1}{\lambda_2}}.
\end{equation}
Este caso da lugar a focos atractores ($\lambda_1=\lambda_2<0$), nodos atractores ($\lambda_1<\lambda_2<0$), líneas atractoras ($\lambda_1<\lambda_2=0$), Silla ($\lambda_1<0<\lambda_2$), líneas repulsoras ($\lambda_1=0<\lambda_2$), nodos repulsores ($0<\lambda_1<\lambda_2$), focos repulsores ($0<\lambda_1=\lambda_2$).\\

\textbf{Una dirección propia:} Existe una matriz $P$ tal que $J=P^{-1}AP$, con

\begin{equation}
J=\begin{pmatrix}
\lambda&1\\
0        &\lambda\\
\end{pmatrix}
\end{equation}
luego el sistema \eqref{sistema lineal} es equivalente a 
\begin{equation}
\dot{y}=Jy
\end{equation}
y el flujo es 

\begin{equation}
e^{Jt}=\begin{pmatrix}
e^{\lambda t}&t\\
0        &e^{\lambda t}\\
\end{pmatrix},
\end{equation}
se tiene un solo autovector $e_1=(1,0)^T$. Si $\lambda \neq 0$ las órbitas estan dadas por $y_1=y_2[\frac{b_1}{b_2}+\frac{1}{\lambda}\ln (\frac{y_2}{y_1})]$ si $b_2\neq 0$. Este caso da lugar a nodos atractores de Jordan ($\lambda<0$), línea neutral ($\lambda=0$), nodos repulsores de Jordan ($\lambda>0$) .\\

\textbf{No hay dirección propia:} Existe una matriz $P$ tal que $J=P^{-1}AP$, con
\begin{equation}
J=\begin{pmatrix}
\alpha&\beta\\
-\beta&\alpha\\
\end{pmatrix}
\end{equation}

luego el sistema \eqref{sistema lineal} es equivalente a 
\begin{equation}
\dot{y}=Jy
\end{equation}

La forma mas facil de encontrar las órbitas es trabajar con coordenadas polares $(r,\theta):$ $y_1= r \cos \theta$ y $y_2=r \sin \theta$. La ecuación se convierte en $r'=\alpha r$ y $\theta '=-\beta$ por tanto se tiene $\frac{dr}{d\theta}=-\frac{\alpha}{\beta}r$ que puede ser integrado fácilmente para obtener la solución $r=r_0 \exp\left[-\frac{\alpha}{\beta}(\theta -\theta _0)\right]$. Se puede asumir que $\beta >0$ porque la ecuación diferencial es invariante ante el cambio $(\beta,y_1)\rightarrow (-\beta,-y_1)$. Este caso da lugar a espirales atractoras ($\alpha <0$), centros ($\alpha=0$), espirales repulsoras ($\alpha>0$)

\begin{definition}{\textbf{Homeomorfismo:}}
Una función $H:\mathbb{R}^n \rightarrow \mathbb{R}^n$ se llama homeomorfismo en $\mathbb{R}^n$ si y solo si es inyectiva, sobreyectiva y continua y además su inversa también lo es.
\end{definition}

\begin{definition}{\textbf{Equivalencia topológica:}}
Dos flujos lineales $e^{At},e^{Bt}$ son topológicamente equivalentes si y solo si existe un Homeomorfismo $H$ y una constante $k>0$ tal que
\begin{equation}
H(e^{At}x)=e^{Bkt}H(x),\quad \textrm{para todo} \quad x\in \mathbb{R}^n \quad \textrm{y para todo}\quad t\in \mathbb{R}.
\end{equation}
\end{definition}

\begin{figure}
    \centering
    \includegraphics[scale=0.4]{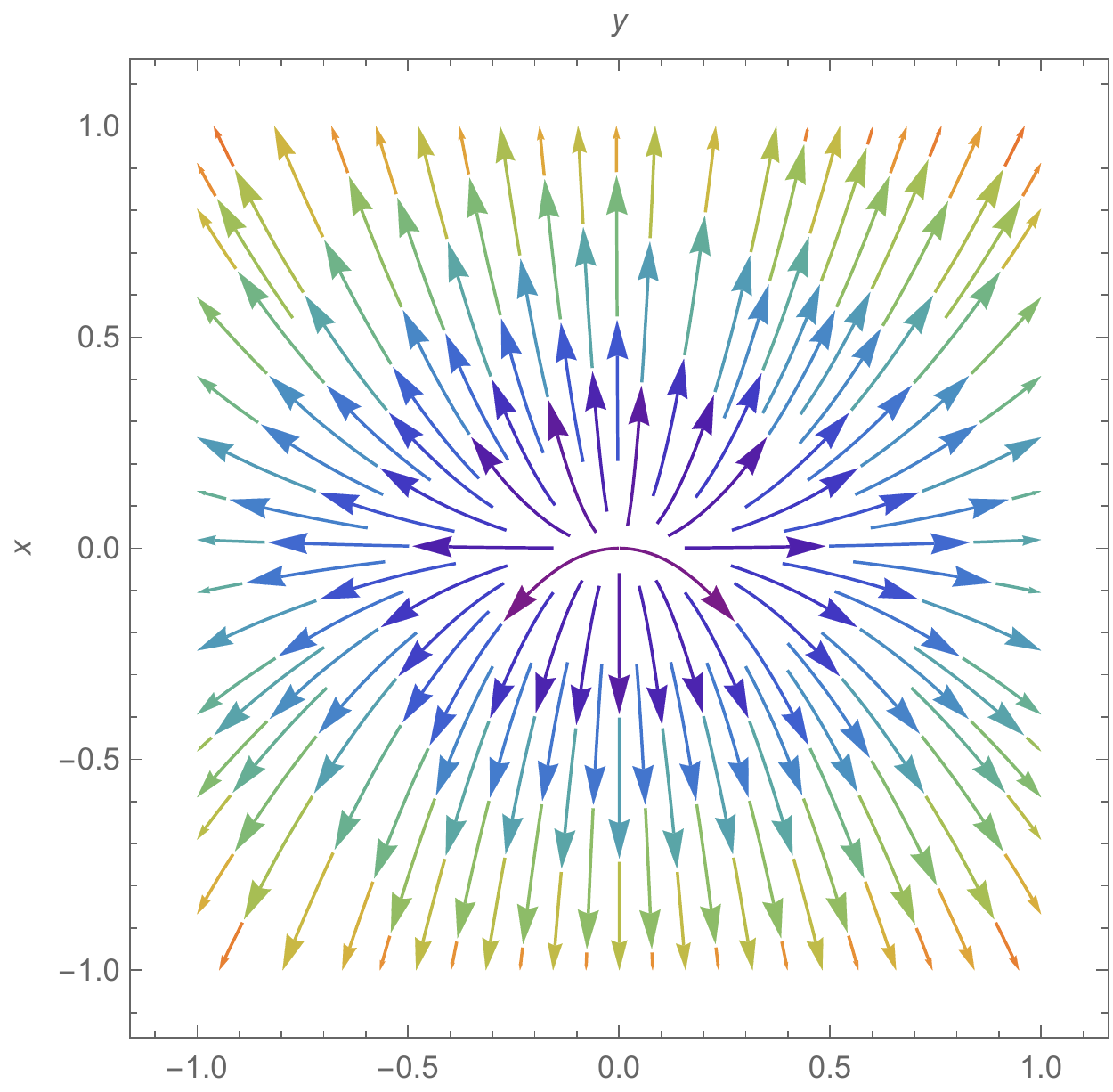}
    \includegraphics[scale=0.4]{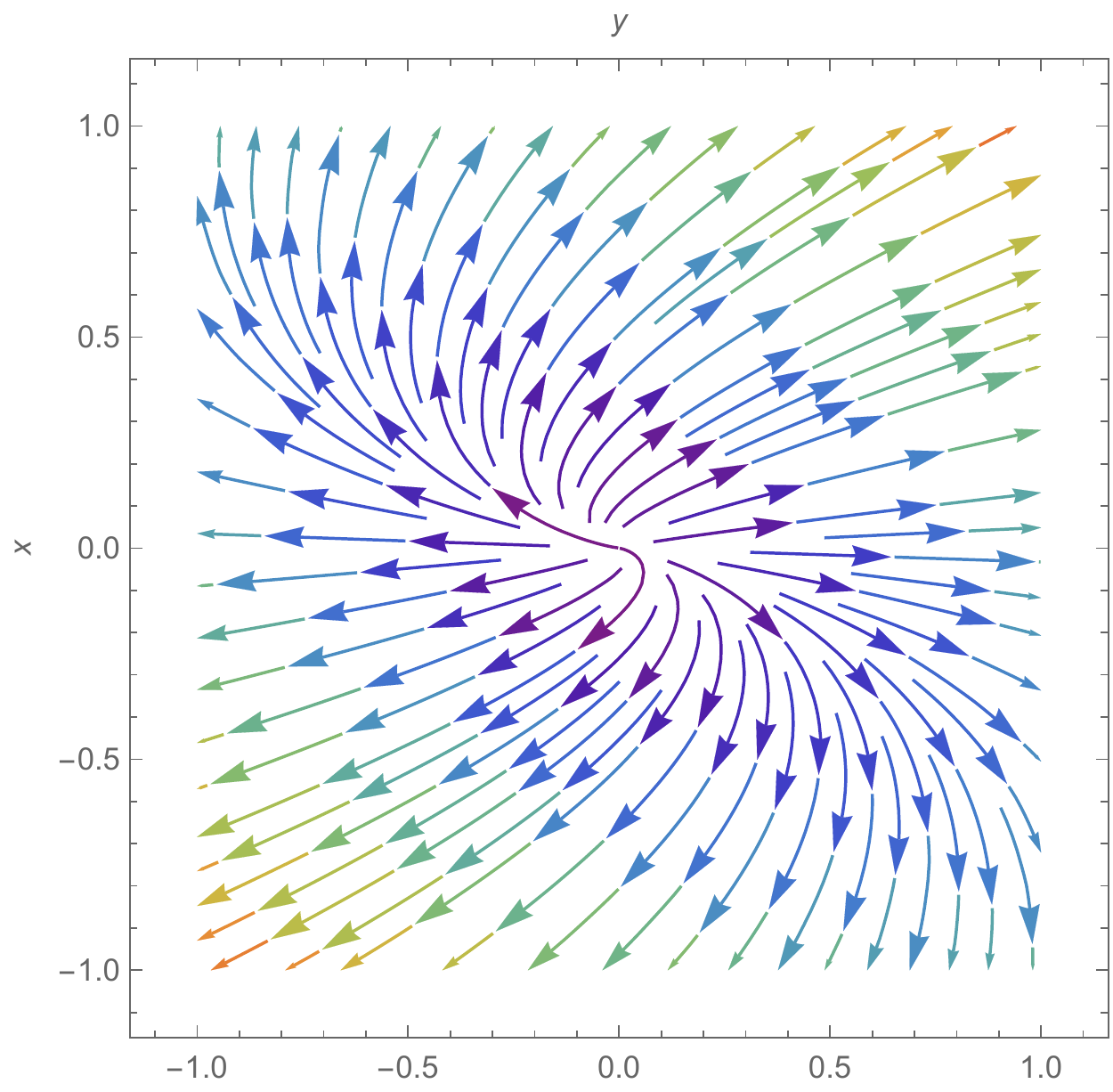}
    \includegraphics[scale=0.4]{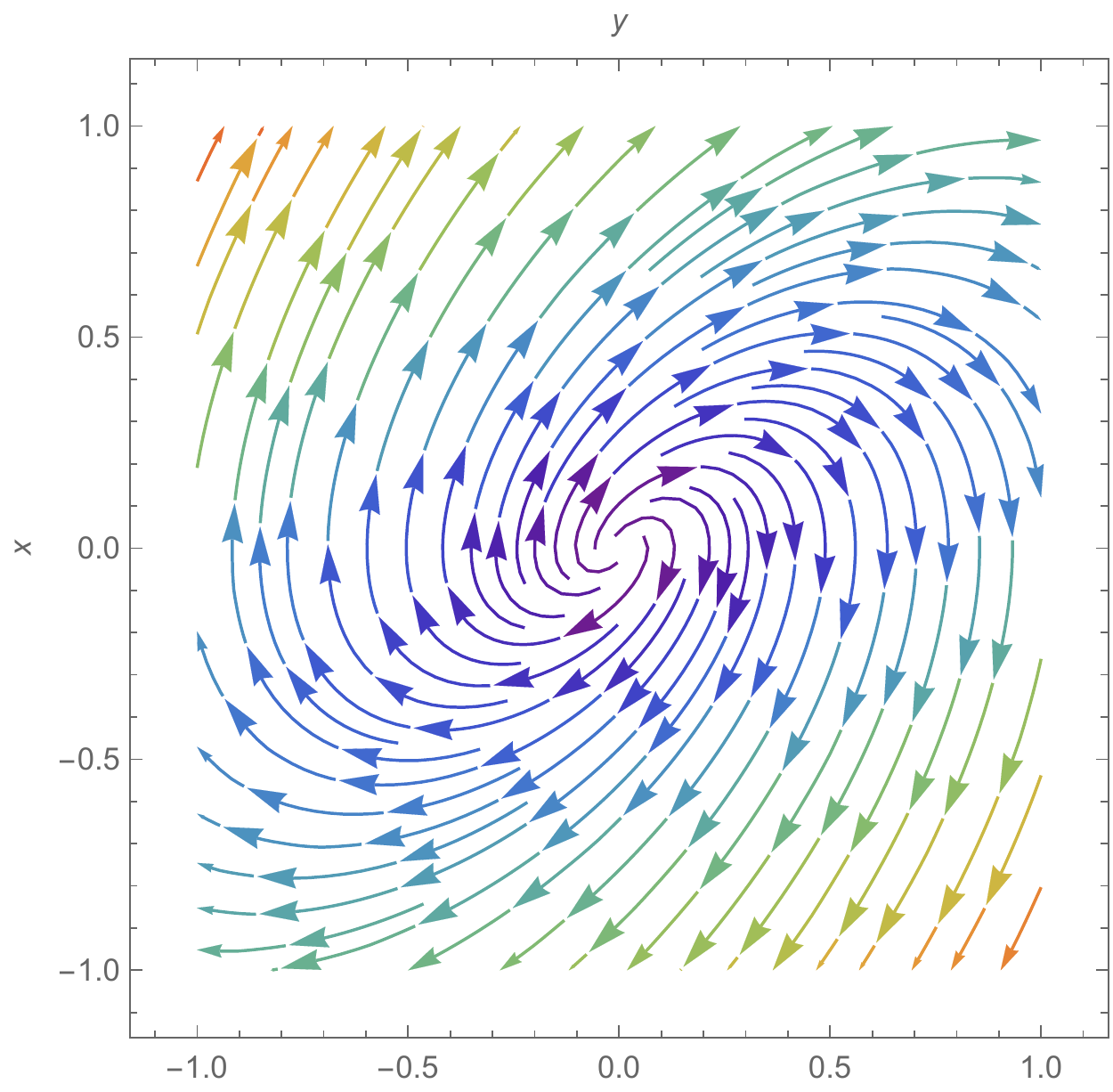}
    \includegraphics[scale=0.4]{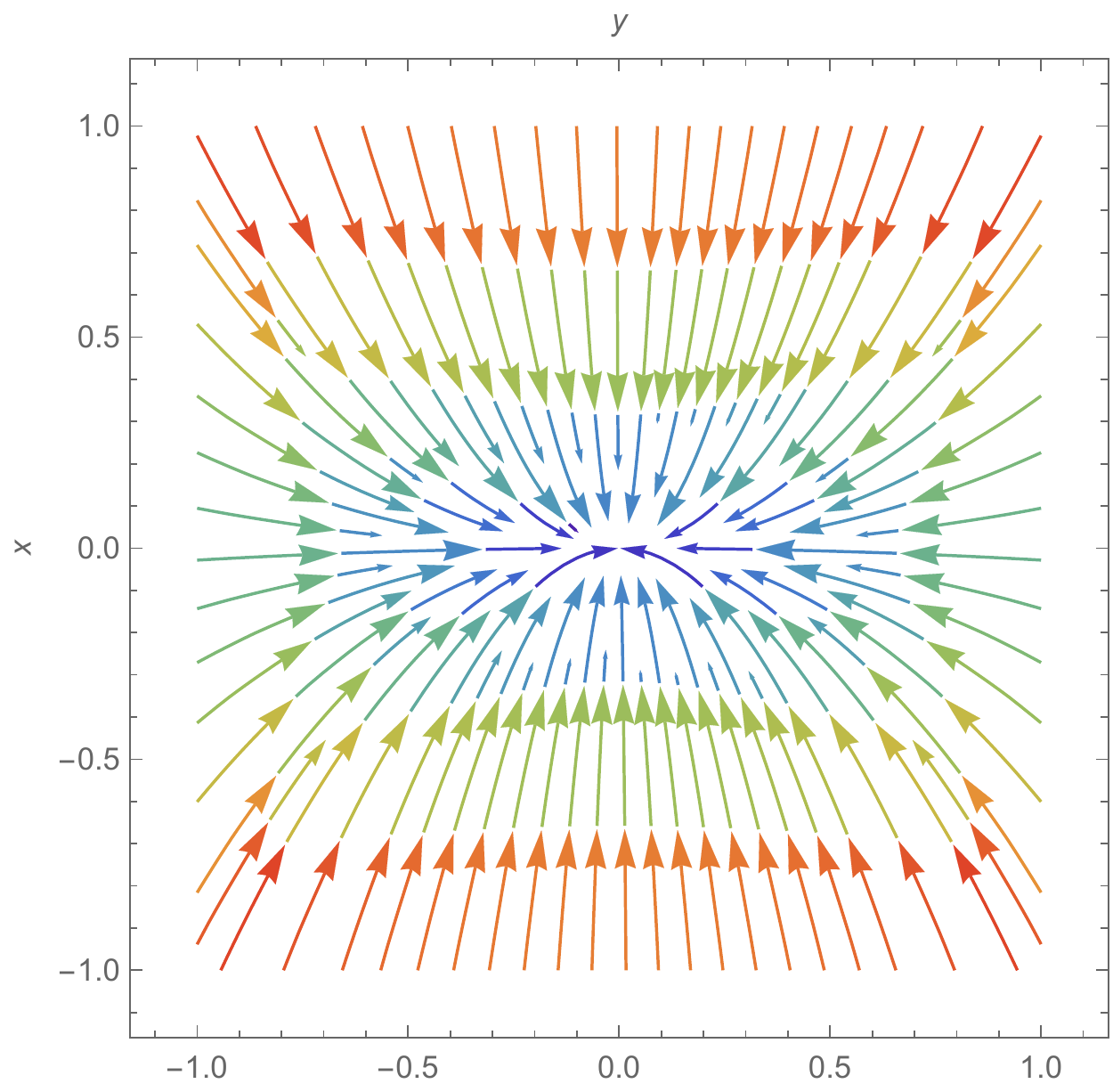}
    \caption{\label{retratos1} Algunas órbitas de sistemas de dimensión 2: foco, nodo, espiral (repulsores) y pozo.}
\end{figure}

\begin{figure}
          \centering
    \includegraphics[scale=0.4]{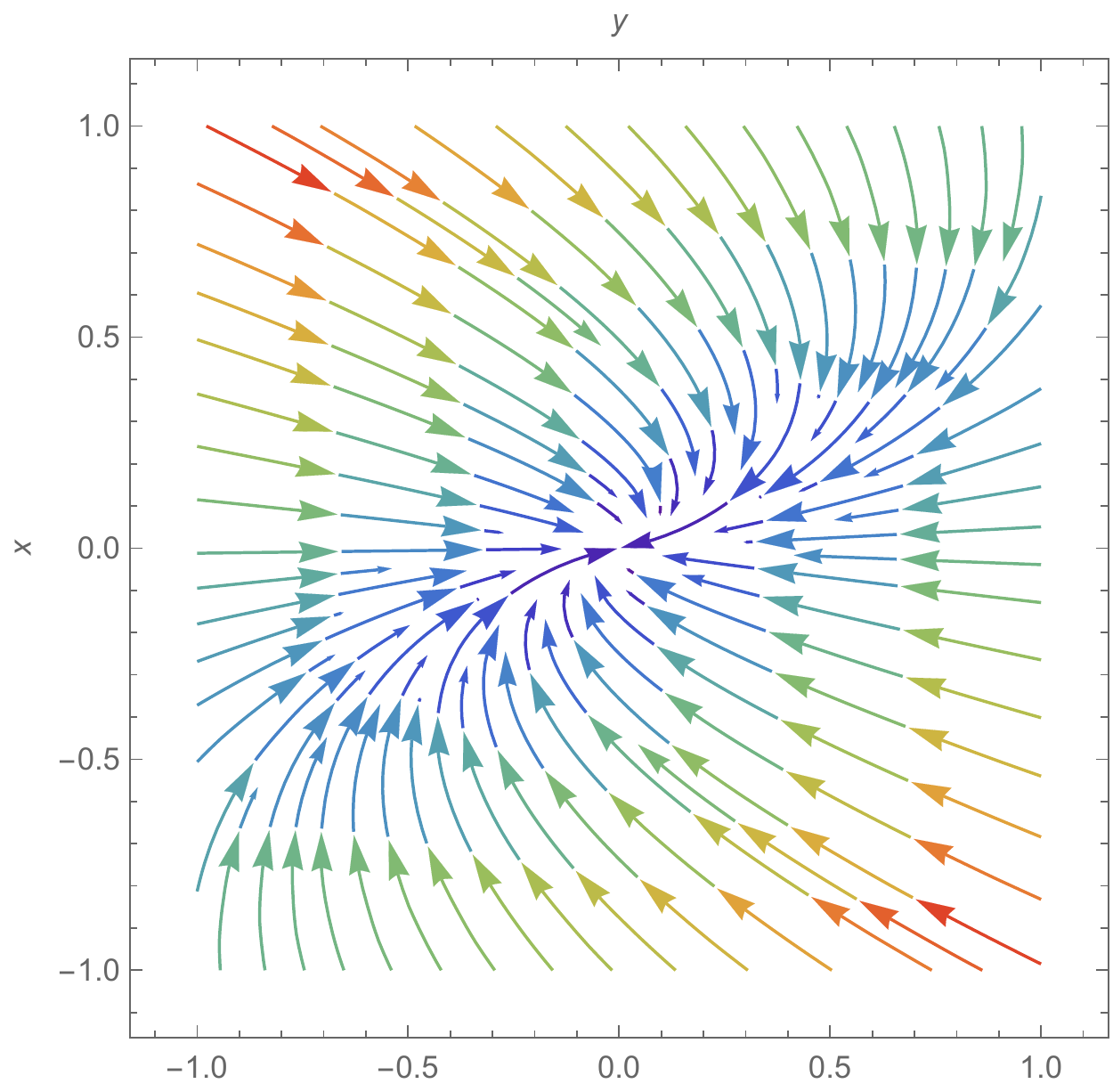}
    \includegraphics[scale=0.4]{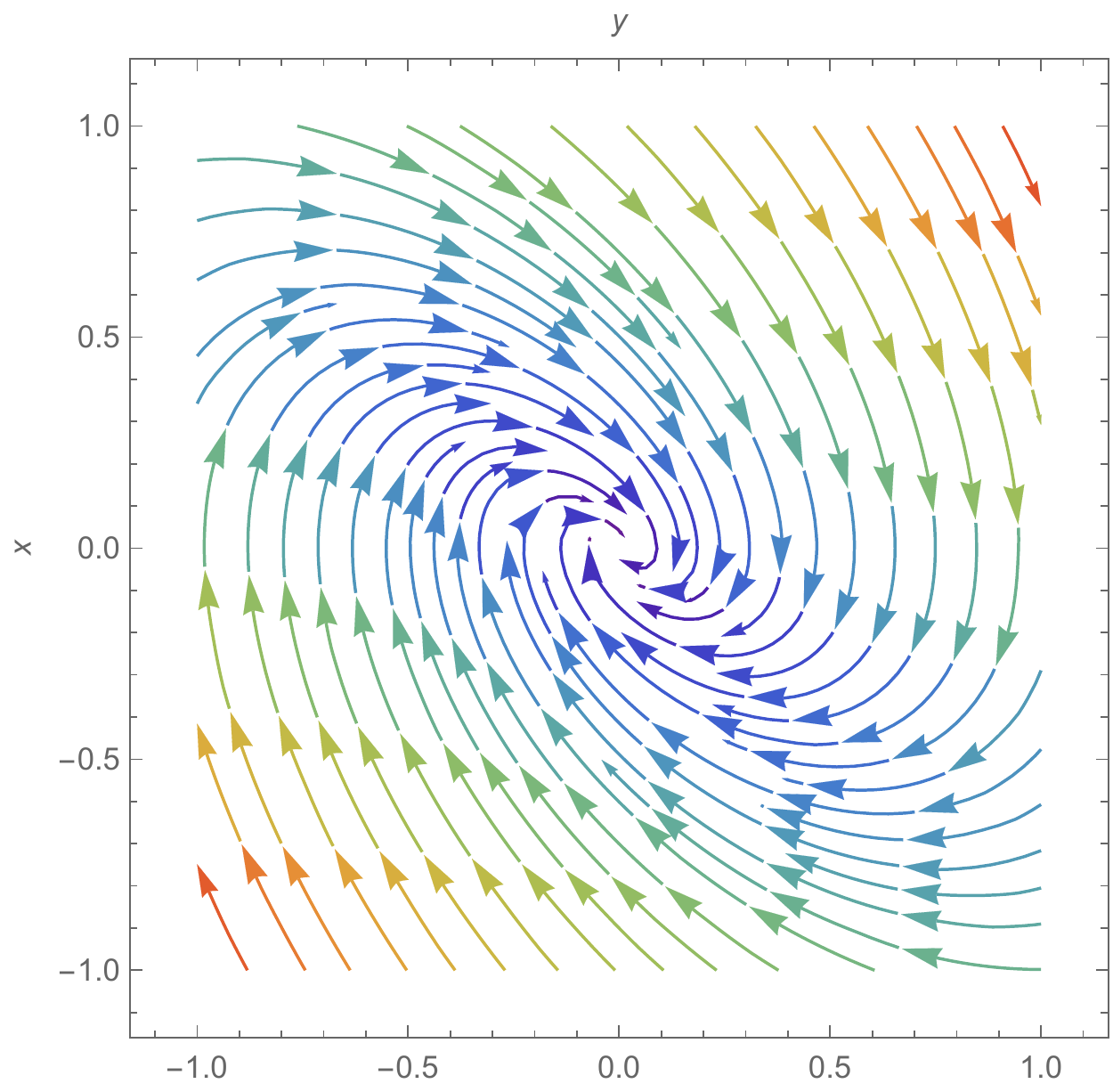}
    \includegraphics[scale=0.4]{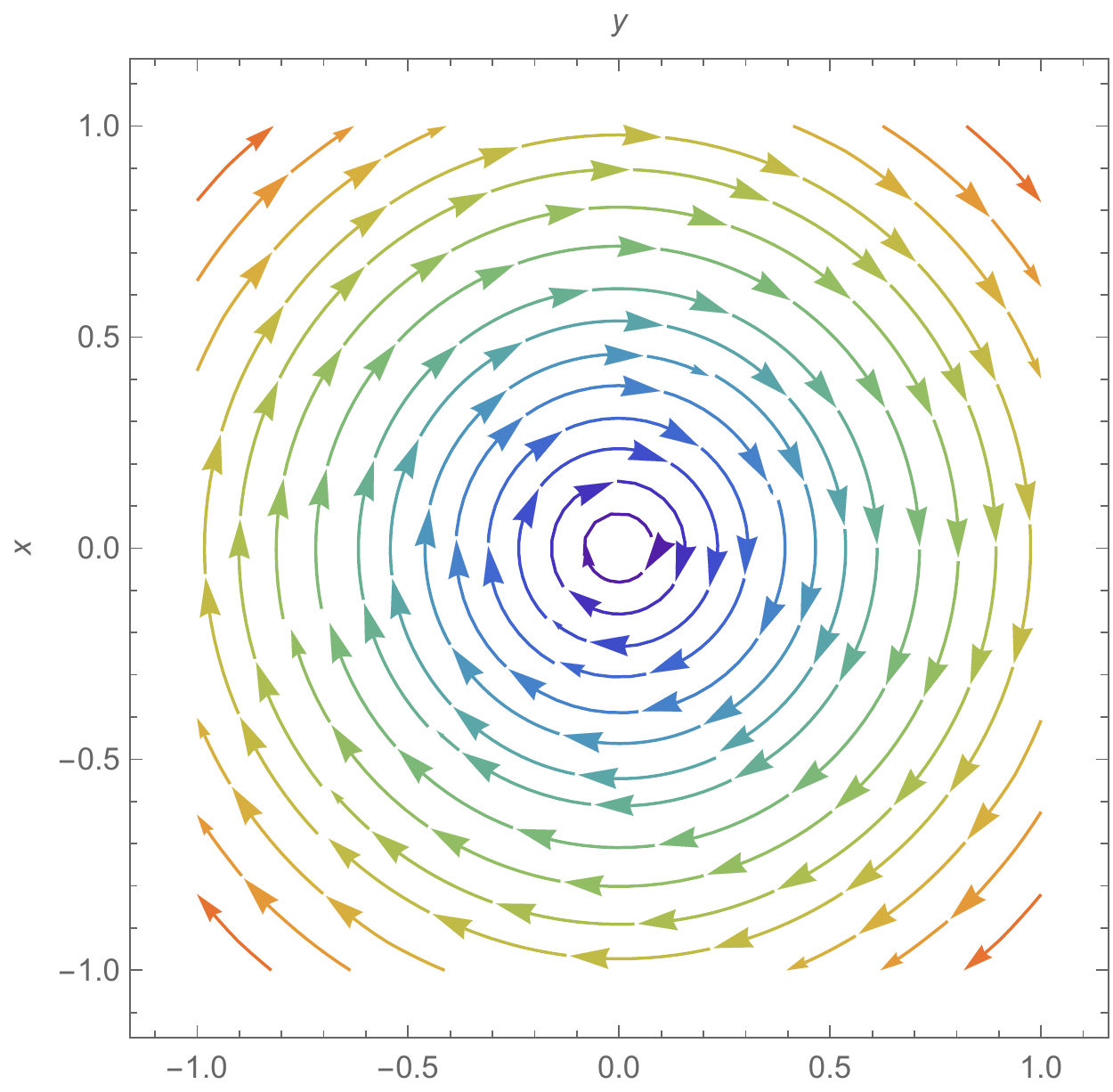}
    \includegraphics[scale=0.4]{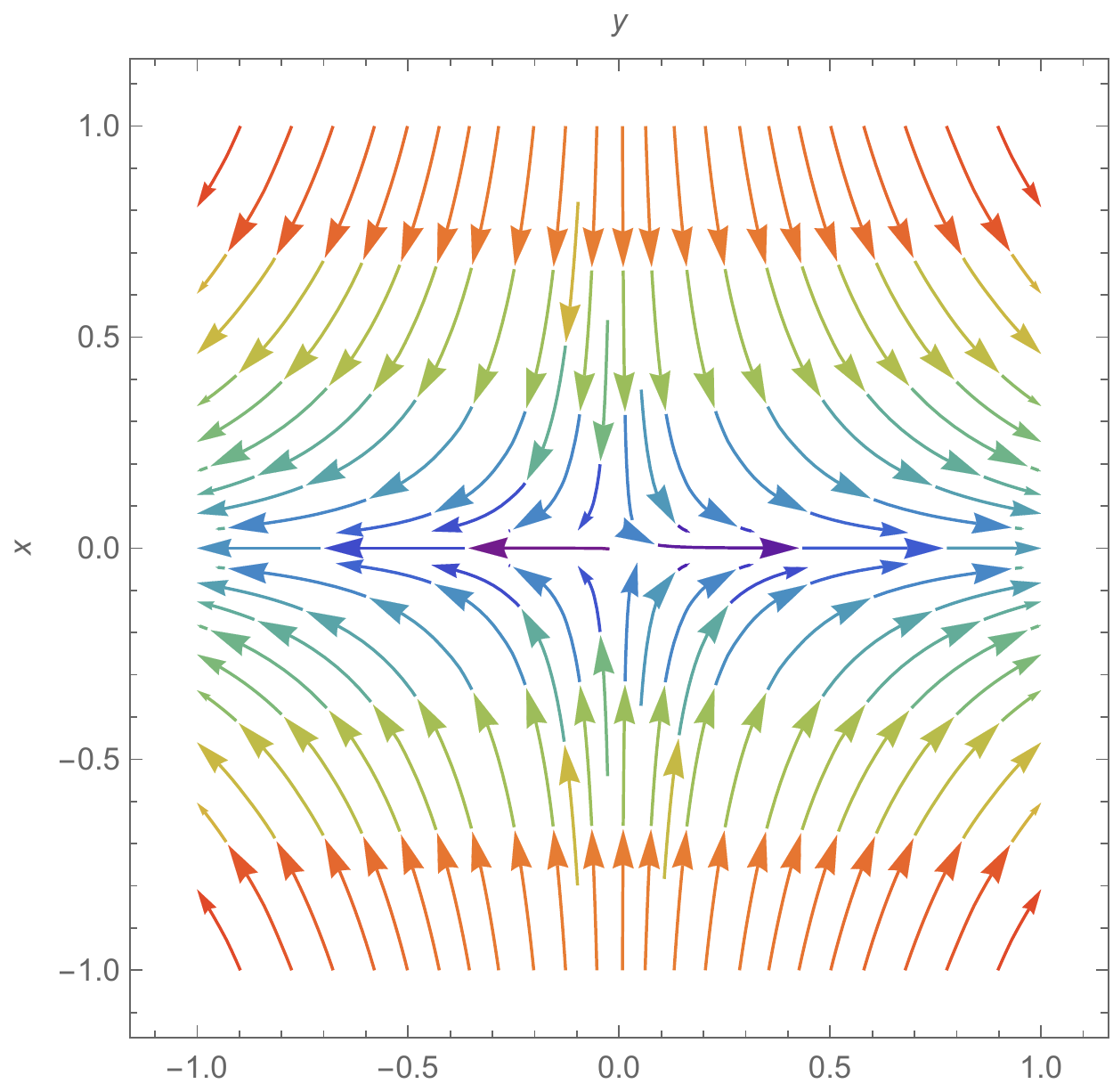}
         \caption{\label{retratos2}Algunas órbitas de sistemas de dimensión 2: nodo, espiral (atractores), centro y silla.}
\end{figure}

\begin{remark} 
Es un resultado conocido que todos los flujos hiperbólicos lineales en $\mathbb{R}^2$ son topológicamente equivalentes  al flujo lineal de una de las siguientes matrices:

\begin{enumerate}
    \item $A=\left(
\begin{array}{cc}
 -1 & 0 \\
 0 & -1 \\
\end{array}
\right)$; pozo estandar.
    \item $A=\left(
\begin{array}{cc}
 1 & 0 \\
 0 & 1 \\
\end{array}
\right)$; fuente estandar.
    \item $A=\left(
\begin{array}{cc}
 -1 & 0 \\
 0 & 1 \\
\end{array}
\right)$; silla estandar.
\end{enumerate}

Los flujos lineales no hiperbólicos en $\mathbb{R}^2$, son equivalentes al flujo lineal de una de las siguientes matrices
\begin{equation*}
    \left(
\begin{array}{cc}
 0 & 0 \\
 0 & 0 \\
\end{array}
 \right),    \left(
\begin{array}{cc}
 0 & -1 \\
 1 & 0 \\
\end{array}
 \right),    \left(
\begin{array}{cc}
 0 & 1 \\
 0 & 0 \\
\end{array}
 \right),    \left(
\begin{array}{cc}
 -1 & 0 \\
 0 & 0 \\
\end{array}
 \right),    \left(
\begin{array}{cc}
 1 & 0 \\
 0 & 0 \\
\end{array}
 \right).
\end{equation*}
Cabe destacar que estos flujos no son topológicamente equivalentes entre ellos. En las figuras \ref{retratos1} y \ref{retratos2}, se presentan algunas órbitas de los casos estándares para sistemas lineales en $\mathbb{R}^2$
\end{remark}

\begin{definition}{\textbf{Conjunto invariante:}}
Un conjunto $S\subset \mathbb{R}^n$ es un conjunto invariante para \eqref{sistema no lineal} si para cualquier punto $a\in S$ la órbita que pasa por $a$ esta totalmente contenida en $S$, es decir $\gamma (a)\subset S.$
\end{definition}

\begin{definition}{\textbf{Subespacios estable, centro e inestable:}}
Considerar el sistema lineal \eqref{sistema lineal} y sea $w_j=u_j+iv_j$ autovector asociado a el autovalor $\lambda_j=a_j+ib_j$ (si $b_j=0, \quad
v_j=0$) y sea
\begin{equation}
B=\{u_1,\cdots ,u_k,u_{k+1},\cdots ,u_m,v_m\}
\end{equation}
base de $\mathbb{R}^n$ con $n=2m-k$. Entonces
\begin{equation}
\begin{split}
E^s=Span\{u_j,v_j|a_j<0\} \\
E^c=Span\{u_j,v_j|a_j<=\} \\
E^u=Span\{u_j,v_j|a_j>0\}
\end{split}
\end{equation}
donde $E^s,E^c,E^u$ son llamados los subespacios estable, centro e inestable del origen.
\end{definition}

\section{Puntos de equilibrio hiperbólicos y Teorema de Hartman Grobman}
\label{seccion1.2}

En esta sección se presentará un teorema que permite decidir la estabilidad local de los puntos de equilibrio de los sistemas que se estudian en esta tesis. El teorema de Hartman-Grobman es útil en el análisis de la estabilidad de los puntos hiperbólicos. Antes de enunciar el teorema se presenta el siguiente caso unidimensional.
\newline
Sea $\dot{x}=f(x)$ donde $x\in \mathbb{R}$ y $x_0$ un punto de equilibrio, su expansión en series de Taylor es 
\begin{equation}\label{aprox lineal}
f(x)\approx f(x_0)+\frac{f'(x_0)}{1!}(x-x_0)+\frac{f''(x_0)}{2!}(x-x_0)^2+\cdots =\sum_{n=0}^\infty \frac{f^{(n)}(x_0)}{n!}(x-x_0)^n.
\end{equation}
\newline
Eliminando los términos de orden mayor o igual a 2 y notando que $x_0$ es punto de equilibrio en (1.25), para valores de $x$ en una vecindad de $x_0$ la ecuación diferencial se comporta como 
\begin{equation}
\dot{x}=f'(x_0)(x-x_0).
\end{equation}   
\newline
Generalizando este proceso a $\mathbb{R}^n$ y ocupando la matriz de \eqref{mtriz linealizacion}, $f$ se puede escribir como 
\begin{equation}
f(x)\approx f(x_0)+Df(x_0)(x-x_0)+\cdots.
\end{equation}
\newline
Luego, para $x$ en una vecindad de $x_0$ el sistema \eqref{sistema no lineal} se comporta como
\begin{equation}
\dot{x}=Df(x_0)(x-x_0).
\end{equation}
Se analizará el sistema \eqref{sistema no lineal} ocupando análisis cualitativo, la idea general es encontrar los puntos de equilibrio del sistema y es de interés estudiar la dinámica en una vecindad de dichos puntos. Esto se logra evaluando la matriz de linealización \ref{mtriz linealizacion} en el punto de equilibrio $x_0$ y estudiando el signo de los autovalores. La estabilidad local de los puntos se clasifican según los siguientes criterios:

\begin{enumerate}
	\item \textbf{Asintóticamente estable:} Cuando la parte real de \textbf{todos} los autovalores tiene signo negativo.
	\item \textbf{Fuente (Inestable):} Cuando la parte real de \textbf{todos} los autovalores tiene signo positivo.
	\item \textbf{Silla (Inestable):} Cuando al menos \textbf{uno} de los autovalores tiene parte real positiva y los demas tienen parte real negativa.
\end{enumerate}

\begin{definition}{\textbf{Conjunto normalmente hiperbólico:}}\label{normhiper}
Un conjunto de puntos de equilibrio no aislados, se llama se tiene normalmente hiperbólico si los  autovalores con parte real cero son aquellos cuyos correspondientes autovectores son tangentes al conjunto.
\end{definition}
\begin{remark}
La estabilidad de un conjunto de puntos de equilibrio normalmente hiperbólico puede ser completamente determinada considerando solo los signos de los autovalores con parte real no nula.
\end{remark}

A continuación se presenta uno de los teoremas centrales en el desarrollo de la tesis.
\begin{theorem}{\textbf{Teorema de Hartman-Grobman:}}
\label{hartgrob}
Sea $f\in C^1(\mathbb{R}^n)$ y sea $x_0$ un punto hiperbólico del sistema diferencial $\dot{x}=f(x)$ entonces existe una vecindad de $x_0$  en el cual el flujo es topológicamente equivalente al flujo de la matriz de linealización \ref{mtriz linealizacion}.
\end{theorem}

Este resultado es válido para puntos de equilibrio hiperbólicos, para el análisis de los puntos no hiperbólicos, se debe utilizar otra estrategia que será presentada en la proxima sección.

\section{Teorema de la variedad estable e inestable}
\label{seccion1.3}

\begin{definition}{\textbf{Variedad diferenciable:}}
Una variedad diferenciable $M$ de dimensión
$n$ (o variedad de clase $C^k$) es un espacio métrico conexo con un cubrimiento por abiertos $\{U_{\alpha}\}$ es decir $M=\cup_{\alpha}U_{\alpha}$ tal que

	\begin{enumerate}
		\item Para todo $\alpha$, existe un Homeomorfismo $h_{\alpha}$  		tal que $h_{\alpha}:U_{\alpha}\rightarrow \textbf{B}=B(x,1)$ en $\mathbb{R}^n$.
		\item Si $U_{\alpha}\cap U_{\beta}\neq \emptyset$ y $h_{\alpha}:U_{\alpha}\rightarrow \textbf{B}$, $h_{\beta}:U_{\beta}\rightarrow \textbf{B}$  son los respectivos homeomorfismos, entonces $h_{\alpha}(U_{\alpha}\cap U_{\beta})$ y $h_{\beta}(U_{\alpha}\cap U_{\beta})$ son subconjutos de $\mathbb{R}^n$ y la función
		\begin{equation*}
		h=h_{\alpha}\circ h_{\beta}^{-1}:h_{\beta}(U_{\alpha}\cap U_{\beta})\rightarrow h_{\alpha}(U_{\alpha}\cap U_{\beta})
		\end{equation*}
		es de clase $C^k$ y para todo $x\in h_{\beta}(U_{\alpha}\cap U_{\beta})$ se tiene $det(Dh(x))\neq 0$.
	\end{enumerate}
\end{definition}

\begin{remark}
El par $(U_{\alpha},h_{\alpha})$ se llama carta de $M$ y el conjunto de todas las cartas se llama atlas de $M$.
\end{remark}

A continuación se presentan dos teoremas importantes sobre variedades que serán de gran uso en esta tesis.

\begin{theorem}{\textbf{Teorema de la variedad estable e inestable:}}
Sea $E$ un abierto de $\mathbb{R}^n$ que contiene al origen, sea $f\in C^1(E)$ y sea $\phi_t$ el flujo del sistema no lineal \eqref{sistema no lineal}. Suponer $f(0)=0$ y $Df(0)$ tiene $k$ autovalores con parte real negativa y $n-k$ autovalores con parte real positiva. Entonces existe una variedad diferenciable $S$ de dimensión $k$ tangente al suebspacio estable $E^s$ del sistema lineal \eqref{sistema lineal} en $0$ tal que para todo $t\geq 0$, $\phi_t(S)\subset S$ y para todo $x_0\in S$,
	\begin{equation}
	\lim_{t\rightarrow \infty}\phi_t(x_0)=0;
	\end{equation}
y existe una variedad diferenciable $U$ de dimensión $n-k$ tangente al subespacio inestable $E^U$ de \eqref{sistema lineal} en $0$ tal que para todo $t\leq 0$, $\phi_t(U)\subset U$ y para todo $x_0\in U$,
	\begin{equation}
	\lim_{t\rightarrow -\infty}\phi_t(x_0)=0.
	\end{equation}
\end{theorem}

\section{Flujos en una dimensión}
\label{Strogatz}

En esta sección seguimos el enfoque de Strogatz \cite{Stro} para ilustrar que vale la pena atacar estos problemas desde el punto de vista cualitativo y geométrico en lugar de buscar la solución exacta de EDO's no lineales, esto será de utilidad al momento de analizar la estabilidad del origen en los cálculos de las variedades invariantes. 

\begin{figure}[h!]
    \centering
    \includegraphics[scale=0.35]{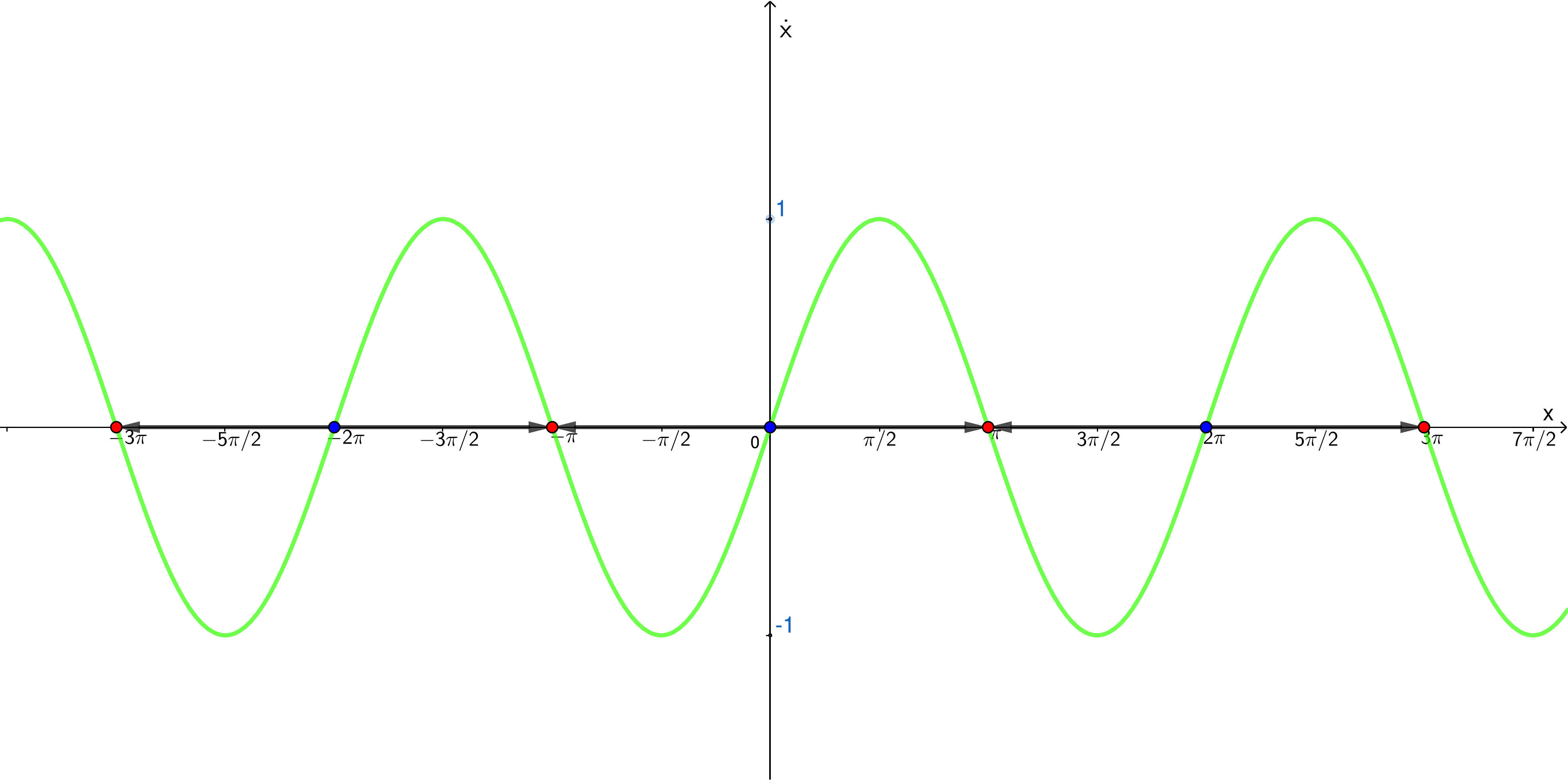}
    \caption{Flujo en la recta para $\dot{x}= \sen{x}$, rojo= pozos (atractores), azul= fuente (repulsores)}
    \label{fig:flujorecta}
\end{figure}

Considerar \begin{equation}
\label{sinx}
    \dot{x}=\sin{x}
\end{equation}
separando variables obtenemos
\begin{equation*}
    \begin{aligned}
    t & =\int \csc{x} dx =-\ln{|\csc{x}+\cot{x}|}+C.
    \end{aligned}
\end{equation*}

Suponiendo que $x=x_0$ en $t=0$. Entonces $C=\ln{|\csc{x_0}+\cot{x_0}|}$. La solución es 
\begin{equation}
    t=\ln{\left|\frac{\csc{x_0}+\cot{x_0}}{\csc{x}+\cot{x}}\right|}
\end{equation}
Esta solución es exacta, pero no es fácil de interpretar. Pensemos en $t$ como el tiempo, $x$ como la posición de una partícula imaginaria que se mueve a lo largo de la recta real y $\dot{x}$ como la velocidad de la particula, entonces la ecuación \eqref{sinx} representa un campo vectorial en la recta real, representa el vector velocidad $\dot{x}$ en cada $x$. Para mostrar esto, se plotea en el los ejes $x,y=\dot{x}$ y se dibujan flechas en el eje $x$ para indicar el vector velocidad correspondiente en cada $x$. La flecha apunta a la derecha cuando $\dot{x}>0$ y a la izquierda cuando $\dot{x}<0$, en la \eqref{fig:flujorecta} se ilustra la discusión previa.

\lhead{Capítulo \ref{ch_2}}
\rhead{Teoría Einstein-æther}
\cfoot{\thepage}
\renewcommand{\headrulewidth}{1pt}
\renewcommand{\footrulewidth}{1pt}

\chapter{Teoría Einstein-æther}\label{ch_2}

En este capítulo se discute el marco teórico, que es la Teoría de la gravedad Einstein-\ae ther, que contiene a la Teoría de la Relatividad General en un caso límite. Se discuten los formalismo $1+3$ y se introduce el formalismo diagonal homotético, que permite escribir las ecuaciones de campo de Einstein --pasando de ecuaciones diferenciales parciales (provenientes de la formulación $1+3$) a ecuaciones diferenciales ordinarias-- como un conjunto equivalente de ecuaciones de evolución más restriciones algebraicas, por lo que se puede usar las técnicas de la teoría cualitativa de los sistemas dinámicos para el análisis de estabilidad de sus soluciones. 

\section{Gravedad Einstein-æther}
\label{aetheory}

La acción para la teoría Einstein-æther es el funcional covariante más general que involucra derivadas parciales de orden a lo sumo dos (sin incluir derivadas totales) de la métrica de espacio-tiempo $g_{ab}$ y del campo vectorial  $u^a$ llamado æther  \cite{Jacobson:2008aj,Garfinkle:2011iw,Carroll:2004ai} dado por: 
\begin{align}
& S=\int d^{4}x\sqrt{-g}\left[  \frac{1}{2}R +\mathcal{L}_{\text{æ}} + M \left(  u^{c}u_{c
} + 1\right) + \mathcal{{L}}_m  \right]  ,\label{action}
\end{align}
donde 
\begin{equation}\label{aeLagrangian}
\mathcal{L}_{\text{æ}} \equiv -K^{a b}{}_{c d}\nabla_{a}u^{c
}\nabla_{b}u^{d},
\end{equation} es el Lagrangiano Einstein-æther \cite{Jacobson:2008aj} con
\begin{equation}
K^{a b}{}_{c d}\equiv c_{1}g^{a b}g_{c d} + c_{2}
\delta_{c}^{a}\delta_{d}^{b} + c_{3}\delta_{d}^{a}
\delta_{c}^{b} + c_{4}u^{a}u^{b}g_{c d}.
\end{equation}
La acción contiene el término de Einstein-Hilbert $ \frac{1}{2}R$, donde $R$ denota el escalar de Ricci; $g_{a b}$ denota el tensor métrico;  $K^{a b}{}_{c d}$ es un tensor de cuatro índices correspondiente a un término cinético para el æther, el cual contiene $4$ coeficientes adimensionales $c_{i}$; $M$ es un multiplicador de Lagrange que fuerza la unitaridad, i.e.,  $u^{c}u_{c
}=-1$, del vector æther, el cual es un vector tipo tiempo \cite{Garfinkle:2011iw}.
Se usará  la signatura  $({-}{+}{+}{+})$, y las unidades se eligen tal que la velocidad de la luz definida por la métrica $g_{ab}$ es la unidad y $\kappa^2\equiv 8\pi G=1.$

Las ecuaciones de  evolución se obtienen variando la acción de la teoría Einstein--æther, \eqref{action}, con respecto a la métrica, al vector æther y al multiplicador de Lagrange \cite{Donnelly:2010cr, Jacobson:2000xp}. Estos modelos contemplan:
\begin{itemize}
\item Los efectos de la anisotropía e inhomogeneidades (por ejemplo, la curvatura) sobre la geometría de los modelos esféricamente simétricos en consideración.
\item La contribución del tensor de energía-momentum $T _{ab}^{\text{æ}}$ del campo æther que depende de los
parámetros adimensionales $ c_i $. En Relatividad General, todos los $ c_i = 0 $, por lo que las ecuaciones de campo de Einstein se generalizan. 
Para estudiar los efectos de la materia, se puede en primera instancia sustituir los valores correspondientes a Relatividad General, o valores próximos a estos.
\item Cuando se estudia la fenomenología de las teorías con un marco preferido, generalmente se supone que este marco
coincide, al menos aproximadamente, con el marco cosmológico de reposo definido por el escalar de expansión de Hubble, $H$, del universo.
En particular, en un universo isotrópico y espacialmente homogéneo el campo æther se alineará con el marco cósmico
(marco de reposo natural preferido del CMB) y por lo tanto está relacionado con la tasa de expansión del universo.
\item 
En principio, en modelos esféricamente simétricos el marco preferido determinado por el
æther puede ser diferente (es decir, inclinado) al marco de reposo del CMB. Esto agrega términos adicionales al tensor de energía-momentum del æther $T _{ab}^{\text{æ}}$, por ejemplo, un ángulo de inclinación hiperbólico, $\nu$, que mide el impulso (\underline{boost}) del æther con respecto al marco de reposo del CMB 
\cite{Kanno:2006ty, Carruthers:2010ii}. En modelos homogéneos pero espacialmente anisotrópicos se espera que dicho ángulo de inclinación hiperbólico, $\nu$ decaiga junto con su derivada cuando $t\rightarrow +\infty$ \cite{Coley:2004jm,Coley:2006nw}.
\end{itemize}
Todos los campos æther esféricamente simétricos son hipersuperficie--ortogonales y, por lo tanto, todas las soluciones esféricamente simétricas  de la teoría æther también serán soluciones en límite infrarojo de la gravedad de Ho\v{r}ava. Lo contrario no es cierto en general, pero se cumple para soluciones con simetría esférica con  centro regular \cite{Jacobson:2008aj} .
\\
Cuando se impone  la simetría esférica, el æther es hipersuperficie--ortogonal,
y entonces tiene rotación (\underline{twist}) nulo. Por lo tanto, sin pérdida de
generalidad es posible hacer $ c_4 $ cero  \cite{Jacobson:2008aj}.
Después de la redefinición de parámetros para eliminar $ c_4 $, el espacio de parámetros es 3-
 dimensional. Los $ c_i $ contribuyen a la constante gravitacional  Newtoniana efectiva $ G $; entonces se puede especificar un parámetro $ c_i $ para hacer $8 \pi G = 1$. Los restantes dos parámetros caraterizan dos magnitudes físicas no triviales (por ejemplo, la masa y radio de Schwarzschild de una distribución de materia). Las otras restricciones impuestas a los $ c_i $ se resumen en \cite{Jacobson:2008aj} (por ejemplo, ver las ecuaciones 43-46 en \cite{Barausse:2011pu}). 

En el caso de simetría esférica en la teoría de Einstein-æther debe ser cuidadoso al elegir
el gauge (una condición de gauge adicional).  Normalmente, en Relatividad General, las coordenadas esféricamente simétricas se eligen para que la métrica se simplifique (por ejemplo, una elección de la función de lapso $ N $) o se puede elegir que el fluido sea comóvil. Aquí se ha elegido el campo vectorial æther alineado con el vector de marco temporal $\e_0$
(gauge æther comóvil) y no se puede simplificar $N(t, x)$ más. Esto puede hacer difícil las comparaciones con
Relatividad General  en algunos casos especiales.

Las ecuaciones de campo que se obtienen variando (\ref{action}) con respecto de 
$g^{ab}$, $u^a$, y $\lambda$ están dadas respectivamente por \cite{Garfinkle:2007bk}:
\begin{subequations}
\bea
{G_{ab}} &=& {T^{TOT}_{ab}}
\label{EFE2}
\\
M {u_b} &=& {\nabla _a} {{J^a}_b}+ c_4 \udot_a \nabla_b u^a
\label{evolveu}
\\
{u^a}{u_a} &=& -1,\label{unit}
 \eea
 \end{subequations}
donde $G_{ab}$ es el tensor de Einstein de la métrica $g_{ab}$,  ${T^{TOT}_{ab}}$
es el tensor de energía-momentum total, ${T^{TOT}_{ab}}=T _{ab}^{\text{æ}}+T^{mat}_{ab}$,
donde $T^{mat}_{ab}$ es la contribución total de todas las fuentes de materia. Se omitirá
$T^{mat}_{ab}$ por el momento (y se añadirá más adelante para modelos con fluido perfecto y con campo escalar). Se comienza con el caso vacío ($\mathcal{{L}}_m=0$). 
Las cantidades ${J^a}_b,\; {\udot_a}$
y el tensor de energía-momento æther $T _{ab}^{\text{æ}}$ están dados por
\begin{subequations}
\begin{align} 
{{J^a}_m} & =
-{{K^{ab}}_{mn}}{\nabla_b}{u^n}
\label{J} \\
{\dot u_a} &= {u^b}{\nabla _b}{u_a}
\label{a} \\
{T _{ab}^{\text{æ}}} &= 2c_{1}(\nabla_{a}u^{c}\nabla_{b}u_{c}-
\nabla^{c}u_{a}\nabla_{c}u_{b})  - 2[\nabla_{c}(u_{(a} J^{c}{}_{b)}) + \nabla_{c}(u^{c
}J_{(a b)}) - \nabla_{c}(u_{(a}J_{b)}{}^{c})] \nonumber\\
&  -2 c_4 \udot_a \udot_b + 2 M u_a u_b + g_{a b}\mathcal{L}_{\text{æ}}. \label{aestress}
\end{align}
\end{subequations}
Tomando la contracción de  \eqref{evolveu} con $u^b$ y con la métrica inducida  $h^{b c}:= g^{b c}+u^b u^c$ se obtienen las ecuaciones
\begin{subequations}
\label{aether_eqs}
\begin{align}
\label{definition:lambda}
&M = - u^b \nabla_a J^a_b-c_4 \udot_a \udot^a,\\
\label{restriction_aether}
& 0 = h^{b c}\nabla_a J^a_b + c_4 h^{b c} \udot_a \nabla_b u^a.
\end{align}
\end{subequations}
Se usará la ecuación \eqref{definition:lambda} como definición del multiplicador de Lagrange, mientras que la segunda ecuación conlleva a un conjunto de condiciones de compatibilidad que el vector æther debe satisfacer.

\section{Descomposición de las Ecuaciones de Campo según el Formalismo $1+3$}
\label{section2.2}
El formalismo $1+3$ \cite{vanElst:1996dr, Coley:2015qqa,wainwrightellis1997}  es  muy apropiado para estudiar modelos esféricamente simétricos con fluido perfecto y para el análisis cualitativo y numérico. En estos modelos la métrica está dada por:
\begin{equation}
ds^2=-N^2dt^2+(\ex)^{-2}dx^2+(\ey)^{-2}(d\y^2+sen^2 \y d\z^2),
\end{equation}
donde $N, \ex$ y $\ey$ son funciones de $t$ y $x$. \newline
Los campos vectoriales de Killing estan dados por \cite{Stephani:2003tm}:
\begin{equation}
\partial_{\z},\quad \cos{\z} \partial_{\y}-\sin{\z} \cot{\y} \partial_{\z},\quad \sin{\z} \partial_{\y}+\cos{\z} \cot{\y} \partial_{\z}.
\end{equation}
Los vectores de marco escritos en forma coordenada son:
\begin{equation}
\e_0=N^{-1}\partial_t,\quad \e_1=\ex \partial_x,\quad \e_2=\ey \partial_{\y},\quad \e_3=\ez \partial_{\z}.
\end{equation}
donde $\ez =\ey / \sin{\y}.$ 
Por tanto, se tienen restricciones en las variables cinemáticas:
\begin{equation}
\sigma_{\alpha \beta}=diag(-2\sigma_+,\sigma_+,\sigma_+),\quad \omega_{\alpha \beta}=0,\quad \udot_\alpha=(\udot_1,0,0),
\end{equation}
donde 
\begin{equation}
\udot_1=\e_1 \ln{N};
\end{equation}
se tienen restricciones en las funciones de conmutación espacial:
\begin{equation}
a_{\alpha}=(a_1,a_2,0),\quad 
n_{\alpha \beta=}\begin{pmatrix}
0 & 0 & n_{13}\\
0 & 0 & 0 \\
n_{13} & 0 & 0
\end{pmatrix},
\end{equation}
donde
\begin{equation}
a_1=\e_{1} \ln{\ey},\quad a_2=n_{13}=-\frac{1}{2}\ey\cot{\y};
\end{equation}
y se tienen restricciones en las componentes de materia:
\begin{equation}
q_{\alpha}=(q_1,0,0),\quad \pi_{\alpha \beta}=diag(-2\pi_+,\pi_+,\pi_+).
\end{equation}
La rotación de marco $\Omega_{\alpha \beta}$  es cero.
La constante cosmológica se hace cero por simplicidad. 
\newline 
$n_{13}$ solo aparece en las ecuaciones junto con $\e_2 n_{13}$ a través de la curvatura espacial de Gauss de las $2$-esferas 
\begin{equation}
{}^2K:=2(\e_2-2n_{13})n_{13},
\end{equation}
que se simplifica a 
\begin{equation}
{}^2K=(\ey)^2.
\end{equation}
Por tanto, la dependencia de $\y$ no aparece explícitamente en las ecuaciones. Se usara ${}^2K$ en lugar de $\ey$ para escribir las ecuaciones de campo.\newline
Las curvaturas espaciales también se simplifican a:
\begin{subequations}
\begin{equation}
{}^3S_{\alpha \beta}=\text{diag}(-2{}^3S_+,{}^3S_+,{}^3S_+),
\end{equation}
con ${}^3R$ y ${}^3S_+$ definidas por:
\begin{equation}
{}^3R=4\e_1a_1-6a_{1}^2+2{}^2K
\end{equation}
\begin{equation}
{}^3S_+=-\frac{1}{3}\e_1a_1+\frac{1}{3}{}^2K.
\end{equation}
\end{subequations}
Las componentes de la curvatura de Weyl se simplifican a:
\begin{equation}
E_{\alpha \beta}=diag(-2E_+,E_+,E_+), \quad H_{\alpha \beta}=0,
\end{equation}
con $E_+$ definido por:
\begin{equation}
E_+=H\sigma_+ + \sigma_{+}^2+{}^3S_+ -\frac{1}{2}\pi_+.
\end{equation}

Por motivos de simplificación, se escribirá
\begin{equation*}
{}^2K=K,\quad \udot_1=\udot,\quad a_1=a
\end{equation*}
Las variables esenciales son:
\begin{equation}
N,\quad \ex,\quad K,\quad H,\quad \sigma_+,\quad a,\quad \mu,\quad q_1,\quad p,\quad \pi_+
\end{equation}
y las variables auxiliares son
\begin{equation}
{}^3 K,\quad {}^3 S_+,\quad \udot.
\end{equation}
Las ecuaciones se escriben como:
\begin{subequations}
\begin{align}
&\e_0 \ex = (-H+2\sigma_+)\ex \\
&\e_0 K = -2(H+\sigma_+)K \\
&\e_0 H = -H^2-2\sigma_{+}^2
\frac{1}{3}(\e_1+\udot-2a)\udot-\frac{1}{6}(\mu+3p) \\
&\e_0 \sigma_+  = -3H\sigma_+-\frac{1}{3}(\e_1+\udot+a)\udot-{}^3S_+ +\pi_+ \\
&\e_0 a = (-H+2\sigma_+)a-(\e_1+\udot)(H+\sigma_+)\\
&\e_0 \mu = -3H(\mu+p)-(\e_1+2\udot-2a)q_1-6\sigma_+\pi_+ \\
&\e_0 q_1 = (-4H+2\sigma_+)q_1-\e_1p-(\mu+p)\udot+2(\e_1+\udot-3a)\pi_+
\end{align}
\end{subequations}
Las ecuaciones de restricción son las restricciones de Gauss y Codazzi junto con la definición de $a$:
\begin{subequations}
\begin{align}
0 & = 3H^2+\frac{1}{2}{}^3R-3\sigma_{+}^2-\mu\\
0 & =  -2\e_1(H+\sigma_+)+6a\sigma_+ +q_1\\
0 & =  (\e_1-2a)K
\end{align}
\end{subequations}
donde las curvaturas espaciales estan dadas por:
\begin{subequations}
\begin{equation}
{}^3R=4\e_1a-6a^2+2K
\end{equation}
\begin{equation}
{}^3S_+=-\frac{1}{3}\e_1 a+\frac{1}{3}K.
\end{equation}
\end{subequations}
Hasta el momento no se tienen ecuaciones de evolución para $N, p$ y $\pi_+$, para ello, se debe especificar la función de lapso $N$ mediante un \emph{gauge} temporal, y se debe especificar un modelo de fluido para las ecuaciones de estado y ecuación de transporte de $p$ y $\pi_+$.

\section{Formulación diagonal homotética}
\label{homotetica}
En la formulación diagonal homotética, el elemento de línea puede ser escrito en forma diagonal, donde una de las coordenadas se adapta a la simetría homotética \cite{Bogoyavlensky}:
\begin{align}
\label{metricTSS}
& d\tilde{s}^2 = e^{2 t} ds^2= e^{2 t}\Big[- b_1^{-2}(x)  dt^2 + dx^2   +b_2^{-2}(x) (d\y^2 + \sin^2 (\y)  d\z^2)\Big].
\end{align}
Ahora, introduciendo las definiciones:
$\theta=\frac{\sqrt{3}}{3}\left(2\alpha -\beta\right), \quad \sigma= \frac{\sqrt{3}}{3}\left(-\alpha + 2 \beta\right), \quad  \alpha=3\widehat{\beta^0}, \quad  \beta= 3\widehat{\beta^+},   b_1^{-1}=e^{\beta^0-2\beta^+},  \quad  b_2^{-1}=e^{\beta^0+\beta^+}$, donde  $\widehat{...}$ denota la derivada con respecto a la variable espacial $x$. Las cantidades $\alpha$ and $\beta$ son respectivamente el escalar de expansión y el escalar de cizalla de la congruencia normal a la superficie de simetría del espacio-tiempo estático $\left(\mathcal{M}, d{s}^2\right)$, el cuál está conformalmente relacionado con el espacio-tiempo físico $\left(\mathcal{M}, d {\tilde{s}}^2\right)$ mediante el factor de homotecia $e^{2 t}$. Se considera un vector æther no inclinado $\mathbf{u}=e^{-t }b_1 \partial_t$. 
\newline 
Si se asume que contenido de materia del  universo físico  $\left(\mathcal{M}, d\tilde{s}^2\right)$ es un fluido perfecto, este se especifica mediante un vector cuadrivelocidad  $\mathbf{v}$ dado por
$\mathbf{v} = \Gamma e^{-t }(-b_1 \partial_t + v \partial_x),\quad   \Gamma = (1-v^2)^{-\frac12}$, donde $v$ es el parámetro de inclinación, que es una función de $x$, con $-1\leq v \leq 1$. En general se elige  una ecuación de estado lineal para el fluido perfecto $p = (\gamma-1) \mu, \quad 1 \leq \gamma < 2$. Por conveniencia, se define   $\mu_t=\frac{e^{-2 t } \left((\gamma -1) v^2+1\right) \mu}{1-v^2}$, 
que también es una función que depende solamente de  $x$. Esta cantidad  $\mu_t$ representa la densidad de energía del fluido, medida por un observador asociado con la simetría homotética. 
Por otra parte, si el contenido de materia es el de un campo escalar no homegéneo, $\phi(t,x)$, con su potencial de auto--interacción  $V(\phi(t,x))$, estos deben respetar la homotecia de la simetría conformalmente estática asociada al elemento de línea \eqref{metricTSS}, por lo que tienen que ser de la forma \cite{Coley:2002je}:$ \phi(t,x)=\psi (x)-\lambda t, \quad  V(\phi(t,x))= e^{-2 t} U(\psi(x)), \quad  U(\psi)=U_0 e^{-\frac{2 \psi}{\lambda}}$,
donde por convenio se asume $\lambda>0$, tal que si $\psi>0$, $U\rightarrow 0$ cuando $\lambda \rightarrow 0$, lo que restringe el tipo de potencial a ser considerado.  

El formalismo diagonal homotético tiene algunas desventajas. Las superficies de simetría en general cambian la causalidad. Luego, en la formulación diagonal homotética el espacio-tiempo debe ser cubierto con dos sistemas de coordenadas (dos cartas); uno cuando el vector de Killing homotético es tipo tiempo, y otro cuando el vector de Killing homotético es tipo espacio. Estas dos regiones tienen que ``empalmarse'' en la región donde el vector de Killing es nulo \cite{Goliath:1998mx}. Sin embargo, la formulación tiene más ventajas que desventajas; la principal es que permite escribir las ecuaciones del campo, que son un sistema bien definido de ecuaciones  en derivadas parciales (PDE) de primer orden en dos variables (provenientes del formalismo $1+3$) como un sistema de ecuaciones diferenciales ordinarias utilizando las simetrías que provienen de los vectores de Killing. Las ecuaciones resultantes son muy similares a las de los modelos con hipersuperficies homogéneas.  A su vez, es posible escribir dichas ecuaciones como un sistema dinámico, lo que hace posible el estudio del modelo usando las técnicas de la teoría cualitativa de los sistemas dinámicos y esto hace posible obtener una descripción en el espacio de fases completa, lo que lleva a una mejor compresión de la dinámica del modelo.

\lhead{Capítulo \ref{ch_3}}
\rhead{Modelos con fluido perfecto}
\cfoot{\thepage}
\renewcommand{\headrulewidth}{1pt}
\renewcommand{\footrulewidth}{1pt}

\chapter{Modelo con fluido perfecto}\label{ch_3}

En este capítulo se estudian métricas temporales, auto-similares y esféricamente simétricas (también conocidas por conformalmente estáticas) en teoría Einsten-æther con fluido perfecto. Se deducen las ecuaciones de evolución, y se discuten las restricciones en el espacio de parámetros de la teoría que son compatibles para las diferentes geometrías. Se introducen variables Hubble-normalizadas, obteniéndose un sistema de ecuaciones diferenciales con respecto a una coordenada ``radial'' independiente. Las ecuaciones se reducen a un sistema dinámico autónomo que satisface restricciones algebraicas. Estas restricciones se pueden utilizar para reducir la dimensión del sistema y permiten el uso de herramientas de la teoría cualitativa de los sistemas dinámicos.

\section{Modelos temporales autosimilares esféricamente simétricos con fluido perfecto.}
\label{model3}
En esta sección, se estudian modelos temporales autosimilares esféricamente simétricos con fluido perfecto usando la descomposición covariante $1+3$ \cite{Coley:2015qqa,wainwrightellis1997}, siguiendo la estrategia usada en \cite{Goliath:1998mx}. Estos  modelos se caracterizan por un grupo simétrico homotético 4-dimensional  $H_4$ actuando multiplemente transitivamente sobre superficies temporales 3-dimensionales. Como se explicó antes,  para la métrica  conformalmente estática dada por elemento de línea \eqref{metricTSS}, se pueden definir los siguientes escalares:
\begin{align}
& \theta=\frac{\sqrt{3}}{3}\left(2\alpha -\beta\right), \quad \sigma= \frac{\sqrt{3}}{3}\left(-\alpha + 2 \beta\right), \quad  \alpha=3\widehat{\beta^0}, \quad  \beta= 3\widehat{\beta^+}, \nonumber\\  &  b_1^{-1}=e^{\beta^0-2\beta^+},  \quad  b_2^{-1}=e^{\beta^0+\beta^+},    
\end{align}
donde  $\widehat{...}$ denota la derivada con respecto a la variable espacial $x$ y las cantidades $\alpha$ and $\beta$ son respectivamente el escalar de expansión y el escalar de cizalla de la congruencia normal a la superficie de simetría del universo estático $\left(\mathcal{M}, d{s}^2\right)$.

Usando la métrica  \eqref{metricTSS}, y los escalares anteriormente dados, el Lagrangiano \eqref{aeLagrangian} se convierte en 
\begin{align}
&\mathcal{L}_{u} = \frac{1}{3} e^{-2 t} \left(( c_{1}- c_{4}) {\sigma}^2-9  b_{1}^2 ( c_{1}+3 c_{2}+ c_{3})\right).
\end{align}
El multiplicador de Lagrange  \eqref{definition:lambda} se calcula como
\begin{align}
& M=-3 e^{-2 t}b_1^2 (c_{1}+c_{2}+ c_{3}) -\frac{1}{3} e^{-2 t} {\sigma}({\sigma}(- c_{1}+ c_{3}+2  c_{4})+2 c_{3} {\theta}) -\frac{ c_{3} e^{-2 t} \widehat{{\sigma}}}{\sqrt{3}}.
\end{align}
Mientras que la ecuación del æther   \eqref{restriction_aether} se reduce a 
\begin{equation}
{e^{-t}  b_{1} {\sigma} (2  c_{1}+3  c_{2}+ c_{3}- c_{4})}=0.
\end{equation}
Además, la traza de la 3-curvatura intríseca de Ricci de las 3-superficies espaciales ortogonales a $\mathbf{u}$ está dada por 
\begin{align}
& {}^{*} R =-\frac{2}{3} e^{-2 t} \left(-3  b_{2}^2+2 \sqrt{3} \left(\widehat{ {\theta}}+\widehat{ {\sigma}}\right)+3 ( {\theta}+ {\sigma})^2\right).
\end{align}

Para simplificar estas expresiones es conveniente hacer una reparametrización de los parámetros del æther que son análogos a la dada en \cite{Jacobson:2013xta}:
\begin{displaymath}
c_\theta = c_2 + (c_1 + c_3)/3,\ c_\sigma = c_1 + c_3,\ c_\omega = c_1 - c_3,\ c_a = c_4 - c_1.
\end{displaymath}
Donde los nuevos parámetros tienen una interpretación física inmediata, correspondiendo a términos en el Lagrangiano relativos a la expansión, al escalar de cizalla, a la aceleración y al  \underline{twist} del æther.  Como los modelos esféricamente simétricos son hipersuperficie ortogonales, el campo æther tiene \underline{twist} nulo y por tanto es independiente del parámetro de  \underline{twist}, $c_\omega$ (el acoplamiento $c_1-c_3$ no aparece en las ecuaciones de campo, solamente aparece $c_1+c_3$) \cite{Jacobson:2013xta}; esto equivale a elegir $c_4=0$  \cite{Jacobson:2008aj}). Puede obtenerse una segunda condición sobre los $c_i$ renormalizando la constante gravitacional efectiva de Newton $G$.
El resto de los parámetros del modelo pueden ser caracterizados por $2$ parámetros constantes no triviales. Para imponer la condición del  æther  \eqref{restriction_aether} se toma $c_a = 3 c_{\theta }$, $(2  c_{1}+3  c_{2}+ c_{3}- c_{4})=0$. 
\newline
Por tanto el espacio de parámetros se reduce a una constante, $c_\theta$. 
\newline
El tensor de energía-momento del æther  se puede expresar por 
\begin{equation}
\label{Taether}
{T^{\text{æ}}}_{a}^{b}=\left(
\begin{array}{cccc}
 e^{-2 t} \mu & -e^{-2 t} q & 0 & 0 \\
-e^{-2 t} q & e^{-2 t} (p-2 \pi) & 0 & 0 \\
 0 & 0 & e^{-2 t} (p+\pi) & 0 \\
 0 & 0 & 0 & e^{-2 t} (p+\pi) \\
\end{array}
\right),
\end{equation}
donde se han incluido las siguientes definiciones que dependen únicamente de  $x$:
 \begin{align}
     & \mu = c_{\theta } \left(9 b_1^2-4  {\theta}  {\sigma}-2 \sqrt{3} \widehat{ {\sigma}}-3  {\sigma}^2\right), \quad p =  \frac{1}{3} c_{\theta } \left(9 b_1^2- {\sigma}^2\right), \\ \nonumber & q =   -2 \sqrt{3} b_1 c_{\theta }  {\sigma}, \quad \pi =   -\frac{2}{3} c_{\theta } {\sigma}^2.
 \end{align}
El tensor de energy-momentum  para la materia está dado por
\begin{equation}
\label{Tm}
{T^{m}}_{a}^{b}=\left(
\begin{array}{cccc}
 e^{-2 t} \mu_{t}  & -\frac{e^{-2 t} \gamma  v  \mu_{t}}{(\gamma -1) v ^2+1} & 0 & 0 \\
 -\frac{e^{-2 t} \gamma  v  \mu_{t}}{(\gamma -1) v ^2+1} & \frac{e^{-2 t} \left(v ^2+\gamma -1\right) \mu_{t}}{(\gamma -1) v ^2+1} & 0 & 0 \\
 0 & 0 & -\frac{e^{-2 t} (\gamma -1) \left(v ^2-1\right) \mu_{t}}{(\gamma -1) v ^2+1} & 0 \\
 0 & 0 & 0 & -\frac{e^{-2 t} (\gamma -1) \left(v ^2-1\right) \mu_{t}}{(\gamma -1) v ^2+1} \\
\end{array}
\right).
\end{equation}
Usando las ecuaciones de Einstein, las identidades de Jacobi y las identidades de Bianchi contraidas, se obtiene un sistema de ecuaciones diferenciales ordinarias para los vectores de marco y las funciones de conmutación, y una ecuación extra para el æther. Se elige el gauge comóvil para æther; quedando como grado de libertad una reparametrización de las variables espaciales y de la variable temporal. Las ecuaciones finales son:\\ 
\noindent{\bf Ecuaciones de propagación:}  
\begin{subequations}
\label{modelouno}
\begin{align}
&\widehat{ {\theta}}=-\sqrt{3} b_{2}^2-\frac{ {\sigma}\left(2 C_2  {\sigma}+ {\theta}\right)}{\sqrt{3}}-\frac{\sqrt{3} \gamma \mu_{t} v^2}{(\gamma -1) v^2+1}, \label{EQ48 }\\ 
&\widehat{ {\sigma}}=-\frac{ {\sigma} (2  {\theta}+ {\sigma})}{\sqrt{3}}+\frac{\sqrt{3} \mu_{t} \left(2-3 \gamma +(\gamma -2)
   v^2\right)}{2 C_2 \left((\gamma -1) v^2+1\right)},\\ 
&\widehat{b_1}= \frac{b_1  {\sigma}}{\sqrt{3}},\\
&\widehat{b_2}= -\frac{b_2 ( {\theta}+ {\sigma})}{\sqrt{3}},\\
&\widehat{v}=  \frac{\left(1-v^2\right)}{\sqrt{3} \gamma  \left(1-\gamma +v^2\right)} \Big\{ \gamma  v (2 (\gamma -1) \theta +\gamma  \sigma) +\sqrt{3} b_1\left((\gamma -1) (3\gamma -2)+(\gamma -2) v^2\right)\Big\}.
\end{align}
\end{subequations}
\noindent{\bf Ecuación para  $\mu_t$:}
	\begin{equation}
	\mu_t= \frac{\left((\gamma -1) v^2+1\right) }{3 \left(1-\gamma -v^2\right)}\left(C_2 \left(3 b_1^2+\sigma ^2\right)+3 b_2^2-\theta^2\right),
	\end{equation}
\noindent{\bf	Ecuación auxiliar:}
\begin{align}
& \widehat{\mu _{t}}=\frac{\mu_{t}}{\sqrt{3} \left(\gamma -v^2-1\right) \left((\gamma -1) v^2+1\right)}
 \Big\{ \gamma  \left(\sigma+(\gamma -1) v^4 (2 \theta +\sigma )-v ^2 ((4 \gamma -6) \theta+\gamma  \sigma )\right)\nonumber \\
& +2 \sqrt{3} b_1 v \left((7-3 \gamma ) \gamma +(\gamma  (2 \gamma -5)+4) v^2-4\right)\Big\}.
\end{align}
\noindent{\bf Restricción:}
\begin{equation}
\gamma  \mu _{t} v-\frac{2 C_2 b_1 \sigma \left((\gamma -1) v^2+1\right)}{\sqrt{3}}=0.\end{equation}
Para simplificar la notación se han definido $C_1=1-2c_\sigma, C_2=1+3c_\theta,C_3=1+c_a$ donde  sustituyendo $C_1=1, C_2=1, C_3=1$  se  recupera  Relatividad  General.
La condición  $c_a = 3 c_{\theta }$, implica $C_2=C_3$, y el parámetro $C_1$ no aparece explícitamente en las  ecuaciones (solo en la definición del multiplicador de Lagrange). 

Para gas ideal con $\gamma=1$  (fluido sin presión; polvo) el tensor de energía-momentum de la materia se reduce a: 
\begin{equation}
\label{Tmgas-ideal}
{T^{m}}_{a}^{b}=    \left(
\begin{array}{cccc}
 e^{-2 t} \mu _{t} & -e^{-2 t} \mu _{t} v & 0 & 0 \\
 -e^{-2 t} \mu _{t} v  & e^{-2 t} \mu _{t} v^2 & 0 & 0 \\
 0 & 0 & 0 & 0 \\
 0 & 0 & 0 & 0 \\
\end{array}
\right),
\end{equation}
donde $v$ mide la inclinación del fluido respecto a la cuadrivelocidad del aether. Las ecuaciones se reducen a 
\begin{subequations}
\label{YYYYY}
\begin{align}
& \widehat{\theta}= -\sqrt{3} b_2^2-\frac{\sigma\left(2 C_2 \sigma+\theta \right)}{\sqrt{3}}-\sqrt{3} \mu_{t} v^2,\\
& \widehat{\sigma}= -\frac{\sqrt{3}\mu_{t}( \left(v^2+1\right)}{2
   C_2}-\frac{\sigma (2 \theta+\sigma)}{\sqrt{3}},\\
& \widehat{b_1}= \frac{b_1 \sigma}{\sqrt{3}},\\
& \widehat{b_2}=  -\frac{b_2 (\theta+\sigma)}{\sqrt{3}},\\
& \widehat{\mu_t}= \mu_{t}\left(\frac{v^2 (\sigma-2 \theta)-\sigma}{\sqrt{3} v^2}-2 b_1 v\right),\\
& \widehat{v}= b_1 \left(v^2-1\right)-\frac{\sigma \left(v^2-1\right)}{\sqrt{3} v},
\end{align}
\end{subequations}
con restricciones 
\begin{subequations}
\label{YYYYY-rest}
\begin{align}
& \mu_{t} v-\frac{2 C_2 b_1 \sigma}{\sqrt{3}}=0,\\
& -C_2 \left(3 b_1^2+\sigma^2\right)-3 b_2^2+\theta^2-3 \mu_{t} v^2=0.
\end{align}
\end{subequations}

Para recuperar la Relatividad General se debe hacer $C_2 = 1$. Por esto es natural considerar $C_1 = O(1)$, luego se puede asumir que  $C_2>0$. Primero se impone la restricción
$\frac{\left(\gamma  v^2-v^2+1\right)}{\gamma +v^2-1}\geq 0$. Esto es, $0<\gamma \leq 1, 1-\gamma \leq v^2\leq 1$, o   $1<\gamma <2, -1\leq v\leq 1$. Esta condición es natural debido a la condición de energía usual para el fluido que se expresa como $1<\gamma <2$. Por lo tanto, se cumple la segunda condición.   La condición $\mu_t\geq 0$ junto con la condición natural de energía, lleva a 
\begin{equation}
C_2 \left(3 b_1^2+\sigma ^2\right)+3 b_2^2 \leq \theta^2.
\end{equation}
Por hipótesis $C_2>0$. Luego, $\theta^2$ es la cantidad dominante, además los términos en el miembro izquierdo de la desigualdad son ambos no negativos. Esto sugiere considerar ecuaciones $\theta$-normalizadas. 

En la siguiente sección se estableceran distintas  condiciones de estabilidad de las soluciones de equilibrio de sistemas dinámicos construidos a partir de  los modelos temporales autosimilares esféricamente simétricos (métricas conformalmente estáticas) con fluido perfecto en la Teoría Einstein \AE ther. Se utilizarán métodos numéricos para apoyar y validar los resultados analíticos.  
En particular, se estudiarán cuatro modelos específicos, estos son: modelos con inclinación extrema \eqref{extreme_tilt_0}; fluido perfecto sin presión \eqref{3.41-3.43}; el sistema reducido \eqref{eq:3.57} en el conjunto invariante $A=v=0$;  y el sistema general \eqref{reducedsyst}.

\section{$\theta$-Normalización de las ecuaciones}
\label{thetaperfectfluid}

En esta sección se definen las siguientes variables normalizadas,
\begin{align}
{\Sigma} =\frac{\sigma}{\theta},\quad  A=\frac{\sqrt{3} b_1}{\theta},\quad K=\frac{3 b_2^2}{\theta^2},\quad {\Omega}_t =\frac{3\mu_t}{\theta^2},
\end{align}
la coordenada radial
$$ f'=\frac{df}{d \eta} := \frac{\sqrt{3}\widehat{f}}{\theta},$$
y se define el parámetro ${r}$, de manera análoga al ``parámetro gradiente de Hubble'' ${r}$, dado por
\begin{equation}
   \widehat{\theta}=-r {\theta}^2,
\end{equation} tal que
\begin{equation}
\label{defnr}
{r}=\frac{{\Sigma} \left(2 C_2 {\Sigma} +1\right)}{\sqrt{3}}+\frac{\gamma  {\Omega}_t v^2}{\sqrt{3} \left((\gamma -1) v^2+1\right)}+\frac{K}{\sqrt{3}},
\end{equation}
luego, se obtienen las ecuaciones del  primer modelo objeto de estudio de esta tesis.\\ 
\newline
\noindent{\bf Ecuaciones de propagación:} 
\begin{small}
\begin{subequations}
\label{system29}
\begin{align}
&{\Sigma}'=-{{\Sigma}} \left({{\Sigma}}-\sqrt{3} {r}+2\right)+\frac{{{\Omega}_t} \left(-3 \gamma +(\gamma -2) v^2+2\right)}{2 C_2 \left((\gamma -1) v^2+1\right)},\\
& A'=A \left({{\Sigma}}+\sqrt{3} {r}\right),\\
& K'=	2 K \left(-{{\Sigma}}+\sqrt{3} {r}-1\right),\\
& v'=\frac{\left(v^2-1\right)}{\gamma  \left(\gamma -v^2-1\right)} \Big\{\gamma  v \left(\gamma  {{\Sigma}}+2 \gamma -2\right) +A \Big[(\gamma -1) (3 \gamma -2)+ (\gamma -2)
   v^2\Big]\Big\}. \label{system29v}
	\end{align}    
\end{subequations}
\end{small}
\newline
\noindent{\bf Ecuación para ${\Omega}_t$:}
\begin{equation}
\label{defnOmegat}
{\Omega}_{t}=\frac{\left(\gamma  v^2-v^2+1\right) \left(1-C_2{{\Sigma}}^2- C_2 A^2 -K\right)}{\gamma +v^2-1}.
\end{equation}
\newline
\noindent{\bf Ecuación auxiliar:}
\begin{align}
& {\Omega}_{t}'= 2 \sqrt{3} {r} {\Omega}_{t}+\frac{{\Omega}_{t}}{\left(\gamma -v^2-1\right) \left(\gamma  v^2-v^2+1\right)} 
\Big\{\gamma  {{\Sigma}}+(\gamma -1) \gamma  \left({{\Sigma}}+2\right) v^4 -\gamma  v^2 \left(\gamma  {{\Sigma}}+4 \gamma -6\right) \nonumber \\
& +2 A v \Big[(\gamma  (2 \gamma -5)+4) v^2-  (\gamma -1) (3 \gamma -4) \Big]\Big\},
	\end{align}
\newline
\noindent{\bf Restricción:}
\begin{equation}
\label{constraintmod1}
\gamma  {\Omega}_t v -2 A C_2 {\Sigma} \left((\gamma -1) v^2+1\right)=0.
	\end{equation}
Es fácil verificar que las ecuaciones anteriores son invariantes bajo la siguiente transformación
\begin{equation}
({\Sigma},A,K,v)\rightarrow({\Sigma},-A,K,-v). 
\end{equation}
Por tanto, es suficiente analizar el caso $A\geq 0$.

Luego de sustituir las expresiones \eqref{defnr} y \eqref{defnOmegat}  sistema \eqref{system29} se reduce a  
\begin{subequations}
\label{Eq:22}
\begin{align}
   & \Sigma'= -\frac{1}{2 C_2 \left(\gamma +v^2-1\right)} \Bigg\{A^2 C_2 \left(-3 \gamma +v^2 (\gamma +2 \gamma  C_2 \Sigma -2)+2\right) \nonumber\\
   & +K \left(-3 \gamma -2 (\gamma -1) C_2 \Sigma +v^2 (\gamma +2 (\gamma -1) C_2 \Sigma -2)+2\right) \nonumber \\
   & +\left(C_2 \Sigma
   ^2-1\right) \left(-3 \gamma -4 (\gamma -1) C_2 \Sigma +(\gamma -2) v^2 (2 C_2 \Sigma +1)+2\right)\Bigg\},
   \\
   & A'= \frac{A }{\gamma +v^2-1} \Bigg\{v^2 \left(2 \Sigma  (C_2 \Sigma +1)-\gamma 
   \left(C_2 \left(A^2+\Sigma ^2\right)-1\right)\right) \nonumber \\
   & +2 (\gamma -1) \Sigma  (C_2 \Sigma +1)-(\gamma -1) K \left(v^2-1\right)\Bigg\}, \\
   & K'= \frac{2 K }{\gamma +v^2-1} \Bigg\{v^2 \left(-C-2 \gamma A^2 +\gamma -(\gamma
   -2) C_2 \Sigma ^2-1\right) \nonumber\\
   & +(\gamma -1) \left(2 C_2 \Sigma ^2-1\right)-(\gamma -1) K \left(v^2-1\right)\Bigg\},\\ 
   & v'=\frac{\left(v^2-1\right) \left(A \left(3 \gamma ^2-5 \gamma +(\gamma -2)
   v^2+2\right)+\gamma  v (\gamma  (\Sigma +2)-2)\right)}{\gamma  \left(\gamma -v^2-1\right)}.
\end{align}
\end{subequations}
Asumimos $v^2\neq \gamma-1$. El caso  $v^2= \gamma-1$ se estudia en la sección \ref{SL}. Los conjuntos invariantes $v=\pm 1$ serán analizados en \ref{tiltfluidoperfecto}. 
En la sección \ref{fluidosinpresion}, será estudiado el modelo de materia con gas ideal ($\gamma=1$). El caso general con $v\neq 0$ (fluido perfecto inclinado) será estudiado en la sección \ref{general}. Los puntos de equilibrio con $v=0, A=0$ serán estudiados en la seccion \ref{vcero}. Finalmente, en la sección \ref{DiscusionC1} se resumen los resultados obtenidos en este capítulo, y se comparan los resultados con aquellos obtenidos en la literatura y se discuten los nuevos resultados.

Notar que el gradiente de la restricción \eqref{constraintmod1} se anula en los tres puntos de equilibrios siguientes: $(\Sigma,A,K,v)=(0,0,1,0),(0,0,1,\pm 1)$. Por esto, para estos puntos se mantienen los cuatro autovalores, ya que la restricción es degenerada en ellos. 

\begin{enumerate}
\item El punto de equilibro $N_1:(\Sigma,A,K,v)=(0,0,1,0)$ tiene autovalores $\{-2,-1,1,2\}$, por tanto es una silla hiperbólica.
\item  Los puntos de equilibrio $N_{2,3}:(\Sigma,A,K,v)=(0,0,1,\pm 1)$ tienen autovalores \newline $\left\{0,-1,1,\frac{4}{\gamma -2}+4\right\}$ por tanto son sillas no hiperbólicas.
\end{enumerate}

Además, tenemos los puntos de equilibrio de \eqref{reducedsyst}\newline  $\widetilde{M}^{\pm}:({\Sigma}, {A}, K, v)=\left(0,1,0, \frac{(\gamma -1) \gamma  \pm \left(\sqrt{(\gamma -1) \left((\gamma
   -1) \gamma ^2+(2-\gamma ) (3 \gamma -2)\right)}\right)}{2-\gamma }\right),\quad \Omega_t=0$, que existen sólo si $C_2=1$. 

Los autovalores de  $\widetilde {M}^{+}$ son \newline
\bigskip
\newline
\begin{doublespace}
\noindent\( {\left\{-2,\frac{1}{(-2+\gamma ) (-1+\gamma )}\left(\gamma  \left(8+(-4+\gamma ) \gamma -\sqrt{(-1+\gamma ) (-4+\gamma  (8+(-4+\gamma
) \gamma ))}\right)+\right.\right.}\\
 {\left.4 \left(-1+\sqrt{(-1+\gamma ) (-4+\gamma  (8+(-4+\gamma ) \gamma ))}\right)\right),}\\
 {\frac{1}{2 (-1+\gamma ) \gamma }\left(-2-\sqrt{(-1+\gamma ) (-4+\gamma  (8+(-4+\gamma ) \gamma ))}+\right.}\\
 {\gamma  \left(6-\gamma  (3+\gamma )+\sqrt{(-1+\gamma ) (-4+\gamma  (8+(-4+\gamma ) \gamma ))}\right)+}\\
 {\surd \left((-1+\gamma ) \left(-4 \left(2+\sqrt{(-1+\gamma ) (-4+\gamma  (8+(-4+\gamma ) \gamma ))}\right)+\right.\right.}\\
 {\gamma  \left(4 \left(5+\sqrt{(-1+\gamma ) (-4+\gamma  (8+(-4+\gamma ) \gamma ))}\right)+\right.}\\
 {\gamma  \left(-28+6 \sqrt{(-1+\gamma ) (-4+\gamma  (8+(-4+\gamma ) \gamma ))}+\gamma  \right.}\\
 {\left.\left.\left.\left.\left.\left(29+\gamma  (-11+2 \gamma )-2 \sqrt{(-1+\gamma ) (-4+\gamma  (8+(-4+\gamma ) \gamma ))}\right)\right)\right)\right)\right)\right),}\\
 {-\frac{1}{2 (-1+\gamma ) \gamma }\left(2+\sqrt{(-1+\gamma ) (-4+\gamma  (8+(-4+\gamma ) \gamma ))}-\right.}\\
 {\gamma  \left(6-\gamma  (3+\gamma )+\sqrt{(-1+\gamma ) (-4+\gamma  (8+(-4+\gamma ) \gamma ))}\right)+}\\
 {\surd \left((-1+\gamma ) \left(-4 \left(2+\sqrt{(-1+\gamma ) (-4+\gamma  (8+(-4+\gamma ) \gamma ))}\right)+\right.\right.}\\
 {\gamma  \left(4 \left(5+\sqrt{(-1+\gamma ) (-4+\gamma  (8+(-4+\gamma ) \gamma ))}\right)+\right.}\\
 {\gamma  \left(-28+6 \sqrt{(-1+\gamma ) (-4+\gamma  (8+(-4+\gamma ) \gamma ))}+\right.}\\
 {\left.\left.\left.\left.\left.\left.\gamma  \left(29+\gamma  (-11+2 \gamma )-2 \sqrt{(-1+\gamma ) (-4+\gamma  (8+(-4+\gamma ) \gamma ))}\right)\right)\right)\right)\right)\right)\right\}}\).
\end{doublespace}
\newpage
\begin{figure*}
    \centering
    \subfigure[$\widetilde{M}^{+}$]{\includegraphics[scale=0.4]{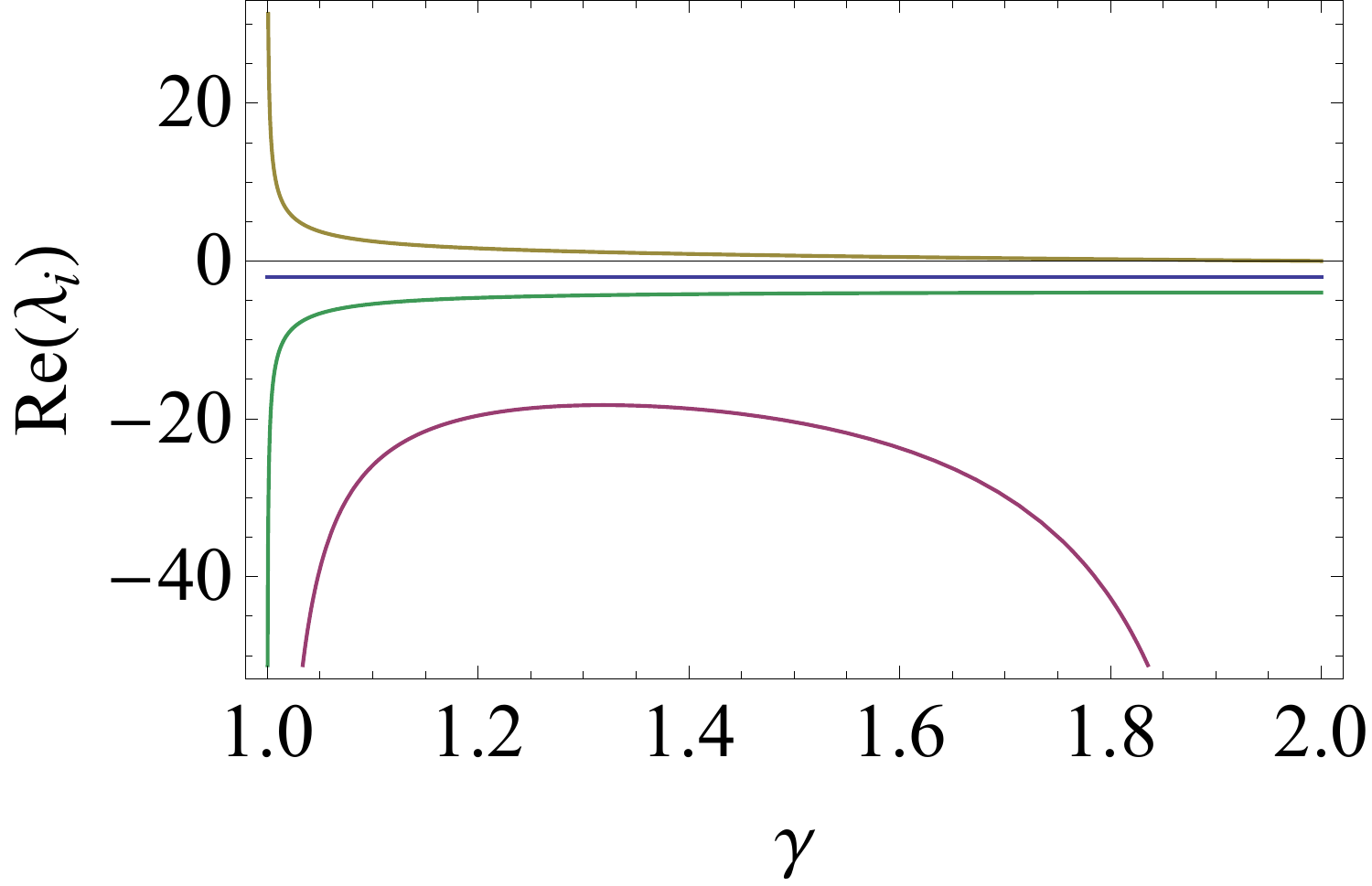}} \hspace{2cm}
    \subfigure[$\widetilde{M}^{-}$]{\includegraphics[scale=0.4]{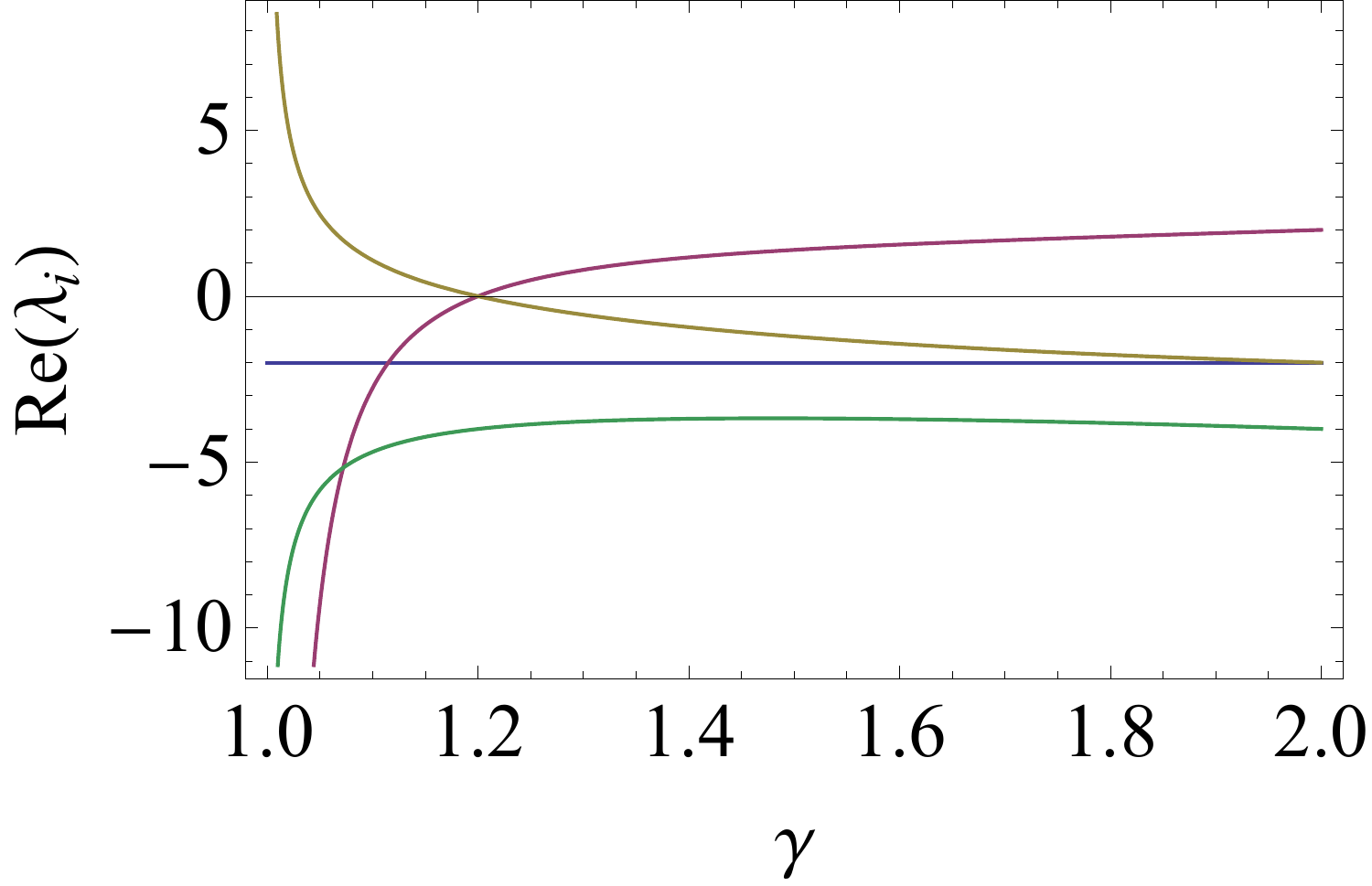}}
    \caption[{Parte real de los autovalores  del punto de equilibrio $\widetilde{M}^{\pm}$ para $C_2=1$.}]{\label{eigenM} Parte real de los autovalores  del punto de equilibrio $\widetilde{M}^{\pm}$ para $C_2=1$.}
   \end{figure*}
   \FloatBarrier
Los autovalores de $\widetilde {M}^{-}$ son \newline
\begin{doublespace}
\noindent\( {\left\{-2,\frac{1}{(-2+\gamma ) (-1+\gamma )}\left(-4 \left(1+\sqrt{(-1+\gamma ) (-4+\gamma  (8+(-4+\gamma ) \gamma ))}\right)+\right.\right.}\\
 {\left.\gamma  \left(8+(-4+\gamma ) \gamma +\sqrt{(-1+\gamma ) (-4+\gamma  (8+(-4+\gamma ) \gamma ))}\right)\right),}\\
 {\frac{1}{2 (-1+\gamma ) \gamma }\left(-2+\sqrt{(-1+\gamma ) (-4+\gamma  (8+(-4+\gamma ) \gamma ))}-\right.}\\
 {\gamma  \left(-6+\gamma  (3+\gamma )+\sqrt{(-1+\gamma ) (-4+\gamma  (8+(-4+\gamma ) \gamma ))}\right)+}\\
 {\surd \left((-1+\gamma ) \left(4 \left(-2+\sqrt{(-1+\gamma ) (-4+\gamma  (8+(-4+\gamma ) \gamma ))}\right)+\right.\right.}\\
 {\gamma  \left(-4 \left(-5+\sqrt{(-1+\gamma ) (-4+\gamma  (8+(-4+\gamma ) \gamma ))}\right)+\right.}\\
 {\gamma  \left(-28-6 \sqrt{(-1+\gamma ) (-4+\gamma  (8+(-4+\gamma ) \gamma ))}+\gamma  \right.}\\
 {\left.\left.\left.\left.\left.\left(29+\gamma  (-11+2 \gamma )+2 \sqrt{(-1+\gamma ) (-4+\gamma  (8+(-4+\gamma ) \gamma ))}\right)\right)\right)\right)\right)\right),}\\
 {-\frac{1}{2 (-1+\gamma ) \gamma }\left(2-\sqrt{(-1+\gamma ) (-4+\gamma  (8+(-4+\gamma ) \gamma ))}+\right.}\\
 {\gamma  \left(-6+\gamma  (3+\gamma )+\sqrt{(-1+\gamma ) (-4+\gamma  (8+(-4+\gamma ) \gamma ))}\right)+}\\
 {\surd \left((-1+\gamma ) \left(4 \left(-2+\sqrt{(-1+\gamma ) (-4+\gamma  (8+(-4+\gamma ) \gamma ))}\right)+\right.\right.}\\
 {\gamma  \left(-4 \left(-5+\sqrt{(-1+\gamma ) (-4+\gamma  (8+(-4+\gamma ) \gamma ))}\right)+\right.}\\
 {\gamma  \left(-28-6 \sqrt{(-1+\gamma ) (-4+\gamma  (8+(-4+\gamma ) \gamma ))}+\right.}\\
 {\left.\left.\left.\left.\left.\left.\gamma  \left(29+\gamma  (-11+2 \gamma )+2 \sqrt{(-1+\gamma ) (-4+\gamma  (8+(-4+\gamma ) \gamma ))}\right)\right)\right)\right)\right)\right)\right\}}\).
\end{doublespace}
En la figura \ref{eigenM} se representa la parte  real de los autovalores  del punto de equilibrio $\widetilde{M}^{\pm}$ para $C_2=1$, mostrando que en general tiene comportamiento de silla. Para $\gamma=\frac{6}{5}$, los autovalores de $\widetilde{M}^{-}$ son $\{-2,0,0,-4\}$ y en este caso es no hiperbólico.

\subsection{Superficie de no extendibilidad de las soluciones}
\label{SL}
Para $\gamma>1$, la superficie definida por $v^2= \gamma-1$  es la superficie de no extendibilidad de las soluciones y es llamada superficie sónica. 
La curva parametrizada por $\Sigma$, 
\begin{equation}
SL_{\pm}: A=-\frac{\gamma 
   \varepsilon  (\gamma  (\Sigma
   +2)-2)}{4 (\gamma
   -1)^{3/2}},  v=\varepsilon\sqrt{\gamma -1},   \varepsilon=\pm 1,
\end{equation}
es llamada línea sónica. Las soluciones divergen en un tiempo finito cuando las soluciones se acercan a la superficie sónica $v^2= \gamma-1$. La única manera en que se puede pasar a través de la superficie sónica es cuando el numerador de la ecuación  \eqref{system29v} también se anula, esto es a través de la línea sónica $SL_{\pm}$. En $SL_{\pm}$ se anulan tanto el denominador como el numerador de \eqref{system29v} esto indica la presencia de una singularidad del sistema \eqref{system29}. El sistema \eqref{reducedsyst} admite los puntos de equilibrio 
\begin{enumerate}
    \item $SL_1: C_2= \frac{\gamma
   ^2}{4 (\gamma -1)^2},\Sigma =
   \frac{2 (\gamma -1)}{\gamma
   },v=\sqrt{\gamma -1},A=
  - \frac{\gamma  (\gamma  (\Sigma
   +2)-2)}{4 (\gamma
   -1)^{3/2}}$, 
    \item $SL_2: C_2= \frac{\gamma
   ^2}{4 (\gamma -1)^2},\Sigma =
   -\frac{2 (\gamma -1)}{\gamma
   },v= -\sqrt{\gamma -1},A=
   \frac{\gamma  (\gamma  (\Sigma
   +2)-2)}{4 (\gamma
   -1)^{3/2}}$,
\end{enumerate}
los cuáles yacen en la línea sónica. Estos puntos no existen en relatividad general cuando $1<\gamma<2$. Cuando $\gamma=2, C_2=1$ dichos puntos existen, y como $\gamma=2$  el fluido se comporta como materia rígida. Adicionalmente, si $\gamma=2, C_2=1$, estos puntos corresponden a modelos con inclinación extrema ($v=\varepsilon$), $SL_1: \Sigma=1, A=-2, v=1$, y $SL_2: \Sigma=-1, A=0, v=-1$.

En la superficie sónica debe satisfacerse la desigualdad 
$C_2(A^2+ \Sigma^2)\leq 1$, que corresponde a $K\geq 0$, lo cual impone condiciones adicionales a los parámetros.

Para estudiar localmente el comportamiento cerca de esta línea sónica, es posible hacer el cambio de variable independiente  a la nueva variable $\xi$ dada por \begin{equation}
    \frac{d \xi}{d \eta}= \frac{1}{(\gamma-1)-v^2},
\end{equation}
obteniéndose el sistema 
\begin{subequations}
\begin{align}
  & \frac{d \Sigma}{d \xi}=  \frac{\Sigma \left(1-\gamma
   +v^2\right) }{\gamma 
   v} \Big\{A^2 \gamma 
   C_2 v+A \left(3 \gamma +2
   (\gamma -1) C_2 \Sigma
   +v^2 (-(\gamma +2 (\gamma -1)
   C_2 \Sigma
   -2))-2\right) \nonumber \\
   & +\gamma  v
   \left(1-C_2 \Sigma
   ^2\right)\Big\},
   \\
  & \frac{d A}{d \xi}= -\frac{A \left(-\gamma
   +v^2+1\right) \left(\gamma  v
   \left(1-A^2 C_2\right)+2
   \Sigma  \left(A (\gamma -1)
   C_2
   \left(v^2-1\right)+\gamma 
   v\right)+\gamma  C_2
   \Sigma ^2 v\right)}{\gamma 
   v},\\
 & \frac{d v}{d \xi}= \frac{\left(v^2-1\right)
   \left(A \left(3 \gamma ^2-5
   \gamma +(\gamma -2)
   v^2+2\right)+\gamma  v (\gamma 
   (\Sigma +2)-2)\right)}{\gamma
   },\\
&  \frac{d \eta}{d \xi}={(\gamma-1)-v^2},   
\end{align}
\end{subequations}
la nueva variable no es monótona dado que $(\gamma-1)-v^2$ puede cambiar el signo, por lo que el sistema no es apropiado para hacer el análisis cualitativo/asintótico del sistema dado que no representa un sistema dinámico. Sin embargo, el sistema es apropiado para la integración numérica en una vecindad de la línea sónica $SL_{\pm}$, y para el análisis de estabilidad local de $SL_{\pm}$ a nivel perturbativo. 
\newline
$SL_{\pm}$ puede parametrizarse según 
\begin{equation}
    \gamma = v_0^2+1, \quad A= -\frac{\left(v_0^2+1\right) \left(\Sigma_0 +(\Sigma_0 +2)
   v_0^2\right)}{4 v_0^3}, \quad \Sigma=\Sigma_0. 
\end{equation}
Definiendo el parámetro de ecuación de estado $\omega=\gamma-1$, se deduce que $\omega=v_0^2$.

Definiendo las perturbaciones lineales
\begin{equation}
\delta_{v}=v-v_0, \quad \delta_{A}=A+\frac{\left(v_0^2+1\right) \left(\Sigma_{0}+(\Sigma_{0}+2)
   v_0^2\right)}{4 v_0^3}, \quad \delta_{\Sigma}= \Sigma-\Sigma_0,
\end{equation}

se deducen las ecuaciones para las perturbaciones lineales
\begin{subequations}
\begin{align}
  & \frac{d \delta_\Sigma }{d \xi}=  \delta_v \Bigg\{\frac{\Sigma_0 \left((C_2+2) v_0^4+2 (C_2-2)
   v_0^2+C_2-2\right)}{2 v_0}  +\frac{C_2 \Sigma_0^3 \left(-3 v_0^4+2
   v_0^2+1\right)^2}{8 v_0^5} \nonumber \\
   & +\frac{\Sigma_0^2 \left((5 C_2+1) v_0^6-(C_2+3)
   v_0^4+(3 C_2-5) v_0^2+C_2-1\right)}{2 v_0^3}\Bigg\},
   \end{align}
  \begin{align}
  & \frac{d \delta_A }{d \xi}= \delta_v
   \Bigg\{-\frac{\left(v_0^2+1\right) \left(C_2 \left(v_0^2+1\right)^2-4 v_0^2\right)}{4
   v_0^2} \nonumber \\
   & -\frac{\Sigma_0 \left(v_0^2+1\right) \left(11 C_2 v_0^6+(C_2-20)
   v_0^4+(9 C_2-4) v_0^2+3 C_2\right)}{8 v_0^4}  -\frac{C_2 \Sigma_0^3
   \left(3 v_0^6+v_0^4-3 v_0^2-1\right)^2}{32 v_0^8} \nonumber \\
   & -\frac{\Sigma_0^2
   \left(v_0^2+1\right) \left(19 C_2 v_0^8-4 (C_2+4) v_0^6+2 (C_2-8)
   v_0^4+12 C_2 v_0^2+3 C_2\right)}{16 v_0^6}\Bigg\},
   \end{align}
   \begin{align}
  & \frac{d \delta_v}{d \xi}= \delta_v
   \left(v_0^4+\frac{\Sigma_0 \left(v_0^2-1\right) \left(v_0^2+1\right)^2}{2
   v_0^2}-1\right) +\frac{v_0 \left(v_0^2-1\right) \left(4 \delta_A v_0^3+\delta_\Sigma  \left(v_0^2+1\right)^2\right)}{v_0^2+1}.
\end{align}
\end{subequations}
Los autovalores correspondientes al punto 
$(\delta_\Sigma, \delta_A, \delta_v)=(0,0,0)$ son $\lambda_1=0$ y las raíces $\lambda_2$ y $\lambda_3$ del polinomio característico 
\begin{align}
  & P(\lambda):=-\lambda ^2+\frac{\lambda  \left(v_0^4-1\right)
   \left(\Sigma_0+(\Sigma_0+2) v_0^2\right)}{2 v_0^2} \nonumber \\
   & -\left(v_0^2-1\right) v_0^2 \left(C_2 \left(v_0^2+1\right)^2-4
   v_0^2\right) \nonumber \\
   & +\frac{\Sigma_0^2 \left(v_0^2-1\right) \left(-C_2 \left(-3 v_0^4+2
   v_0^2+1\right)^2-12 v_0^2+2 \left(v_0^2+6\right) v_0^6-2\right)}{4
   v_0^2} \nonumber \\
   & -\Sigma_0 \left(v_0^2-1\right) \left((5 C_2-1) v_0^6-(C_2+9)
   v_0^4+(3 C_2+1) v_0^2+C_2+1\right).  
\end{align}
Esto es, 
\begin{doublespace}
\noindent\( {\left\{\lambda_{2}=\frac{\left(-1+ v_0^2\right) \left(1+ v_0^2\right)^2 \Sigma_0}{4  v_0^2}-\right.}\\
 {\frac{1}{4  v_0^{13} \left(1+ v_0^2\right)}\left(2  v_0^{13}+2  v_0^{15}-2  v_0^{17}-2  v_0^{19}+\surd \left(- v_0^{22}
\left(-1+ v_0^2\right) \left(1+ v_0^2\right)^2 \right.\right.}\\
 {\left(\Sigma_0^2+(-1+4 C_2)  v_0^{10} (2+\Sigma_0) (2+9 \Sigma_0)+ v_0^2 \Sigma_0 (4+(11+4
C_2) \Sigma_0)-2  v_0^6 \right.}\\
 {(-2+(-8+\Sigma_0) \Sigma_0+4 C_2 (-2+(-6+\Sigma_0) \Sigma_0))- v_0^8 (68+\Sigma_0 (152+51
\Sigma_0)+}\\
 {\left.\left.\left.\left.16 C_2 \left(-2+\Sigma_0+3 \Sigma_0^2\right)\right)+2  v_0^4 (2+\Sigma_0 (12+25
\Sigma_0+8 C_2 (1+\Sigma_0)))\right)\right)\right),}\\\\\\
 {\lambda_{3}=\frac{\left(-1+ v_0^2\right) \left(1+ v_0^2\right)^2 \Sigma_0}{4  v_0^2}+\frac{1}{4  v_0^{13}
\left(1+ v_0^2\right)}\left(-2  v_0^{13}-2  v_0^{15}+2  v_0^{17}+2  v_0^{19}+\right.}\\
 {\surd \left(- v_0^{22} \left(-1+ v_0^2\right) \left(1+ v_0^2\right)^2 \left(\Sigma_0^2+(-1+4 C_2)  v_0^{10}
(2+\Sigma_0) (2+9 \Sigma_0)+ v_0^2 \Sigma_0 \right.\right.}\\
 {(4+(11+4 C_2) \Sigma_0)-2  v_0^6 (-2+(-8+\Sigma_0) \Sigma_0+4 C_2 (-2+(-6+\Sigma_0) \text{$\Sigma
$0}))-}\\
 { v_0^8 \left(68+\Sigma_0 (152+51 \Sigma_0)+16 C_2 \left(-2+\Sigma_0+3 \Sigma_0^2\right)\right)+}\\
 {\left.\left.\left.\left.2  v_0^4 (2+\Sigma_0 (12+25 \Sigma_0+8 C_2 (1+\Sigma_0)))\right)\right)\right)\right\}}\).
\end{doublespace}
Aparece un autovalor cero por estar en presencia de una curva de puntos de equilibrio. 
Además,  la curva 
\begin{equation}
\delta_{\Sigma}(\Sigma_0)= \Sigma-\Sigma_0,\quad \delta_{A}(\Sigma_0)=A+\frac{\left(v_0^2+1\right) \left(\Sigma_{0}+(\Sigma_{0}+2)
   v_0^2\right)}{4 v_0^3}, \quad \delta_{v}(\Sigma_0)=v-v_0, 
\end{equation}
tiene vector tangente en el punto $(\delta_\Sigma, \delta_A, \delta_v)=(0,0,0)$:  
$$\frac{d}{d \Sigma_0}\left(\delta_{\Sigma}(\Sigma_0),
\delta_{A}(\Sigma_0),
\delta_{v}(\Sigma_0)\right)=\left(1,-\frac{\left(v_0^2+1\right)^2}{4 v_0^3},0\right),$$ y el autovector asociado al autovalor cero, que está dado por 
$$\mathbf{v}:(\delta_\Sigma, \delta_A, \delta_v)=\left(-\frac{4 v_0^3}{\left(v_0^2+1\right)^2},1,0\right),$$ es paralelo al vector tangente a la curva en el punto. Luego la curva es normalmente hiperbólica. De acuerdo a lo comentado a continuación de la definición \ref{normhiper}, sección \ref{seccion1.2}, se puede estudiar la estabilidad considerando solo los signos de los autovalores no nulos.

\begin{figure*}
    \centering
    \subfigure[$C_2=0.5$]{\includegraphics[scale=0.24]{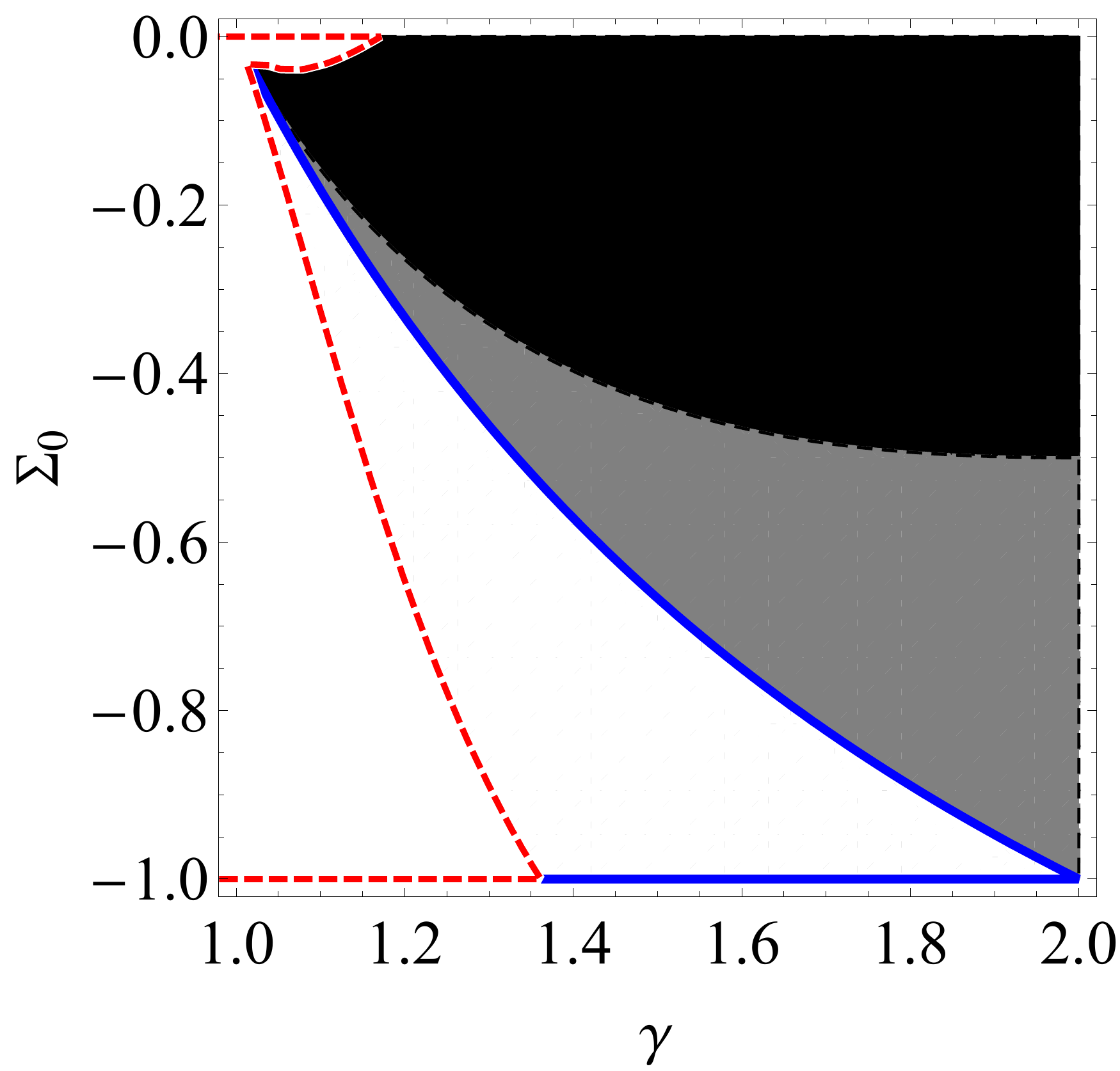}}\hspace{1cm}
	\subfigure[\label{SL-stabilitya} $C_2=1$]{\includegraphics[scale=0.24]{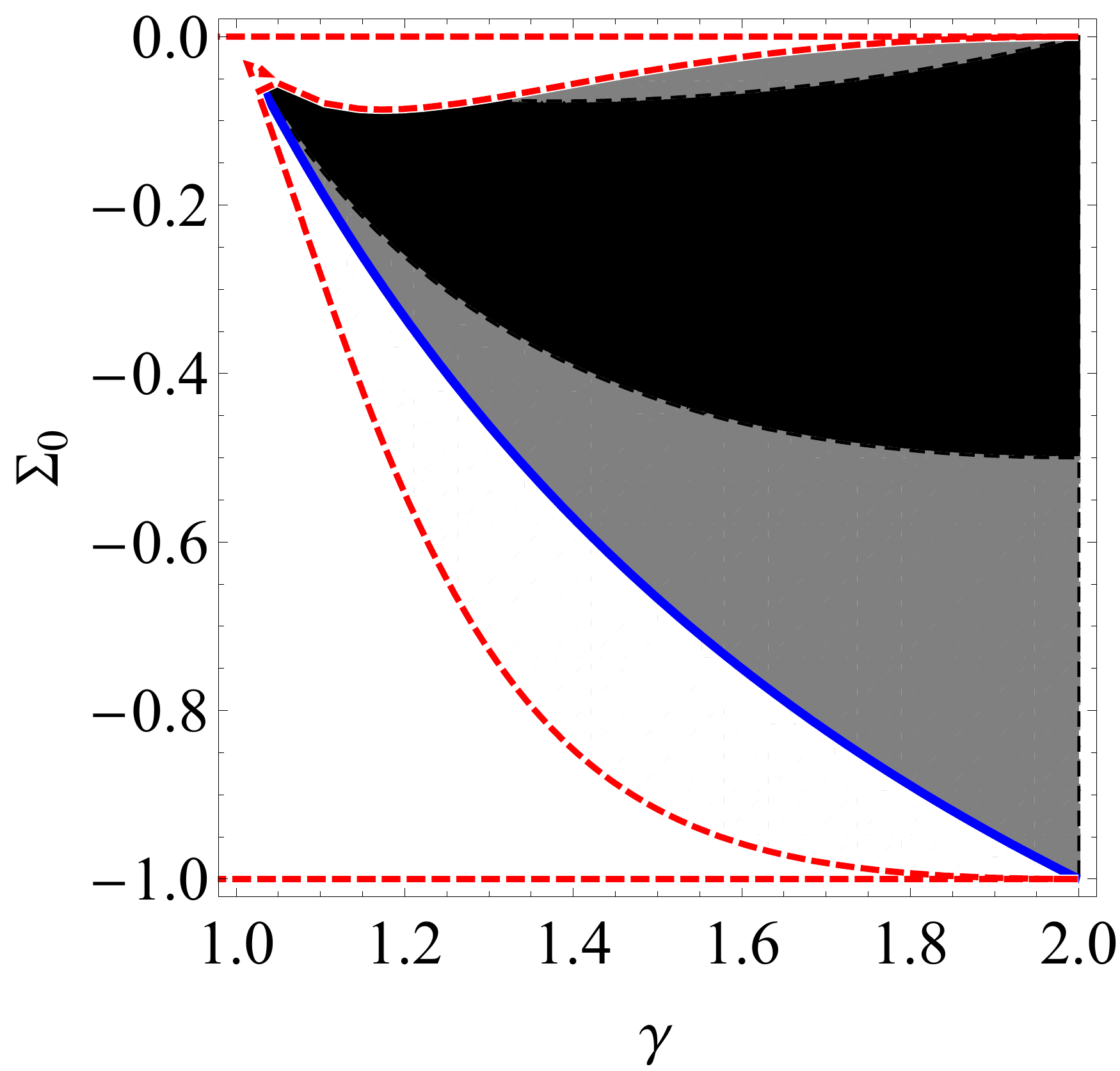}}\hspace{1cm}
	\subfigure[$C_2=2$]{\includegraphics[scale=0.24]{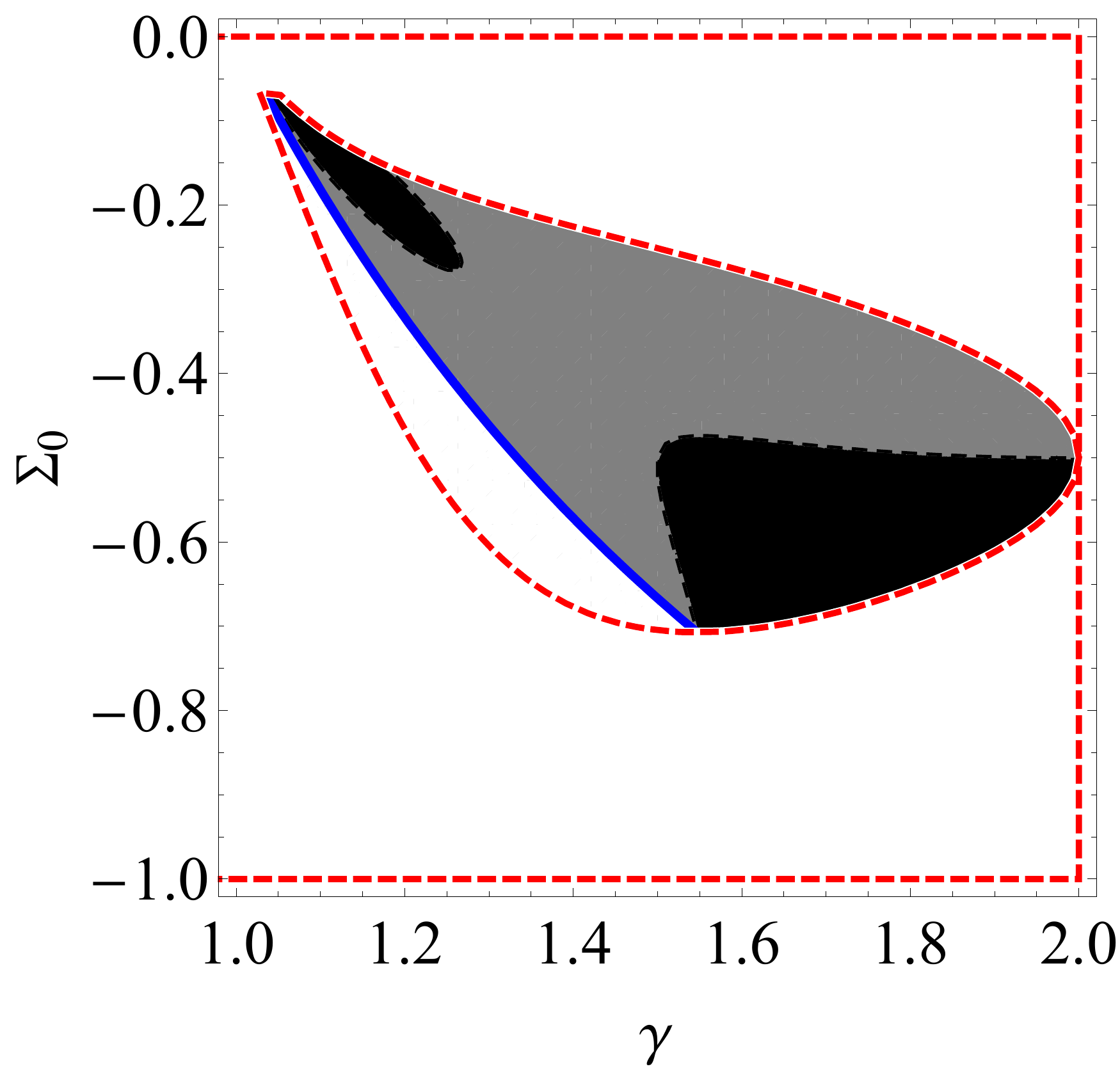}}\hspace{1cm}
    \caption[{\label{SL-stability} Regiones de estabilidad para $SL_{-}: v_0=-\sqrt{\gamma-1}, A=\frac{\gamma 
   (\gamma  (\Sigma
   +2)-2)}{4 (\gamma
   -1)^{3/2}}$, para $C_2=0.5, 1, 2$  y $\gamma\in[1, 2]$.}]{\label{SL-stability} Regiones de estabilidad para $SL_{-}: v=-\sqrt{\gamma-1}, A=\frac{\gamma 
   (\gamma  (\Sigma
   +2)-2)}{4 (\gamma
   -1)^{3/2}}$, para $C_2=0.5, 1, 2$ y $\gamma\in[1, 2]$, donde $v_0=-\sqrt{\gamma-1}, \Sigma_0$ denotan los valores de $v, \Sigma$ en un punto arbitario fijado en la curva sónica.}
   \end{figure*}
Para el análisis de estabilidad de la línea sónica $SL_{\pm}$, se identifican algunos conjuntos invariantes:
\begin{enumerate}
    \item $A_0=0$:  $\frac{\left(v_0^2+1\right) \left(\Sigma_0+(\Sigma_0+2) v_0^2\right)}{4
   v_0^3}=0$, 
    \item $K_0=0$: $1-C_2 \left(\Sigma_0^2+\frac{\left(v_0^2+1\right)^2 \left(\Sigma_0+(\Sigma_0+2) v_0^2\right)^2}{16 v_0^6}\right)=0$,
\end{enumerate}
ambos determinan curvas en el espacio $v_0,\Sigma_0$ que al ser invariantes no pueden ser atravesados por órbitas. 
En la discusión se impondrán la condiciones de existencia:
\begin{align*}
   &\frac{\left(v_0^2+1\right) \left(\Sigma_0+(\Sigma_0+2) v_0^2\right)}{4
   v_0^3}\geq 0, \quad  1-C_2 \left(\Sigma_0^2+\frac{\left(v_0^2+1\right)^2 \left(\Sigma_0+(\Sigma_0+2) v_0^2\right)^2}{16 v_0^6}\right)\geq 0.
\end{align*}
En la figura \ref{SL-stability} se representan regiones de estabilidad para $SL_{-}: v_0=-\sqrt{\gamma-1}, A=\frac{\gamma 
   (\gamma  (\Sigma_0   +2)-2)}{4 (\gamma   -1)^{3/2}}$, para (a) $C_2=0.5$, (b) $C_2=1$ y (c) $C_2=2$ y $\gamma\in[1, 2]$, donde $v_0=-\sqrt{\gamma-1}, \Sigma_0$ denotan los valores de $v, \Sigma$ en un punto fijado en la curva sónica. La region sin sombrear representa la región donde $A_0<0$ ó $K_0<0$, que es la región no física. La línea roja puenteada corresponde a $K=0$ y la línea azul gruesa corresponde a $A=0$. La región representada en color gris corresponde a la región donde $(\delta_\Sigma, \delta_A, \delta_v)=(0,0,0)$ es silla. La region representada en color negro corresponde a la región donde $(\delta_\Sigma, \delta_A, \delta_v)=(0,0,0)$ es  estable. 
   La figura \ref{SL-stabilitya} reproduce los resultados de estabilidad que se muestran en la figura 1 de  \cite{Goliath:1998mx} (con la excepción de una franja en el espacio de parámetros, representada en color gris, donde el punto de equilibrio es silla $(\delta_\Sigma, \delta_A, \delta_v)=(0,0,0)$, ver esquina superior derecha en la figura \ref{SL-stabilitya}, cuyo análisis fue omitido en \cite{Goliath:1998mx}). Para $SL_{+}$ las condiciones de existencia  $\frac{\left(v_0^2+1\right) \left(\Sigma_0+(\Sigma_0+2) v_0^2\right)}{4
   v_0^3}\geq 0,$
no se verifica, por lo que se concluye que están fuera de la región física y se omite su análisis.

\subsection{Conjuntos invariantes $v=\pm 1$}
\label{tiltfluidoperfecto}
En esta sección, se estudia $v^2=1$, i.e., $v=\pm 1$, que corresponde al caso de inclinación extrema. Entonces, de las ecuaciones. \eqref{constraintmod1}, \eqref{defnOmegat}, y \eqref{defnr} es posible deducir 
\begin{subequations}
\begin{align}
& {\Omega}_t = \pm 2  A C_2 {\Sigma},\\
& K=1- C_2 A^2  -C_2 \Sigma ^2 \mp  2  A C_2 {\Sigma},\\
& \sqrt{3} r=  1+\Sigma+C_2 {\Sigma}^2- C_2 A^2.   
\end{align}
\end{subequations}

Para luego obtener el sistema reducido 2-dimensional, 
\begin{subequations}
\label{extreme_tilt_0}
\begin{align}
&{\Sigma}'=-{{\Sigma}} \left(1 \pm 2  A-C_2( {\Sigma}^2- A^2)\right), \\
 & A'=A \left( 1+ 2{\Sigma}+C_2( {\Sigma}^2- A^2)\right),
	\end{align}
\end{subequations}	
donde $\pm 1$ denota el signo de $v$. 

\begin{figure*}
\centering
	\subfigure[]{\includegraphics[scale=0.5]{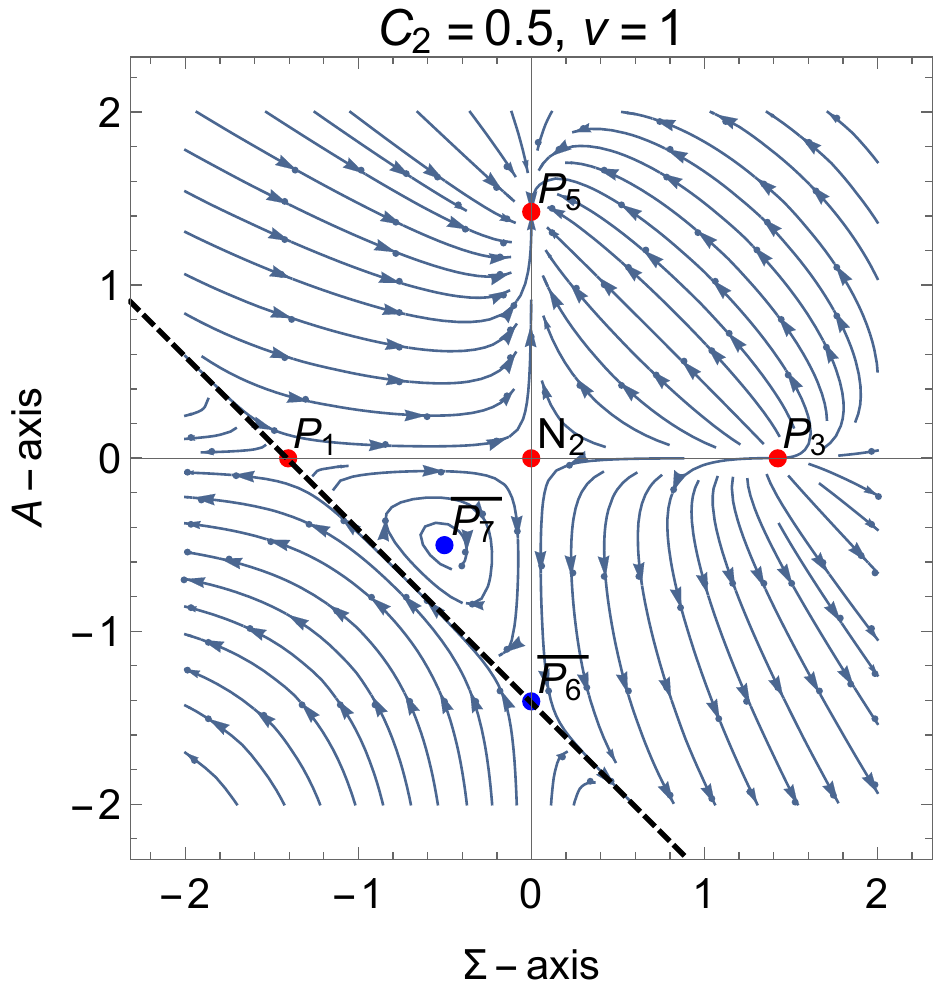}}
	\subfigure[]{\includegraphics[scale=0.5]{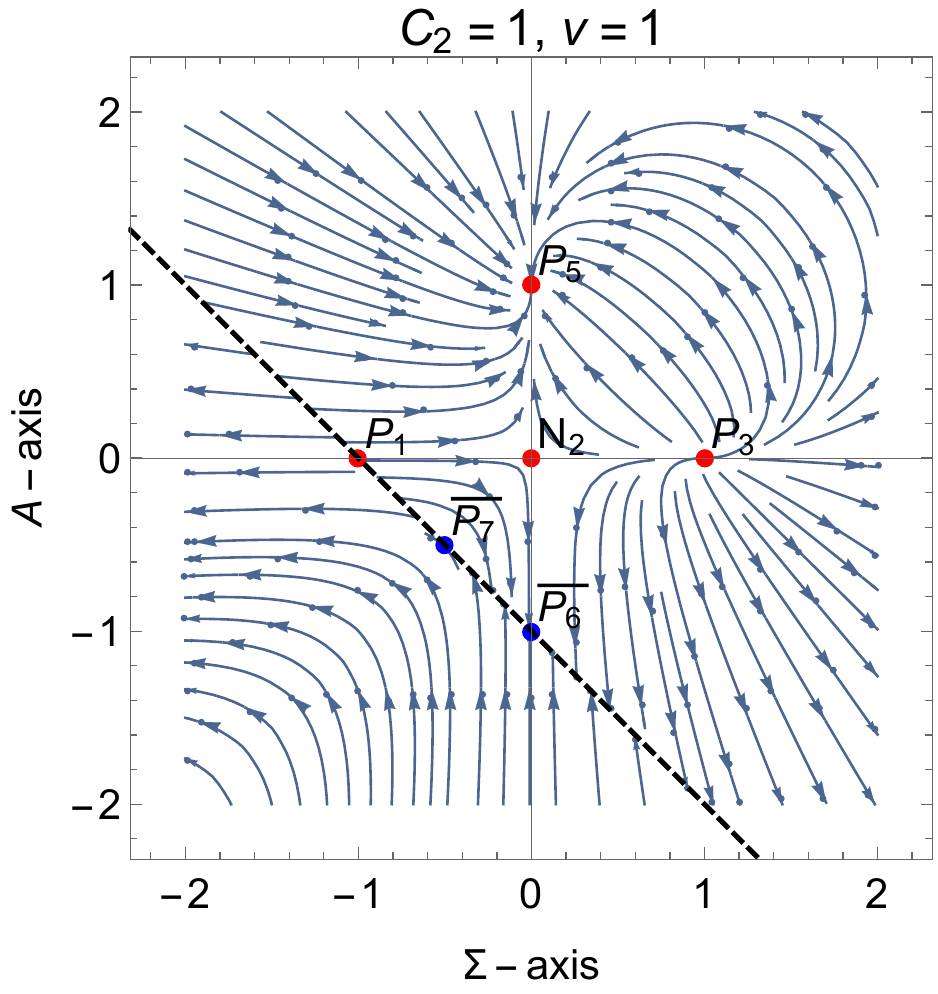}}
	\subfigure[]{\includegraphics[scale=0.5]{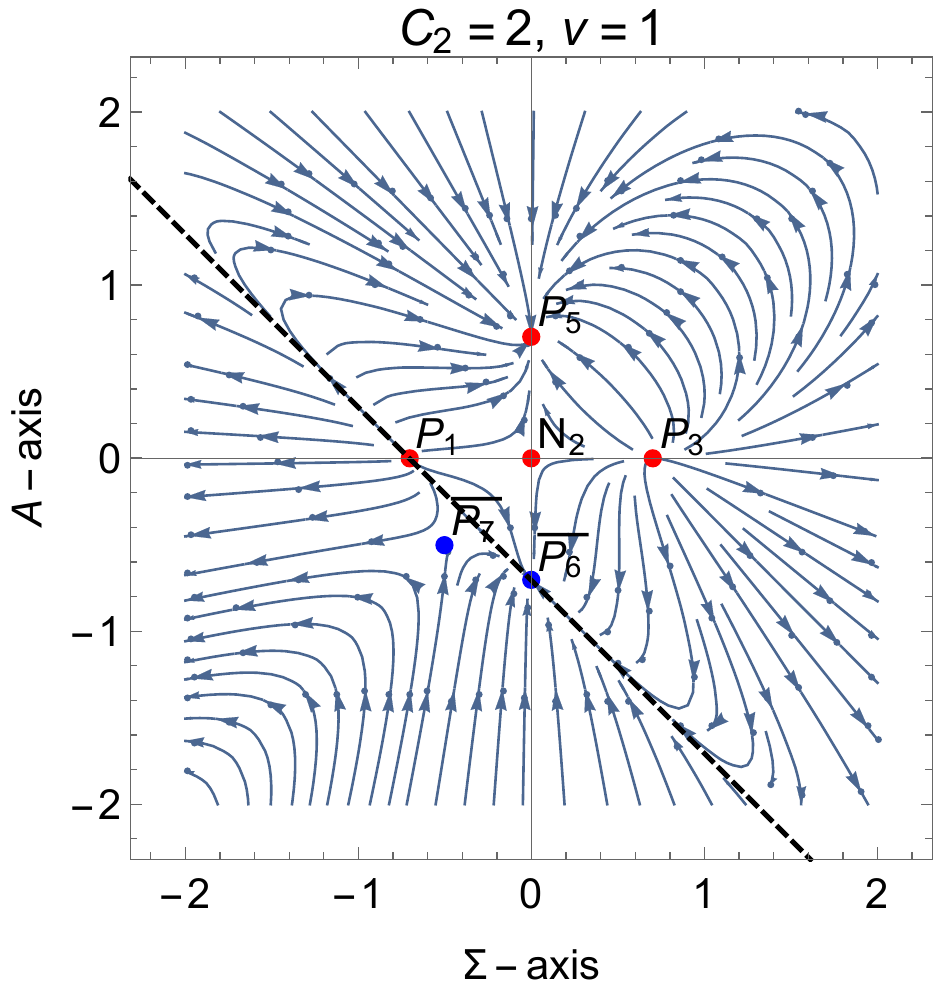}}
	\subfigure[]{\includegraphics[scale=0.5]{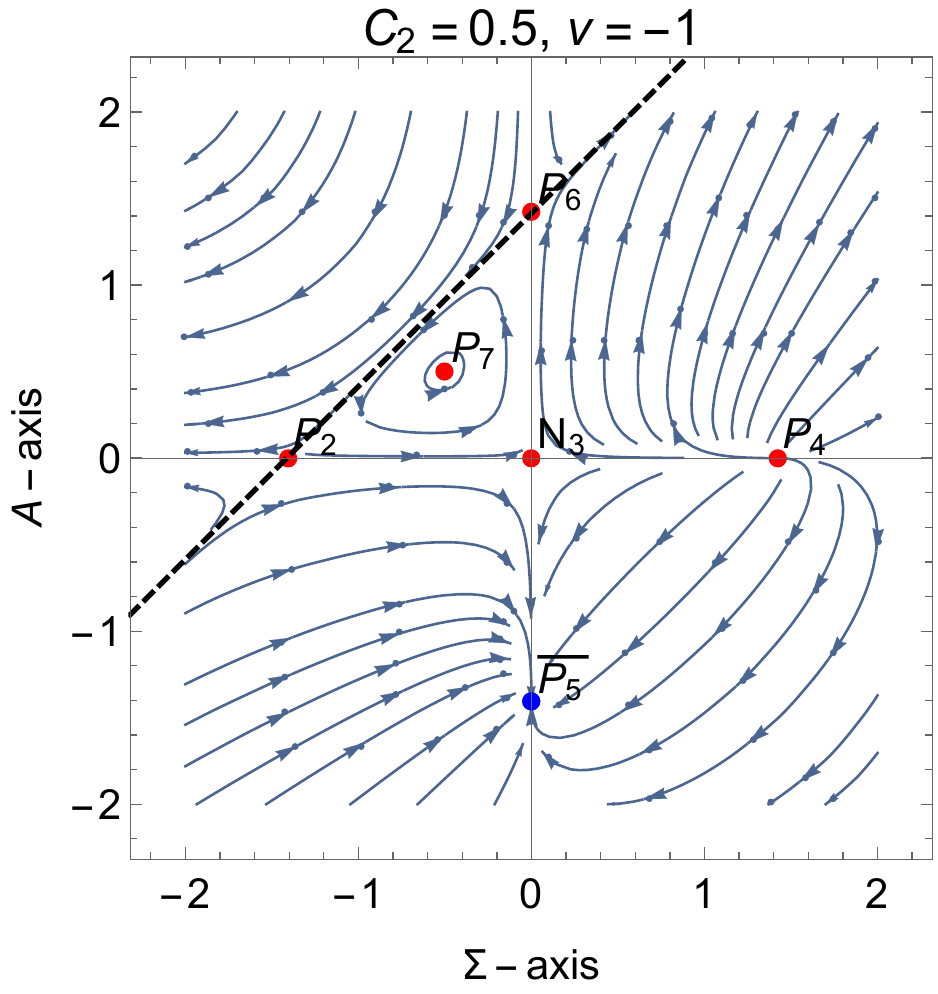}}
	\subfigure[]{\includegraphics[scale=0.5]{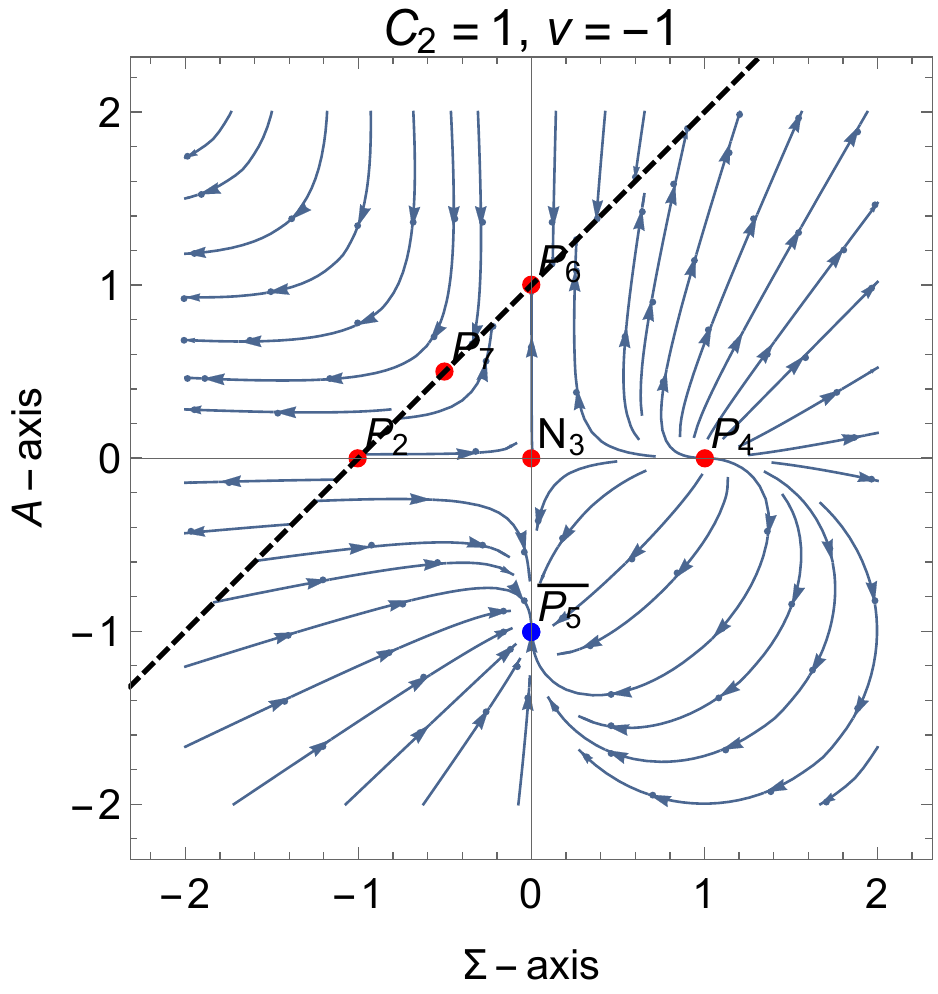}}
	\subfigure[]{\includegraphics[scale=0.5]{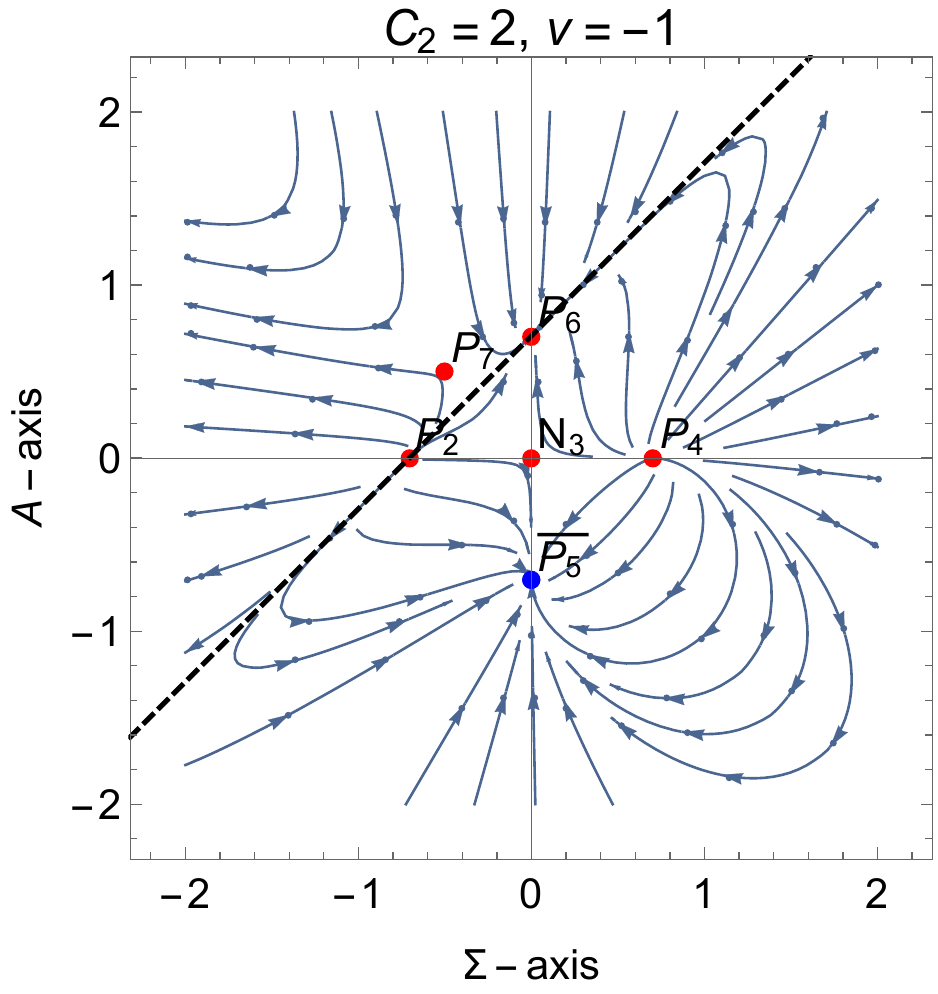}}
	\caption[Algunas órbitas  de los sistemas  \eqref{extreme_tilt_0} con $v=1,-1$, para diferentes elecciones de parámetros.]{\label{FIG1a-d} Algunas órbitas  de los sistemas  \eqref{extreme_tilt_0} con $v=1,-1$, para diferentes elecciones de parámetros. Se incluyen los puntos $\overline{P_5},\overline{P_6},\overline{P_7}$. Donde $\overline{P_i}$ denota los puntos simétricos de los ${P_i}$. La línea punteada representa $H_{-}$ (resp. $\overline{H_{-}}=H_{+}$) para $C_2=1$.}
\end{figure*}

Los sistemas \eqref{extreme_tilt_0} se relacionan a través del cambio simultáneo de  $A \rightarrow -A$, y la rama ``$+$'' por la rama ``$-$''. Por tanto, sin pérdida de generalidad, se estudia la rama positiva ``$+$'', con $A\geq 0$. 
\begin{table*}[b!]
\caption{\label{Tab1} Análisis cualitativo de los puntos de equilibrio de los sistemas de los sistemas  \eqref{extreme_tilt_0} correspondiente a los casos de inclinación extrema $v=1,-1$.}
\begin{tabular}{|m{0.7cm}|m{2.6cm}|m{2.8cm}|m{5.5cm}|m{1.3cm}| }
\hline
Etiq. & $(\Sigma,A),(v)$& Autovalores & Estabilidad &  $(K,\Omega_t)$\\
 \hline
  $N_{2,3}$ & $(0,0),(\pm 1)$& $\left\{-1,1\right\}$  & Silla hiperbólica & $(1,0)$\\\hline
	$P_{1,2}$ & $(-\frac{1}{\sqrt{C_2}},0), (\pm 1)$  & $\left\{2-\frac{2}{\sqrt{C_2}},2\right\}$ & Silla hiperbólica para $0<C_2<1$ \newline No hiperbólico para $C_2=1$ \newline Fuente hiperbólica para $C_2>1$& (0,0)\\\hline
 $P_{3,4}$ & $(\frac{1}{\sqrt{C_2}},0),( \pm1)$ & $\left\{\frac{2}{\sqrt{C_2}}+2,2\right\}$ & Fuente hiperbólica para $C_2>0$  & $(0,0)$\\\hline
  $P_5$ & $\left(0, \frac{1}{\sqrt{C_2}}\right), (1) $& $\left\{-\frac{2}{\sqrt{C_2}}-2,-2\right\}$  & Pozo hiperbólico para $C_2>0$ & $(0,0)$\\\hline
	 $P_6$ & $\left(0, \frac{1}{\sqrt{C_2}}\right),(- 1) $& $\left\{\frac{2}{\sqrt{C_2}}-2,-2\right\}$  & Silla hiperbólica para $0<C_2<1$ \newline No hiperbólico para $C_2=1$ \newline Pozo hiperbólico para $C_2>1$ & $(0,0)$\\\hline
	$P_{7}$ & $\left(-\frac{1}{2}, \frac{1}{2}\right),(- 1)$ & $\left\{\pm\sqrt{C_2-1}\right\}$ & Silla hiperbólica para $C_2>1$ \newline No hiperbólico $C_2\leq 1$  &$\scriptstyle{(1-C_2, \frac{C_2}{2})}$\\
 \hline
\end{tabular}
\end{table*}
En la tabla \ref{Tab1} se presenta el análisis cualitativo de los puntos de equilibrio de los sistemas  \eqref{extreme_tilt_0} correspondientes a los casos de inclinación extrema $v=1,-1$, que son los siguientes:
\begin{enumerate}
    \item $N_{2,3}:(\Sigma,A)=(0,0)$ (para $v=\pm 1$) con autovalores $\{-1,1\}$ es una silla hiperbólica.
    \item $P_{1,2}:(\Sigma,A)=(-\frac{1}{\sqrt{C_2}},0)$ (para $v=\pm 1$) con autovalores $\left\{2-\frac{2}{\sqrt{C_2}},2\right\}$ son 
    \begin{enumerate}
        \item Sillas hiperbólicas para $0<C_2<1$. 
        \item No hiperbólicos para $C_2=1$.
        \item Fuentes hiperbólicas  $C_2>1$.
    \end{enumerate}
    \item $P_{3,4}:(\Sigma,A)=(\frac{1}{\sqrt{C_2}},0)$ (para $v=\pm 1$) con autovalores $\left\{2+\frac{2}{\sqrt{C_2}},2\right\}$ son fuentes hiperbólicas para $C_2>0$.
    \item $P_5:(\Sigma,A)=\left(0, \frac{1}{\sqrt{C_2}}\right)$ (para $v=1$) con autovalores $\left\{-\frac{2}{\sqrt{C_2}}-2,-2\right\}$ es pozo hiperbólico para $C_2>0$.
    
    \item $P_6:(\Sigma,A)=\left(0, \frac{1}{\sqrt{C_2}}\right)$ (para $v=-1$) con autovalores $\left\{\frac{2}{\sqrt{C_2}}-2,-2\right\}$ es
    \begin{enumerate}
        \item Silla hiperbólica para $0<C_2<1$. 
        \item No hiperbólico para $C_2=1$.
        \item Pozo hiperbólica  $C_2>1$.
    \end{enumerate}
    \item $P_7:(\Sigma,A)=\left(-\frac{1}{2}, \frac{1}{2}\right)$ (para $v=-1$) con autovalores $\left\{-\sqrt{C_2-1},\sqrt{C_2-1}\right\}$ es 
    \begin{enumerate}
        \item Silla hiperbólica para $C_2>1$.
        \item No hiperbólico para $C_2\leq 1$.
    \end{enumerate}
\end{enumerate}
En la figura \ref{FIG1a-d}  se presentan algunas órbitas  de los sistemas  \eqref{extreme_tilt_0} con $v=1,-1$, para diferentes elecciones de parámetros. Se incluyen los puntos $\overline{P_5},\overline{P_6},\overline{P_7}$, donde $\overline{P_i}$ denota los puntos simétricos de los ${P_i}$.  La línea invariante punteada representa $H_{-}$ (resp. $H_{+}$) para $C_2=1$. Se confirman los resultados analíticos discutidos anteriormente. 
\FloatBarrier

\subsection{Cálculo de variedades invariantes para $P_6$}
Para el punto $P_6$ con $v=-1$ se obtiene la línea invariante  $A-\Sigma-\frac{1}{\sqrt{C_2}}$ que corresponde a la variedad inestable de $P_6$ cuando $C_2<1$, es una línea estable para $C_2=1$, y para $C_2>1$ pertenece a la variedad estable 2D de $P_6$. Introduciendo el cambio de variables 
\begin{subequations}
\begin{align}
& x= \Sigma, \\
& y= A-\Sigma-\frac{1}{\sqrt{C_2}},   
\end{align}
se obtienen las ecuaciones
\end{subequations}
\begin{subequations}
\label{variedad_extreme_tilt}
\begin{align}
&{x}'=x \left(-1+x^2 (1+\mu )^2-(1+x+y+(x+y) \mu )^2+2 \left(x+y+\frac{1}{1+\mu }\right)\right),\\
 & y'=-y (2+y+y \mu ) (1+y+y \mu +2 x(1+\mu )),
	\end{align}
\end{subequations}
donde se ha introducido el parámetro
\begin{equation}
    \mu= \sqrt{C_2}-1, \quad C_2=(\mu +1)^2. 
\end{equation}
Los autovalores de  $P_6$ son $\left\{-\frac{2 \mu }{\mu
   +1},-2\right\}$. Esto es, para $\mu>0$,  i.e., para $C_2>1$, el punto $P_6$ es un pozo en el conjunto invariante  $v=-1$. Para $\mu<0$,  i.e., para $C_2<1$, el punto $P_6$ tiene una variedad inestable invariante tangente al eje $x$. 
   La variedad local inestable de $P_6$ se puede expresar como el gráfico
\begin{equation}
    \left\{(x, y): y=h(x) , |x|<\delta\right\},
\end{equation}
donde $h$ satisface el problema con condición inicial
\begin{align}
& h (2+(1+\mu ) h) (1+2 x (1+\mu )+(1+\mu ) h)\nonumber \\
& +x \left(-1+x^2 (1+\mu )^2+2 \left(x+\frac{1}{1+\mu }+h\right)-(1+x+h+\mu
 (x+h))^2\right) h'=0, \\ & h(0)=0, \quad h'(0)=0.  
\end{align}
La ecuación diferencial anterior admite primera integral 
\begin{align}
    \frac{-x (1+\mu ) h^{p}(1+x+x \mu +(1+\mu ) h)}{ (2+(1+\mu ) h)^{q}} =C_1, \quad p=-\frac{\mu }{1+\mu }>0, \nonumber\\ q=\frac{2+\mu }{1+\mu }<0, \quad -1< \mu<0. 
\end{align} 
donde $C_1$ es una constante de integración. 
	\begin{figure*}
\centering
	\subfigure[]{\includegraphics[scale=0.5]{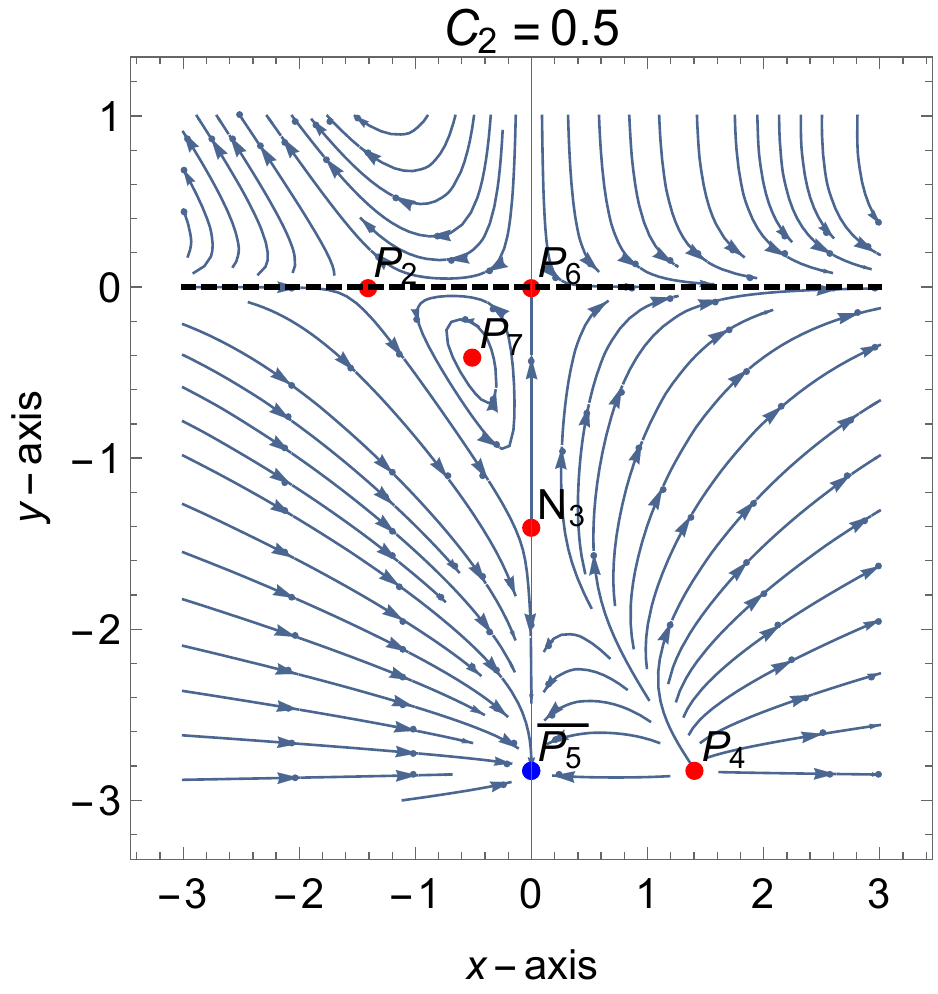}}
	\subfigure[]{\includegraphics[scale=0.5]{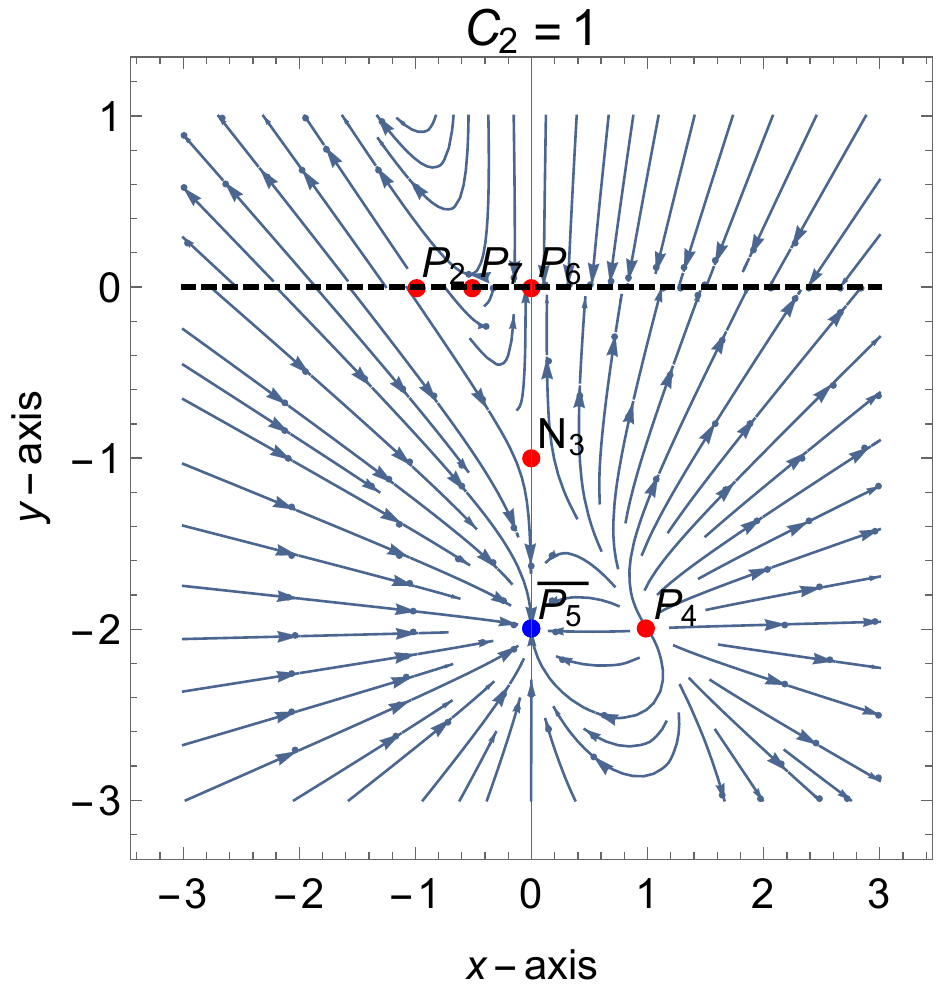}}
	\subfigure[]{\includegraphics[scale=0.5]{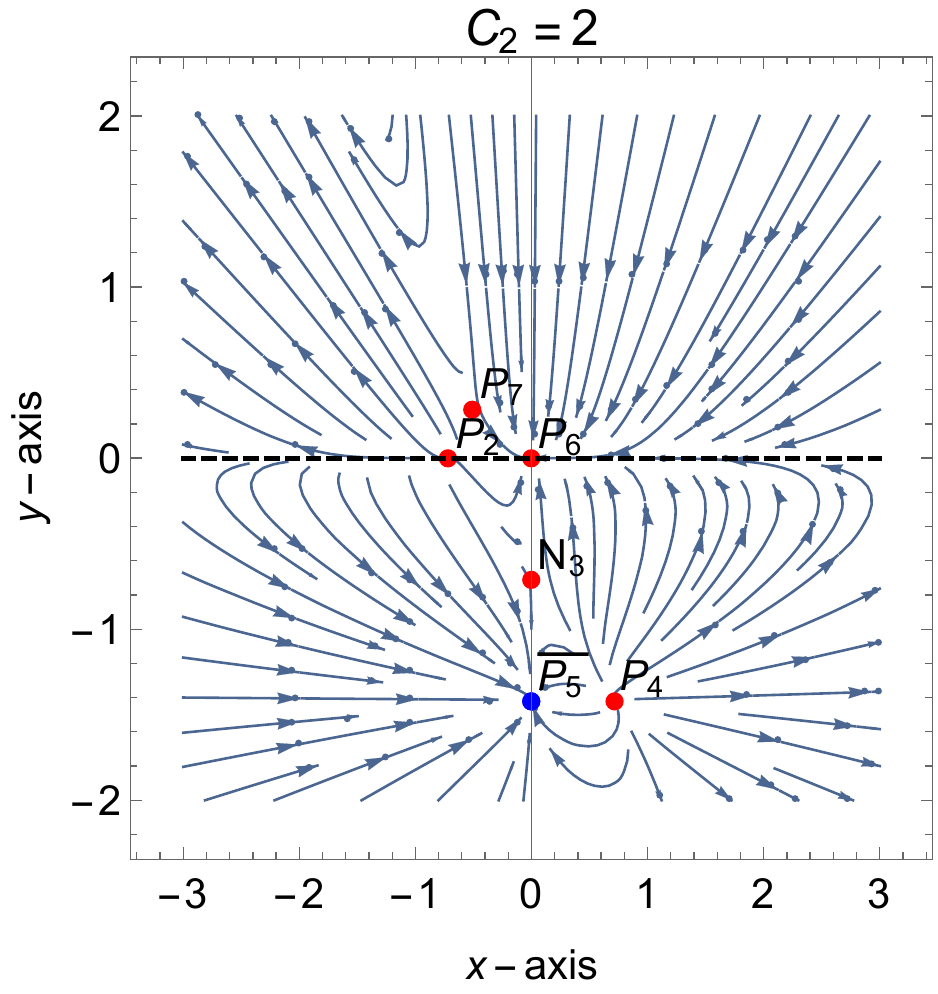}}
	\caption{\label{FIG2a-c} Algunas órbitas  del sistema   \eqref{variedad_extreme_tilt} para distintas elecciones de parámetros.}
\end{figure*}
\begin{figure}
    \includegraphics[scale=0.32]{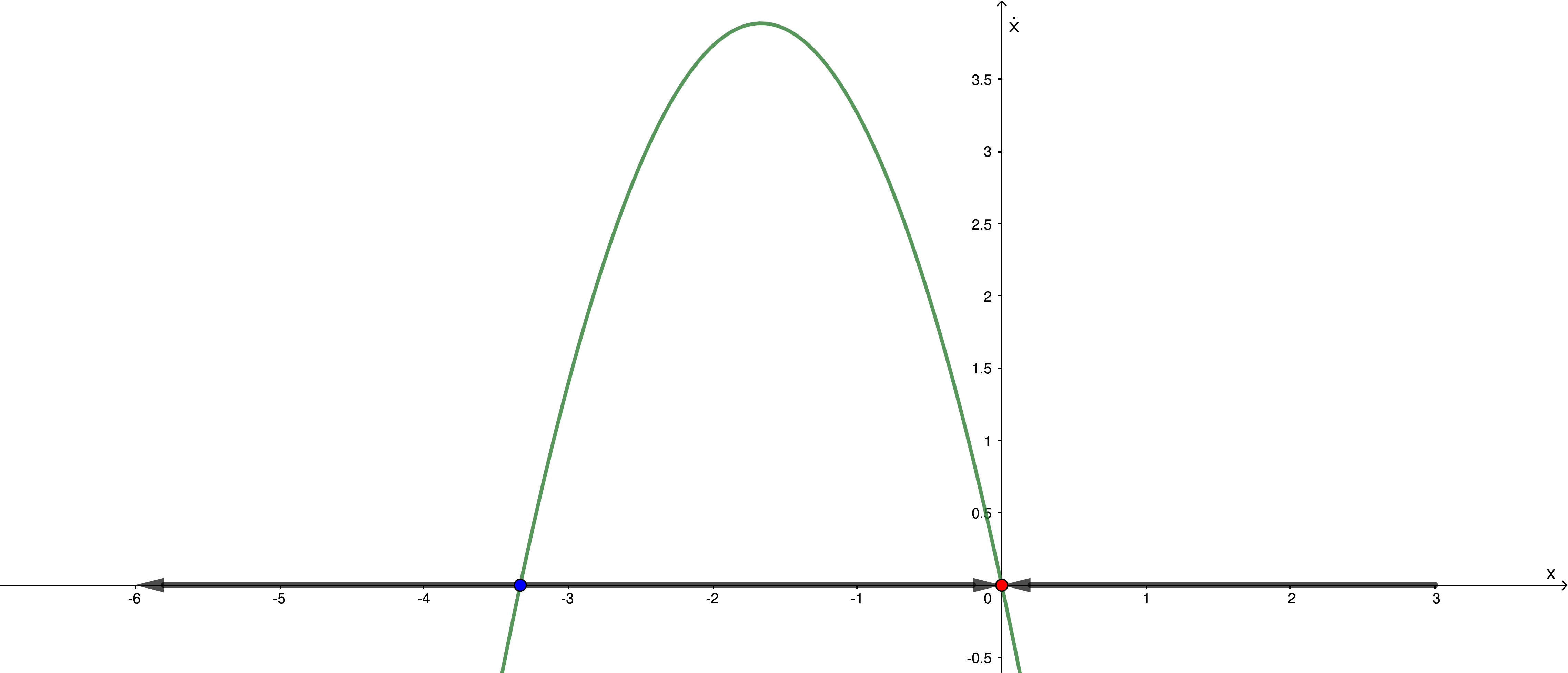}
    \caption{Flujo en una dimensión para el sistema \eqref{L3.49}, para $\mu =-0.7$. El análisis de estabilidad no cambia para los valores de $\mu$ en $-1 < \mu < 0$.}
    \label{fig:x-mu}
\end{figure}
Imponiendo la condición  $h(0)=0$, se obtiene $C_1=0$. Resolviendo para $h$ la ecuación resultante, se obtienen dos soluciones  $$h(x)=0,$$ y 
$$h(x)= -x-\frac{1}{1+\mu }.$$\
La última solución no es aceptable ya que 
$h'(x)=-1 \neq 0$, por tanto la condición de tangencialidad  $h'(0)=0$ no se cumple. Esto implica que la variedad inestable esta dada por la solución trivial:
\begin{equation}
    \left\{(x, y): y=0 , |x|<\delta\right\}.
\end{equation}
La dinámica en la variedad inestable está dada por 
\begin{equation}
\label{L3.49}
 {x}'=-\frac{2 \mu  x (\mu  x+x+1)}{\mu +1},   
\end{equation}
cuya solución es
\begin{equation}
    x(\eta )= \frac{x_0}{e^{\frac{2 \eta  \mu }{\mu +1}} (\mu  x_0+x_0+1)-(\mu +1) x_0}, 
\end{equation}
esta pasa por  $x(0)=x_0$, con $-1 < \mu < 0$. Entonces, $\lim_{\eta\rightarrow -\infty} x(\eta)=0$. Esto dice que las soluciones son asintóticas al origen cuando $\eta\rightarrow -\infty$, por tanto tienden a $P_6$.

En la figura \ref{fig:x-mu} se presenta el flujo 1-dimensional de \eqref{L3.49} para $-1< \mu<0$, donde se ilustra que el origen del sistema \eqref{L3.49} es estable. Luego, aplicando el Teorema de la Variedad Inestable se confirma que  $P_6$ es una silla, como fue comentado con anterioridad en la tabla \ref{Tab1}.  
En la figura \ref{FIG2a-c} se representan algunas órbitas  del sistema   \eqref{variedad_extreme_tilt} para distintas elecciones de parámetros.
\FloatBarrier

\subsection{Gas ideal $\gamma=1$}
\label{fluidosinpresion}
Para gas ideal con $\gamma=1$  (fluido sin presión; polvo) las ecuaciones \eqref{YYYYY} y restricciones \eqref{YYYYY-rest} se transforman en
\begin{subequations}
\label{dust-1}
\begin{align}
&{\Sigma}'=\frac{A^2 C_2 \left(v^2 (1-2 C_2 \Sigma )+1\right) +\left(C_2 \Sigma ^2-1\right) \left(2 C_2 \Sigma  v^2+v^2+1\right)+K
   \left(v^2+1\right)}{2 C_2 v^2},\\
& A'=-C_2 A^3 +A \Sigma  (C_2 \Sigma +2)+A,\\
& K'=	2 C_2 K \left(\Sigma ^2-A^2\right),\\
& v'=\frac{\left(v^2-1\right) (A v-\Sigma )}{v},
	\end{align}
\end{subequations}
donde hemos usado las expressiones de $\Omega_{t}$ y $r$ dadas por
\begin{subequations}
\begin{align}
\Omega_{t}=\frac{1-C_2 A^2 -C_2 \Sigma ^2-K}{v^2}, \quad r=\frac{1-C_2 A^2 +C_2 \Sigma ^2+\Sigma}{\sqrt{3}}.
	\end{align}
\end{subequations}
La restricción \eqref{constraintmod1} se reduce a  
\begin{equation}
-\frac{C_2 \left(A^2+2 A \Sigma  v+\Sigma ^2\right)+K-1}{v}=0,
\end{equation}
que puede ser resuelta para $K$ obteniéndose  
\begin{equation}
K=-C_2 A^2 -2 A C_2 \Sigma  v-C_2 \Sigma ^2+1.
\end{equation}
Esto nos permite estudiar el sistema 3-dimensional reducido para las variables $\Sigma, A, v$: 
\begin{subequations}
\label{3.41-3.43}
\begin{align}
&{\Sigma}'=-\frac{\Sigma  \left(A \left(A C_2 v+v^2+1\right)+v\right)}{v},\\
& A'=-C_2 A^3 +A \Sigma  (C_2 \Sigma +2)+A,\\
& v'=\frac{\left(v^2-1\right) (A v-\Sigma )}{v},
	\end{align}
\end{subequations}
definido en el espacio de estados
\begin{align}
    & \Bigg\{(\Sigma, A, v)\in\mathbb{R}^3:  C_2(A^2 + \Sigma ^2)\leq 1, \nonumber \\
    & \quad C_2 \left(A^2 +2 A \Sigma  v+ \Sigma ^2\right) \leq 1,  \quad v \in [-1, 0)\cup (0, 1]\Bigg\}.
\end{align}

Los puntos de equilibrio del sistema \eqref{3.41-3.43} son los siguientes:
\begin{enumerate}
    \item  $N:(\Sigma,A,v)=(0,0,v)$  con autovalores  $\{0,-1,1\}$ es silla no hiperbólica.
    \item $N_{2,3}:(\Sigma,A,v)=(0,0,\pm 1) $ con autovalores $\{-1,1,0\}$ son sillas no hiperbólicas.
    \item $P_{1,2}:(\Sigma,A,v)=\left(-\frac{1}{\sqrt{C_2}},0,\pm 1\right)$ con autovalores  $\left\{\frac{2}{\sqrt{C_2}},2-\frac{2}{\sqrt{C_2}},2\right\}$ son \begin{enumerate}
        \item Fuentes hiperbólicas para $C_2>1$.
        \item Sillas hiperbólicas para $C_2<1$. 
        \item No hiperbólicos para  $C_2=1$.
    \end{enumerate} 
    \item $P_{3,4}:(\Sigma,A,v)=\left(\frac{1}{\sqrt{C_2}},0,\pm 1\right)$ con autovalores $\left\{-\frac{2}{\sqrt{C_2}},2+\frac{2}{\sqrt{C_2}},2\right\}$ son sillas hiperbólicas para $C_2>0$.
    \item $P_5:(\Sigma,A,v)=\left(0, \frac{1}{\sqrt{C_2}}, 1\right)$ con autovalores  $\left\{\frac{2}{\sqrt{C_2}},-\frac{2}{\sqrt{C_2}}-2,-2\right\}$  es silla hiperbólica para $C_2>0$. 
    \item $P_6:(\Sigma,A,v)=\left(0, \frac{1}{\sqrt{C_2}},- 1\right)$ con autovalores  $\left\{-\frac{2}{\sqrt{C_2}},\frac{2}{\sqrt{C_2}}-2,-2\right\}$ es
    \begin{enumerate}
        \item Pozo hiperbólico para $C_2>1$.
        \item Silla hiperbólica para $C_2<1$.
        \item No hiperbólico para $C_2=1$.
    \end{enumerate}
    \item $P_{7}:(\Sigma,A,v)=\left(-\frac{1}{2},\frac{1}{2},- 1\right)$ con autovalores $\left\{0,\pm \sqrt{C_2-1}\right\}$ es 	
    \begin{enumerate}
	\item silla no hiperbólica para $C_2>1$.
	\item no hiperbólico con tres autovalores iguales a cero para $C_2=1$.
	\item no hiperbólico con un autovalor igual a cero y dos autovalores imaginarios puros para $0\leq
   C_2<1$.
	\end{enumerate}
    \item $P_8:(\Sigma,A,v)={\left(-\frac{1}{2 C_2}, \frac{\sqrt{4 C_2-3}}{2 C_2}, - \frac{1}{\sqrt{4 C_2-3}}\right)}$ con autovalores  $\left\{k \lambda_1,k \lambda_2, k \lambda_3\right\}$, con  $k=\frac{1}{2 C_2^2 (4 C_2-3)^{3/2}}$ and $\lambda_1, \lambda_2$ y $\lambda_3$ son las raíces del  polinomio en $\lambda$:  $P(\lambda)=-16 (C_2-1)^2 C_2^3 (4   C_2-3)^{11/2}-4 (C_2-1)  C_2^2 (8 C_2-7) (4
   C_2-3)^3 \lambda +\lambda ^3$. 
   \begin{figure*}[ht!]
    \centering
    \includegraphics[scale=0.4]{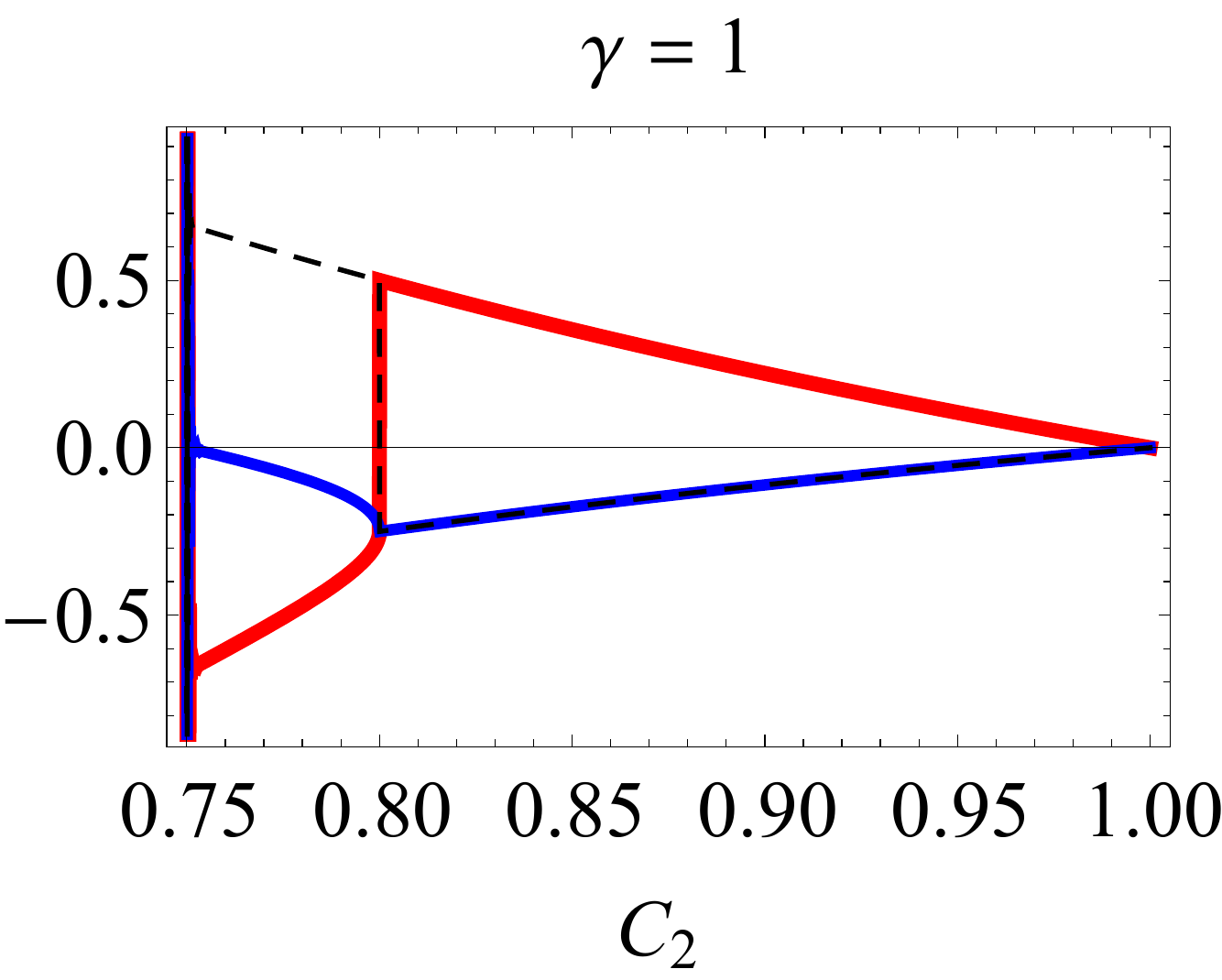} \hspace{2cm}
    \includegraphics[scale=0.4]{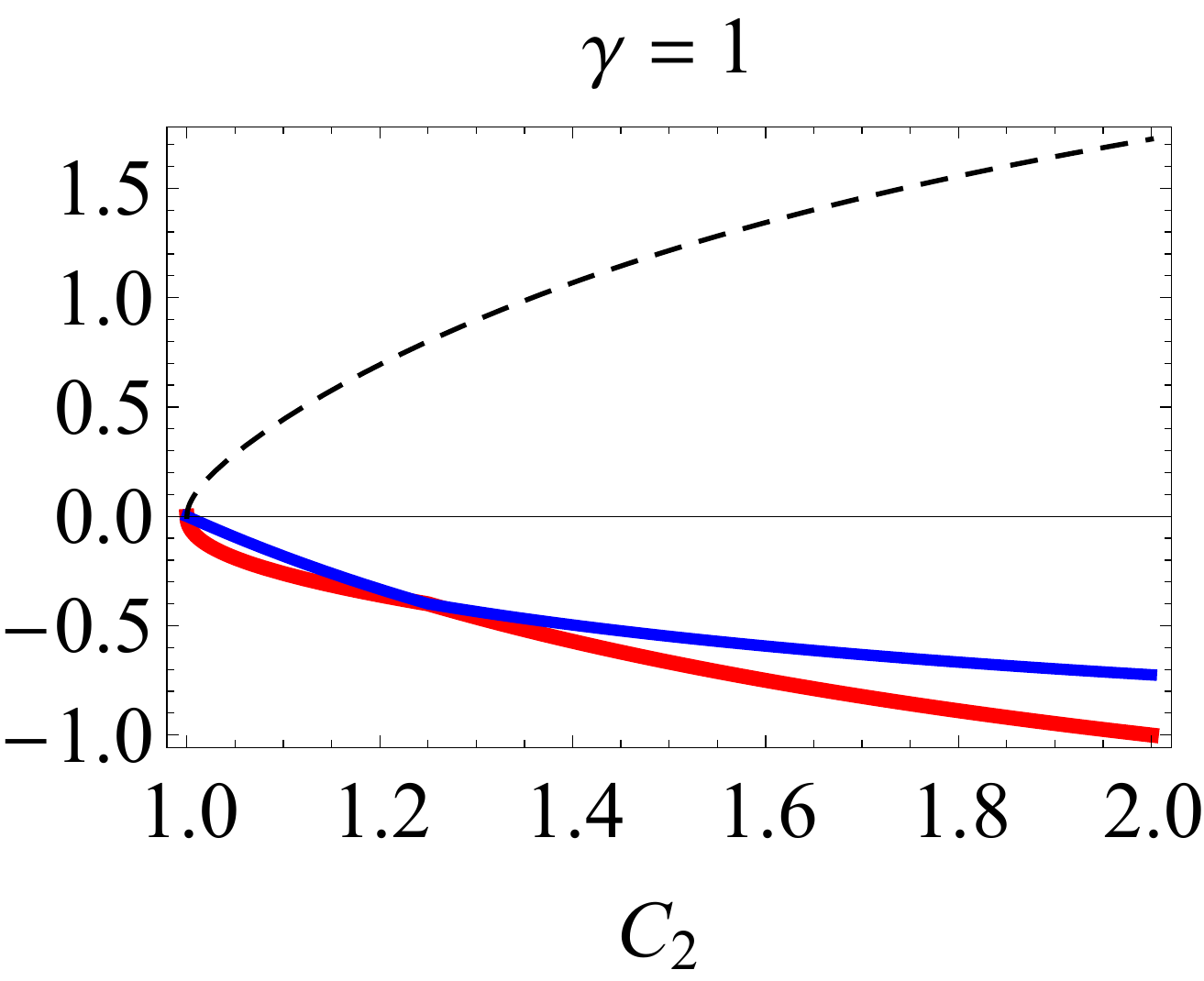}
    \caption[{Parte real de los autovalores  del punto de equilibrio $P_8$ para $C_2\geq \frac{3}{4}$ y $\gamma=1$.}]{\label{eigenP8gamma1} Parte real de $\lambda_i$ correspondientes al punto de equilibrio $P_8$ para $C_2\geq \frac{3}{4}$ y $\gamma=1$, mostrando que en general tiene comportamiento de silla.}
   \end{figure*}
   En la figura \ref{eigenP8gamma1} se
   representa gráficamente la parte real $\lambda_i$ correspondientes al punto de equilibrio $P_8$ para $C_2\geq \frac{3}{4}$ y $\gamma=1$, mostrando que en general tiene comportamiento de silla.
   Como antes, se ha prescindido de un factor multiplicativo $k$ en los autovalores para determinar su estabilidad, lo cual es aplicable dado que el punto de equilibrio es silla o es no hiperbólico y  se requiere sólo mostrar que al menos dos autovalores tienen partes reales con signo distinto o tienen partes reales cero, para verificar que el punto de equilibrio es silla o es no hiperbólico, por lo que el signo de un factor multiplicativo es irrelevante. 
   \item $H_{\pm}:(\Sigma,A,v)=\left(\Sigma_0,-\varepsilon(1+\Sigma_0), \varepsilon\right)$, $\varepsilon=\pm 1$, existe para $C_2=1$. Los autovalores son $\left\{0,-4 \Sigma _c-2\right\}$. 
\end{enumerate}
Los criterio de estabilidad de los puntos de equilibrio del sistema \eqref{3.41-3.43} para fluido sin presión ($\gamma =1$) y $v\neq 0$  se resumen en la tabla \ref{Tab2}. En este caso particular, los puntos $N_1$, $N_2$ and $N_3$ se unen en una línea que llamada $N$. 
\begin{table*}
\caption{\label{Tab2} Análisis cualitativo de los puntos de equilibrio del sistema \eqref{3.41-3.43} para fluido sin presión ($\gamma =1$) y $v\neq 0$. Se incluye la línea $N$.}
\begin{tabular}{|m{0.7cm}|m{3.5cm}|m{3.5cm}|m{3.4cm}|m{1.7cm}|}
\hline
Etiq. &  $(\Sigma,A,v)$& Autovalores & Estabilidad  & ${(K,\Omega_t)}$\\
 \hline
 $N$ & $(0,0,v)$  & $\{0,-1,1\}$    & Silla no hiperbólica & (1,0)\\\hline
$N_{2,3}$ & $(0,0,\pm 1) $& $\{-1,1,0\}$  & Silla no hiperbólica & $(1,0)$\\\hline
  $P_{1,2}$ & $\left(-\frac{1}{\sqrt{C_2}},0,\pm 1\right)$ & $\left\{\frac{2}{\sqrt{C_2}},2-\frac{2}{\sqrt{C_2}},2\right\}$ & Fuente hiperbólica para $C_2>1$ \newline Silla hiperbólica para $C_2<1$ \newline no hiperbólico para  $C_2>1$  & $(0,0)$\\\hline
$P_{3,4}$ & $\left(\frac{1}{\sqrt{C_2}},0,\pm 1\right)$&$\left\{-\frac{2}{\sqrt{C_2}},2+\frac{2}{\sqrt{C_2}},2\right\}$ & Silla hiperbólica para $C_2>0$ & $(0,0)$\\\hline
$P_5$ & $\left(0, \frac{1}{\sqrt{C_2}}, 1\right)$& $\left\{\frac{2}{\sqrt{C_2}},-\frac{2}{\sqrt{C_2}}-2,-2\right\}$ &  Silla hiperbólica  para $C_2>0$ &$(0,0)$\\\hline
$P_6$ & $\left(0, \frac{1}{\sqrt{C_2}},- 1\right)$& $\left\{-\frac{2}{\sqrt{C_2}},\frac{2}{\sqrt{C_2}}-2,-2\right\}$  & Pozo hiperbólico para $C_2>1$ \newline Silla hiperbólica para $C_2<1$ \newline No hiperbólico para $C_2=1$ &$(0,0)$\\\hline
$P_{7}$ & $\left(-\frac{1}{2},\frac{1}{2},- 1\right)$& $\left\{0,\pm \sqrt{C_2-1}\right\}$ & Silla no hiperbólica  para $C_2> 1$ & $\scriptstyle{(1-C_2,\newline \frac{C_2}{2})}$\\\hline
$P_8$ &${\scriptscriptstyle\left(-\frac{1}{2 C_2}, \frac{\sqrt{4 C_2-3}}{2 C_2}, - \frac{1}{\sqrt{4 C_2-3}}\right)}$ & ver texto & ver texto  & ${\scriptscriptstyle\left(0,\newline \frac{4 C_2-3}{2 C_2}\right)}$\\ \hline
$H_{-}$ &  ${\scriptscriptstyle\left(\Sigma_0,1+\Sigma_0, - 1\right), C_2=1}$ & $\left\{0,-2(1+ 2\Sigma _c)\right\}$ & No hiperbólico &  ${\scriptscriptstyle\left(0, -2 \Sigma_0 (1+\Sigma_0)\right)}$\\\hline
$H_{+}$ &  ${\scriptscriptstyle\left(\Sigma_0,-1-\Sigma_0,  1\right), C_2=1}$ & $\left\{0,-2(1+ 2\Sigma _c)\right\}$ & No hiperbólico &  ${\scriptscriptstyle\left(0, -2 \Sigma_0 (1+\Sigma_0)\right)}$\\
 \hline
\end{tabular}
\end{table*}
\FloatBarrier

\subsection{Caso general $v\neq 0$}
\label{general}

En el caso general $v\neq 0$ es posible reducir la dimensión del sistema cuando la restricción \eqref{constraintmod1} no es degenerada, procediendo de la siguiente manera. Las ecuaciones  \eqref{constraintmod1} y  \eqref{defnOmegat} se pueden resolver globalmente para ${\Omega}_{t}$ y $K$ (suponiendo que  $\gamma v \neq 0$ y $\gamma  v^2-v^2+1 \neq 0$). Esto es, 
\begin{subequations}
\begin{align}
& {\Omega}_t = \frac{2 A C_2 {\Sigma} \left((\gamma -1) v^2+1\right)}{\gamma
 v}, \\
& K=- C_2 A^2-\frac{2 A C_2 \Sigma  \left(\gamma
   +v^2-1\right)}{\gamma  v}-C_2 \Sigma ^2+1.
\end{align} 
\end{subequations}
Luego, se obtiene el sistema 3-dimensional para las variables $\Sigma,A,v$ siguiente: 
\begin{subequations}
\label{reducedsyst}
\begin{align}
&{\Sigma}'=\Sigma  \left(-C_2 A^2+\frac{A\left(-3 \gamma -2 (\gamma -1) C_2 \Sigma +v^2(\gamma +2 (\gamma -1) C_2 \Sigma -2)+2\right)}{\gamma  v}+C_2 \Sigma^2-1\right),\\
& A'=A \left(-C_2 A^2+\frac{2 A (\gamma -1) C_2 \Sigma \left(v^2-1\right)}{\gamma v}+\Sigma (C_2 \Sigma +2)+1\right),\\
& v'=\frac{\left(v^2-1\right)\left(A \left(3 \gamma ^2-5\gamma +(\gamma -2) v^2+2\right)+\gamma  v (\gamma (\Sigma +2)-2)\right)}{\gamma \left(\gamma -v^2-1\right)}.
	\end{align}
\end{subequations}

  \begin{table}
\caption{\label{Tab0} Análisis cualitativo de los puntos de equilibrio del sistema  \eqref{system29} con $v\neq 0$.}
\begin{tabularx}{\textwidth}{|c|c|c|X|}
\hline
Etiq &  $(\Sigma,A,v, K)$& Autovalores & Estabilidad  \\
 \hline
 $N$ & $(0,0,v,1)$, $\gamma=1$ & $\{0,-1,1,2\}$ & Silla no hiperbólica. \\\hline
 $N_{1}$ & $(0,0,0,1)$& $\{-2,-1,1,2\}$  & Silla hiperbólica. \\\hline
$N_{2,3}$ & $(0,0,\pm 1,1) $ & $\left\{0,-1,1,\frac{4}{\gamma -2}+4\right\}$  & Silla no hiperbólica. \\\hline
  $P_{1,2}$ & $(-\frac{1}{\sqrt{C_2}},0,\pm 1,0)$  & $\left\{2-\frac{2}{\sqrt{C_2}},2,\frac{2 \gamma \left(\frac{1}{\sqrt{C_2}}-2\right)+4}{2-\gamma }\right\}$  &   Fuente hiperbólica para  $1<\gamma <2,  1<C_2<\frac{\gamma
   ^2}{4 \gamma ^2-8 \gamma +4}$. \newline
   	  Silla hiperbólica para 
	   $ 1<\gamma <2,0<C_2<1$,  $1<\gamma <2, C_2>\frac{\gamma ^2}{4 \gamma ^2-8 \gamma +4}$. \newline
	  No hiperbólico para for $1<\gamma <2, C_2=\frac{4\gamma ^2}{(\gamma-1)^2}$, o   $1<\gamma <2, C_2=1$. \\\hline
$P_{3,4}$ & $\left(\frac{1}{\sqrt{C_2}},0,\pm 1,0\right)$ & $\left\{\frac{2}{\sqrt{C_2}}+2,2,\frac{-4 \gamma -\frac{2 \gamma}{\sqrt{C_2}}+4}{2-\gamma }\right\}$  & Silla hiperbólica  $1<\gamma <2,  C_2>0$. \\\hline
$P_5$ & $\left(0, \frac{1}{\sqrt{C_2}}, 1, 0\right)$& $ {\scriptscriptstyle \left\{-\frac{2}{\sqrt{C_2}}-2,-2,\frac{6 \gamma +4 (\gamma -1)
   \sqrt{C_2}-8}{(\gamma -2) \sqrt{C_2}}\right\}}$  &    Pozo hiperbólico  (ver texto). \newline
	  Silla hiperbólica (ver texto).
	   \newline
      No hiperbólico  (ver texto).\\
	 \hline 
	 \end{tabularx}
\end{table}

\begin{table}[ht!]
\caption{\label{Tab00} Análisis cualitativo de los puntos de equilibrio del sistema  \eqref{system29} con $v\neq 0$ (cont.).}
\begin{tabularx}{\textwidth}{|c|c|X|X|}
\hline
Etiq &  $(\Sigma,A,v, K)$& Autovalores & Estabilidad  \\
 \hline
$P_6$ & $ \left(0, \frac{1}{\sqrt{C_2}}, - 1, 0\right)$& ${\scriptscriptstyle\left\{\frac{2}{\sqrt{C_2}}-2,-2,\frac{-6 \gamma +4 (\gamma -1)
   \sqrt{C_2}+8}{(\gamma -2) \sqrt{C_2}}\right\}}$ &    Pozo hiperbólico para $1<\gamma <2,  C_2>1$.
	\newline 
	  Silla hiperbólica para
	 $1<\gamma <2, 0<C_2<1 $, o  
	      $\frac{4}{3}<\gamma <2,  0<C_2<\frac{(4-3\gamma )^2}{4 (\gamma -1)^2}$, o   $\frac{4}{3}<\gamma <2, C_2>1$.
	\newline 
	  No hiperbólico para 
	  $1<\gamma \leq \frac{4}{3}, C_2=1$, o   $\frac{4}{3}<\gamma <2,  C_2=\frac{9 \gamma^2-24 \gamma +16}{4 \gamma^2-8 \gamma +4}$, o  
	    $\frac{4}{3}<\gamma <2,  C_2=1$. \\\hline
$P_{7}$ & $\left(-\frac{1}{2},\frac{1}{2},- 1, 1-C_2\right)$& $\left\{0,-\sqrt{C_2-1},\sqrt{C_2-1}\right\}$ & 
  Silla no hiperbólica para $C_2>1, 1<\gamma <2$. \newline
  No hiperbólico con tres autovalores nulos para $C_2=1,  \gamma \neq 0$. \newline
  No hiperbólico con un autovalor cero y dos imaginarios putos para $1<\gamma <2,  0\leq
   C_2\leq 1$. \\\hline
$P_8$ & $\left(-\frac{1}{2 C_2},-\frac{1}{2 C_2 \Delta },\Delta \right)$ & 
$\left\{k \lambda_1,k \lambda_2, k \lambda_3\right\}$ si $\gamma=1$ & Silla.  \\\hline
$P_9$ &$(0, \lambda_{+} v_{-}, v_{-})$  & $\{\mu_1, \mu_2, \mu_3\}$  & Silla.  \\
 \hline
 $P_{10}$ &$(0, \lambda_{-} v_{+}, v_{+})$  & $\{\nu_1, \nu_2, \nu_3\}$ &   Pozo hiperbólico (ver texto).
\newline 
  Silla hiperbólica  para
 (ver texto). \\
 \hline
\end{tabularx}
\end{table}
El sistema dinámico \eqref{reducedsyst} admite algunos conjuntos invariantes, estos son: $v=\pm 1$, correspondiente al caso de inclinación extrema, y los conjuntos invariantes $A=0$ y $\Sigma=0$. 
Los puntos de equilibrio del sistema \eqref{reducedsyst} son los siguientes.
\begin{enumerate}
\item  $P_{1,2}:(\Sigma,A,v)=(-\frac{1}{\sqrt{C_2}},0,\pm 1)$, con autovalores $\left\{2-\frac{2}{\sqrt{C_2}},2,\frac{2 \gamma \left(\frac{1}{\sqrt{C_2}}-2\right)+4}{2-\gamma }\right\}$. Estos son:
	\begin{enumerate}
	\item Fuentes hiperbólicas para $1<\gamma <2, 
   1<C_2<\frac{\gamma
   ^2}{4 \gamma ^2-8 \gamma +4}$.
   	\item Sillas hiperbólicas para
	   \begin{enumerate}
	       \item $1<\gamma <2,0<C_2<1$, o 
	       \item $1<\gamma <2,C_2>\frac{\gamma ^2}{4 \gamma ^2-8 \gamma +4}$.
	   \end{enumerate}
	\item No hiperbólicos para
	 \begin{enumerate}
	     \item $1<\gamma <2, C_2=\frac{4\gamma ^2}{(\gamma-1)^2}$, o 
         \item $1<\gamma <2, C_2=1$. 
	 \end{enumerate}
	\end{enumerate}
\item  $P_{3,4}:(\Sigma,A,v)=(\frac{1}{\sqrt{C_2}},0,\pm 1)$, con autovalores $\left\{\frac{2}{\sqrt{C_2}}+2,2,\frac{-4 \gamma -\frac{2 \gamma}{\sqrt{C_2}}+4}{2-\gamma }\right\}$. Estos son sillas hiperbólicas para $1<\gamma <2,  C_2>0$.
\item $P_5:(\Sigma,A,v)=(0,\frac{1}{\sqrt{C_2}},1)$, con autovalores $\left\{-\frac{2}{\sqrt{C_2}}-2,-2,\frac{6 \gamma +4 (\gamma -1)
   \sqrt{C_2}-8}{(\gamma -2) \sqrt{C_2}}\right\}$. Estos son
	\begin{enumerate}
	\item pozo hiperbólico para
	  \begin{enumerate}
	      \item $1<\gamma \leq \frac{4}{3}, C_2>\frac{9 \gamma^2-24 \gamma +16}{4 \gamma^2-8 \gamma +4}$, o 
	      \item $\frac{4}{3}<\gamma <2, C_2>0$. 
	  \end{enumerate} 
	\item silla hiperbólica para $1<\gamma <\frac{4}{3},  0<C_2<\frac{(4-3 \gamma )^2}{4 (\gamma -1)^2}$
	\item no hiperbólico para $1<\gamma <\frac{4}{3},  C_2=\frac{9 \gamma^2-24 \gamma +16}{4 \gamma^2-8 \gamma +4}$.
	\end{enumerate}
	\item $P_6:(\Sigma,A,v)=(0,\frac{1}{\sqrt{C_2}},-1)$, con autovalores $\left\{\frac{2}{\sqrt{C_2}}-2,-2,\frac{-6 \gamma +4 (\gamma -1)
   \sqrt{C_2}+8}{(\gamma -2) \sqrt{C_2}}\right\}$
	\begin{enumerate}
	\item pozo hiperbólico para $1<\gamma <2,  C_2>1$.
	\item silla hiperbólica para \begin{enumerate}
	    \item $1<\gamma <2, 0<C_2<1$, o 
	    \item $\frac{4}{3}<\gamma <2,  0<C_2<\frac{(4-3 \gamma )^2}{4 (\gamma -1)^2}$, o 
	    \item $1<\gamma \leq \frac{4}{3},  0<C_2<1$.
	\end{enumerate}
	\item no hiperbólico para
	  \begin{enumerate}
	      \item $1<\gamma \leq \frac{4}{3}, C_2=1$, o  
	      \item $\frac{4}{3}<\gamma <2,  C_2=\frac{9 \gamma^2-24 \gamma +16}{4 \gamma^2-8 \gamma +4}$, o 
	      \item $\frac{4}{3}<\gamma <2,  C_2=1$.
	  \end{enumerate}
	\end{enumerate}

 \item $P_{7}:(\Sigma,A,v)=(-\frac{1}{2},\frac{1}{2},-1)$, con autovalores $\left\{0,-\sqrt{C_2-1},\sqrt{C_2-1}\right\}$ es 
	\begin{enumerate}
	\item silla no hiperbólica para $C_2>1$. 
	\item no hiperbólico con tres autovalores iguales a cero para $C_2=1$.
	\item no hiperbólico con un autovalor igual a cero y dos autovalores imaginarios puros para $0\leq
   C_2<1$.
	\end{enumerate}

\item$P_8:(\Sigma,A,v)=\left(-\frac{1}{2 C_2},-\frac{1}{2 C_2 \Delta },\Delta \right)$, donde $\Delta=-\sqrt{\frac{2-3 \gamma }{\gamma -4 \gamma  C_2+2}}$.
Para este punto se verfica la ecuación $\Sigma =A v=-\frac{1}{2 C_2}$. Este punto existe para  $1\leq \gamma <2,  C_2>\frac{\gamma +2}{4 \gamma }$.
En el caso $\gamma=1$, los autovalores son $\left\{k \lambda_1,k \lambda_2, k \lambda_3\right\}$, con  $k=\frac{1}{2 C_2^2 (4 C_2-3)^{3/2}}$ y $\lambda_1, \lambda_2$ y $\lambda_3$ son las raíces del  polinomio en $\lambda$:  
$P(\lambda)=-16 (C_2-1)^2 C_2^3 (4   C_2-3)^{11/2}-4 (C_2-1)  C_2^2 (8 C_2-7) (4
   C_2-3)^3 \lambda +\lambda ^3$, estas raíces deben ser evaluadas numéricamente debido a que sus expresiones son complicadas. Para $C_2=1+\delta + \mathcal{O}(\delta^2)$, se tiene $\lambda_1= -2 \delta +\mathcal{O}\left(\delta ^2\right)$, $\lambda_2=-\frac{\left(3 \gamma ^2-8 \gamma +4\right) \sqrt{\delta }}{(\gamma
   -2) (3 \gamma -2)}+\frac{\left(-3 \gamma ^2+6 \gamma -4\right) \delta }{(\gamma -2) (3 \gamma
   -2)}+\mathcal{O}\left(\delta ^{3/2}\right)$, y 
   $\lambda_3=\frac{\left(3 \gamma ^2-8 \gamma +4\right) \sqrt{\delta }}{(\gamma -2) (3
   \gamma -2)}+\frac{\left(-3 \gamma ^2+6 \gamma -4\right) \delta }{(\gamma -2) (3 \gamma -2)}+\mathcal{O}\left(\delta
   ^{3/2}\right)$, esto muestra un comportamiento de silla  para valores de los parámetros cercanos a los valores de Relatividad General, por ejemplo, para fluido sin presión ($\gamma=1$) los $\lambda_i$ son aproximadamente $\left\{-2 \delta ,\delta +\sqrt{\delta },\delta -\sqrt{\delta }\right\}$. Para $\delta<0$ se tienen dos autovalores imaginarios complejos conjugados con parte real negativa y un autovalor real positivo, mientras que, para $\delta>0$, se tiene un autovalor negativo y los demás tienen signos diferentes. En la figura \ref{eigenP83D} se representa gráficamente la parte real de $\lambda_i$ (que difieren de los autovalores asociados al punto de equilibrio $P_8$ en el factor $k$) para $C_2\geq \frac{2+\gamma}{4 \gamma}$ y $\gamma\in[1, 2]$. La figura ilustra que el punto de equilibrio es en general una silla o es no hiperbólico. Dicho procedimiento en que se prescinde de un factor multiplicativo en los autovalores para determinar su estabilidad solo es aplicable cuando el punto de equilibrio es silla o no hiperbólico y la razón es simple, se requiere sólo mostrar que al menos dos autovalores tienen partes reales con signo distinto o partes reales cero, por lo que el signo de un factor multiplicativo es irrelevante.  

\item $P_9:(\Sigma,A,v)=(0, \lambda_{+} v_{-}, v_{-})$, donde \newline
   $v_{-}=\frac{\sqrt{(\gamma -1) \left(\gamma  \left(2 \gamma ^2 C_2-\gamma  (2 C_2+3)-2 \sqrt{\gamma
   -1} \sqrt{C_2} \sqrt{\gamma  (\gamma  ((\gamma -1) C_2-3)+8)-4}+8\right)-4\right)}}{\gamma
   -2}$,
y \newline $\lambda_{+}= \frac{\gamma +\frac{\sqrt{\gamma  (\gamma  ((\gamma -1) C_2-3)+8)-4}}{\sqrt{\gamma -1}
   \sqrt{C_2}}}{2-3 \gamma }$, tal que $C_2=\frac{1}{\lambda_{+}^2 v_{-}^2}$. Los autovalores son $\mu_1=-2, \mu_2=-\gamma + \frac{\sqrt{8 \gamma +\gamma ^3 C_2-\gamma ^2 (C_2+3)-4}}{\sqrt{\gamma -1} \sqrt{C_2}}$ (este siempre es positivo para $1<\gamma<2, C_2>0$), y el tercer autovalor es 
   $\mu_3=\frac{2 (\gamma -1) \gamma  \left(-\gamma -v_{-}^4+3 \gamma  v_{-}^2-4 v_{-}^2+1\right)-2 (\gamma -2) \lambda  v_{-}^2 \left(-3
   \gamma +v_{-}^2+3\right) \left(\gamma +v_{-}^2-1\right)}{\gamma  \left(-\gamma +v_{-}^2+1\right)^2}$.
Al notar que al menos existe cambio de signo en dos autovalores, es decir $\mu_1 \mu_2<0$, para  $1<\gamma<2, C_2>0$, se concluye que es silla. 
\begin{figure*}
    \centering
    \includegraphics[scale=0.45]{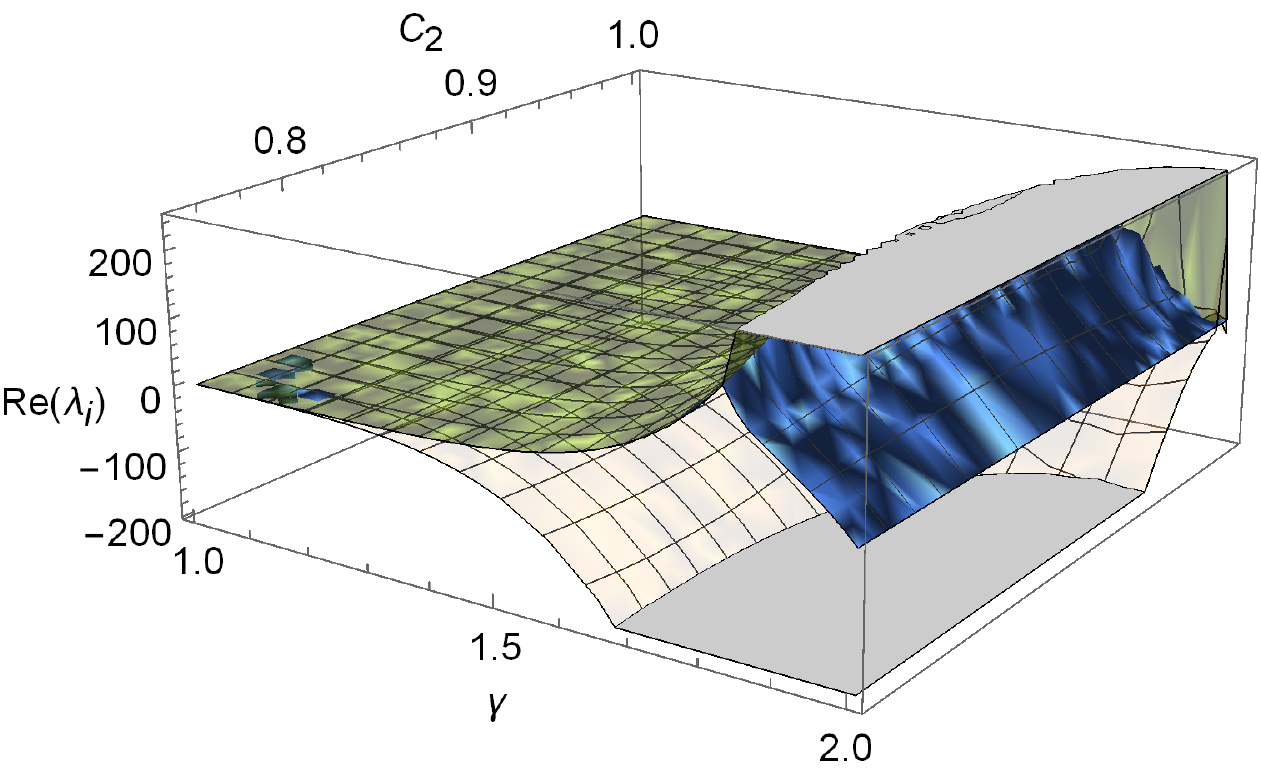} \hspace{2cm}
    \includegraphics[scale=0.45]{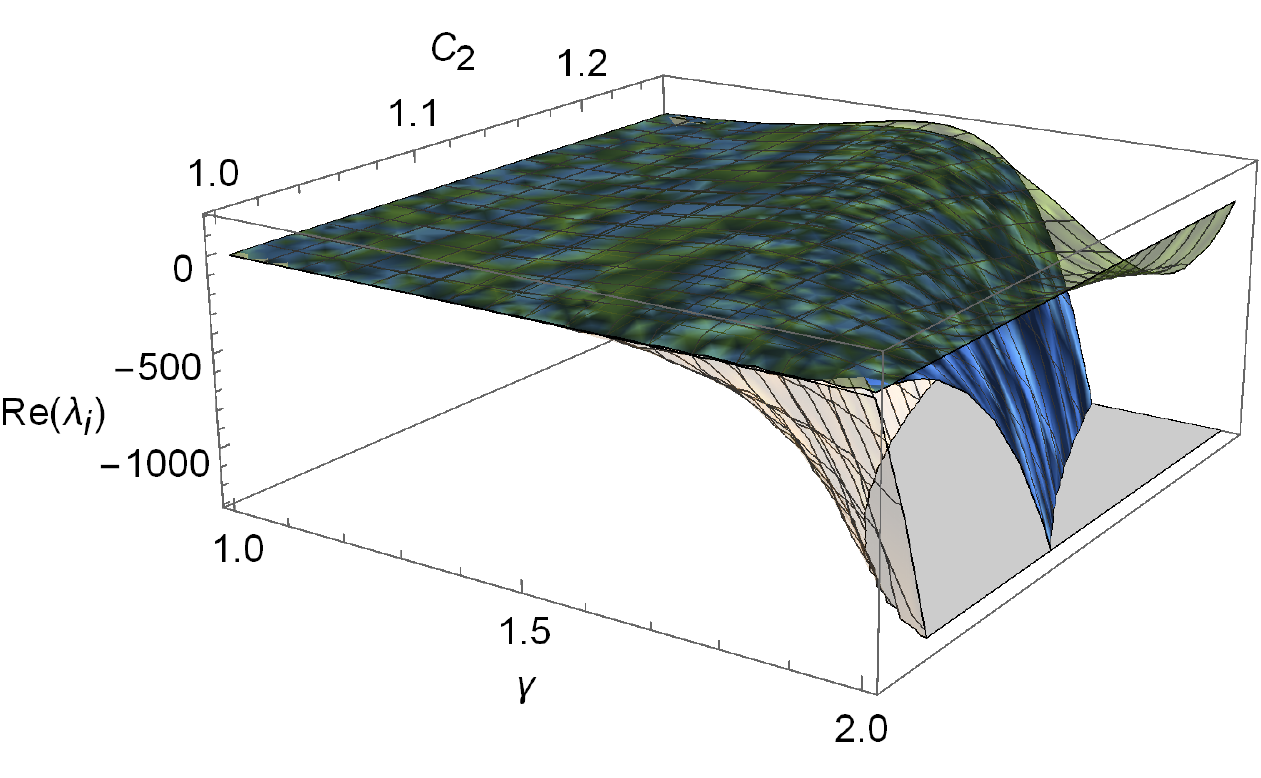}
    \caption[{Parte real de los autovalores del punto de equilibrio $P_8$ para $C_2\geq \frac{2+\gamma}{4 \gamma}$ y $\gamma\in[1, 2]$.}]{\label{eigenP83D} Parte real de $\lambda_i$ correspondiente al punto de equilibrio $P_8$ para $C_2\geq \frac{2+\gamma}{4 \gamma}$ y $\gamma\in[1, 2]$.}
   \end{figure*}
\item  
$P_{10}:(\Sigma,A,v)=(0, \lambda_{-} v_{+}, v_{+})$, donde \newline
   $v_{+}=\frac{\sqrt{(\gamma -1) \left(\gamma  \left(2 \gamma ^2 C_2-\gamma  (2 C_2+3)+2 \sqrt{\gamma -1} \sqrt{C_2} \sqrt{\gamma 
   (\gamma  ((\gamma -1) C_2-3)+8)-4}+8\right)-4\right)}}{2-\gamma }$,
y \newline $\lambda_{-}=\frac{\gamma -\frac{\sqrt{\gamma  (\gamma  ((\gamma -1) C_2-3)+8)-4}}{\sqrt{\gamma -1} \sqrt{C_2}}}{2-3 \gamma }$, tal que  $C_2=\frac{1}{\lambda_{-}^2 v_{+}^2}$. Los autovalores son $\nu_1=-2, \nu_2=-\gamma -\frac{\sqrt{\gamma  (\gamma  ((\gamma -1) C_2-3)+8)-4}}{\sqrt{\gamma -1} \sqrt{C_2}}$ (este siempre es negativo para $1<\gamma <2,  C_2>0$), y el tercer autovalor es  $\nu_3= \frac{-2 (\gamma -2) \lambda  v_{+}^2 \left(-3 \gamma +v_{+}^2+3\right) \left(\gamma +v_{+}^2-1\right)-2 (\gamma -1) \gamma  \left(\gamma +v_{+}^4+(4-3 \gamma )
   v_{+}^2-1\right)}{\gamma  \left(-\gamma +v_{+}^2+1\right)^2}$.
   
El punto es un pozo hiperbólico para 
\begin{enumerate}
    \item $1<\gamma \leq \gamma_0, 0<C_2<\frac{4 \gamma -4}{\gamma ^2}$, o   
    \item $1<\gamma <\gamma_0, \frac{4 \gamma -4}{\gamma
   ^2}<C_2<\frac{(4-3 \gamma )^2}{4 (\gamma -1)^2}$, o   
 
   \item $\gamma_0<\gamma <\frac{4}{3}, 0<C_2<\frac{(4-3 \gamma )^2}{4 (\gamma -1)^2}$,
\end{enumerate}
donde $\gamma_0=-\frac{2}{27} \left(-11-\frac{14}{\sqrt[3]{27 \sqrt{57}-197}}+\sqrt[3]{27 \sqrt{57}-197}\right)\approx 1.22033$.
Es silla hiperbólica para 
 \begin{enumerate}
     \item $1<\gamma \leq \gamma_0, C_2>\frac{(4-3 \gamma )^2}{4 (\gamma -1)^2}$, o  
     \item $\gamma_0<\gamma \leq \frac{4}{3}, 
   \frac{(4-3 \gamma )^2}{4 (\gamma -1)^2}<C_2<\frac{4 (\gamma -1)}{\gamma ^2}$, o  
   \item $\gamma_0<\gamma \leq \frac{4}{3}, 
   C_2>\frac{4 (\gamma -1)}{\gamma ^2}$, o   
   \item $\frac{4}{3}<\gamma <2,  0<C_2<\frac{4 (\gamma -1)}{\gamma ^2}$, o  
   \item $\frac{4}{3}<\gamma <2,  C_2>\frac{4 (\gamma -1)}{\gamma ^2}$.
 \end{enumerate}
 \end{enumerate}

En las tablas \ref{Tab0} y \ref{Tab00} se presenta un resumen del análisis cualitativo de los puntos de equilibrio del sistema \eqref{system29} donde se usan las notaciones
\newline $v_{\pm}=\frac{\sqrt{(\gamma -1) \left(\gamma  \left(2 \gamma ^2 C_2-\gamma  (2 C_2+3)\pm 2 \sqrt{\gamma
   -1} \sqrt{C_2} \sqrt{\gamma  (\gamma  ((\gamma -1) C_2-3)+8)-4}+8\right)-4\right)}}{\gamma
   -2}$,\newline 
 $\lambda_{\pm}= \frac{\gamma \pm \frac{\sqrt{\gamma  (\gamma  ((\gamma -1) C_2-3)+8)-4}}{\sqrt{\gamma -1}
   \sqrt{C_2}}}{2-3 \gamma }$, $\gamma_0=-\frac{2}{27} \left(-11-\frac{14}{\sqrt[3]{27 \sqrt{57}-197}}+\sqrt[3]{27 \sqrt{57}-197}\right)\approx 1.22033$, $\Delta=-\sqrt{\frac{2-3 \gamma }{\gamma -4 \gamma  C_2+2}}$,  $k=\frac{1}{2 C_2^2 (4 C_2-3)^{3/2}}$, y $\lambda_1, \lambda_2$ además $\lambda_3$ cumplen $P(\lambda_i)=0$, donde
$P(\lambda)=-16 (C_2-1)^2 C_2^3 (4   C_2-3)^{11/2}-4 (C_2-1)  C_2^2 (8 C_2-7) (4
   C_2-3)^3 \lambda +\lambda ^3$.

\FloatBarrier

\subsection{Conjunto invariante $v= A=0$.}
\label{vcero}

En esta sección se presenta el análisis de estabilidad de los puntos de equilibrio del sistema \eqref{Eq:22} en el conjunto invariante $A=v=0$. 
En la siguiente lista se hace el análisis de estabilidad conservando los cuatro autovalores.  
\begin{enumerate}
    \item $N_1:(\Sigma,A,K,v)=(0,0,1,0)$ tiene autovalores $\{-2,-1,1,2\}$, por tanto es una silla hiperbólica.
    
    \item $P_{11}:(\Sigma,A,K,v)=\left(-\frac{1}{\sqrt{C_2}}, 0, 0, 0\right)$.  Tiene autovalores \\
    $\left\{2-\frac{2}{\sqrt{C_2}},2,\frac{\gamma }{(\gamma -1) \sqrt{C_2}}-2,\frac{-3 \gamma
   +4 (\gamma -1) \sqrt{C_2}+2}{(\gamma -1) \sqrt{C_2}}\right\}$.
    \begin{enumerate}
        \item Es una fuente si $\gamma >1,  \frac{(2-3 \gamma )^2}{16 (\gamma -1)^2}<C_2<\frac{\gamma ^2}{4 (\gamma -1)^2}$. 
        \item Es silla si 
         \begin{enumerate}
             \item $1<\gamma <2,   C_2>\frac{\gamma ^2}{4(\gamma -1)^2}$, o 
             \item $1<\gamma <2, 
   0<C_2<\frac{(2-3 \gamma )^2}{16 (\gamma -1)^2}$
         \end{enumerate}
         \item No hyperbólico si: 
         \begin{enumerate}
             \item $1<\gamma <2,  C_2=1$, o  
             \item $1<\gamma <2,  C_2=\frac{(2-3 \gamma )^2}{16 (\gamma -1)^2}$, o 
             \item $1<\gamma <2,  C_2=\frac{\gamma ^2}{4 (\gamma -1)^2}$.
         \end{enumerate}
    \end{enumerate}
    
    \item $P_{12}:(\Sigma,A,K,v)=\left(\frac{1}{\sqrt{C_2}}, 0, 0,  0\right)$, Tiene autovalores \\$\left\{\frac{2}{\sqrt{C_2}}+2,2,-\frac{\gamma }{(\gamma -1) \sqrt{C_2}}-2,\frac{3 \gamma
   +4 (\gamma -1) \sqrt{C_2}-2}{(\gamma -1) \sqrt{C_2}}\right\}$. 
   
   Es una silla si $
   1<\gamma <2,  C_2>0$. 
   
    \item $P_{13}:(\Sigma,A,K,v)=\left(-\frac{2-3 \gamma }{4
   C_2(1- \gamma)}, 0, 0, 0\right)$. Satisface $\Omega_t\geq 0$ para ${ C_2}\geq \frac{(2-3 \gamma )^2}{16 (\gamma -1)^2}$.

   Tiene autovalores \\
   $\left\{-\frac{(\gamma -2) (3 \gamma -2)}{8 (\gamma -1)^2 C_2},\frac{(2-3 \gamma )^2}{4 (\gamma
   -1)^2 C_2}-2,\frac{\gamma  (3 \gamma -2)}{4 (\gamma -1)^2 C_2}-2,\frac{(2-3 \gamma
   )^2}{8 (\gamma -1)^2 C_2}-2\right\}$. 
    \begin{enumerate}
        \item Fuente para $1<\gamma <2,  0<C_2<\frac{(2-3 \gamma )^2}{16 (\gamma -1)^2}$. 
        \item Silla para 
         \begin{enumerate}
             \item $1<\gamma <2,  \frac{(2-3 \gamma )^2}{16 (\gamma -1)^2}<C_2<\frac{\gamma  (3 \gamma -2)}{8(\gamma -1)^2}$, o  
             \item $1<\gamma <2,  \frac{\gamma  (3 \gamma -2)}{8 (\gamma -1)^2}<C_2<\frac{(2-3 \gamma )^2}{8(\gamma -1)^2}$
             \item $1<\gamma <2,  C_2>\frac{(2-3 \gamma )^2}{8 (\gamma -1)^2}$.
         \end{enumerate}
    \end{enumerate}

   \item $P_{14}:(\Sigma,A,K,v)=\left(-\frac{2 (\gamma -1)}{3 \gamma -2}, 0,  \frac{(2-3 \gamma )^2-8 (\gamma -1)^2 C_2}{(2-3 \gamma )^2}, 0 \right)$. Tiene autovalores
   $\{\lambda_1,  \lambda_2, \lambda_3, \lambda_4\}=\Bigg\{\frac{2-\gamma }{3 \gamma -2},-\frac{4 (\gamma -1)}{3 \gamma -2},-\frac{1}{2}+\frac{\sqrt{64 (\gamma
   -1)^2 C_2-7 (2-3 \gamma )^2}}{4-6 \gamma }, -\frac{1}{2}-\frac{\sqrt{64 (\gamma -1)^2
   C_2-7 (2-3 \gamma )^2}}{6 \gamma -4}\Bigg\}$.  
   
   \begin{enumerate}
       \item No hiperbólico para 
        \begin{enumerate}
            \item $C_2=\frac{(2-3 \gamma )^2}{8 (\gamma -1)^2}, 1<\gamma \leq 2$, o
            \item $C_2\geq  0,  \gamma =1$, o
            \item $C_2\geq  0, \gamma =2$.
        \end{enumerate}
        \item Silla en los demás casos. 
   \end{enumerate}
   En la figura \ref{fig:my_label} se representan las partes reales de los autovalores $\lambda_i$ para el punto de equilibrio $(\Sigma, A,K,v)=\left(-\frac{2 (\gamma -1)}{3 \gamma -2}, 0,  \frac{(2-3 \gamma )^2-8 (\gamma -1)^2 C_2}{(2-3 \gamma )^2}, 0 \right)$ es cual representa soluciones estáticas, para $1\leq \gamma\leq 2$ y $C_2\geq 0$, ilustrando que dicho punto de equilibrio es no hiperbólico en los casos (a)-1,2,3 anteriormente descritos o  es silla. 
   \begin{figure}
       \centering
       \includegraphics[scale=0.45]{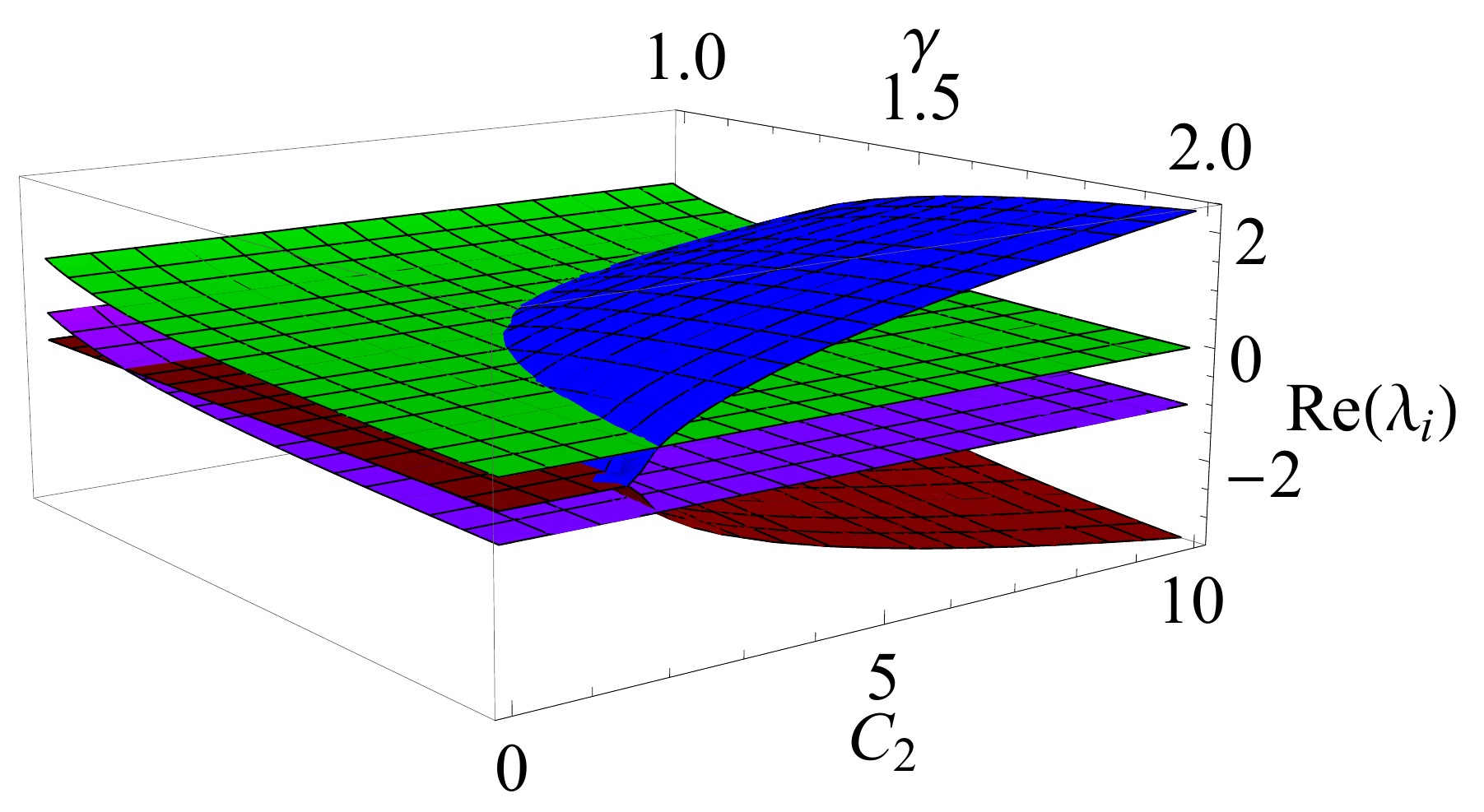}
       \caption{ \label{fig:my_label} Partes reales de los autovalores $\lambda_i$ para el punto de equilibrio $(\Sigma, A,K,v)=\left(-\frac{2 (\gamma -1)}{3 \gamma -2}, 0,  \frac{(2-3 \gamma )^2-8 (\gamma -1)^2 C_2}{(2-3 \gamma )^2}, 0 \right)$ para $1\leq \gamma\leq 2$ y $C_2\geq 0$.}
         \end{figure}
\end{enumerate}

\subsubsection{Sistema reducido}
\label{sect:3.7.1}
 En este caso particular, donde $A=v=0$, la restricción \eqref{constraintmod1} se satisface trivialmente. Además, de la ecuación \eqref{defnOmegat} resulta
\begin{equation}
(\gamma-1){\Omega}_{t}= \left(1-C_2{{\Sigma}}^2 -K\right).
\end{equation}
Imponiendo las condiciones de energía  para $\gamma\in(1,2)$, $\Omega_t\geq 0$, obtenemos el sistema dinámico reducido 
\begin{subequations}
\label{eq:3.57}
\begin{align}
& \Sigma'= \frac{\left(C_2 \Sigma ^2-1\right) (3 \gamma +4 (\gamma -1) C_2 \Sigma -2)+K (3
   \gamma +2 (\gamma -1) C_2 \Sigma -2)}{2 (\gamma -1) C_2},\\
& K'= 2 K \left(2 C_2 \Sigma^2+K-1\right), 
\end{align}
\end{subequations}
definido en el plano de fase 
\begin{equation}
    \left\{(\Sigma, K)\in \mathbb{R}^2 :  C_2{{\Sigma}}^2 +K \leq 1, K \geq 0\right\}.
\end{equation}
En la tabla \ref{Tab2Avcero} se hace un resumen de los resultados del análisis cualitativo de los puntos de equilibrio del sistema \eqref{Eq:22} con $v= A=0$. Se incluye la línea $N_1$.

\begin{figure*}
    \centering
    \includegraphics[scale=0.45]{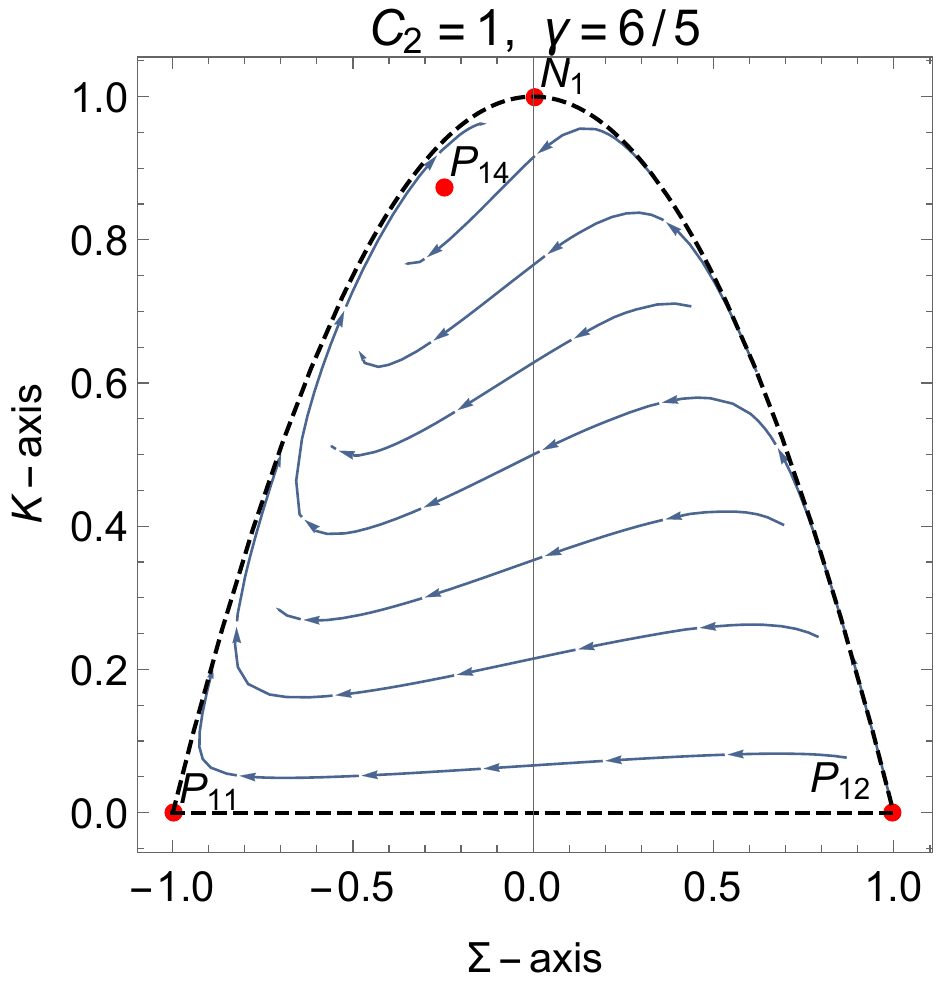} 
    \includegraphics[scale=0.45]{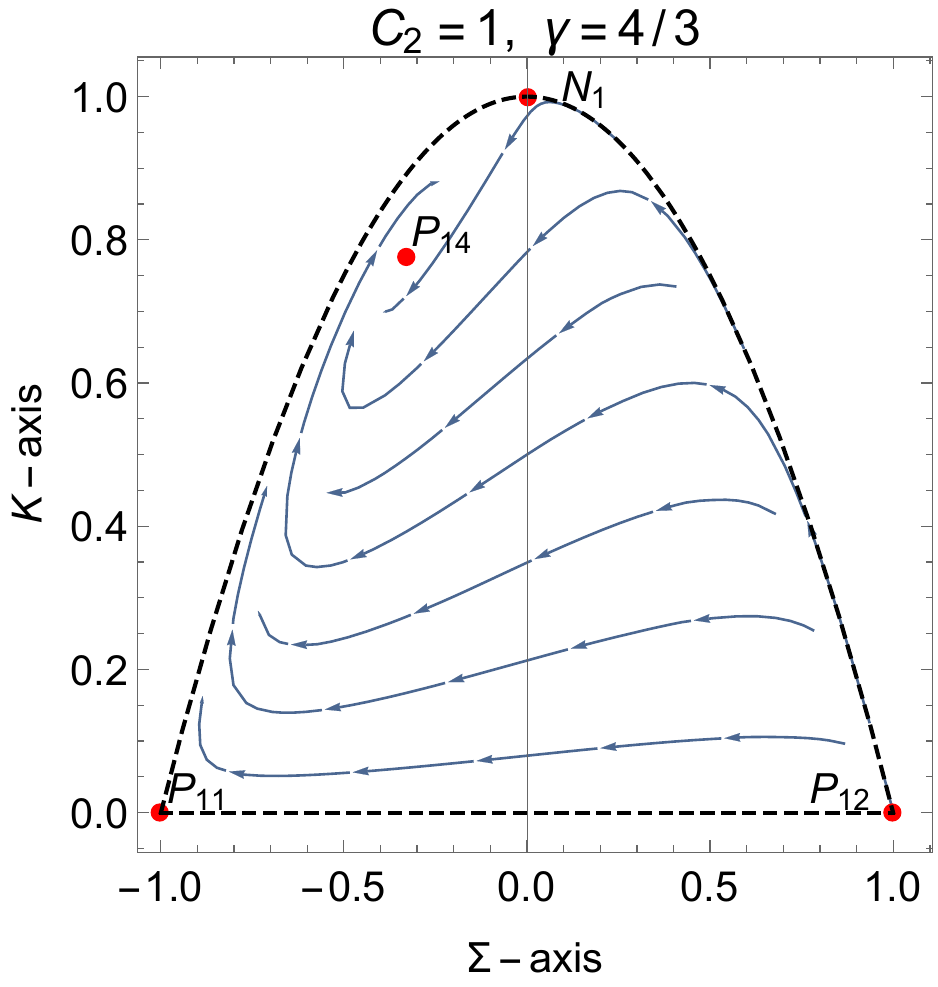} 
    \includegraphics[scale=0.45]{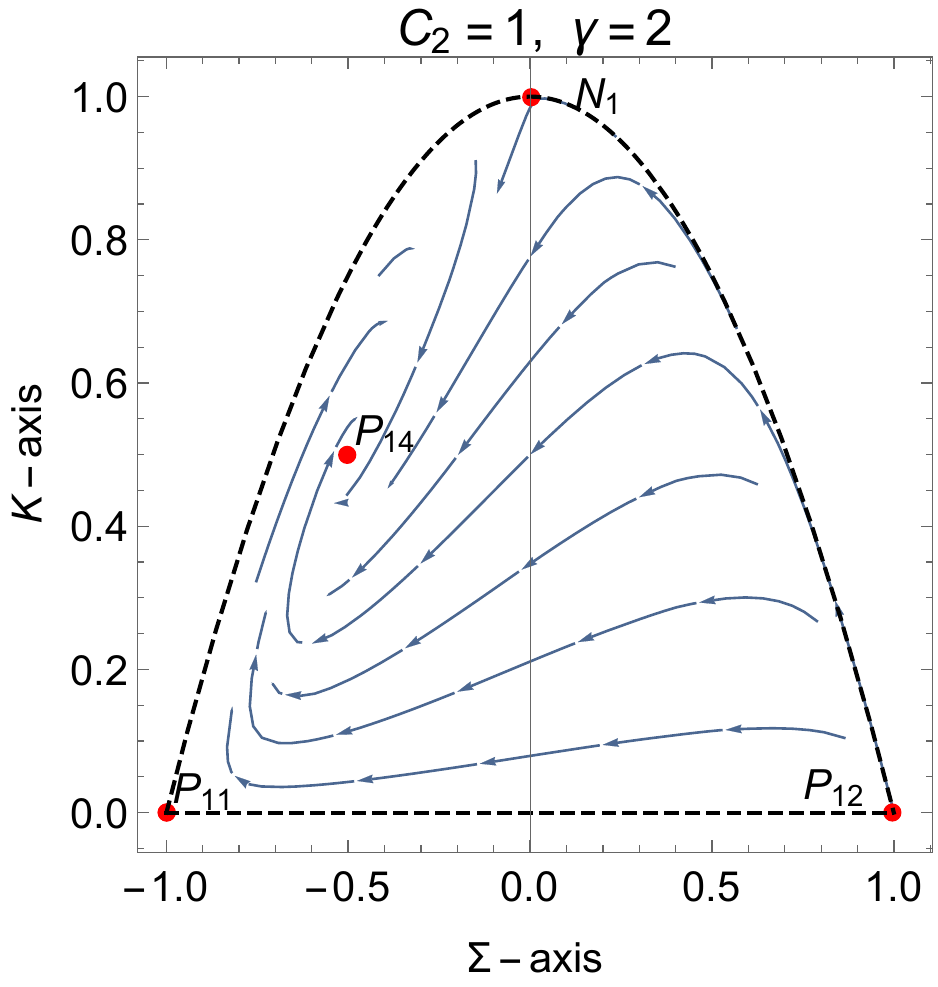} 
    \includegraphics[scale=0.45]{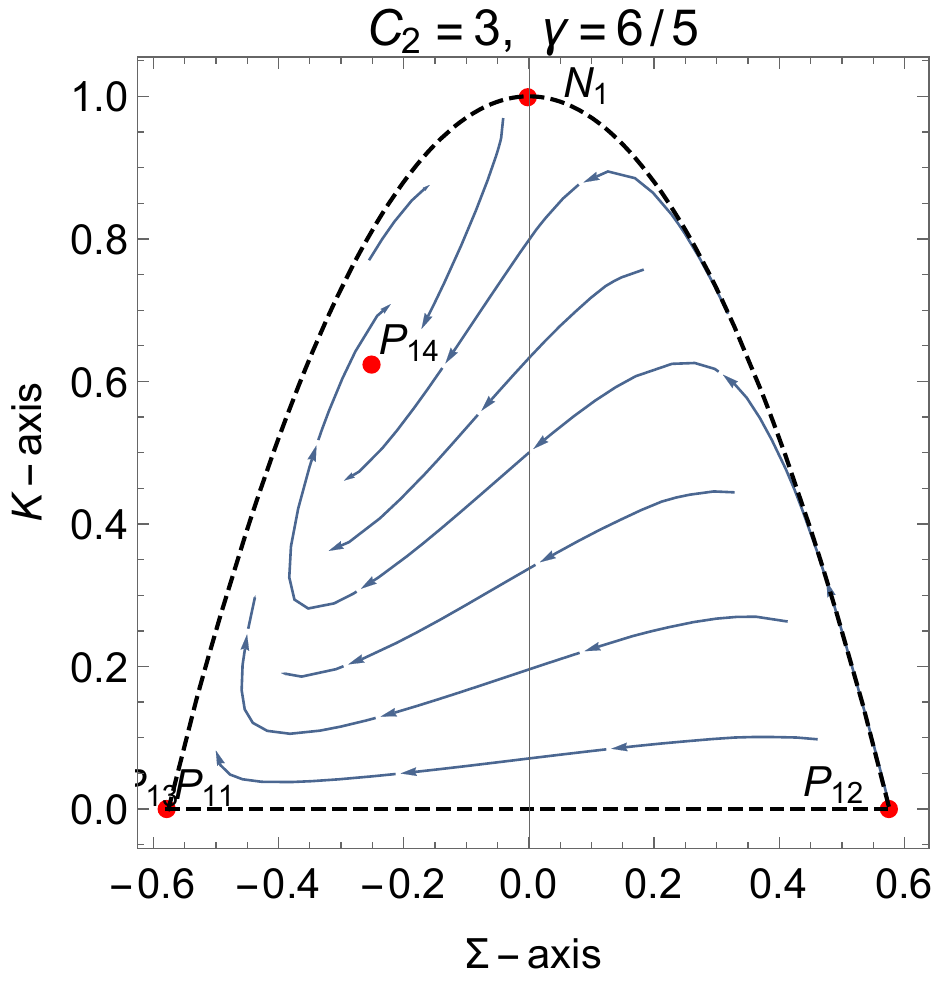} 
    \includegraphics[scale=0.45]{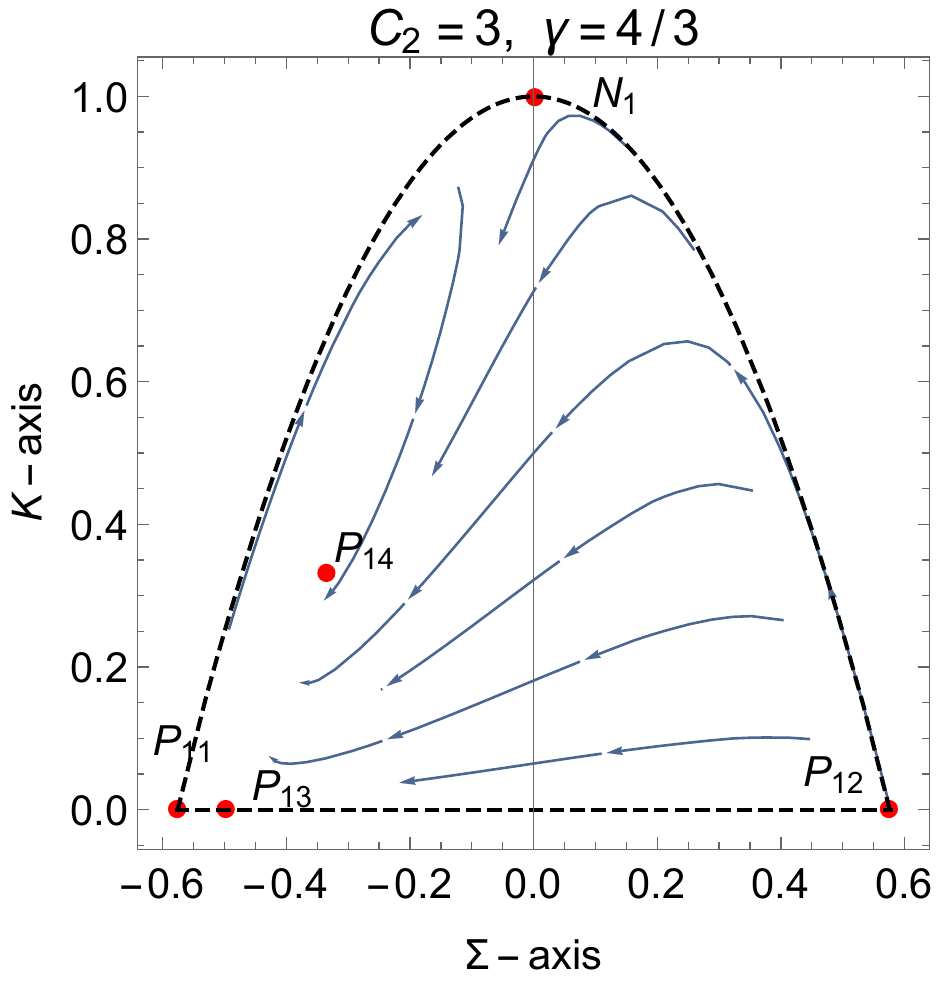} 
    \includegraphics[scale=0.34]{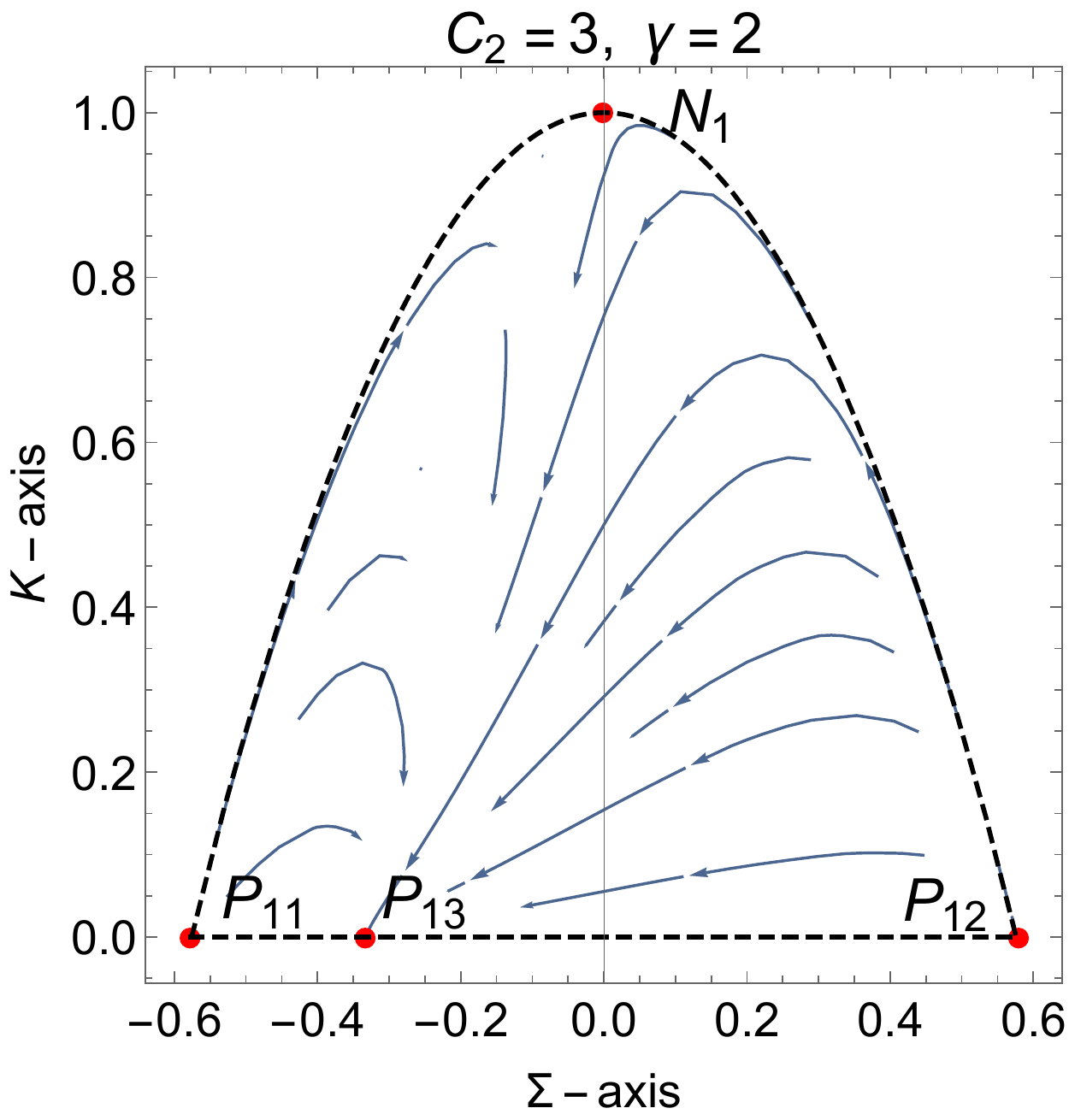}
    \includegraphics[scale=0.45]{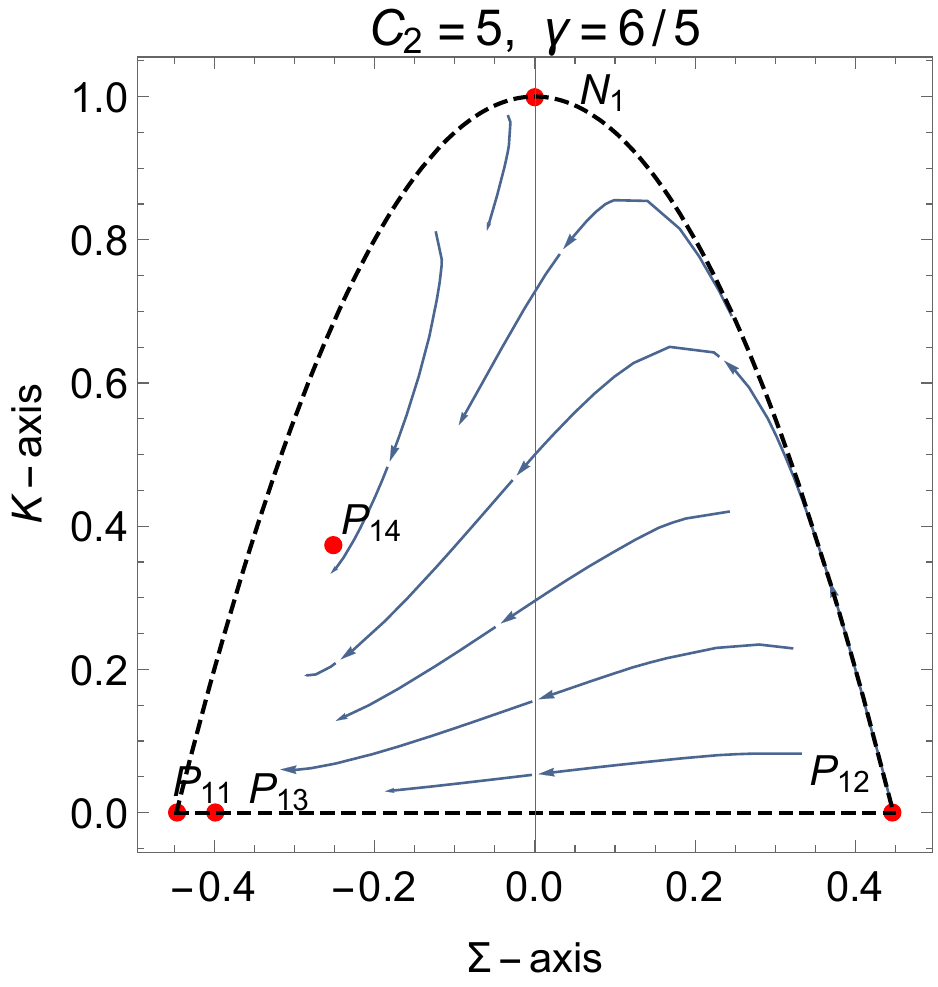} 
    \includegraphics[scale=0.45]{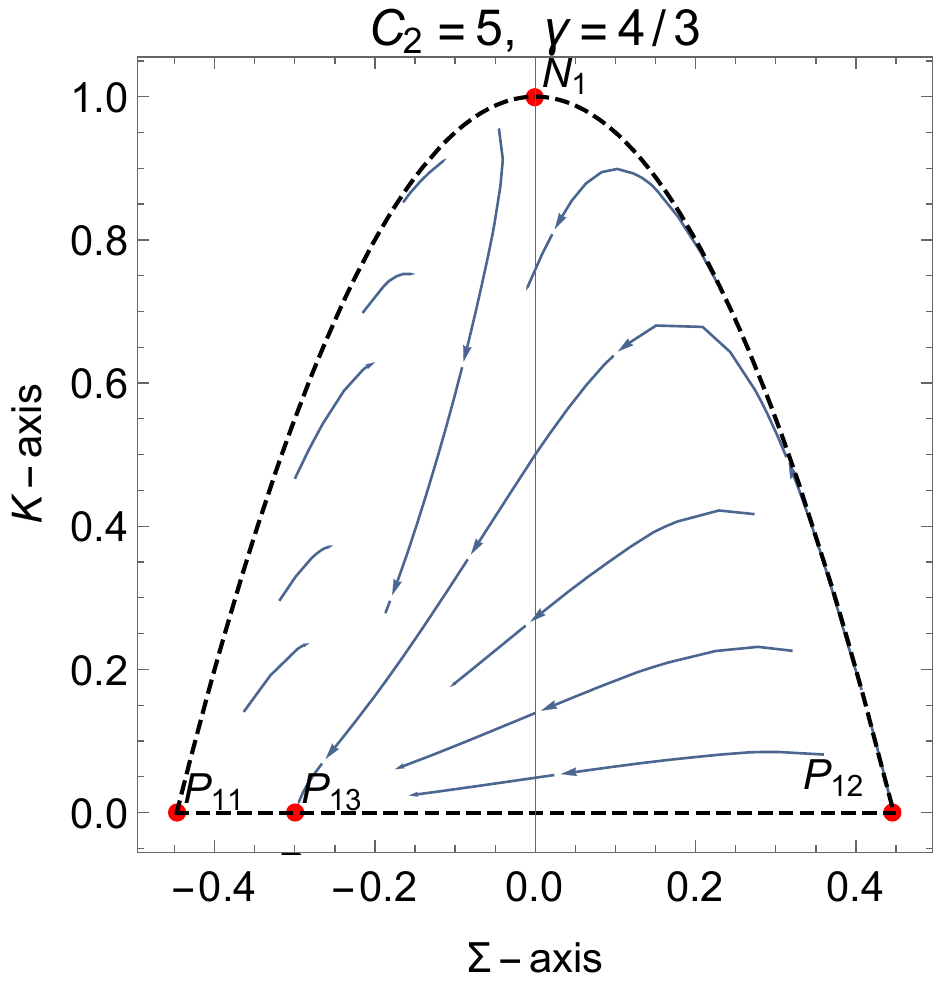} 
    \includegraphics[scale=0.45]{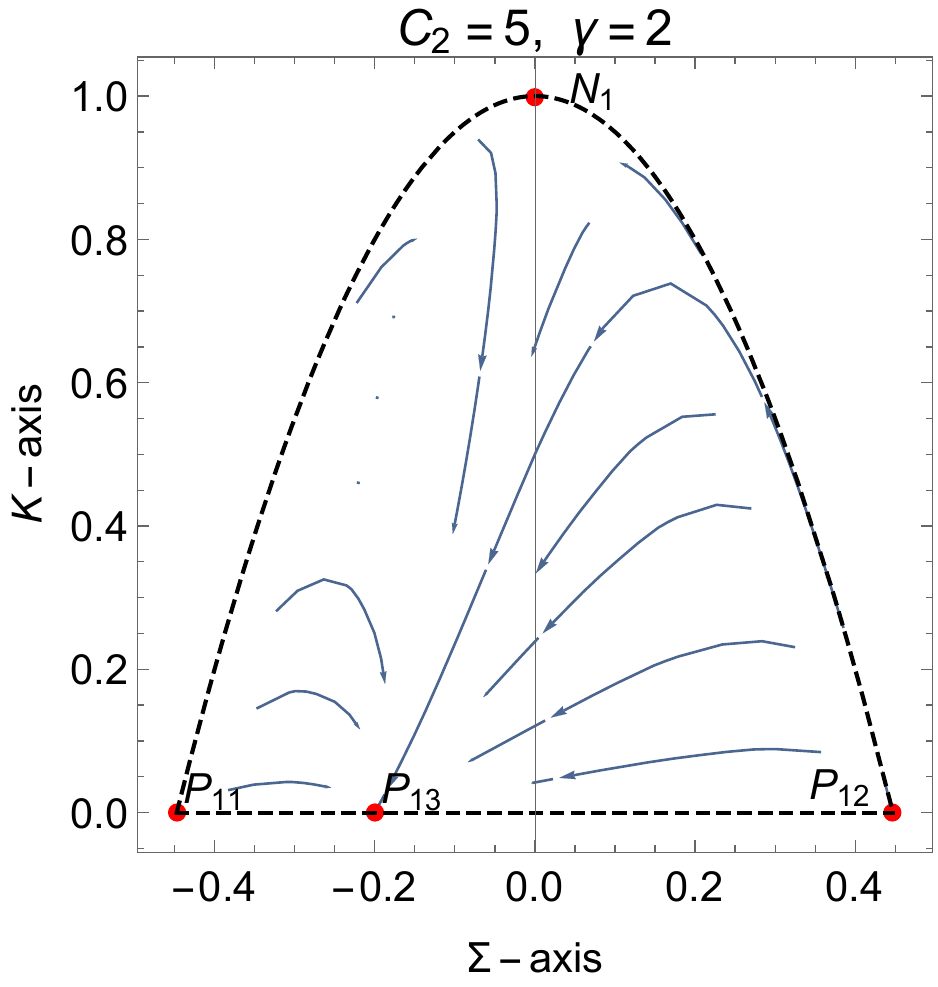} 
    \caption[{Órbitas del sistema \eqref{eq:3.57} con $v=0, \quad A=0$ con $1<\gamma \leq 2$}]{\label{Bla} Órbitas del sistema \eqref{eq:3.57} con $v=0, \quad A=0$ con $1<\gamma \leq 2$.}
   \end{figure*}

En la figura \ref{Bla} se muestran algunas órbitas en el retato de fase del sistema \eqref{eq:3.57} con $v= A=0$ y $1<\gamma \leq 2$.
\begin{table*}
\caption{\label{Tab2Avcero} Análisis cualitativo de los puntos de equilibrio del sistema \eqref{eq:3.57} con $v=0, \quad A=0$ con $1<\gamma \leq 2$. Se incluye la línea $N_1$. }
\begin{tabular}{|m{0.7cm}|m{2.8cm}|m{4.0cm}|m{3.5cm}|m{2.3cm}|}
\hline
Etiq. &  $(\Sigma, K)$ & Autovalores \newline (plano $\Sigma$--$K$)  & Estabilidad  (plano $\Sigma$--$K$)  & ${ \Omega_t}$\\
 \hline
$N_1$ & $(0, 1)$ & $\{-1,2\}$  & Silla & $0$\\ \hline
$P_{11}$ & $\left(-\frac{1}{\sqrt{C_2}}, 0, \right)$ & $\left\{2,\frac{2-3 \gamma }{(\gamma -1) \sqrt{C_2}}+4\right\}$  &
Silla para \newline 
$\scriptscriptstyle 0<C_2<\frac{(2-3 \gamma )^2}{16 (\gamma -1)^2}.$ \newline
No hiperbólico para \newline $\scriptscriptstyle C_2=\frac{(2-3 \gamma )^2}{16 (\gamma -1)^2}$. \newline
\newline  Fuente local para \newline
$\scriptscriptstyle C_2>\frac{(2-3 \gamma )^2}{16 (\gamma -1)^2}$. & $0$ \\\hline
$P_{12}$ & $\left(\frac{1}{\sqrt{C_2}}, 0 \right)$ & $\left\{2,\frac{3 \gamma -2}{(\gamma -1) \sqrt{C_2}}+4\right\}$ & Fuente local para $C_2>0$. & $0$ \\\hline
$P_{13}$ & $\left(-\frac{2-3 \gamma }{4 C_2(1- \gamma)}, 0\right)$ & $\Bigg\{\frac{(2-3 \gamma )^2-16 (\gamma -1)^2 C_2}{8 (\gamma -1)^2 C_2}, \newline \frac{(2-3 \gamma
   )^2-8 (\gamma -1)^2 C_2}{4 (\gamma -1)^2 C_2}\Bigg\}$  & Fuente local para \newline $\scriptscriptstyle 0<C_2<\frac{(2-3 \gamma )^2}{16 (\gamma -1)^2}$. \newline 
   Silla para \newline
   $\scriptscriptstyle \frac{(2-3 \gamma )^2}{16 (\gamma -1)^2}<C_2<\frac{(2-3 \gamma )^2}{8 (\gamma
   -1)^2}$.
   \newline
   No hiperbólico para \newline
   $\scriptscriptstyle   C_2=\frac{(2-3 \gamma )^2}{16 (\gamma -1)^2}$, o  \newline $\scriptscriptstyle   C_2=\frac{(2-3 \gamma )^2}{8 (\gamma
   -1)^2}$. 
   \newline
    Atractor local para \newline 
    $\scriptscriptstyle C_2>\frac{(2-3 \gamma )^2}{8 (\gamma -1)^2}$ & $\scriptscriptstyle\frac{16 (\gamma -1)^2 C_2-(2-3 \gamma )^2}{16 (\gamma -1)^3 C_2}$\\\hline
   $P_{14}$ & $\scriptscriptstyle\Bigg(-\frac{2 (\gamma -1)}{3 \gamma -2}, \newline \frac{(2-3 \gamma )^2-8 (\gamma -1)^2 C_2}{(2-3 \gamma )^2}\Bigg)$. & $\scriptscriptstyle \Bigg\{-\frac{1}{2}+\frac{\sqrt{64 (\gamma
   -1)^2 C_2-7 (2-3 \gamma )^2}}{4-6 \gamma }, \newline -\frac{1}{2}-\frac{\sqrt{64 (\gamma -1)^2
   C_2-7 (2-3 \gamma )^2}}{6 \gamma -4}\Bigg\}$ & Atractor local si \newline $\scriptscriptstyle 0<C_2\leq \frac{7 (2-3 \gamma )^2}{64 (\gamma -1)^2}$, o 
   \newline $\scriptscriptstyle \frac{7 (2-3 \gamma )^2}{64 (\gamma -1)^2}<C_2<\frac{(2-3 \gamma
   )^2}{8 (\gamma -1)^2}$. \newline No hiperbólico si \newline 
   $\scriptscriptstyle C_2=\frac{(2-3 \gamma )^2}{8 (\gamma -1)^2}$.
   \newline Silla si \newline $\scriptscriptstyle C_2>\frac{(2-3 \gamma )^2}{8 (\gamma -1)^2}$ & $\scriptscriptstyle \frac{4 (\gamma -1) C_2}{(2-3 \gamma )^2}$. \\\hline
\end{tabular}
\end{table*}

\FloatBarrier

\section{Discusión}
\label{DiscusionC1}
En este capítulo, se analizaron cualitativamente sistemas de ecuaciones diferenciales asociados a un modelo con métricas temporales autosimilares esféricamente simétricos con fluido perfecto dado por \eqref{modelouno} utilizando las herramientas presentadas en el capítulo \ref{ch_1}. Se presentaron las ecuaciones normalizadas con la variable $\theta$ y se estudiaron cuatro modelos específicos, estos son: modelos con inclinación extrema \eqref{extreme_tilt_0}; fluido perfecto sin presión \eqref{3.41-3.43}; el sistema reducido \eqref{eq:3.57} en el conjunto invariante $A=v=0$;  y el sistema general \eqref{reducedsyst}. Los puntos hiperbólicos fueron clasificados de acuerdo a sus condiciones de estabilidad de acuerdo al teorema \ref{hartgrob} mientras que los puntos no hiperbólicos fueron clasificados como sillas.

Además, fue posible recuperar los resultados obtenidos en \cite{Goliath:1998mx}. En la siguiente lista se presentan los puntos obtenidos en \cite{Goliath:1998mx}  y su correspondencia con los puntos discutidos en este capítulo: 
\begin{enumerate}

    \item $SL_{\pm}$: Líneas sónicas definidas por ${A}=-\frac{\gamma 
   \varepsilon  (\gamma  (\Sigma
   +2)-2)}{4 (\gamma
   -1)^{3/2}}$,
$v=\varepsilon\sqrt{\gamma -1}$, fueron analizadas en la sección \ref{SL}. A diferencia de relatividad general, para  cuando $1<\gamma<2$ el sistema \eqref{reducedsyst} admite los puntos de equilibrio 
\begin{enumerate}
    \item $SL_1: C_2= \frac{\gamma
   ^2}{4 (\gamma -1)^2},\Sigma =
   \frac{2 (\gamma -1)}{\gamma
   },v=\sqrt{\gamma -1},A=
  - \frac{\gamma  (\gamma  (\Sigma
   +2)-2)}{4 (\gamma
   -1)^{3/2}}$, 
    \item $SL_2: C_2= \frac{\gamma
   ^2}{4 (\gamma -1)^2},\Sigma =
   -\frac{2 (\gamma -1)}{\gamma
   },v= -\sqrt{\gamma -1},A=
   \frac{\gamma  (\gamma  (\Sigma
   +2)-2)}{4 (\gamma
   -1)^{3/2}}$,
\end{enumerate}
los cuáles yacen en la línea sónica.  Si $\gamma=2, C_2=1$ dichos puntos existen, y como $\gamma=2$  el fluido se comporta como materia rígida. Adicionalmente, si $\gamma=2, C_2=1$, estos puntos corresponden a modelos con inclinación extrema ($v=\varepsilon$), $SL_1: \Sigma=1, A=-2, v=1$, y $SL_2: \Sigma=-1, A=0, v=-1$. 

    \item $C^0=({\Sigma}, {A}, v)=(0,0,0),\quad ({K},\Omega_t)=(1,0)$, corresponde a $N_1$. 
    
    \item $C^{\pm }=({\Sigma}, {A}, v)=(0,0,\pm 1),\quad ({K},\Omega_t)=(1,0)$, corresponden a $N_{2,3}$. 
    
    \item $K^{\pm}_-=({\Sigma}, {A}, v)=(-1,0,\pm 1),\quad ({K},\Omega_t)=(0,0)$, corresponden a $P_{1,2}$ cuando $C_2=1$.
    
    \item $K^{\pm}_+=({\Sigma}, {A}, v)=(1,0,\pm 1),\quad ({K},\Omega_t)=(0,0)$, corresponden a $P_{3,4}$ cuando $C_2=1$.
    
    \item $M^+=({\Sigma}, {A}, v)=(0,1,1),\quad ({K},\Omega_t)=(0,0)$, corresponde a  $P_5$ cuando $C_2=1$.
    
    \item $M^-=({\Sigma}, {A}, v)=(0,1,-1),\quad ({K},\Omega_t)=(0,0)$, corresponde a $P_6$ cuando $C_2=1$.
    
    \item $H^-$: La recta definida por $ {A}( {\Sigma})= {\Sigma}  +1$, $v( {\Sigma})=-1$,$\quad (0,-2 {\Sigma} {A})$. 
    
    \item $K^0_-=({\Sigma}, {A}, v)=(-1,0,0),\quad ({K},\Omega_t)=(0,0)$, corresponde a $P_{11}$ cuando $C_2=1$.
    
    \item $K^0_+=({\Sigma}, {A}, v)=(1,0,0),\quad ({K},\Omega_t)=(0,0)$,  corresponde a $P_{12}$ cuando $C_2=1$. 
    
    \item $T=({\Sigma}, {A}, v)=\left(-2\frac{\gamma-1}{3\gamma-2},0,0\right),\quad ({K},\Omega_t)=\left(\frac{\gamma^2+4(\gamma-1)}{(3\gamma -2},\frac{4(\gamma-1)}{(3\gamma -2}\right)$,  corresponde a $P_{13}$ cuando $C_2=1$. 
    
    \item $\widetilde{M}^{\pm}=({\Sigma}, {A}, v)=\left(0,1,\frac{(\gamma -1) \gamma  \pm \left(\sqrt{(\gamma -1) \left((\gamma
   -1) \gamma ^2+(2-\gamma ) (3 \gamma -2)\right)}\right)}{2-\gamma }\right),\quad ({K},\Omega_t)=(0,0)$, existe si $C_2=1$.
    
   \end{enumerate}
Los autores de \cite{Goliath:1998mx} utilizan la notación $\text{Kernel}^{\text{sgn}(v)}_{\text{sgn}({\Sigma})}$, cuando no hay confusión se puede omitir $\text{sgn}(v)$ o $\text{sgn}({\Sigma})$ el kernel indica la interpretación del punto: $M,C$ representan el espacio tiempo de Minkowski; $K$ representa una solución de Kasner; $T$ corresponde a soluciones estáticas; $SL_{\pm}$ corresponde a un espacio plano FLRW y orbitas estáticas dependiendo del parámetro $\gamma$; La línea  de puntos de equilibrio $H^-$ se asocia con un cambio de causalidad del campo vectorial homotético. Los puntos $\widetilde {M}^{\pm}$ son puntos de equilibrio de  \eqref{reducedsyst}, sólo si $C_2=1$. 

\lhead{Capítulo \ref{ch_4}}
\rhead{Modelos con campo escalar}
\cfoot{\thepage}
\renewcommand{\headrulewidth}{1pt}
\renewcommand{\footrulewidth}{1pt}

\chapter{Modelo con campo escalar}\label{ch_4}

En este capítulo se estudian métricas temporales, auto-similares y esféricamente simétricas (o sea, conformalmente estáticas) en teoría Einsten-æther con fluido perfecto y con campo escalar con potencial exponencial que respeta la simetría homotética. Se analizan cualitativamente cuatro sistemas de ecuaciones diferenciales ordinarias (y varios subsistemas) que modelan distintas situaciones de interés físico. Se procede de la misma manera que en el capítulo anterior, presentando condiciones de estabilidad de los puntos de equilibrio.

\section{Modelos temporales autosimilares esféricamente simétricos con campo escalar}
\label{SECT:4.1}
Como hemos comentado con anterioridad,  para  que un campo escalar no homogéneo $\phi(t,x)$ con potencial  $V(\phi(t,x))$ cumpla la simetría conformal estática tienen que tener la forma \cite{Coley:2002je}:
\begin{align}
\label{phi_respect_symm}
& \phi(t,x)=\psi (x)-\lambda t, \quad  V(\phi(t,x))= e^{-2 t} U(\psi(x)), \quad  U(\psi)=U_0 e^{-\frac{2 \psi}{\lambda}},    
\end{align}
donde hemos asumido, por conveniencia, $\lambda>0$, tal que para $\psi>0$, $U\rightarrow 0$ cuando $\lambda \rightarrow 0$.
\newline lUego, el tensor de energía-momentum del campo escalar estará dado por
\begin{equation}
{T^{\psi}}_{a}^{b}=\left(
\begin{array}{cccc}
 \mu_\phi & q_\phi  & 0 & 0 \\
 q_\phi & p_\phi-2\pi_\phi & 0 & 0 \\
 0 & 0 & p_\phi+\pi_\phi & 0 \\
 0 & 0 & 0 & p_\phi+\pi_\phi \\
\end{array}
\right),
\end{equation}
donde $\widehat{...}$ denota la derivada con respecto a la variable espacial $x$ y 
\begin{subequations}
\begin{small}
\begin{align}
& \mu_\phi:= \frac{1}{2} \e_0(\phi)^2 + \frac{1}{2}\e_1(\phi)^2 +V(\phi) =\frac{1}{2} \lambda ^2 e^{-2 t} {b_1}^2+U_0 e^{-2 t-\frac{2 \psi (x)}{\lambda }}+\frac{1}{2} e^{-2 t} {\widehat{\psi}}^2,\\
& p_\phi:= \frac{1}{2} \e_0(\phi)^2 - \frac{1}{6}\e_1(\phi)^2 -V(\phi) = \frac{1}{2} \lambda ^2 e^{-2 t} {b_1}^2-U_0 e^{-2 t-\frac{2 \psi (x)}{\lambda
   }}-\frac{1}{6} e^{-2 t} {\widehat{\psi}}^2,\\
& q_\phi:=  -\e_0(\phi)  \e_1(\phi)=\lambda  e^{-2 t} {b_1} {\widehat{\psi}},\\
&\pi_\phi:= - \frac{1}{3}\e_1(\phi)^2=-\frac{1}{3} e^{-2 t} {\widehat{\psi}}^2.   
\end{align}
\end{small}
\end{subequations}

\noindent{\bf Ecuaciones de propagación:} 
\label{modelodos}
\begin{small}
\begin{subequations}
\begin{align}
&\widehat{ {\theta}}= -\sqrt{3} {b_2}^2-\frac{\sigma \left(2 C_2 \sigma+\theta\right)}{\sqrt{3}}- \sqrt{3} \Psi^2-\frac{\sqrt{3} \gamma \mu_{t} v^2}{(\gamma -1) v^2+1},
\\
&\widehat{ {\sigma}}=-\frac{\sqrt{3} \lambda ^2 {b_1}^2}{C_2}+\frac{\sqrt{3} U_0 e^{-\frac{2 \psi}{\lambda
   }}}{C_2}-\frac{\sigma  (2 \theta +\sigma)}{\sqrt{3}}  +\frac{\sqrt{3} \mu _{t} \left(-3 \gamma +(\gamma -2) v^2+2\right)}{2 C_2 \left((\gamma -1) v^2+1\right)}, \\
&\widehat{b_1}= \frac{b_1  {\sigma}}{\sqrt{3}}, \\
&\widehat{b_2}= -\frac{b_2 ( {\theta}+ {\sigma})}{\sqrt{3}},\\
&\widehat{v}=  \frac{\left(v^2-1\right)}{\sqrt{3} \gamma  \left(\gamma -v^2-1\right)} \Big\{\gamma 
   v (2 (\gamma -1) \theta +\gamma  \sigma )  + \sqrt{3} b_1 \left(3 \gamma ^2-5 \gamma +(\gamma -2) v^2+2\right)\Big\},
 \\
&\widehat{\Psi}= 2 \lambda  {b_1}^2-\frac{(2 \theta+\sigma) \Psi}{\sqrt{3}}-\frac{2 U_0 e^{-\frac{2 \psi}{\lambda }}}{\lambda },
\\
& \widehat{\psi}=\Psi.
\end{align}
\end{subequations}
\end{small}
\noindent{\bf	Ecuación auxiliar:}
\begin{subequations}
\begin{align}
& \widehat{\mu _{t}}=\frac{\mu_{t}}{\sqrt{3} \left(\gamma -v^2-1\right) \left((\gamma -1) v^2+1\right)}  \Big\{\gamma  \left(\sigma+(\gamma -1) v^4 (2 \theta +\sigma )-v ^2 ((4 \gamma -6) \theta+\gamma  \sigma )\right) \nonumber \\
& +2 \sqrt{3} b_1 v \left((7-3 \gamma ) \gamma +(\gamma  (2 \gamma -5)+4) v^2-4\right)\Big\}.
\end{align}
\end{subequations}

\noindent{\bf Restricción:}
\begin{equation}
3 \gamma  \mu_{t}  v- b_1\left((\gamma -1) v^2+1\right) \left(2 \sqrt{3} C_2 \sigma+3 \lambda  \Psi\right)=0.\end{equation}
\noindent{\bf Ecuación para  $\mu_t$:}
	\begin{align}
	&{b_1}^2 \left(C_2+\frac{\lambda ^2}{2}\right)+{b_2}^2+\frac{1}{3} C_2 \sigma^2+\frac{1}{2} \Psi^2=\frac{\theta^2}{3}+U_0 e^{-\frac{2 \psi}{\lambda }}-\frac{\mu_{t} \left(\gamma +v^2-1\right)}{(\gamma -1) v^2+1}   \leq \theta^2+U_0 e^{-\frac{2 \psi}{\lambda }}.
	\end{align}

En general, para $\mu_t\geq 0, 0<\gamma \leq 1, 1-\gamma \leq v^2\leq 1$, o    $\mu_t\geq 0,1<\gamma <2, -1\leq v\leq 1$ asumiendo que $C_2\geq 0$, el término $\theta^2$ es dominante, lo que sugiere usar $\theta$-normalización. 

En la sección \ref{Sect:4.2} se discute la $\theta$-normalizacion de las ecuaciones. Dada la dificultad computacional de la obtención de los puntos de equilibrio del sistema general, en diferentes subsecciones se estudian algunos casos específicos de interés físico. En particular en la sección \ref{gamma=2/3} se discute el caso especial $\Omega_t=v=0$ y $\gamma=2/3$, correspondiente a
un fluido cosmológico en forma de gas ideal con ecuación de estado $p_m =(\gamma-1)\mu_m$, con $\gamma=2/3$ que mimetiza un espacio-tiempo FLRW  con curvatura no nula. 
En la sección \ref{Isotrop} se estudia el  conjunto invariante $\Sigma=0$ (no confundir con los modelos isotrópicos). En la sección \ref{tilt2} se estudian los conjuntos invariantes $v=\pm 1$ que corresponden a inclinación extrema. Finalmente, en la sección \ref{Section4.7} se estudia el conjunto invariante $A=v=0$. En todos los casos se procede de la misma manera que en el capítulo anterior, presentando condiciones de estabilidad de los puntos de equilibrio.  Los resultados parciales se discuten en la sección \ref{SECT:4-3} y en la sección 
\ref{progreso} se comenta brevemente sobre trabajo en progreso. 

\section{$\theta$-Normalizacion de las ecuaciones.}
\label{Sect:4.2}
En esta sección, se consideran las variables normalizadas
\begin{equation}
{\Sigma} =\frac{\sigma}{\theta},\quad  A=\frac{\sqrt{3} b_1}{\theta},\quad K=\frac{3 b_2^2}{\theta^2},\quad {\Omega}_t =\frac{3\mu_t}{\theta^2}, \quad u= \sqrt{\frac{3}{2}} \frac{\Psi}{\theta}, \quad w=\frac{ e^{-\frac{ \psi}{\lambda }}\sqrt{3 U_0}}{\theta},
\end{equation}
\newline junto con la coordenada radial
$$\frac{df}{d \eta} := \frac{\sqrt{3}\widehat{f}}{\theta}.$$
Se define el parámetro ${r}$, de manera análoga al  ``parámetro gradiente de Hubble'' ${r}$, por   $$\widehat{\theta}=-r {\theta}^2,$$ donde 
\begin{equation}
  \sqrt{3} r=  2 C_2 \Sigma ^2+K+\frac{\gamma  {\Omega_t}  v^2}{(\gamma -1) v^2+1}+\Sigma +2 u^2.
\end{equation}
En estas variables, el sistema dinámico está dado por: 
\begin{subequations}
\begin{align}
& \Sigma'= -\frac{A^2 \lambda ^2}{C_2}+2 C_2 \Sigma ^3+\frac{{\Omega_t}  \left(-3 \gamma +2 \gamma  C_2 \Sigma  v^2+(\gamma -2) v^2+2\right)}{2 C_2 \left((\gamma -1) v^2+1\right)}+\frac{w^2}{C_2}+\Sigma  \left(K+2 u^2-2\right),\\
& A'=A \left(2 C_2 \Sigma ^2+K+2 \left(\Sigma +u^2\right)+\frac{\gamma  v^2 {\Omega_t} }{(\gamma -1) v^2+1}\right),\\
& K'=2 K \left(2 C_2 \Sigma ^2+K+2 u^2+\frac{\gamma  v^2 {\Omega_t} }{(\gamma -1) v^2+1}-1\right),\\
& u'= \sqrt{2} A^2 \lambda +2 C_2 \Sigma ^2 u+u \left(K+\frac{\gamma  v^2 {\Omega_t} }{(\gamma -1) v^2+1}-2\right)+2 u^3-\frac{\sqrt{2} w^2}{\lambda },\\
& w'= w\left(2 C_2 \Sigma ^2+K+\Sigma +2 u^2-\frac{\sqrt{2} u}{\lambda }+\frac{\gamma  v^2 {\Omega_t} }{(\gamma -1) v^2+1}\right),\\
& v'=\frac{\left(v^2-1\right) \left(A \left(3 \gamma ^2-5 \gamma +(\gamma -2) v^2+2\right)+\gamma  v (\gamma  (\Sigma +2)-2)\right)}{\gamma  \left(\gamma -v^2-1\right)},\\
& {\Omega_t}'={\Omega_t}  \Bigg(4 C_2 \Sigma
   ^2+2 K+2 \Sigma +4 u^2+\frac{2 \gamma  v^2 {\Omega_t} }{(\gamma -1) v^2+1}+\Bigg) \nonumber\\
   & +{\Omega_t}\Bigg(\frac{2 A v \left((7-3
   \gamma ) \gamma +(\gamma  (2 \gamma -5)+4) v^2-4\right)}{\left((\gamma -1) v^2+1\right) \left(\gamma -v^2-1\right)}\Bigg)\\ \nonumber & +{\Omega_t}\Bigg(\frac{\gamma  \left(\Sigma +v^2 \left(-\gamma  (\Sigma +4)+(\gamma -1) (\Sigma +2) v^2+6\right)\right)}{\left((\gamma -1) v^2+1\right) \left(\gamma -v^2-1\right)}\Bigg).
\end{align}
\end{subequations}
Las restricciones son
\begin{subequations}
\begin{align}
& C_2 \left(A^2+\Sigma ^2\right)+\frac{A^2 \lambda ^2}{2}+K+u^2-w^2 +\frac{{\Omega_t} \left(\gamma +v^2-1\right)}{(\gamma -1) v^2+1} =1,\\
&\gamma  {\Omega_t}  v -A \left((\gamma -1) v^2+1\right) \left(2 C_2 \Sigma +\sqrt{2} \lambda  u\right)=0. 
\end{align}
\end{subequations}
 estas pueden ser resueltas globalmente para  $\Omega_t$ y $K$ para obtener 
\begin{subequations}
\begin{align}
&\Omega_t= \frac{A \left(\gamma  v^2-v^2+1\right) \left(2 C_2 \Sigma +\sqrt{2} \lambda  u\right)}{\gamma  v},\\
& K= \frac{1}{2} \left(-2 A^2 C_2 -\lambda ^2A^2 -2 C_2 \Sigma ^2-2 u^2+2 w^2+2\right) \nonumber\\ & -\frac{A \left(\gamma +v^2-1\right) \left(\gamma  v^2-v^2+1\right) \left(2 C_2 \Sigma +\sqrt{2} \lambda 
   u\right)}{\gamma  v \left((\gamma -1) v^2+1\right)}.
\end{align}
\end{subequations}
Finalmente se obtiene el sistema reducido 5-dimensional 
\begin{subequations}
\label{reducedsystSF}
\begin{align}
& \Sigma'=-\frac{A^2 \lambda ^2 \left(C_2 \Sigma +2\right)}{2 C_2}+C_2 \left(\Sigma ^3-A^2 \Sigma \right)+\frac{w^2}{C_2}+\Sigma  \left(u^2+w^2-1\right) \nonumber \\
   & +\frac{A \left(2 C_2 \Sigma +\sqrt{2} \lambda  u\right) \left(-3 \gamma +2 (\gamma -1) C_2 \Sigma  \left(v^2-1\right)+(\gamma -2) v^2+2\right)}{2 \gamma  C_2 v},\\
& A'= -\frac{1}{2} A^3 \lambda ^2+A \left(C_2 \left(\Sigma ^2-A^2\right)+2 \Sigma +u^2+w^2+1\right)\\ \nonumber &+\frac{A^2 (\gamma -1) \left(v^2-1\right) \left(2
   C_2 \Sigma +\sqrt{2} \lambda  u\right)}{\gamma  v},\\
& v'=\frac{\left(v^2-1\right) \left(A \left(3 \gamma ^2-5 \gamma +(\gamma -2) v^2+2\right)+\gamma  v (\gamma  (\Sigma +2)-2)\right)}{\gamma  \left(\gamma -v^2-1\right)},\end{align}
\begin{align}
& u'= u \left(C_2 \left(\Sigma ^2-A^2\right)+u^2+w^2-1\right)+\sqrt{2} A^2 \lambda -\frac{1}{2} A^2 \lambda ^2 u\\ \nonumber & +\frac{A (\gamma -1) u \left(v^2-1\right) \left(2 C_2 \Sigma +\sqrt{2} \lambda  u\right)}{\gamma  v}-\frac{\sqrt{2} w^2}{\lambda },\\
& w'=-\frac{1}{2} A^2 \lambda ^2
   w+\frac{\sqrt{2} A (\gamma -1) \lambda  u \left(v^2-1\right) w}{\gamma  v}-\frac{\sqrt{2} u w}{\lambda }\nonumber \\
 & + \frac{w \left(C_2 \left(A^2 \gamma  (-v)+2 A (\gamma -1) \Sigma  \left(v^2-1\right)+\gamma  \Sigma ^2 v\right)+\gamma  v \left(\Sigma +u^2+w^2+1\right)\right)}{\gamma  v}.
\end{align}
\end{subequations}
Si se toman  $u=w=0$ y se toma el límite cuando  $\lambda \rightarrow 0$ se recupera el sistema \eqref{reducedsyst}.
Dada la dificultad computacional que enfrentamos al intentar obtener analíticamente todos los puntos de equilibrio del sistema \eqref{reducedsystSF} para hacer un análisis exhaustivo (analítico o numérico) de sus condiciones de estabilidad, en las siguientes secciones estudiaremos algunos casos particulares de interés físico. 

\subsection{Caso especial $\Omega_t=v=0$ y $\gamma=2/3$}
\label{gamma=2/3}
Un fluido cosmologico en forma de gas ideal con ecuación de estado $p_m =(\gamma-1)\mu_m$, con $\gamma=2/3$ describe un espacio-tiempo
FLRW  con curvatura no nula. Las ecuaciones son: 
\begin{subequations}
\begin{align}
& \Sigma'=-\frac{\left(A^2 \lambda ^2-w^2\right)}{C_2}+2 C_2 \Sigma ^3+\Sigma  \left(K+2 u^2-2\right),\\ 
& A'=A \left(2 C_2 \Sigma ^2+K+2 \left(\Sigma +u^2\right)\right), \\
& K'=2 K \left(2 C_2 \Sigma ^2+K+2 u^2-1\right),\\
& u'= \frac{\sqrt{2} \left(A^2 \lambda ^2-w^2\right)}{\lambda }+u \left(2 C_2 \Sigma ^2+K-2\right)+2 u^3,\\
& w'= w \left(2 C_2 \Sigma ^2+K+\Sigma +2 u^2-\frac{\sqrt{2} u}{\lambda }\right),  \end{align}
\end{subequations}
con restricciones 
\begin{subequations}
\begin{align}
& -\frac{1}{3} A \left(2 C_2 \Sigma +\sqrt{2} \lambda  u\right)=0,\\
& \frac{1}{6} \left(-2 C_2 \left(A^2+\Sigma ^2\right)-A^2 \lambda ^2-2 K-2 u^2+2 w^2+2\right)=0.
\end{align}
\end{subequations}
Estas se pueden resolver para obtener
\begin{subequations}
\begin{align}
   & u= -\frac{\sqrt{2} C_2 \Sigma }{\lambda }, \\
   & K=  -C_2 \left(A^2+\Sigma ^2\right)-\frac{A^2 \lambda ^2}{2}-\frac{2 C_2^2 \Sigma ^2}{\lambda ^2}+w^2+1,
\end{align}
\end{subequations}
esto permite obtener el sistema reducido 3-dimensional
\begin{subequations}
\label{scalar-field-A}
\begin{align}
&\Sigma'=C_2 \left(\Sigma ^3-A^2 \Sigma \right)+\frac{1}{2} \Sigma  \left(-A^2 \lambda ^2+2 w^2-2\right)+\frac{(w-A \lambda ) (A \lambda +w)}{C_2}+\frac{2 C_2^2 \Sigma ^3}{\lambda ^2},\\
& A'= -\frac{1}{2} A^3 \lambda ^2+A C_2 \left(\Sigma
   ^2-A^2\right)+\frac{2 A C_2^2 \Sigma ^2}{\lambda ^2}+A \left(2 \Sigma +w^2+1\right), \\
& w'=w \left(\frac{C_2 \left(\lambda ^2 \left(\Sigma ^2-A^2\right)+2 C_2 \Sigma ^2+2 \Sigma \right)}{\lambda ^2}-\frac{A^2 \lambda ^2}{2}+\Sigma
   +w^2+1\right).
\end{align}
\end{subequations}
Los puntos de equilibrio del sistema \eqref{scalar-field-A} son los siguientes. 
\begin{enumerate}
   \item $N_1:(\Sigma,A,w)=(0,0,0)$ con autovalores $\{-1,1,1\}$ es silla hiperbólica.
    \item $Q_1:(\Sigma,A,w)=\left(-\frac{1}{2}, \frac{\sqrt{C_2}}{\sqrt{2} \lambda },0\right)$, con autovalores \\$\left\{\frac{1}{2}-\frac{{C_2}}{\lambda ^2},\frac{1}{2} \left(-\frac{\sqrt{8
   {C_2}^2+(4 {C_2}-7) \lambda ^2}}{\lambda }-1\right),\frac{1}{2}
   \left(\frac{\sqrt{8 {C_2}^2+(4 {C_2}-7) \lambda ^2}}{\lambda
   }-1\right)\right\}$. Es
   \begin{enumerate}
       \item pozo hiperbólico para 
       \begin{enumerate}
           \item $0<\lambda \leq \frac{\sqrt{7}}{2}, \; \frac{\lambda}{4} \sqrt{\lambda ^2+14}-\frac{\lambda^2}{4}<{C_2}<
           \frac{\lambda}{4}\sqrt{\lambda ^2+16}-\frac{\lambda ^2}{4}$, o   
            \item $\frac{\sqrt{7}}{2}<\lambda <\sqrt{2}, \; \frac{\lambda ^2}{2}<{C_2}<\frac{\lambda}{4} \sqrt{\lambda ^2+16}-\frac{\lambda ^2}{4}$, o   
             \item $0<\lambda <\frac{\sqrt{7}}{2}, \; \frac{\lambda ^2}{2}<{C_2}\leq \frac{1}{4} \sqrt{\lambda ^4+14
   \lambda ^2}-\frac{\lambda ^2}{4}$.
       \end{enumerate}
       \item silla hiperbólica para
       \begin{enumerate}
           \item  $0<\lambda \leq \sqrt{2}, \; {C_2}>\frac{\lambda}{4} \sqrt{\lambda ^2+16 }-\frac{\lambda ^2}{4}$, o   
           \item $\lambda >\sqrt{2}, \; {C_2}>\frac{\lambda ^2}{2}$, o   
           \item $\lambda >\sqrt{2}, \; \frac{\lambda}{4} \sqrt{\lambda ^2+16}-\frac{\lambda
   ^2}{4}<{C_2}<\frac{\lambda ^2}{2}$, o   
            \item $\frac{\sqrt{7}}{2}<\lambda \leq \sqrt{2}, \; \frac{\lambda}{4} \sqrt{\lambda ^2+14}-\frac{\lambda
   ^2}{4}<{C_2}<\frac{\lambda ^2}{2}$, o   
            \item $\lambda >\sqrt{2}, \; \frac{\lambda}{4} \sqrt{\lambda ^2+14}-\frac{\lambda
   ^2}{4}<{C_2}<\frac{\lambda}{4} \sqrt{\lambda ^2+16}-\frac{\lambda ^2}{4}$, o   
            \item $0<\lambda \leq \frac{\sqrt{7}}{2}, \; 0<{C_2}<\frac{\lambda ^2}{2}$, o   
            \item $\lambda >\frac{\sqrt{7}}{2}, \; 0<{C_2}\leq \frac{\lambda}{4} \sqrt{\lambda ^2+14}-\frac{\lambda ^2}{4}$.
       \end{enumerate}
       \item no hiperbólico para
       \begin{enumerate}
           \item $\lambda >0, \; {C_2}=\frac{\lambda ^2}{2}$, o   
           \item $\lambda >0, \; {C_2}=\frac{\lambda}{4} \sqrt{\lambda ^2+16}-\frac{\lambda ^2}{4}$.
       \end{enumerate}
   \end{enumerate}
   \item $Q_2:(\Sigma,A,w)=\left(-\frac{\lambda ^2}{C_2 \left(2 C_2+\lambda ^2\right)}, \frac{\sqrt{4 C_2^2+2
   (C_2-1) \lambda ^2}}{\sqrt{C_2} \left(2 C_2+\lambda ^2\right)},0\right)$ con autovalores \\ $\left\{\frac{1}{C_2}-\frac{4}{2 C_2+\lambda ^2},-\frac{8}{2 C_2+\lambda
   ^2}+\frac{4}{C_2}-2,-\frac{4}{2 C_2+\lambda ^2}+\frac{2}{C_2}-2\right\}$ existe para  $\lambda >0,\\   {C_2}\geq \frac{1}{4} \lambda 
   \sqrt{\lambda ^2+8}-\frac{\lambda ^2}{4}$ y es
    \begin{enumerate}
        \item pozo hiperbólico para 
        \begin{enumerate}
            \item $0<\lambda \leq \sqrt{2}, C_2>\frac{\lambda}{4} \sqrt{\lambda ^2+16}-\frac{\lambda ^2}{4}$, o   
            \item $\lambda >\sqrt{2}, C_2>\frac{\lambda ^2}{2}$.
        \end{enumerate}
        \item silla hiperbólica para 
        \begin{enumerate}
            \item $0<\lambda \leq 1, \frac{\lambda}{4} \sqrt{\lambda ^2+8}-\frac{\lambda
   ^2}{4}<C_2<\frac{\lambda}{4} \sqrt{\lambda ^2+16}-\frac{\lambda ^2}{4}$, o   
            \item $1<\lambda <\sqrt{2}, \frac{\lambda ^2}{2}<C_2<\frac{\lambda}{4} \sqrt{\lambda ^2+16}-\frac{\lambda ^2}{4}$, o   
            \item $1<\lambda \leq \sqrt{2}, \frac{\lambda}{4} \sqrt{\lambda ^2+8}-\frac{\lambda
   ^2}{4}<C_2<\frac{\lambda ^2}{2}$, o   
            \item $\lambda >\sqrt{2}, \frac{\lambda}{4} \sqrt{\lambda ^2+8}-\frac{\lambda
   ^2}{4}<C_2<\frac{\lambda}{4} \sqrt{\lambda ^2+16}-\frac{\lambda ^2}{4}$, o   
            \item $\lambda >\sqrt{2}, \frac{\lambda}{4} \sqrt{\lambda ^2+16}-\frac{\lambda
   ^2}{4}<C_2<\frac{\lambda ^2}{2}$.
        \end{enumerate}
        \item no hiperbólico para
        \begin{enumerate}
            \item $\lambda \geq 1, C_2=\frac{\lambda ^2}{2}$, o   
            \item $C_2=\frac{1}{4} \left(-\lambda ^2-\sqrt{\lambda ^2+16} \lambda \right)$, o   
            \item $C_2=\frac{1}{4} \left(\lambda  \sqrt{\lambda ^2+16}-\lambda ^2\right)$, o   
            \item $C_2=\frac{1}{4} \left(-\lambda ^2-\sqrt{\lambda ^2+8} \lambda \right)$, o   
            \item $C_2=\frac{1}{4} \left(\lambda  \sqrt{\lambda ^2+8}-\lambda ^2\right)$.
        \end{enumerate}
    \end{enumerate}
    \item $Q_3:(\Sigma,A,w)=\left(-\frac{\lambda }{\sqrt{C_2 \left(2 C_2+\lambda ^2\right)}},0,0\right)$ con autovalores\\ $\left\{2,2-\frac{2 \lambda }{\sqrt{C_2 \left(2 C_2+\lambda
   ^2\right)}},2-\frac{\sqrt{C_2 \left(2 C_2+\lambda ^2\right)}}{C_2 \lambda }\right\}$ y es 
   \begin{enumerate}
       \item fuente hiperbólica para
       \begin{enumerate}
           \item $\frac{1}{\sqrt{2}}<\lambda \leq 1, C_2>\frac{\lambda ^2}{4 \lambda ^2-2}$, o   
           \item $\lambda >1, C_2>\frac{\lambda}{4} \sqrt{\lambda ^2+8}-\frac{\lambda ^2}{4}$.
       \end{enumerate}
       \item silla hiperbólica para
       \begin{enumerate}
           \item $\lambda >0, 0<C_2<\frac{\lambda}{4} \sqrt{\lambda ^2+8}-\frac{\lambda ^2}{4}$, o   
           \item $0<\lambda \leq \frac{1}{\sqrt{2}},  {C_2}>0$, o   
           \item $\lambda >1, \frac{\lambda ^2}{4 \lambda ^2-2}<C_2<\frac{1}{4}
   \sqrt{\lambda ^4+8 \lambda ^2}-\frac{\lambda ^2}{4}$, o  
           \item $0<\lambda \leq \frac{1}{\sqrt{2}}, C_2>\frac{1}{4} \sqrt{\lambda ^4+8
   \lambda ^2}-\frac{\lambda ^2}{4}$, o  
           \item $\frac{1}{\sqrt{2}}<\lambda <1, \frac{1}{4} \sqrt{\lambda ^4+8 \lambda
   ^2}-\frac{\lambda ^2}{4}<C_2<\frac{\lambda ^2}{4 \lambda ^2-2}$, o  
           \item $\lambda >\frac{1}{\sqrt{2}}, 
   0<{C_2}<\frac{\lambda ^2}{4 \lambda ^2-2}$.
       \end{enumerate}
       \item no hiperbólico para
       \begin{enumerate}
           \item $0<\lambda \leq \frac{1}{\sqrt{2}}, C_2=\frac{1}{4} \sqrt{\lambda ^4+8
   \lambda ^2}-\frac{\lambda ^2}{4}$, o  
           \item $\frac{1}{\sqrt{2}}<\lambda <1, C_2=\frac{1}{4} \sqrt{\lambda ^4+8
   \lambda ^2}-\frac{\lambda ^2}{4}$, o  
           \item $\frac{1}{\sqrt{2}}<\lambda <1, C_2=\frac{\lambda ^2}{4 \lambda ^2-2}$, o  
           \item $\lambda =1, C_2=\frac{1}{2}$, o  
           \item $\lambda >1, C_2=\frac{\lambda ^2}{4 \lambda ^2-2}$, o  
           \item $\lambda >1, C_2=\frac{1}{4} \sqrt{\lambda ^4+8 \lambda
   ^2}-\frac{\lambda ^2}{4}$.
       \end{enumerate}
   \end{enumerate}

   \item $Q_4:(\Sigma,A,w)=\left(\frac{\lambda }{\sqrt{C_2 \left(2 C_2+\lambda ^2\right)}},0,0\right)$ con autovalores \\ $\left\{2,\frac{2 \lambda }{\sqrt{C_2 \left(2 C_2+\lambda ^2\right)}}+2,\frac{\sqrt{C_2
   \left(2 C_2+\lambda ^2\right)}}{C_2 \lambda }+2\right\}$
  es fuente hiperbólica para $\lambda >0, C_2>0$.

   \item $Q_{5,6}:(\Sigma,A,w)=\left(-\frac{1}{2 {C_2}},0,\pm \frac{\sqrt{{C_2} \left(2-4 \lambda ^2\right)+\lambda ^2}}{2
   \sqrt{{C_2}} \lambda }\right)$ con autovalores \\ $\left\{\frac{1}{\lambda ^2}-\frac{1}{2 {C_2}},\frac{1}{{C_2}}+\frac{2}{\lambda ^2}-2,\frac{1}{2
   {C_2}}+\frac{1}{\lambda ^2}-2\right\}$ existe para $0<\lambda \leq \frac{1}{\sqrt{2}},  {C_2}>0$ o $\lambda >\frac{1}{\sqrt{2}},  0<{C_2}\leq \frac{\lambda ^2}{4
   \lambda ^2-2}$ y es
   \begin{enumerate}
       \item fuente hiperbólica para
       \begin{enumerate}
           \item $0<\lambda \leq \frac{1}{\sqrt{2}}, \; {C_2}>\frac{\lambda ^2}{2}$, o   
           \item $\frac{1}{\sqrt{2}}<\lambda <1, \; \frac{\lambda ^2}{2}<{C_2}<\frac{\lambda ^2}{4 \lambda ^2-2}$.
       \end{enumerate}
       \item silla hiperbólica para 
       \begin{enumerate}
           \item $0<\lambda \leq 1, 0<C_2<\frac{\lambda ^2}{2}$, o   
           \item $\lambda >1, 0<C_2<\frac{\lambda ^2}{4 \lambda ^2-2}$. 
       \end{enumerate}
       \item no hiperbólico para 
       \begin{enumerate}
           \item $C_2=\frac{1}{2}, \lambda =1$, o   
           \item $C_2=\frac{\lambda ^2}{2}, 0<\lambda <1$, o   
           \item $C_2=\frac{\lambda ^2}{4 \lambda ^2-2}, \lambda >1$, o   
           \item $C_2=\frac{\lambda ^2}{4 \lambda ^2-2}, \frac{1}{\sqrt{2}}<\lambda <1$.
       \end{enumerate}
   \end{enumerate}

   \item $Q_{7,8}:(\Sigma,A,w)=\left(0, \frac{1}{\sqrt{C_2-\frac{\lambda ^2}{2}}},\pm \frac{\lambda }{\sqrt{C_2-\frac{\lambda
   ^2}{2}}}\right)$ con autovalores\\  $\left\{-\frac{\mu_1(\lambda,C_2)}{C_2 \lambda  \left(2 C_2-\lambda ^2\right)^{3/2}}  ,-\frac{\mu_2(\lambda,C_2)}{C_2 \lambda  \left(2 C_2-\lambda ^2\right)^{3/2}}  ,-\frac{\mu_3(\lambda,C_2)}{C_2 \lambda  \left(2 C_2-\lambda ^2\right)^{3/2}}  \right\}$ donde las expresiones para los $\mu_i$ dependen de $\lambda, C_2$. Existe para $C_2\geq \frac{\lambda ^2}{2}$. En la figura \ref{parte real}, se presenta un plot 3D donde se muestran las partes reales de los autovalores $\mu_i$. Esto ilustra que los puntos de equilibrio son típicamente sillas o no hiperbólicos con tres autovalores nulos dependiendo de la elección de los parámetros. 
\end{enumerate}
     
\begin{figure*}[!t]
    \centering
    \includegraphics[scale=0.5]{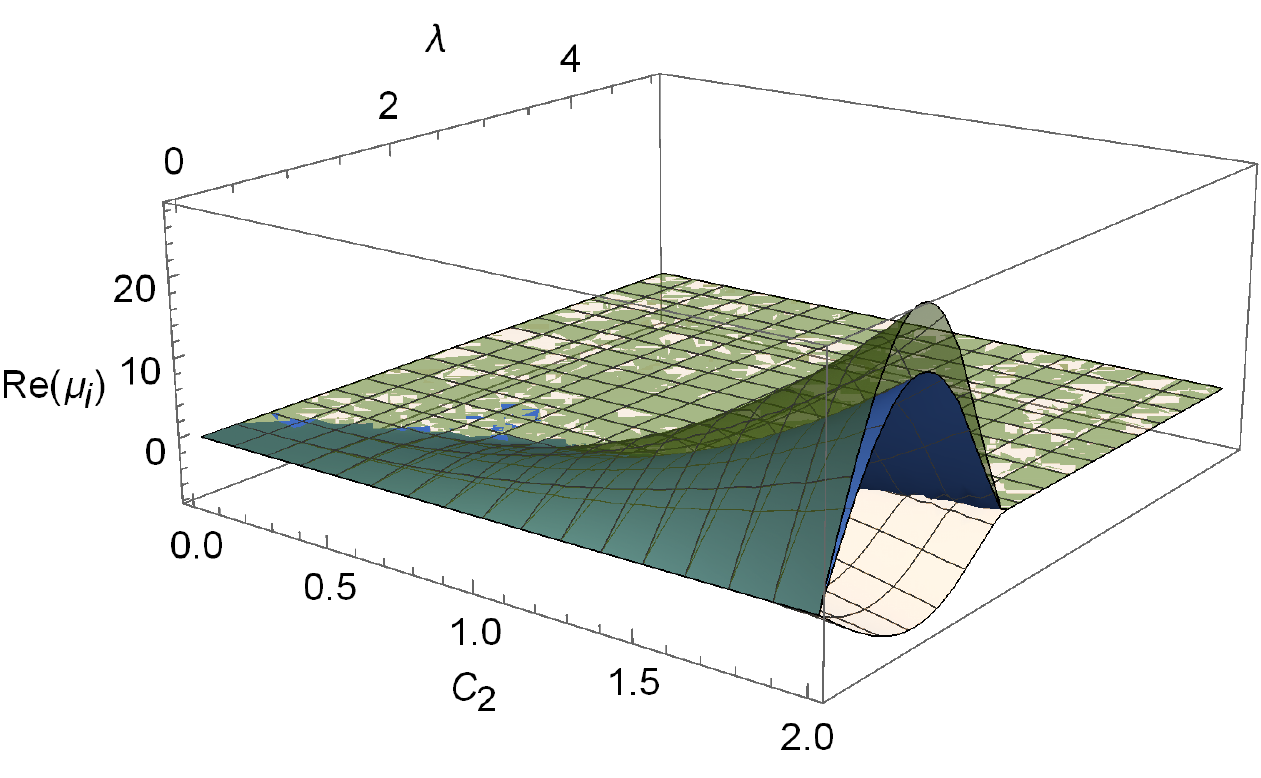}  \caption[{Parte real de los  $\mu_i$ correspondientes a los puntos de equilibrio $(\Sigma,A,w)=\left(0, \frac{1}{\sqrt{C_2-\frac{\lambda ^2}{2}}},\pm \frac{\lambda }{\sqrt{C_2-\frac{\lambda
   ^2}{2}}}\right)$.}]{\label{parte real} Parte real de los  $\mu_i$ se aprecia que los puntos de equilibrio $(\Sigma,A,w)=\left(0, \frac{1}{\sqrt{C_2-\frac{\lambda ^2}{2}}},\pm \frac{\lambda }{\sqrt{C_2-\frac{\lambda
   ^2}{2}}}\right)$ son tipicamente sillas o no hiperbólicos con tres autovalores nulos.}
   \end{figure*}
   
\subsection{Conjunto invariante $\Sigma=0$}
\label{Isotrop}

Imponiendo la condición  $\Sigma=0$ se deducen las restricciones: 
\begin{subequations}
\begin{align}
& \Omega = -\frac{2 \left(\gamma  v^2-v^2+1\right) \left(w^2-A^2 \lambda ^2\right)}{-3 \gamma +\gamma  v^2-2 v^2+2}, \\
& K=\frac{1}{2} \left(A^2 \left(-\left(2 C_2+\lambda ^2\right)\right)+\frac{4 \left(\gamma +v^2-1\right) (w-A \lambda ) (A \lambda +w)}{-3 \gamma +(\gamma -2) v^2+2}-2 u^2+2 w^2+2\right), \\
& 2 A^2 \gamma  \lambda ^2 v+\sqrt{2} A \lambda  u \left(3 \gamma -(\gamma -2) v^2-2\right)-2 \gamma  v w^2=0.     
\end{align}
\end{subequations}
Finalmente la dinámica en el conjunto  invariante $\Sigma=0$ esta determinada por el sistema dinámico reducido 
\begin{subequations}
\label{scalar-field-B}
\begin{align}
& A'= \frac{1}{2} A \left(-2 A^2 C_2+A^2 \lambda ^2+\frac{\sqrt{2} A \lambda  u \left(v^2+1\right)}{v}+2 u^2+2\right), \\
& v'= \frac{\left(v^2-1\right) \left(A \left(3 \gamma ^2-5 \gamma +(\gamma -2) v^2+2\right)+2 (\gamma -1) \gamma  v\right)}{\gamma  \left(\gamma -v^2-1\right)}, \\
& u'=\frac{\gamma  u \left(-2 A^2 C_2 v+A \left(A \lambda ^2 v+\sqrt{2} \lambda  u \left(v^2+1\right)+2 v^2-6\right)+2 \left(u^2-1\right) v\right)-4 A u \left(v^2-1\right)}{2 \gamma v}.
\end{align}
\end{subequations}

Los puntos de equilibrio del sistema  \eqref{scalar-field-B} son los siguientes. 

\begin{enumerate}
\item $N_{2,3}:(A,v,u)=(0,\pm 1,0)$ con autovalores $\left\{-1,1,\frac{4}{\gamma -2}+4\right\}$ son sillas hiperbólicas porque dos autovalores tienen signos opuestos y es silla no hiperbólica para $\lambda >0,  \gamma =1,  C_2>0$.
\item $M_{1,2}:(A,v,u)=(0,1,\pm 1)$ con autovalores $\left\{2,2,\frac{4}{\gamma -2}+4\right\}$ son
	\begin{enumerate}
	\item fuentes hiperbólicas para $\lambda >0,  0<\gamma <1,  C_2>0$, o   
	\item sillas hiperbólicas para $\lambda >0,  1<\gamma <2,  C_2>0$, o   
	\item no hiperbólicos para $\lambda >0,  \gamma =1,  C_2>0$.
	\end{enumerate}
\item $M_{3,4}:(A,v,u)=(0,-1,\pm 1)$ con autovalores $\left\{2,2,\frac{4}{\gamma -2}+4\right\}$ las condiciones de estabilidad son las mismas del punto anterior.

\item $Q_9:(A,v,u)=\left(\frac{1}{\sqrt{C_2-\frac{\lambda ^2}{2}}},1,0\right)$ con autovalores \\ $\left\{-2,-\frac{2}{\sqrt{C_2-\frac{\lambda ^2}{2}}}-2,\frac{-4 \gamma
   -\frac{6 \gamma }{\sqrt{C_2-\frac{\lambda
   ^2}{2}}}+\frac{8}{\sqrt{C_2-\frac{\lambda ^2}{2}}}+4}{2-\gamma
   }\right\}$ existe para  $\lambda >0,  C_2\geq \frac{\lambda ^2}{2}$ y es
	\begin{enumerate}
	\item pozo hiperbólico para 
	\begin{enumerate}
	\item $\lambda >0,  1<\gamma \leq \frac{4}{3},  C_2>\frac{1}{4}
   \left(\frac{(4-3 \gamma )^2}{(\gamma -1)^2}+2 \lambda ^2\right)$, o   
	\item $\lambda >0,  \frac{4}{3}<\gamma <2,  2 C_2>\lambda ^2$.
	\end{enumerate}
	\item silla hiperbólica para 	\begin{enumerate}
	    \item  $1<\gamma <\frac{4}{3},  \lambda >0,  C_2=\frac{1}{4}
   \left(\frac{(4-3 \gamma )^2}{(\gamma -1)^2}+2 \lambda ^2\right)$, o  
   \item $\lambda >0,  0<\gamma \leq 1,  2 C_2>\lambda ^2$.
	\end{enumerate}
	\item no hiperbólico para $1<\gamma <\frac{4}{3},  \lambda >0,  C_2=\frac{1}{4} \left(\frac{(4-3
   \gamma )^2}{(\gamma -1)^2}+2 \lambda ^2\right)$.
	\end{enumerate}

\item $Q_{10}:(A,v,u)=\left(\frac{1}{\sqrt{C_2-\frac{\lambda ^2}{2}}},-1,0\right)$ con autovalores \\ $\left\{-2,\frac{2}{\sqrt{C_2-\frac{\lambda ^2}{2}}}-2,\frac{-4 \gamma
   +\frac{6 \gamma }{\sqrt{C_2-\frac{\lambda
   ^2}{2}}}-\frac{8}{\sqrt{C_2-\frac{\lambda ^2}{2}}}+4}{2-\gamma
   }\right\}$ existe para $C_2\geq \frac{\lambda ^2}{2}$ y es 
	\begin{enumerate}
	\item pozo hiperbólico para
	\begin{enumerate}
	\item $\lambda >0,  2 C_2>\lambda ^2+2,  1\leq \gamma <2$, o   
	\item $\lambda >0,  0<\gamma <1,  \frac{1}{2} \left(\lambda
   ^2+2\right)<C_2<\frac{1}{4} \left(\frac{(4-3 \gamma )^2}{(\gamma
   -1)^2}+2 \lambda ^2\right)$.
	\end{enumerate}
	\item silla hiperbólica para 
	\begin{enumerate}
	\item $\lambda >0,  0<\gamma <2,  0<2 C_2-\lambda ^2<2$, o   
	\item $\lambda >0,  0<\gamma <1,  C_2>\frac{1}{4} \left(\frac{(4-3 \gamma
   )^2}{(\gamma -1)^2}+2 \lambda ^2\right)$, o   
	\item $\lambda >0,  \frac{4}{3}<\gamma <2,  \frac{\lambda
   ^2}{2}<C_2<\frac{1}{4} \left(\frac{(4-3 \gamma )^2}{(\gamma -1)^2}+2
   \lambda ^2\right)$, o  
   \item $\lambda >0,  0<\gamma \leq \frac{4}{3},  0<2 C_2-\lambda ^2<2$, o   
   \item $\lambda >0,  3 \gamma >4,  \frac{1}{4} \left(\frac{(4-3 \gamma )^2}{(\gamma
   -1)^2}+2 \lambda ^2\right)<C_2<\frac{1}{2} \left(\lambda ^2+2\right)$. 
	\end{enumerate}
	\item no hiperbólico para 
	\begin{enumerate}
	\item $\lambda >0,  0<\gamma <2,  2 C_2=\lambda ^2+2$, o    
	\item $C_2=\frac{1}{4} \left(\frac{(4-3 \gamma )^2}{(\gamma -1)^2}+2 \lambda
   ^2\right),  0<\gamma <1, 
   \lambda >0$, o   
   \item $C_2=\frac{1}{4} \left(\frac{(4-3 \gamma )^2}{(\gamma -1)^2}+2 \lambda
   ^2\right),  \frac{4}{3}<\gamma <2, 
   \lambda >0$.
	\end{enumerate}
	\end{enumerate}

\item $Q_{11}:(A,v,u)=\left(1,-1,\frac{\lambda }{\sqrt{2}}-\sqrt{C_2-1}\right)$ con autovalores $$\left\{-2,-\sqrt{-2 \sqrt{2} \sqrt{{C_2}-1} \lambda +4 {C_2}-3}-1,\sqrt{-2 \sqrt{2} \sqrt{{C_2}-1} \lambda +4 {C_2}-3}-1\right\}$$ existe para $C_2\geq 1$ y es
	\begin{enumerate}
	\item pozo hiperbólico para 
	\begin{enumerate}
	\item $1<C_2<2,  \lambda =\sqrt{2},  0<\gamma <2$, o   
	\item $C_2>1,  0<\lambda <\sqrt{2},  0<\gamma <2,  2 C_2<\lambda
   ^2+2$, o   	
	\item $C_2>1,  \lambda >\sqrt{2},  0<\gamma <2,  4 C_2+\lambda 
   \sqrt{\lambda ^2-2}<\lambda ^2+3$, o   
	\item $\lambda >\sqrt{2},  0<\gamma <2,  2 C_2<\lambda ^2+2,  \lambda 
   \left(\sqrt{\lambda ^2-2}+\lambda \right)+3<4 C_2$, o   
	\item $\lambda >\sqrt{2},  0<\gamma <2,  \lambda  \left(\sqrt{\lambda
   ^2-2}+\lambda \right)+3\geq 4 C_2,  4 C_2+\lambda 
   \sqrt{\lambda ^2-2}\geq \lambda ^2+3$.
	\end{enumerate}
	\item silla hiperbólica para $\lambda >0,  0<\gamma <2,  C_2>\frac{1}{2} \left(\lambda ^2+2\right)$.
	\item no hiperbólico para 
	\begin{enumerate}
	\item $C_2=1 ,  0<\gamma <2,  \lambda >0$, o   
	\item $2 C_2=\lambda ^2+2 ,  0<\gamma <2,   \lambda >0$.
	\end{enumerate}
	\end{enumerate}
	\item $Q_{12}:(A,v,u)=\left(1,-1,\frac{\lambda }{\sqrt{2}}+\sqrt{C_2-1}\right)$ con autovalores $$\left\{-2,-\sqrt{2 \sqrt{2} \sqrt{{C_2}-1} \lambda +4 {C_2}-3}-1,\sqrt{2 \sqrt{2} \sqrt{{C_2}-1} \lambda +4 {C_2}-3}-1\right\}$$ existe para $C_2\geq 1$ y es 
	\begin{enumerate}
	\item silla hiperbólica para $\lambda >0,  0<\gamma <2,  {C_2}>1$.
	\item no hiperbólico para $\lambda >0,  0<\gamma <2,  {C_2}=1$.
	\end{enumerate}
	\item $Q_{13}:(A,v,u)=\left\{\frac{\Delta  \left(\Gamma +(\gamma -1)^2 \gamma ^2 \left(2
   {C_2}-\lambda ^2\right)\right)}{(\gamma -1)^2 \gamma  (3 \gamma -2)
   \left(2 {C_2}-\lambda ^2\right)},-\Delta ,0\right\}$ donde\\ $\Delta=\sqrt{\frac{-\Gamma +\gamma  \left(\gamma  \left((\gamma -1)^2 \left(-\lambda
   ^2\right)-3 \gamma +2 (\gamma -1)^2 {C_2}+11\right)-12\right)+4}{(\gamma
   -2)^2}}$ y \\$\Gamma=\sqrt{(\gamma -1)^3 \gamma ^2 \left(2 {C_2}-\lambda ^2\right) \left(\gamma 
   \left(\gamma  \left(-(\gamma -1) \lambda ^2+2 (\gamma -1)
   {C_2}-6\right)+16\right)-8\right)}$ con autovalores\\ $\left\{-2,\frac{\Gamma }{(\gamma -1)^2 \gamma  \left(2 {C_2}-\lambda
   ^2\right)}-\gamma ,\gamma +\frac{(\gamma -2) \gamma  \left(4 \gamma  (\gamma
   -1)^2+\Gamma \right)}{2 (\gamma -1)^3 \left(\gamma  \left(-\gamma  \lambda
   ^2+2 \gamma  {C_2}-8\right)+8\right)}+\frac{(3 \gamma -4) \Gamma }{2
   (\gamma -1)^3 \gamma  \left(\lambda ^2-2 {C_2}\right)}-2\right\}$ \\ existe para 
	\begin{enumerate}
		\item $\lambda >0, \gamma =1, C_2>0$, o   
		\item $\lambda >0, 2 C_2\geq \lambda ^2, 0<\gamma \leq \frac{2}{3}$, o   
		\item $\lambda >0, 2 C_2>\lambda ^2, 1<\gamma <2$ o
        \item $\lambda >0, \frac{2}{3}<\gamma <1, \gamma  \left(\gamma  \left(-(\gamma
   -1) \lambda ^2+2 (\gamma -1) C_2-6\right)+16\right)<8$
	\end{enumerate} y es
	
	\begin{enumerate}
		\item pozo hiperbólico para $\lambda >0,  \frac{2}{3}<\gamma <1,  {C_2}>\frac{\gamma ^3 \lambda
   ^2-\gamma ^2 \lambda ^2+6 \gamma ^2-16 \gamma +8}{2 \gamma ^3-2 \gamma ^2}$.

		\item silla hiperbólica para 
		\begin{enumerate}
		\item $\lambda >0,  1<\gamma <2,  2 {C_2}>\lambda ^2$, o   
		\item $\lambda >0,  0<\gamma <\frac{2}{3},  2 {C_2}>\lambda ^2$, o   
		\item $\lambda >0,  1<\gamma \leq \frac{4}{3},  \frac{\lambda
   ^2}{2}<{C_2}<\frac{\gamma ^2 \lambda ^2+8 \gamma -8}{2 \gamma ^2}$, o   
		\item $\lambda >0,  \frac{4}{3}<\gamma <2,  \frac{2 \gamma ^2 \lambda ^2+9 \gamma
   ^2-4 \gamma  \lambda ^2-24 \gamma +2 \lambda ^2+16}{4 \gamma ^2-8 \gamma
   +4}<{C_2}<\frac{\gamma ^2 \lambda ^2+8 \gamma -8}{2 \gamma ^2}$.
		\end{enumerate}
		\item no hiperbólico para 
		\begin{enumerate}
		    \item $\lambda >0,  \frac{4}{3}<\gamma <2,  {C_2}=\frac{2 \gamma ^2 \lambda
   ^2+9 \gamma ^2-4 \gamma  \lambda ^2-24 \gamma +2 \lambda ^2+16}{4 \gamma ^2-8
   \gamma +4}$
             \item $\lambda >0, \gamma =\frac{2}{3}, C_2>\frac{\lambda ^2}{2}$
		\end{enumerate}
	\end{enumerate}
	\item $Q_{14}:(A,v,u)=\left\{\frac{\Delta  \left(-\Gamma +(\gamma -1)^2 \gamma ^2 \left(2
   {C_2}-\lambda ^2\right)\right)}{(\gamma -1)^2 \gamma  (3 \gamma -2)
   \left(2 {C_2}-\lambda ^2\right)},-\Delta ,0\right\}$ donde \\ $\Delta=\sqrt{\frac{-\Gamma +\gamma  \left(\gamma  \left((\gamma -1)^2 \left(-\lambda
   ^2\right)-3 \gamma +2 (\gamma -1)^2 {C_2}+11\right)-12\right)+4}{(\gamma
   -2)^2}}$ y \\ $\Gamma=\sqrt{(\gamma -1)^3 \gamma ^2 \left(2 {C_2}-\lambda ^2\right) \left(\gamma 
   \left(\gamma  \left(-(\gamma -1) \lambda ^2+2 (\gamma -1)
   {C_2}-6\right)+16\right)-8\right)}$ con autovalores\\ $\left\{-2,\frac{-\Gamma }{(\gamma -1)^2 \gamma  \left(2 {C_2}-\lambda
   ^2\right)}-\gamma ,\gamma +\frac{(\gamma -2) \gamma  \left(4 (\gamma -1)^2 \gamma -\Gamma
   \right)}{2 (\gamma -1)^3 \left(\gamma  \left(-\gamma  \lambda ^2+2 \gamma 
   {C_2}-8\right)+8\right)}-\frac{(3 \gamma -4) \Gamma }{2 (\gamma -1)^3
   \gamma  \left(\lambda ^2-2 {C_2}\right)}-2\right\}$\\ existe para 
	\begin{enumerate}
		\item $\lambda >0, C_2>0, \gamma =1$  o,
		\item $\lambda >0, C_2>0, 3 \gamma =2$, o   
		\item $\lambda >0, 2 C_2\geq \lambda ^2, 0<\gamma <\frac{2}{3}$  o, 
		\item $\lambda >0, 2 C_2\geq \lambda ^2, 1<\gamma <2$  o,  
		\item $\lambda >0, \frac{2}{3}<\gamma <1, \gamma  \left(\gamma  \left(-(\gamma
   -1) \lambda ^2+2 (\gamma -1) C_2-6\right)+16\right)<8$.
	\end{enumerate} y es 
	\begin{enumerate}
	    \item pozo hiperbólico para 
	    \begin{enumerate}
	        \item $\lambda >0, 0<\gamma <\frac{2}{3}, C_2>\frac{1}{4} \left(\frac{(4-3
   \gamma )^2}{(\gamma -1)^2}+2 \lambda ^2\right)$  o, 
	        \item $\lambda >0, 3 \gamma =2, 2 C_2>\lambda ^2+18$  o, 
	        \item $\lambda >0, \frac{2}{3}<\gamma <1, C_2>\frac{1}{4} \left(\frac{(4-3
   \gamma )^2}{(\gamma -1)^2}+2 \lambda ^2\right)$  o, 
	        \item $\lambda >0, 3 \gamma +\sqrt{13}=7, 2 C_2<\lambda
   ^2+\sqrt{13}+7, \left(7 \sqrt{13}-31\right) \left(2 C_2-\lambda
   ^2\right)<12 \left(\sqrt{13}-4\right)$  o, 
	        \item $\lambda >0, 3 \gamma +\sqrt{13}=7, 18 C_2+\sqrt{13}<9 \lambda
   ^2+11, \lambda ^2<2 C_2$  o, 
	        \item $\lambda >0, C_2<\frac{1}{4} \left(\frac{(4-3 \gamma )^2}{(\gamma
   -1)^2}+2 \lambda ^2\right), \frac{4 (\gamma -1)}{\gamma ^2}+\frac{\lambda
   ^2}{2}<C_2, \gamma >1, 3 \gamma +\sqrt{13}<7$  o, 
	        \item $\lambda >0, C_2<\frac{1}{4} \left(\frac{(4-3 \gamma )^2}{(\gamma
   -1)^2}+2 \lambda ^2\right), 3 \gamma +\sqrt{13}>7, \gamma +\mu _1<0$ (donde $\mu _1\approx -1.22033$ es la raíz real de $P(\mu)=9 \mu ^3+22 \mu ^2+20 \mu +8$)  o, 
	        \item $\lambda >0, C_2<\frac{1}{4} \left(\frac{(4-3 \gamma )^2}{(\gamma
   -1)^2}+2 \lambda ^2\right), 3 \gamma <4, \lambda ^2<2 C_2,
   \gamma +\mu _1>0$ (donde $\mu _1\approx -1.22033$ es la raíz real de $P(\mu)=9 \mu ^3+22 \mu ^2+20 \mu +8$)  o, 
    \item $\lambda >0, 2 C_2>\lambda ^2, \frac{8}{\gamma ^2}+2
   C_2<\frac{8}{\gamma }+\lambda ^2, 3 \gamma +\sqrt{13}>7, \gamma
   +\mu _1\leq 0$ (donde $\mu _1\approx -1.22033$ es la raíz real de $P(\mu)=9 \mu ^3+22 \mu ^2+20 \mu +8$)  o, 
        \item $\lambda >0, 2 C_2>\lambda ^2, \frac{8}{\gamma ^2}+2
   C_2<\frac{8}{\gamma }+\lambda ^2, 3 \gamma +\sqrt{13}<7$.
	    \end{enumerate}
	    \item silla hiperbólica para 
	    \begin{enumerate}
	        \item $\lambda >0, \frac{2}{3}<\gamma <1, \frac{(\gamma -2) (3 \gamma
   -2)}{(\gamma -1) \gamma ^2}+\frac{\lambda ^2}{2}<C_2<\frac{1}{4}
   \left(\frac{(4-3 \gamma )^2}{(\gamma -1)^2}+2 \lambda ^2\right)$  o, 
	        \item $\lambda >0, 3 \gamma +\sqrt{13}=7, 2 C_2>\lambda ^2+\sqrt{13}+7$  o, 
	        \item $\lambda >0, C_2>\frac{1}{4} \left(\frac{(4-3 \gamma )^2}{(\gamma
   -1)^2}+2 \lambda ^2\right), \gamma >1, 3 \gamma +\sqrt{13}<7$  o, 
	        \item $\lambda >0, C_2>\frac{1}{4} \left(\frac{(4-3 \gamma )^2}{(\gamma
   -1)^2}+2 \lambda ^2\right), 3 \gamma +\sqrt{13}>7, \gamma +\mu _1<0$ (donde $\mu _1\approx -1.22033$ es la raíz real de $P(\mu)=9 \mu ^3+22 \mu ^2+20 \mu +8$)  o,  
	        \item $\lambda >0, C_2<\frac{4 (\gamma -1)}{\gamma ^2}+\frac{\lambda
   ^2}{2}, 3 \gamma <4, \frac{1}{4} \left(\frac{(4-3 \gamma )^2}{(\gamma
   -1)^2}+2 \lambda ^2\right)<C_2$  o,  
	        \item $\lambda >0, \gamma <2, 2 C_2>\lambda ^2, 3 \gamma \geq 4,
   \frac{8}{\gamma ^2}+2 C_2<\frac{8}{\gamma }+\lambda ^2$  o,  
	        \item $\lambda >0, \gamma <2, \frac{8}{\gamma ^2}+2 C_2>\frac{8}{\gamma
   }+\lambda ^2, \gamma +\mu _1\geq 0$  o,  
	        \item $\lambda >0, 0<\gamma <\frac{2}{3}, \frac{\lambda
   ^2}{2}<C_2<\frac{1}{4} \left(\frac{(4-3 \gamma )^2}{(\gamma -1)^2}+2
   \lambda ^2\right)$  o, 
	        \item $\lambda >0, 3 \gamma =2, 0<2 C_2-\lambda ^2<18$.
	    \end{enumerate}
	    \item no hiperbólico para 
	    \begin{enumerate}
	    \item $C_2=\frac{1}{4} \left(\frac{(4-3 \gamma )^2}{(\gamma -1)^2}+2 \lambda
   ^2\right), 0<\gamma <1$
	        \item $C_2=\frac{1}{4} \left(\frac{(4-3 \gamma )^2}{(\gamma -1)^2}+2 \lambda
   ^2\right), 1<\gamma <\chi _1$  (donde  $\chi _1\approx 1.22033$ es la raíz real de $P(\chi)=9 \chi ^3-22 \chi ^2+20 \chi -8$)  o, 
	        \item $\lambda >0, \chi _1 <\gamma <\frac{4}{3}$ (donde  $\chi _1\approx 1.22033$ es la raíz real de $P(\chi)=9 \chi ^3-22 \chi ^2+20 \chi -8$).
	    \end{enumerate}
	\end{enumerate}
\end{enumerate}

\subsection{Conjuntos invariantes $v=\pm 1$}
\label{tilt2}
En el caso $v=\varepsilon=\pm 1$ el sistema se reduce a 
\begin{subequations}
\label{scalar-field-C}
\begin{align}
&\Sigma'= \frac{C_2 \Sigma  \left(2 C_2 \left(\Sigma ^2-A^2\right)-A \left(A \lambda ^2+4 \varepsilon \right)+2 u^2+2 w^2-2\right)-2 A \lambda  \left(A \lambda +\sqrt{2} u \varepsilon \right)+2 w^2}{2 C_2},\\
& A'=-\frac{1}{2} A^3 \lambda ^2+A
   C_2 \left(\Sigma ^2-A^2\right)+A \left(2 \Sigma +u^2+w^2+1\right),\\
& u'= C_2 u \left(\Sigma ^2-A^2\right)+u \left(-\frac{A^2 \lambda ^2}{2}+w^2-1\right)+\frac{\sqrt{2} \left(A^2 \lambda ^2-w^2\right)}{\lambda }+u^3,\\
& w'=w \left(C_2\left(\Sigma ^2-A^2\right)-\frac{A^2 \lambda ^2}{2}+\Sigma +u^2-\frac{\sqrt{2} u}{\lambda }+w^2+1\right).
\end{align}
\end{subequations}

Los puntos de equilibrio del sistema  \eqref{scalar-field-C} para $\varepsilon=\pm 1$ son los siguientes:
\begin{enumerate}
    \item $N_1:(\Sigma,A,u,w)=(0,0,0,0)$ con autovalores $\{-1,-1,1,1\}$ es a silla hiperbólica.
   \item $Q_{15,16}:(\Sigma,A,u,w)=\left(\Sigma_0 ,0,\varepsilon \sqrt{1-C_2 \Sigma_0 ^2},0\right)$ con autovalores\\ $\left\{0,2,2 (\Sigma_0 +1),-\varepsilon \frac{\sqrt{2-2 C_2 \Sigma_0 ^2}}{\lambda }+\Sigma_0
   +2\right\}$.  Estas curva de puntos no hiperbólicos contienen a los puntos $P_1,P_2,P_3,P_4$ estudiados en el capítulo \ref{ch_3} y existen para  $\lambda >0, 0<C_2\leq \frac{1}{\Sigma ^2}$ y $\Sigma \in \mathbb{R}$. Estas curvas corresponden a  conjuntos normalmente hiperbólicos. En efecto, la curvas paramétricas definidas por $$\Sigma (\Sigma_0)=\Sigma_0,\quad A(\Sigma_0)=0,\quad u(\Sigma_0)=\varepsilon \sqrt{1-C_2 \Sigma_0 ^2},\quad w(\Sigma_0)=0,$$  tiene como vector tangente en $\Sigma_0$ a $$ \dot{\textbf{r}}(\Sigma_0)=\left(1,0,-\frac{\varepsilon C_2 \Sigma_0}{\sqrt{1-C_2 \Sigma_0 ^2}},0\right)$$ es paralelo al autovector asociado al autovalor nulo dado por $$\textbf{v}(\Sigma_0)=\left(-\frac{\sqrt{1-C_2 \Sigma_0 ^2}}{\varepsilon C_2 \Sigma_0},0,1,0\right).$$ En este caso particular, de acuerdo a lo comentado a continuación de la definición \ref{normhiper}, sección \ref{seccion1.2}, se puede estudiar la estabilidad considerando solo los signos de la partes reales de los autovalores no nulos. De esta manera se concluye que:
      \begin{enumerate}
       \item $Q_{15}$ es fuente para 
       \begin{enumerate}
           \item $C_2>0,  \Sigma_0=\frac{1}{\sqrt{C_2}},  \lambda >0,  1<\gamma <2$, o
           \item $C_2>1,  \Sigma_0=-\frac{1}{\sqrt{C_2}},  \lambda >0,  1<\gamma <2$, o
           \item $0<C_2\leq 1,  -1<\Sigma_0<\sqrt{\frac{1}{C_2}},  \lambda >\sqrt{2} \sqrt{\frac{1-C_2 \Sigma_0^2}{(\Sigma_0+2)^2}},  1<\gamma <2$, o
           \item $C_2>1,    -\sqrt{\frac{1}{C_2}}<\Sigma_0<\sqrt{\frac{1}{C_2}},  \lambda >\sqrt{2} \sqrt{\frac{1-C_2 \Sigma_0^2}{(\Sigma_0+2)^2}},  1<\gamma <2$, o
           \item $C_2\leq 0, 
   \Sigma_0>-1,  \lambda >\sqrt{2} \sqrt{\frac{1-C_2 \Sigma_0^2}{(\Sigma_0+2)^2}},  1<\gamma <2$
       \end{enumerate}
       \item $Q_{15}$ es silla para 
       \begin{enumerate}
           \item $\frac{1}{4}<C_2<1, \Sigma_0=-\frac{1}{\sqrt{C_2}}, \lambda >0, 1<\gamma <2$,  o 
           \item $0<C_2\leq \frac{1}{4}, -2<\Sigma_0<-1, \lambda >\sqrt{2}
   \sqrt{\frac{1-C_2 \Sigma_0^2}{(\Sigma_0+2)^2}}, 1<\gamma <2$,  o 
   \item $\frac{1}{4}<C_2<1, -\sqrt{\frac{1}{C_2}}<\Sigma_0<-1, \lambda >\sqrt{2}
   \sqrt{\frac{1-C_2 \Sigma_0^2}{(\Sigma_0+2)^2}}, 1<\gamma <2$,  o 
   \item $0<C_2\leq 1, -1<\Sigma_0<\sqrt{\frac{1}{C_2}}, 0<\lambda <\sqrt{2} \sqrt{\frac{1-C_2
   \Sigma_0^2}{(\Sigma_0+2)^2}}, 1<\gamma <2$,  o 
   \item $C_2>1, -\sqrt{\frac{1}{C_2}}<\Sigma_0<\sqrt{\frac{1}{C_2}}, 0<\lambda <\sqrt{2} \sqrt{\frac{1-C_2
   \Sigma_0^2}{(\Sigma_0+2)^2}}, 1<\gamma <2$,  o 
   \item $0<C_2<\frac{1}{4}, -\sqrt{\frac{1}{C_2}}\leq \Sigma_0\leq -2, \lambda >0, 1<\gamma <2$,  o 
   \item $\frac{1}{4}\leq C_2<1, -\sqrt{\frac{1}{C_2}}<\Sigma_0<-1, 0<\lambda <\sqrt{2} \sqrt{\frac{1-C_2 \Sigma_0^2}{(\Sigma_0+2)^2}}, 1<\gamma <2$,  o 
   \item $0<C_2<\frac{1}{4}, -2<\Sigma_0<-1, 0<\lambda <\sqrt{2} \sqrt{\frac{1-C_2 \Sigma_0^2}{(\Sigma_0+2)^2}}, 1<\gamma <2$.
       \end{enumerate}
       \item $Q_{15}$ es no hiperbólico para 
       \begin{enumerate}
           \item $C_2=\frac{1}{4},  \Sigma_0=-2,  \lambda >0,  1<\gamma <2$, o 
           \item $C_2=\frac{1}{4},  \Sigma_0=-1,  \lambda >0,  1<\gamma <2$, o 
           \item $0<C_2<\frac{1}{4},  \Sigma_0=-1,  \lambda >0,  1<\gamma <2$, o 
   \item $\frac{1}{4}<C_2\leq 1,  \Sigma_0=-1,  \lambda >0,  1<\gamma <2$, o 
   \item $C_2>1,  -\sqrt{\frac{1}{C_2}}<\Sigma_0<\sqrt{\frac{1}{C_2}},  \lambda =\sqrt{\frac{2(1-C_2 \Sigma_0^2)}{(\Sigma_0+2)^2}},  1<\gamma <2$, o 
   \item $C_2=\frac{1}{4},  -2<\Sigma_0<-1,  \lambda =\frac{1}{2} \sqrt{\frac{8}{\Sigma_0+2}-2},  1<\gamma <2$, o 
   \item $C_2=\frac{1}{4},  -1<\Sigma_0<2,  \lambda
   =\frac{1}{2} \sqrt{\frac{8}{\Sigma_0+2}-2},  1<\gamma <2$, o 
   \item $0<C_2<\frac{1}{4},  -1<\Sigma_0<\sqrt{\frac{1}{C_2}},  \lambda =\sqrt{\frac{2(1-C_2 \Sigma_0^2)}{(\Sigma_0+2)^2}},  1<\gamma <2$, o 
   \item $0<C_2<\frac{1}{4},  -2<\Sigma_0<-1,  \lambda =\sqrt{\frac{2(1-C_2 \Sigma_0^2)}{(\Sigma_0+2)^2}},  1<\gamma
   <2$, o 
   \item $\frac{1}{4}<C_2\leq 1,  -1<\Sigma_0<\sqrt{\frac{1}{C_2}},  \lambda =\sqrt{\frac{2(1-C_2 \Sigma_0^2)}{(\Sigma_0+2)^2}},  1<\gamma <2$, o 
   \item $\frac{1}{4}<C_2<1,  -\sqrt{\frac{1}{C_2}}<\Sigma_0<-1,  \lambda =\sqrt{\frac{2(1-C_2 \Sigma_0^2)}{(\Sigma_0+2)^2}},  1<\gamma <2$.
       \end{enumerate}
   \end{enumerate}
Además,
   \begin{enumerate}
       \item $Q_{16}$ es fuente para 
       \begin{enumerate}
           \item $C_2\leq 0, \Sigma_0>-1, \lambda >0, 1<\gamma <2$, o 
           \item $0<C_2\leq 1, -1<\Sigma_0\leq \sqrt{\frac{1}{C_2}}, \lambda >0, 1<\gamma <2$, o 
           \item $C_2>1,
   -\sqrt{\frac{1}{C_2}}\leq \Sigma_0 \leq \sqrt{\frac{1}{C_2}}, \lambda >0, 1<\gamma <2$.
       \end{enumerate}
       \item $Q_{16}$ es silla para 
       \begin{enumerate}
           \item $C_2=\frac{1}{4}, -2<\Sigma_0<-1, \lambda >0, 1<\gamma <2$, o 
           \item $0<C_2<\frac{1}{4}, -\sqrt{\frac{1}{C_2}}<\Sigma_0<-2, 0<\lambda <\sqrt{2}
   \sqrt{\frac{1-C_2 \Sigma_0^2}{(\Sigma_0+2)^2}}, 1<\gamma <2$, o \item $\frac{1}{4}<C_2<1, -\sqrt{\frac{1}{C_2}}\leq \Sigma_0<-1, \lambda >0, 1<\gamma
   <2$, o 
   \item $0<C_2<\frac{1}{4}, -2\leq \Sigma_0<-1, \lambda >0, 1<\gamma <2$, o 
   \item $0<C_2<\frac{1}{4}, \Sigma_0=-\frac{1}{\sqrt{C_2}}, \lambda >0,
   1<\gamma <2$, o 
   \item $0<C_2<\frac{1}{4}, -\sqrt{\frac{1}{C_2}}<\Sigma_0<-2, \lambda >\sqrt{2} \sqrt{\frac{1-C_2 \Sigma_0^2}{(\Sigma_0+2)^2}}, 1<\gamma <2$.
       \end{enumerate}
       \item $Q_{16}$ es no hiperbólico para 
       \begin{enumerate}
           \item $C_2=\frac{1}{4},  \Sigma_0=-2,  \lambda >0,  1<\gamma <2$, o 
           \item $0<C_2\leq 1,  \Sigma_0=-1,  \lambda >0,  1<\gamma <2$, o 
           \item $0<C_2<\frac{1}{4}, 
   -\sqrt{\frac{1}{C_2}}<\Sigma_0<-2,  \lambda =\sqrt{\frac{2-2 C_2 \Sigma_0^2}{(\Sigma_0+2)^2}},  1<\gamma <2$.
       \end{enumerate}
   \end{enumerate} 
   \item $Q_{17,18}:(\Sigma,A,u,w)=\left(-\frac{1}{2}, -\varepsilon \frac{1}{2},\frac{\lambda }{2 \sqrt{2}},0\right)$ donde un auovalor es cero y los otros tres son las raices del polinomio en $\mu$: $P(\mu)=\mu ^3+\mu ^2+\mu  \left(\frac{\lambda ^2}{2 C_2}-C_2-\frac{\lambda
   ^2}{2}+1\right)-\frac{\lambda ^2}{2 C_2}-C_2+\frac{\lambda
   ^2}{2}+1$. Es silla no hiperbólica o centro, dependiendo de la elección de los parámetros. Ver figura \ref{P4eigen}. 
   \item $Q_{19}:(\Sigma,A,u,w)=\left(-\frac{1}{2 C_2},0,\frac{1}{\sqrt{2} \lambda },  \frac{1}{2} 
   \sqrt{\frac{1}{C_2}+\frac{2}{\lambda ^2}-4}\right)$ con autovalores \\$\left\{\frac{1}{\lambda ^2}-\frac{1}{2 C_2}, \frac{1}{2} \left(\frac{1}{C_2}+\frac{2}{\lambda
   ^2}-4\right), \frac{1}{2} \left(\frac{1}{C_2}+\frac{2}{\lambda
   ^2}-4\right), \frac{1}{C_2}+\frac{2}{\lambda ^2}-2\right\}$ \\ existe para $0<\lambda \leq \frac{1}{\sqrt{2}}, C_2>0$ o para $\lambda
   >\frac{1}{\sqrt{2}}, 0<C_2\leq \frac{\lambda ^2}{4 \lambda ^2-2}$.
   \begin{enumerate}
   \item fuente hiperbólica para 
   \begin{enumerate}
       \item $0<\lambda \leq \frac{1}{\sqrt{2}}, C_2>\frac{\lambda ^2}{2}$, o  
       \item $\frac{1}{\sqrt{2}}<\lambda <1, \frac{\lambda ^2}{2}<C_2<\frac{\lambda
   ^2}{4 \lambda ^2-2}$.
   \end{enumerate}
   \item silla hiperbólica para 
   \begin{enumerate}
       \item $\lambda >1, 0<C_2<\frac{\lambda ^2}{4 \lambda ^2-2}$, o  
       \item $0<\lambda \leq 1,  0<C_2<\frac{\lambda ^2}{2}$.
   \end{enumerate}
   \item no hiperbólico para 
   \begin{enumerate}
       \item $C_2=\frac{1}{2},  \lambda =1$, o  
       \item $C_2=\frac{\lambda ^2}{2},  0<\lambda <1$, o  
       \item $C_2=\frac{\lambda ^2}{4 \lambda ^2-2},  \frac{1}{\sqrt{2}}<\lambda <1$, o   
       \item $C_2=\frac{\lambda ^2}{4 \lambda ^2-2},  \lambda >1$.
   \end{enumerate}
   \end{enumerate}
   \item $Q_{20,21}:(\Sigma,A,u,w)=\left(-\frac{1}{2}, \varepsilon \frac{\sqrt{C_2}}{\sqrt{2} \lambda },\frac{C_2}{\sqrt{2}
   \lambda },0\right)$ con autovalores \\$\left\{\frac{1}{2}-\frac{C_2}{\lambda ^2}, \frac{\sqrt{2} \sqrt{C_2}}{\lambda
   }-1, -\frac{1}{2} \left(1+\frac{\sqrt{8 C_2^2+4 C_2 \lambda ^2-7 \lambda
   ^2}}{\lambda }\right), -\frac{1}{2} \left(1-\frac{\sqrt{8 C_2^2+4 C_2 \lambda ^2-7 \lambda
   ^2}}{\lambda }\right)\right\}$ estos puntos son
   \begin{enumerate}
       \item silla hiperbólica para 
       \begin{enumerate}
       \item $0<\lambda \leq \frac{\sqrt{7}}{2}, \frac{1}{4} \sqrt{\lambda ^4+14 \lambda
   ^2}-\frac{\lambda ^2}{4}<C_2<\frac{1}{4} \sqrt{\lambda ^4+16 \lambda
   ^2}-\frac{\lambda ^2}{4}$, o  
       \item $\frac{\sqrt{7}}{2}<\lambda <\sqrt{2}, \frac{\lambda
   ^2}{2}<C_2<\frac{1}{4} \sqrt{\lambda ^4+16 \lambda ^2}-\frac{\lambda
   ^2}{4}$, o  
       \item $0<\lambda <\frac{\sqrt{7}}{2}, \frac{\lambda ^2}{2}<C_2\leq
   \frac{1}{4} \sqrt{\lambda ^4+14 \lambda ^2}-\frac{\lambda ^2}{4}$, o   
        \item $\lambda >\sqrt{2}, C_2>\frac{\lambda ^2}{2}$
        \item $0<\lambda \leq \sqrt{2}, C_2>\frac{1}{4} \sqrt{\lambda ^4+16 \lambda
   ^2}-\frac{\lambda ^2}{4}$, o   
        \item $\lambda >\sqrt{2}, \frac{1}{4} \sqrt{\lambda ^4+16 \lambda ^2}-\frac{\lambda
   ^2}{4}<C_2<\frac{\lambda ^2}{2}$, o  
        \item $\frac{\sqrt{7}}{2}<\lambda \leq \sqrt{2}, \frac{1}{4} \sqrt{\lambda ^4+14
   \lambda ^2}-\frac{\lambda ^2}{4}<C_2<\frac{\lambda ^2}{2}$, o   
        \item $\lambda >\sqrt{2}, \frac{1}{4} \sqrt{\lambda ^4+14 \lambda ^2}-\frac{\lambda
   ^2}{4}<C_2<\frac{1}{4} \sqrt{\lambda ^4+16 \lambda ^2}-\frac{\lambda
   ^2}{4}$o, 
         \item $0<\lambda \leq \frac{\sqrt{7}}{2}, 0<C_2<\frac{\lambda ^2}{2}$
        \item $\lambda >\frac{\sqrt{7}}{2}, 0<C_2\leq \frac{1}{4} \sqrt{\lambda ^4+14
   \lambda ^2}-\frac{\lambda ^2}{4}$.
       \end{enumerate}
       \item no hiperbólico para 
       \begin{enumerate}
           \item $\lambda >0, C_2=\frac{\lambda ^2}{2}$, o  
           \item $\lambda >0, C_2=\frac{1}{4} \sqrt{\lambda ^4+16 \lambda
   ^2}-\frac{\lambda ^2}{4}$.
       \end{enumerate}
   \end{enumerate}
    \item $Q_{22,23}:(\Sigma,A,u,w)=\left(-\frac{1}{2}, -\varepsilon \frac{\sqrt{C_2}}{\sqrt{2} \lambda },\frac{C_2}{\sqrt{2}
   \lambda },0\right)$ con autovalores \\$\left\{\frac{1}{2}-\frac{C_2}{\lambda ^2}, -\frac{\sqrt{2} \sqrt{C_2}}{\lambda
   }-1, -\frac{1}{2} \left(1+\frac{\sqrt{8 C_2^2+4 C_2 \lambda ^2-7 \lambda
   ^2}}{\lambda }\right), -\frac{1}{2} \left(1-\frac{\sqrt{8 C_2^2+4 C_2 \lambda ^2-7 \lambda
   ^2}}{\lambda }\right)\right\}$ estos puntos son
   \begin{enumerate}
   \item Pozo hiperbólico para 
   \begin{enumerate}
       \item $0<\lambda \leq \frac{\sqrt{7}}{2}, \frac{1}{4} \sqrt{\lambda ^4+14 \lambda
   ^2}-\frac{\lambda ^2}{4}<C_2<\frac{1}{4} \sqrt{\frac{\lambda ^6+14
   \lambda ^4+2}{\lambda ^2}}-\frac{\lambda ^2}{4}$, o  
       \item $\frac{\sqrt{7}}{2}<\lambda <1.35095, \frac{\lambda
   ^2}{2}<C_2<\frac{1}{4} \sqrt{\frac{\lambda ^6+14 \lambda ^4+2}{\lambda
   ^2}}-\frac{\lambda ^2}{4}$, o  
       \item $0<\lambda <\frac{\sqrt{7}}{2}, \frac{\lambda ^2}{2}<C_2\leq \frac{1}{4}
   \sqrt{\lambda ^4+14 \lambda ^2}-\frac{\lambda ^2}{4}$.
   \end{enumerate}
       \item silla hiperbólica para
       \begin{enumerate}
           \item $0<\lambda \leq \frac{\sqrt{7}}{2}, \frac{1}{4} \sqrt{\lambda ^4+14 \lambda
   ^2}-\frac{\lambda ^2}{4}<C_2<\frac{1}{4} \sqrt{\lambda ^4+16 \lambda
   ^2}-\frac{\lambda ^2}{4}$, o  
           \item $\frac{\sqrt{7}}{2}<\lambda <\sqrt{2}, \frac{\lambda
   ^2}{2}<C_2<\frac{1}{4} \sqrt{\lambda ^4+16 \lambda ^2}-\frac{\lambda
   ^2}{4}$, o  
           \item $0<\lambda <\frac{\sqrt{7}}{2}, \frac{\lambda ^2}{2}<C_2\leq
   \frac{1}{4} \sqrt{\lambda ^4+14 \lambda ^2}-\frac{\lambda ^2}{4}$, o  
           \item $0<\lambda \leq \sqrt{2}, C_2>\frac{1}{4} \sqrt{\lambda ^4+16 \lambda
   ^2}-\frac{\lambda ^2}{4}$, o  
            \item $\lambda >\sqrt{2}, C_2>\frac{\lambda ^2}{2}$, o  
            \item $\lambda >\sqrt{2}, \frac{1}{4} \sqrt{\lambda ^4+16 \lambda ^2}-\frac{\lambda
^2}{4}<C_2<\frac{\lambda ^2}{2}$, o  
            \item $\frac{\sqrt{7}}{2}<\lambda \leq \sqrt{2}, \frac{1}{4} \sqrt{\lambda ^4+14
   \lambda ^2}-\frac{\lambda ^2}{4}<C_2<\frac{\lambda ^2}{2}$, o  
            \item $\lambda >\sqrt{2}, \frac{1}{4} \sqrt{\lambda ^4+14 \lambda ^2}-\frac{\lambda
   ^2}{4}<C_2<\frac{1}{4} \sqrt{\lambda ^4+16 \lambda ^2}-\frac{\lambda
   ^2}{4}$, o  
             \item $0<\lambda \leq \frac{\sqrt{7}}{2}, 0<C_2<\frac{\lambda ^2}{2}$, o  
             \item $\lambda >\frac{\sqrt{7}}{2}, 0<C_2\leq \frac{1}{4} \sqrt{\lambda ^4+14
   \lambda ^2}-\frac{\lambda ^2}{4}$.
             \end{enumerate}
             \item no hiperbólico para 
                \begin{enumerate}
                    \item $\lambda >0, C_2=\frac{\lambda ^2}{2}$, o  
                    \item $\lambda >0, C_2=\frac{1}{4} \sqrt{\lambda ^4+16 \lambda
   ^2}-\frac{\lambda ^2}{4}$.
                \end{enumerate}
   \end{enumerate}
   \item $Q_{24,25}:(\Sigma,A,u,w)=\left(-\frac{\lambda ^2}{2 C_2^2+C_2 \lambda ^2},-\varepsilon \delta,\frac{\sqrt{2} \lambda }{2 C_2+\lambda ^2},0\right)$ con autovalores \\ $\left\{\frac{1}{C_2}-\frac{4}{2
   C_2+\lambda ^2},-\frac{8}{2
   C_2+\lambda
   ^2}+\frac{4}{C_2}-2,-\frac{4}{2
   C_2+\lambda ^2}+\frac{2}{C_2}-2,2
  \delta+\frac{2 \lambda ^2}{C_2
   \left(2 C_2+\lambda ^2\right)}-2\right\}$,\\ donde $\delta=\sqrt{\frac{4 C_2^2+2 (C_2-1)
   \lambda ^2}{C_2 \left(2 C_2+\lambda
   ^2\right)^2}}$ existe para $\lambda >0, C_2\geq \frac{1}{4} \sqrt{\lambda ^4+8 \lambda ^2}-\frac{\lambda
   ^2}{4}$ y son  
   \begin{enumerate}
       \item pozo hiperbólico para 
       \begin{enumerate}
           \item $0<\lambda \leq \sqrt{2},  C_2>1$, o 
           \item $\lambda >\sqrt{2},  C_2>\frac{\lambda ^2}{2}$.
       \end{enumerate}
       \item silla hiperbólica para
       \begin{enumerate}
           \item $0<\lambda <\sqrt{2},  \frac{1}{4} \sqrt{\lambda ^4+16 \lambda
   ^2}-\frac{\lambda ^2}{4}<C_2<1$, o  
   \item $0<\lambda \leq 1,  \frac{1}{4} \sqrt{\lambda ^4+8 \lambda ^2}-\frac{\lambda
   ^2}{4}<C_2<\frac{1}{4} \sqrt{\lambda ^4+16 \lambda ^2}-\frac{\lambda
   ^2}{4}$, o  
   \item $1<\lambda <\sqrt{2},  \frac{\lambda ^2}{2}<C_2<\frac{1}{4} \sqrt{\lambda
   ^4+16 \lambda ^2}-\frac{\lambda ^2}{4}$o, 
   \item $\lambda >\sqrt{2},  \frac{1}{4} \sqrt{\lambda ^4+16 \lambda ^2}-\frac{\lambda
   ^2}{4}<C_2<\frac{\lambda ^2}{2}$, o  
   \item $\lambda >\sqrt{2},  1<C_2<\frac{1}{4} \sqrt{\lambda ^4+16 \lambda
   ^2}-\frac{\lambda ^2}{4}$.
       \end{enumerate}
       \item no hiperbólico para
       \begin{enumerate}
           \item $\lambda \geq 1,  C_2=\frac{\lambda ^2}{2}$, o  
           \item $\lambda >0,  C_2=\frac{1}{4} \sqrt{\lambda ^4+8 \lambda
   ^2}-\frac{\lambda ^2}{4}$, o  
   \item $\lambda >0, C_2=\frac{1}{4} \sqrt{\lambda ^4+16 \lambda
   ^2}-\frac{\lambda ^2}{4}$, o  
   \item $\lambda >0,  C_2=1$.
       \end{enumerate}
   \end{enumerate}
   \item $Q_{26,27}:(\Sigma,A,u,w)=\left(-\frac{\lambda ^2}{2 C_2^2+C_2 \lambda ^2},\varepsilon \delta,\frac{\sqrt{2} \lambda }{2 C_2+\lambda ^2},0\right)$  con autovalores \\ $\left\{\frac{1}{C_2}-\frac{4}{2
   C_2+\lambda ^2},-\frac{8}{2
   C_2+\lambda
   ^2}+\frac{4}{C_2}-2,-\frac{4}{2
   C_2+\lambda ^2}+\frac{2}{C_2}-2,-2
   \delta+\frac{2 \lambda ^2}{C_2
   \left(2 C_2+\lambda ^2\right)}-2\right\}$,\\donde $\delta=\sqrt{\frac{4 C_2^2+2 (C_2-1)
   \lambda ^2}{C_2 \left(2 C_2+\lambda
   ^2\right)^2}}$ existe para $\lambda >0, C_2\geq \frac{1}{4} \sqrt{\lambda ^4+8 \lambda ^2}-\frac{\lambda
   ^2}{4}$ y son 
      \begin{enumerate}
       \item pozo hiperbólico para  
       \begin{enumerate}
           \item $0<\lambda \leq \sqrt{2},  C_2>\frac{1}{4} \sqrt{\lambda ^4+16 \lambda
   ^2}-\frac{\lambda ^2}{4}$, o  
           \item $\lambda >\sqrt{2},  C_2>\frac{\lambda ^2}{2}$. \end{enumerate}
       \item silla hiperbólica para
       \begin{enumerate}
           \item $0<\lambda \leq 1,  \frac{1}{4} \sqrt{\lambda ^4+8 \lambda ^2}-\frac{\lambda
   ^2}{4}<C_2<\frac{1}{4} \sqrt{\lambda ^4+16 \lambda ^2}-\frac{\lambda
   ^2}{4}$, o   
           \item $1<\lambda <\sqrt{2},  \frac{\lambda ^2}{2}<C_2<\frac{1}{4} \sqrt{\lambda
   ^4+16 \lambda ^2}-\frac{\lambda ^2}{4}$, o  
           \item $\lambda >\sqrt{2},  \frac{1}{4} \sqrt{\lambda ^4+16 \lambda ^2}-\frac{\lambda
   ^2}{4}<C_2<\frac{\lambda ^2}{2}$, o  
           \item $1<\lambda \leq \sqrt{2},  \frac{1}{4} \sqrt{\lambda ^4+8 \lambda
   ^2}-\frac{\lambda ^2}{4}<C_2<\frac{\lambda ^2}{2}$, o  
           \item $\lambda >\sqrt{2},  \frac{1}{4} \sqrt{\lambda ^4+8 \lambda ^2}-\frac{\lambda
   ^2}{4}<C_2<\frac{1}{4} \sqrt{\lambda ^4+16 \lambda ^2}-\frac{\lambda
   ^2}{4}$.
       \end{enumerate}
       \item no hiperbólico para 
       \begin{enumerate}
           \item $\lambda \geq 1,  C_2=\frac{\lambda ^2}{2}$, o  
           \item $\lambda >0, C_2=\frac{1}{4} \sqrt{\lambda ^4+16 \lambda
   ^2}-\frac{\lambda ^2}{4}$, o  
           \item $\lambda >0,  C_2=\frac{1}{4} \sqrt{\lambda ^4+8 \lambda
   ^2}-\frac{\lambda ^2}{4}$.
       \end{enumerate}
   \end{enumerate}
   \item $Q_{28}:(\Sigma,A,u,w)=\left(\frac{C_2-1}{\lambda ^2}-\frac{1}{2},\frac{1-C_2}{\lambda
   ^2}-\frac{1}{2},\frac{-2 C_2+\lambda ^2+2}{2 \sqrt{2} \lambda },\frac{ 
   \sqrt{(C_2-1) \left(2 C_2+\lambda ^2-2\right)}}{\lambda }\right)$ existe para $\lambda >0, C_2>0, 2 C_2+\lambda ^2\leq 2$ or $\lambda >0, C_2\geq 1$ con autovalores\\ $\left\{\frac{4-4 C_2}{\lambda ^2},\frac{e_1(\lambda,C_2)}{32 C_2 \lambda ^{12}},\frac{e_2(\lambda,C_2)}{32 C_2 \lambda ^{12}},\frac{e_3(\lambda,C_2)}{32 C_2 \lambda ^{12}}\right\}$. En la figura \ref{P11eigen} se muestran las partes reales de los $e_i$, donde se aprecia que el punto tiene comportamiento de silla o no hiperbólico.
   
   \item $Q_{29}:(\Sigma,A,u,w)=\left(-\frac{\lambda ^2}{2 C_2+\lambda ^2},0,\frac{\sqrt{2} C_2 \lambda }{2
   C_2+\lambda ^2}, \sqrt{-\frac{(1-4 C_2)^2 C_2 \lambda
   ^2 \left(2 (4 C_2-1) \lambda ^2-4
   C_2+\lambda ^4\right)^2}{2
   C_2+\lambda ^2}}\right)$ existe para $C_2=\frac{1}{4}, \lambda >0$ o para $C_2=\frac{\lambda ^2 \left(\lambda ^2-2\right)}{4-8 \lambda ^2}, 1<\lambda
   <\sqrt{2}$ o para $C_2=\frac{\lambda ^2 \left(\lambda ^2-2\right)}{4-8 \lambda ^2},
   \frac{1}{\sqrt{2}}<\lambda <1$. Las partes reales de los autovalores se muestran la figura \ref{P12eigen}.
   
   \item $Q_{30}:(\Sigma,A,u,w)=\left(0,  \frac{1}{\sqrt{C_2-\frac{\lambda ^2}{2}}},0, \sqrt{2} \sqrt{-\frac{\lambda ^2
   \left(-2 C_2+\lambda ^2+8\right)^2 \left(2 (2 C_2+1) \lambda ^2+4
   (C_2-1) C_2+\lambda ^4\right)^2}{\lambda ^2-2 C_2}}\right)$ con $\varepsilon=1$ existe para $\lambda >0, C_2\geq \frac{\lambda ^2}{2}$,  con autovalores \\ $\left\{-\frac{f_1(\lambda,C_2)}{2 C_2 \lambda  \left(\lambda ^2-2 C_2\right)^2}, -\frac{f_2(\lambda,C_2)}{2 C_2 \lambda  \left(\lambda ^2-2 C_2\right)^2}, -\frac{f_3(\lambda,C_2)}{2 C_2 \lambda  \left(\lambda ^2-2 C_2\right)^2}, -\frac{f_4(\lambda,C_2)}{2 C_2 \lambda  \left(\lambda ^2-2 C_2\right)^2}\right\}$. 
   La parte real de $f_i$  se muestra en la figura  \ref{P13P14eigen} (izq.).
   
   \item $Q_{31}:(\Sigma,A,u,w)=\left(0, \frac{1}{\sqrt{C_2-\frac{\lambda ^2}{2}}},0, \sqrt{2} \sqrt{-\frac{\lambda ^2
   \left(-2 C_2+\lambda ^2+8\right)^2 \left(2 (2 C_2+1) \lambda ^2+4
   (C_2-1) C_2+\lambda ^4\right)^2}{\lambda ^2-2 C_2}}\right)$ con $\varepsilon=-1$ existe para $\lambda >0, C_2\geq \frac{\lambda ^2}{2}$ con autovalores  \\
   $\left\{ \frac{g_1(\lambda,C_2)}{C_2 \lambda  \left(2 C_2-\lambda ^2\right)^{3/2}},  \frac{g_2(\lambda,C_2)}{C_2 \lambda  \left(2 C_2-\lambda ^2\right)^{3/2}},  \frac{g_3(\lambda,C_2)}{C_2 \lambda  \left(2 C_2-\lambda ^2\right)^{3/2}},  \frac{g_4(\lambda,C_2)}{C_2 \lambda  \left(2 C_2-\lambda ^2\right)^{3/2}}\right\}$.  Las partes reales de los   $g_i$  se muestran en la figura \ref{P13P14eigen} (der.).
\end{enumerate}

\begin{figure*}[!htb]
    \centering
    \includegraphics[scale=0.45]{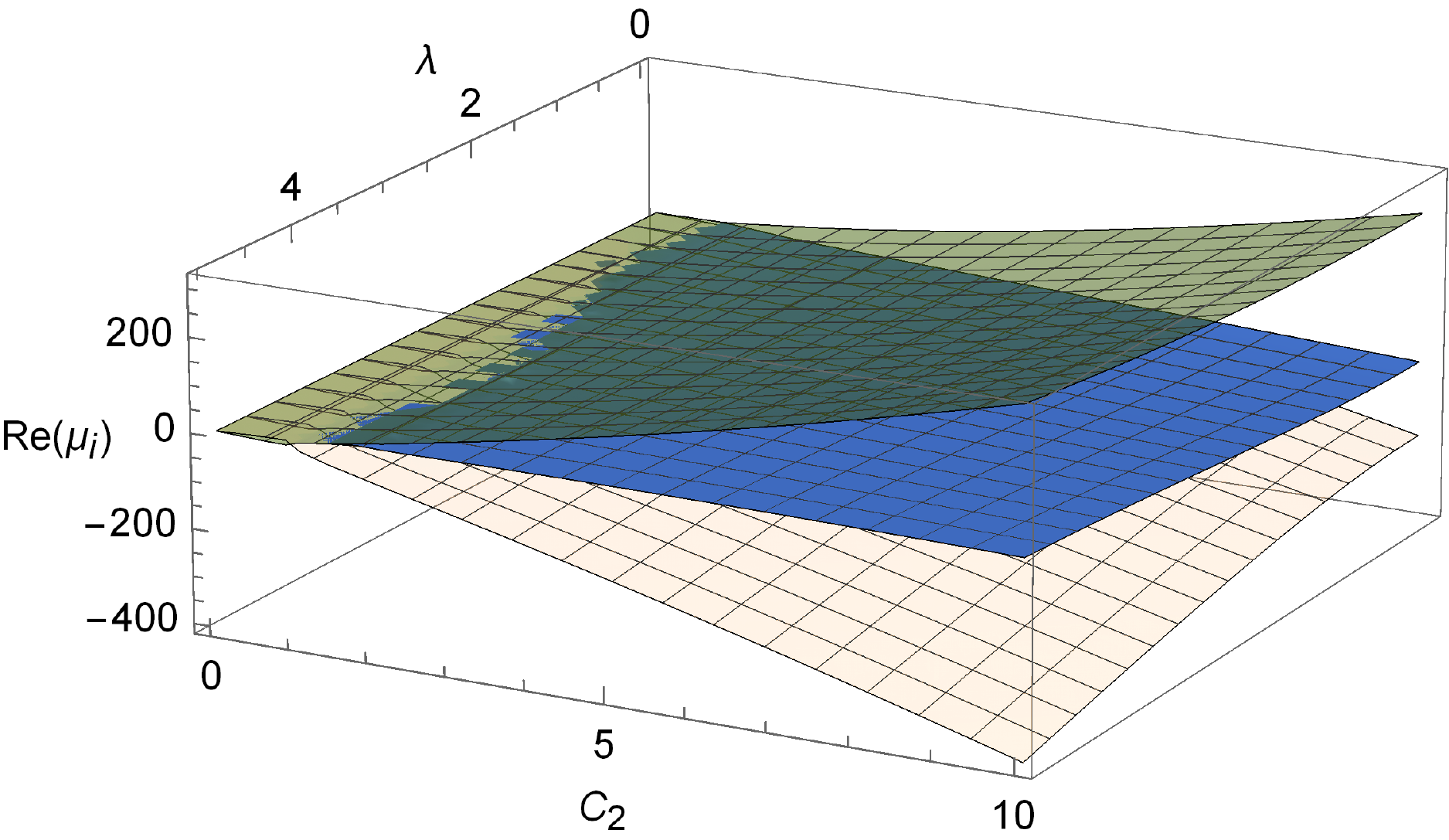}  \caption[Parte real de los $\mu_i$ correspondientes a $(\Sigma,A,u,w)=\left(-\frac{1}{2},\frac{1}{2} \varepsilon ,\frac{\lambda }{2 \sqrt{2}},0\right)$.]{\label{P4eigen} Parte real de los  $\mu_i$ correspondientes al punto de equilibrio $(\Sigma,A,u,w)=\left(-\frac{1}{2},\frac{1}{2} \varepsilon ,\frac{\lambda }{2 \sqrt{2}},0\right)$.}
   \end{figure*}
   
  En la figura \ref{P4eigen} se representa gráficamente la parte real de los  $\mu_i$ correspondientes al punto de equilibrio $(\Sigma,A,u,w)=\left(-\frac{1}{2},\frac{1}{2} \varepsilon ,\frac{\lambda }{2 \sqrt{2}},0\right)$. Dicha figura ilustra que los puntos de equilibrio tienen comportamiento de silla o son no hiperbólicos.
  
\begin{figure*}[!htb]
    \centering
    \includegraphics[scale=0.33]{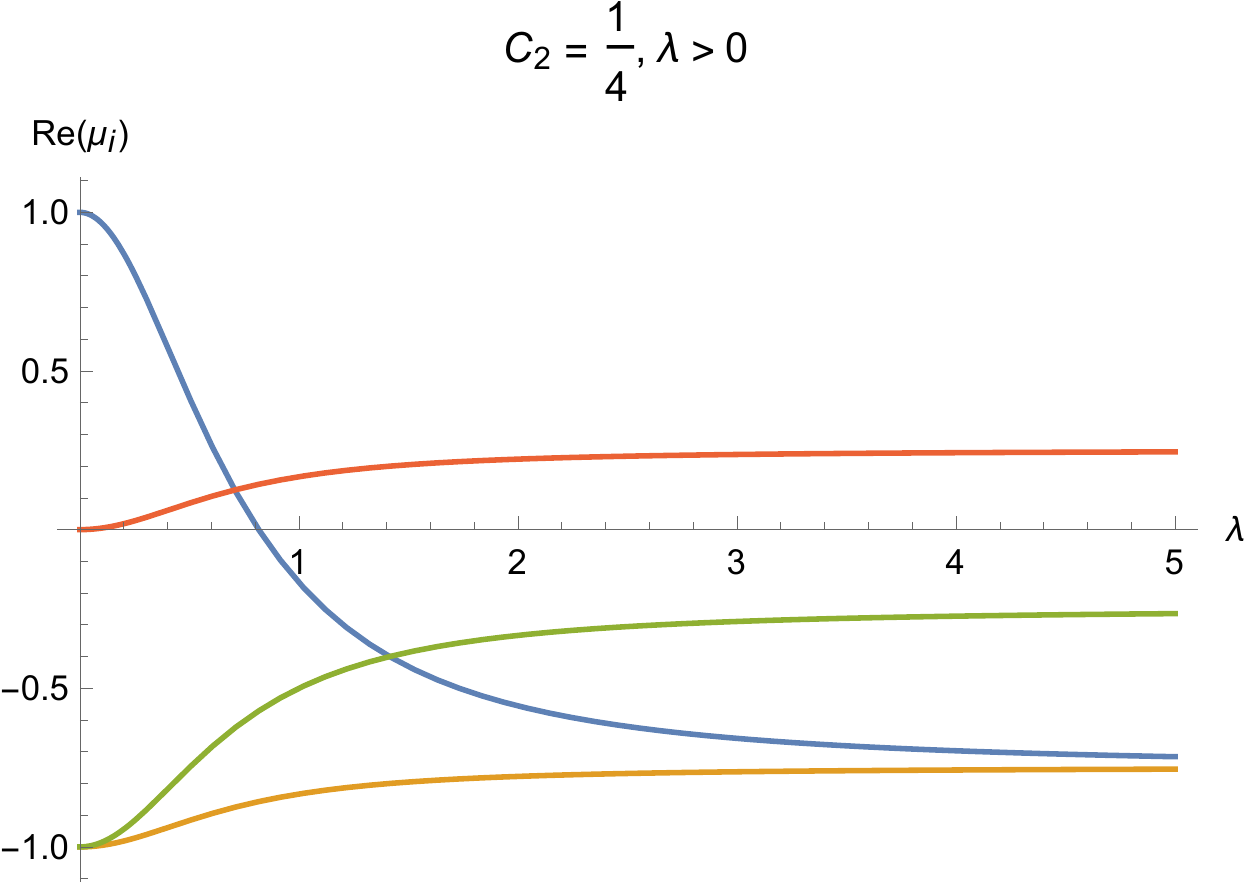} \hspace{1cm}
    \includegraphics[scale=0.33]{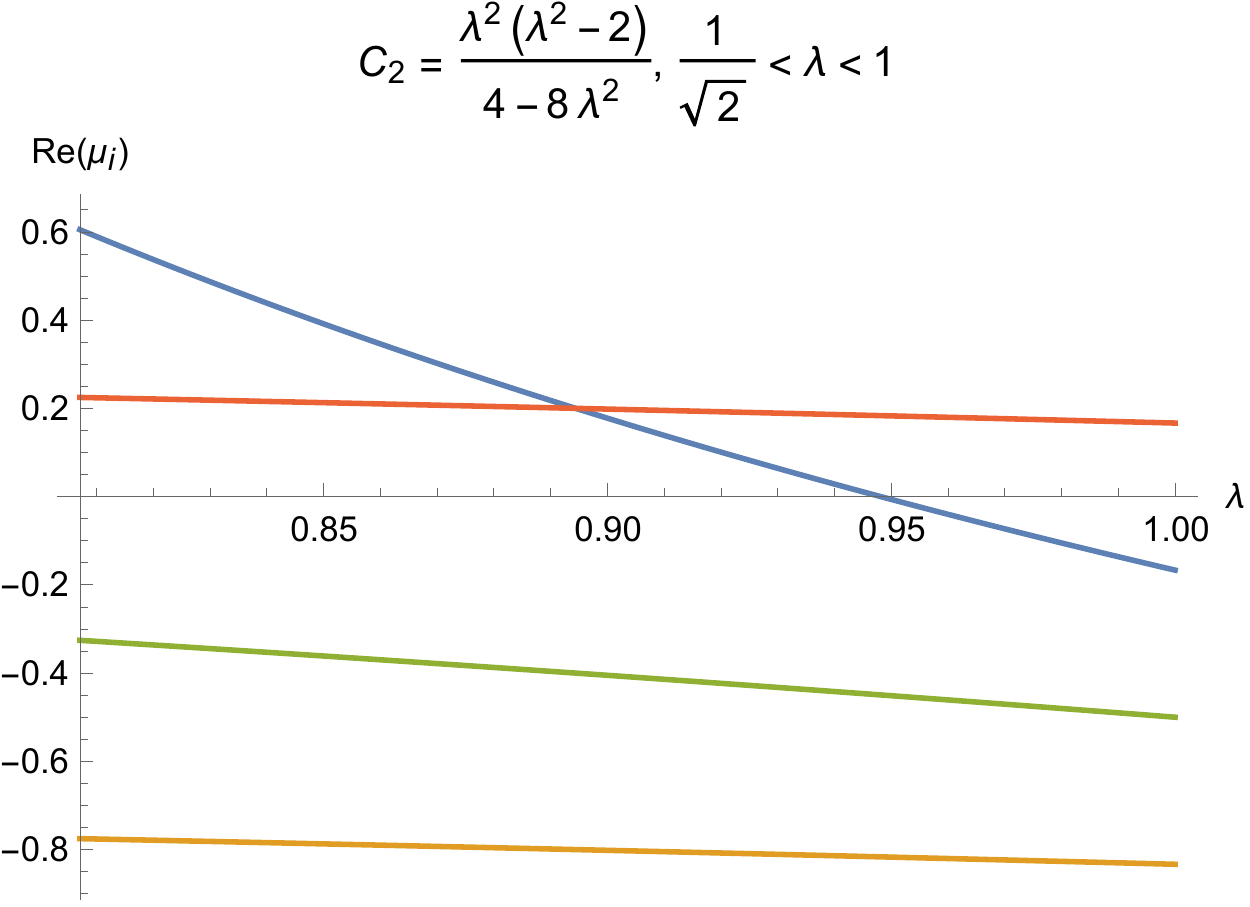} \hspace{1cm}
    \includegraphics[scale=0.33]{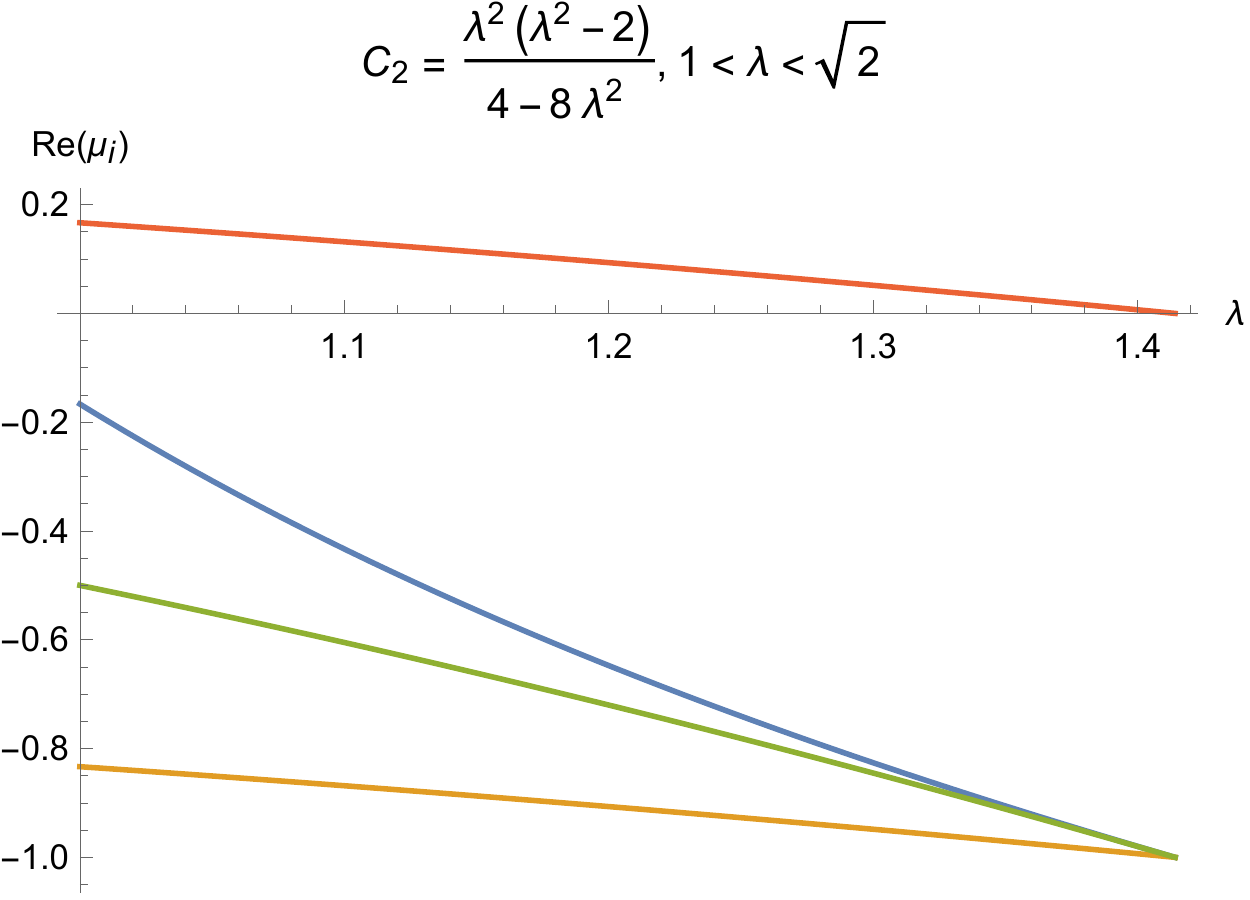}
    \caption[{Parte real de los autovalores del punto  $(\Sigma,A,u,w)=\left(-\frac{\lambda ^2}{2 C_2+\lambda ^2},0,\frac{\sqrt{2} C_2 \lambda }{2
   C_2+\lambda ^2}, 0\right)$, $C_2 \in \left\{\frac{1}{4}, \frac{\lambda ^2 \left(\lambda ^2-2\right)}{4-8 \lambda ^2}\right\}$.}]{\label{P12eigen} Parte real de los autovalores del punto  $(\Sigma,A,u,w)=\left(-\frac{\lambda ^2}{2 C_2+\lambda ^2},0,\frac{\sqrt{2} C_2 \lambda }{2
   C_2+\lambda ^2}, 0\right)$, $C_2 \in \left\{\frac{1}{4}, \frac{\lambda ^2 \left(\lambda ^2-2\right)}{4-8 \lambda ^2}\right\}$, para diferentes elecciones de los parámetro $\lambda$.}
   \end{figure*}
   
En la figura \ref{P12eigen} se representa gráficamente la parte real de los autovalores del punto  $(\Sigma,A,u,w)=\left(-\frac{\lambda ^2}{2 C_2+\lambda ^2},0,\frac{\sqrt{2} C_2 \lambda }{2
   C_2+\lambda ^2}, 0\right)$, $C_2 \in \left\{\frac{1}{4}, \frac{\lambda ^2 \left(\lambda ^2-2\right)}{4-8 \lambda ^2}\right\}$, para diferentes elecciones del parámetro $\lambda$. En dicha figura se ilustra que el punto tiene comportamiento general de silla.

\begin{figure*}
    \centering
    \includegraphics[scale=0.5]{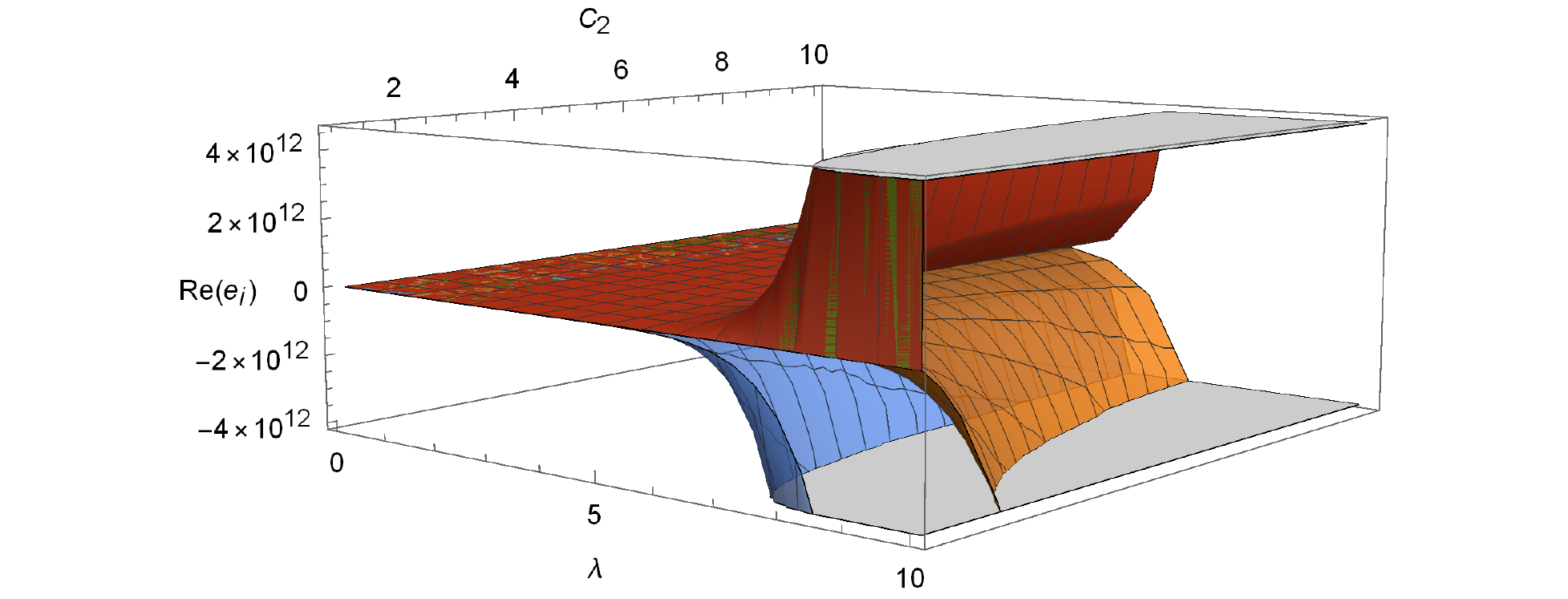}  \caption[Parte real de $e_i$ correspondiente al punto de equilibrio \newline $(\Sigma,A,u,w)=\left(\frac{C_2-1}{\lambda ^2}-\frac{1}{2},\frac{1-C_2}{\lambda
   ^2}-\frac{1}{2},\frac{-2 C_2+\lambda ^2+2}{2 \sqrt{2} \lambda },\frac{ 
   \sqrt{(C_2-1) \left(2 C_2+\lambda ^2-2\right)}}{\lambda }\right)$.]{\label{P11eigen} Parte real de $e_i$ correspondiente al punto de equilibrio $(\Sigma,A,u,w)=\left(\frac{C_2-1}{\lambda ^2}-\frac{1}{2},\frac{1-C_2}{\lambda
   ^2}-\frac{1}{2},\frac{-2 C_2+\lambda ^2+2}{2 \sqrt{2} \lambda },\frac{ 
   \sqrt{(C_2-1) \left(2 C_2+\lambda ^2-2\right)}}{\lambda }\right)$.}
   \end{figure*}
   
   En la figura \ref{P11eigen} se representa gráficamente la parte real de $e_i$ correspondientes al punto de equilibrio $(\Sigma,A,u,w)=\left(\frac{C_2-1}{\lambda ^2}-\frac{1}{2},\frac{1-C_2}{\lambda
   ^2}-\frac{1}{2},\frac{-2 C_2+\lambda ^2+2}{2 \sqrt{2} \lambda },\frac{ 
   \sqrt{(C_2-1) \left(2 C_2+\lambda ^2-2\right)}}{\lambda }\right)$. La figura ilustra que el punto de equilibrio tiene comportamiento de silla o es no hiperbólico.
   
   \begin{figure*}
    \centering
    \includegraphics[scale=0.35]{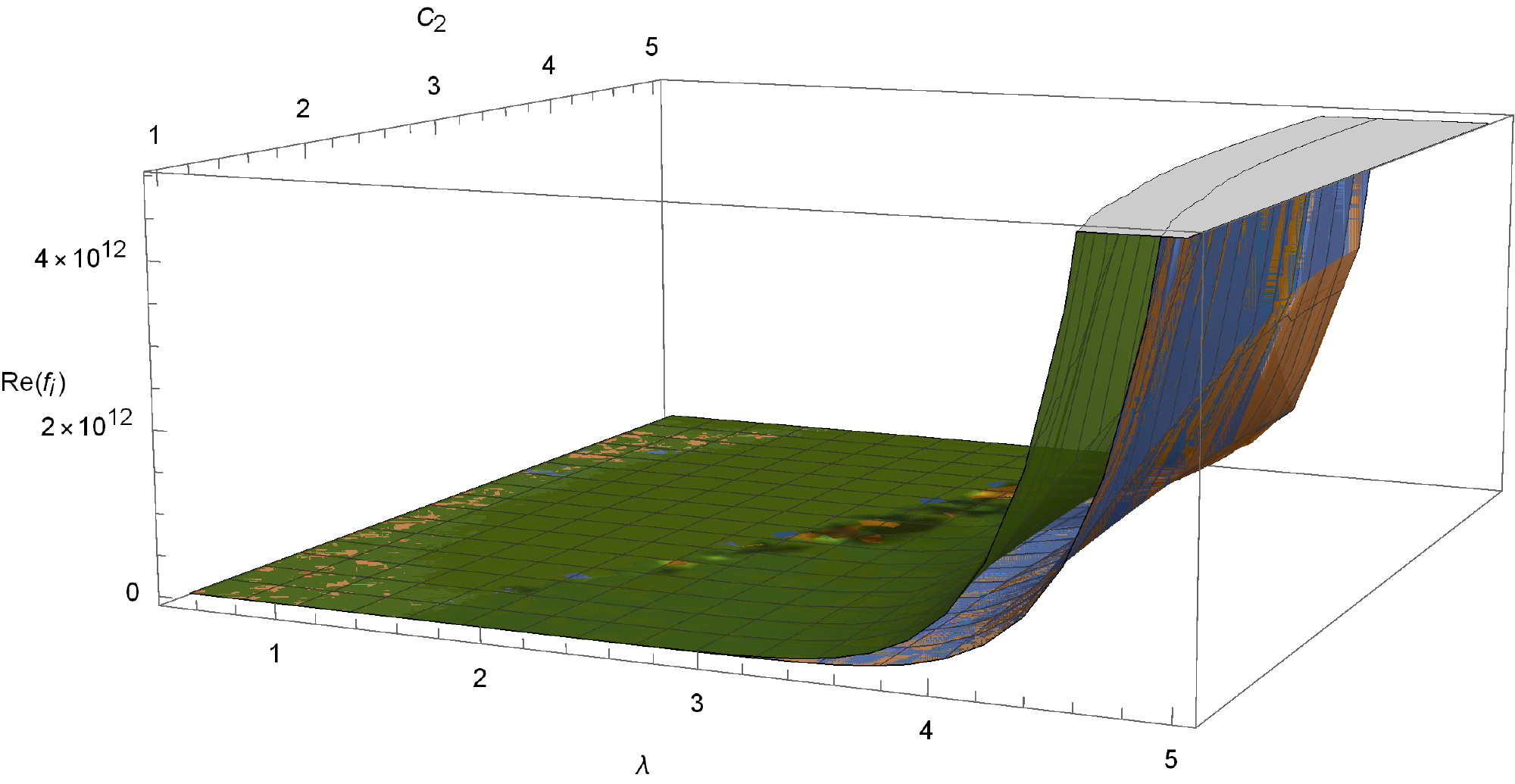} \hspace{0.1cm}
    \includegraphics[scale=0.35]{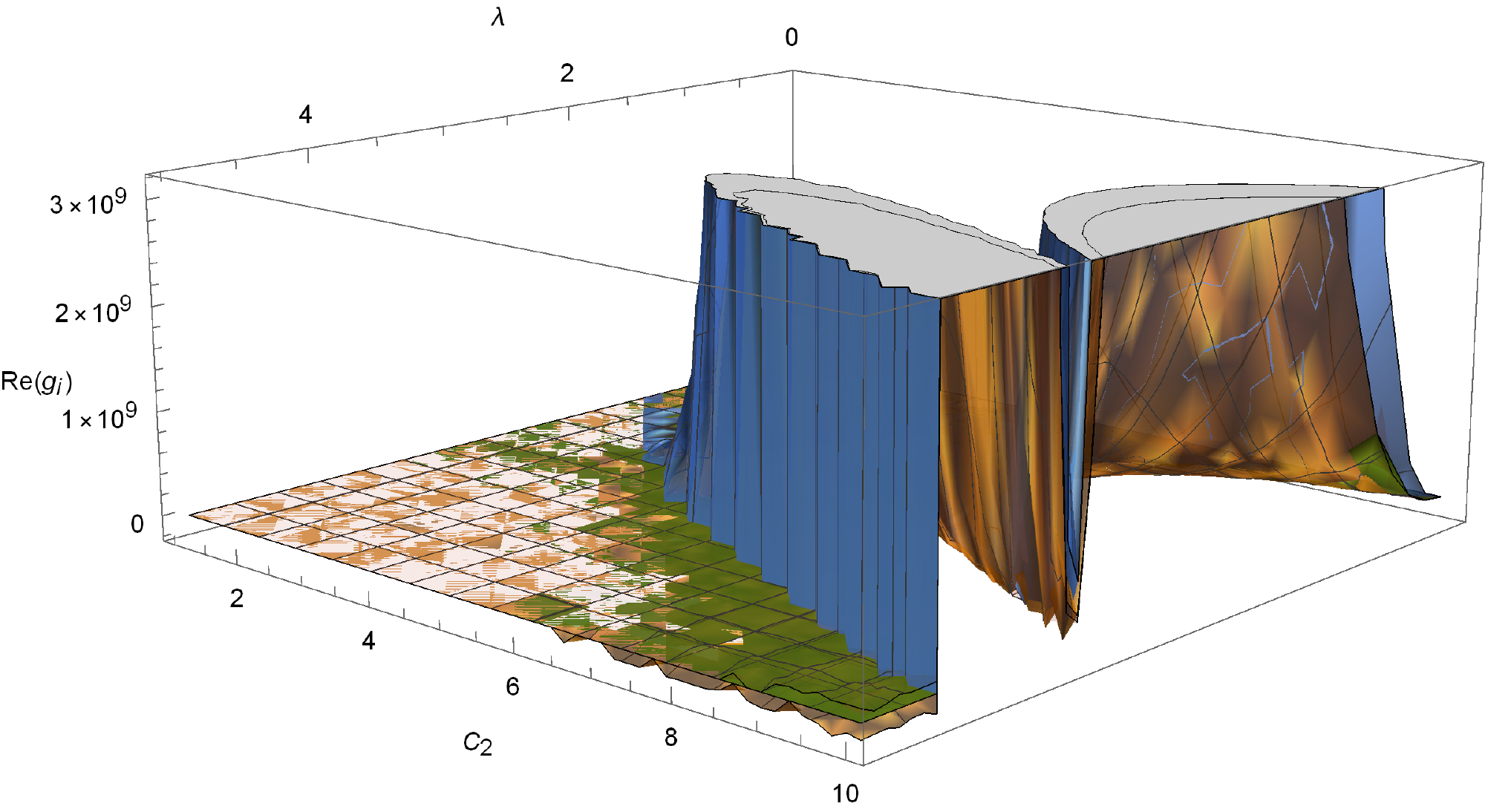}
    \caption[{\small{Parte real de $f_i$ (izq.) y  $g_i$ (der.) correspodiente a los puntos de equilibrio \newline $(\Sigma,A,u,w)=\left(0,  \frac{1}{\sqrt{C_2-\frac{\lambda ^2}{2}}},0, \sqrt{2} \sqrt{-\frac{\lambda ^2
   \left(-2 C_2+\lambda ^2+8\right)^2 \left(2 (2 C_2+1) \lambda ^2+4
   (C_2-1) C_2+\lambda ^4\right)^2}{\lambda ^2-2 C_2}}\right), \varepsilon=1$ y \newline  $(\Sigma,A,u,w)=\left(0,  \frac{1}{\sqrt{C_2-\frac{\lambda ^2}{2}}},0, \sqrt{2} \sqrt{-\frac{\lambda ^2
   \left(-2 C_2+\lambda ^2+8\right)^2 \left(2 (2 C_2+1) \lambda ^2+4
   (C_2-1) C_2+\lambda ^4\right)^2}{\lambda ^2-2 C_2}}\right), \varepsilon=-1$, respectivamente.}}]{\label{P13P14eigen} Parte real de $f_i$ (izq.) y  $g_i$ (der.) para los puntos de equilibrio $(\Sigma,A,u,w)=\left(0,  \frac{1}{\sqrt{C_2-\frac{\lambda ^2}{2}}},0, \sqrt{2} \sqrt{-\frac{\lambda ^2
   \left(-2 C_2+\lambda ^2+8\right)^2 \left(2 (2 C_2+1) \lambda ^2+4
   (C_2-1) C_2+\lambda ^4\right)^2}{\lambda ^2-2 C_2}}\right), \varepsilon=1$ y $(\Sigma,A,u,w)=\left(0,  \frac{1}{\sqrt{C_2-\frac{\lambda ^2}{2}}},0, \sqrt{2} \sqrt{-\frac{\lambda ^2
   \left(-2 C_2+\lambda ^2+8\right)^2 \left(2 (2 C_2+1) \lambda ^2+4
   (C_2-1) C_2+\lambda ^4\right)^2}{\lambda ^2-2 C_2}}\right), \varepsilon=-1$, respectivamente. Tienen comportamiento de fuentes o no hiperbólicos con cuatro autovalores imaginarios puros.}
   \end{figure*}
   
   En la figura \ref{P13P14eigen} se representa la parte real de $f_i$ (izq.) y  $g_i$ (der.) correspondientes a los puntos de equilibrio 
   \newline
   $(\Sigma,A,u,w)=\left(0,  \frac{1}{\sqrt{C_2-\frac{\lambda ^2}{2}}},0, \sqrt{2} \sqrt{-\frac{\lambda ^2
   \left(-2 C_2+\lambda ^2+8\right)^2 \left(2 (2 C_2+1) \lambda ^2+4
   (C_2-1) C_2+\lambda ^4\right)^2}{\lambda ^2-2 C_2}}\right), \varepsilon=1$ y 
   \newline
   $(\Sigma,A,u,w)=\left(0,  \frac{1}{\sqrt{C_2-\frac{\lambda ^2}{2}}},0, \sqrt{2} \sqrt{-\frac{\lambda ^2
   \left(-2 C_2+\lambda ^2+8\right)^2 \left(2 (2 C_2+1) \lambda ^2+4
   (C_2-1) C_2+\lambda ^4\right)^2}{\lambda ^2-2 C_2}}\right), \varepsilon=-1$, respectivamente. De acuerdo a la figura dichos puntos de equilibrio son fuentes o son no hiperbólicos (con cuatro autovalores imaginarios puros).
   
\FloatBarrier

\subsection{Conjunto invariante $A=v=0$}
\label{Section4.7}

En el conjunto invariante $A=v=0$, las ecuaciones se reducen a 
\begin{subequations}
\label{campo-escalar-D}
\begin{align}
    & \Sigma'= 2 C_2 \Sigma ^3+\frac{-\frac{3 \gamma  \Omega }{2}+w^2+\Omega }{C_2}+\Sigma  \left(K+2 u^2-2\right), \\
    & K'= 2 K \left(2 C_2 \Sigma ^2+K+2 u^2-1\right), \\
    & u'=u \left(2 C_2 \Sigma ^2+K-2\right)+2 u^3-\frac{\sqrt{2} w^2}{\lambda }, \\
    & w'= w \left(2 C_2 \Sigma ^2+K+\Sigma +2 u^2-\frac{\sqrt{2} u}{\lambda }\right),
\end{align}
con restricción 
\begin{equation}
\label{eq4:72}
  C_2 \Sigma^2+K+u^2 -(1-\gamma)\Omega_t-w^2 =1.  
\end{equation}
Para $1<\gamma\leq 2$, tenemos la ecuación auxiliar
\begin{equation}
 \Omega_t'= \Omega_t  \left(\left(\frac{1}{\gamma -1}+3\right) \Sigma +4 C_2 \Sigma ^2+2 K+4 u^2\right).
\end{equation}
\end{subequations}

\subsubsection{Sistema reducido}
Para $1<\gamma\leq 2$, la restriccion \eqref{eq4:72} puede ser resuelta globalmente para $\Omega_t$, resultado el sistema reducido 
\begin{subequations}
\label{eq:4.74}
\begin{align}
& \Sigma'=\frac{(3 \gamma -2) \left(C_2 \Sigma ^2+K+u^2-w^2-1\right)}{2 (\gamma -1) C_2}+2 C_2 \Sigma ^3+\frac{w^2}{C_2}+\Sigma  \left(K+2 u^2-2\right) \\
& K'= 2 K \left(2 C_2 \Sigma ^2+K+2 u^2-1\right) \\
& u'= u \left(2 C_2 \Sigma ^2+K+2 u^2-2\right)-\frac{\sqrt{2} w^2}{\lambda } \\
& w'= w \left(2 C_2 \Sigma ^2+K+\Sigma +2 u^2-\frac{\sqrt{2} u}{\lambda }\right). 
\end{align}
\end{subequations}
Para $u=w=0$ y $\lambda \rightarrow 0$ se obtiene el sistema 
\eqref{eq:3.57} se recuperan los puntos de equilibrio de la sección \ref{sect:3.7.1}. 
En la siguiente lista se enumeran los puntos de equilibrio del sistema reducido \eqref{eq:4.74}. Por definición $K\geq 0$, y $w\geq 0$ (si $\theta>0$).  Dado que el sistema \eqref{eq:4.74} es invariante ante el cambio simultáneo $(u, \lambda) \rightarrow (-u, -\lambda)$, y hemos asumido $\lambda>0$, en la discusión siguiente se restringe el análisis a $\lambda>0, u\geq 0$ (el signo de $u$ corresponde al signo de $\Psi$, si $\theta>0$). 

\begin{enumerate}

   \item $N_1: (\Sigma, K, u, w)= \left(0 ,  1 ,  0 ,  0\right)$, $\Omega_t= 0$. Tiene autovalores $\{-1,-1,1,2\}$, por tanto es una silla. 
    
    \item  $L: (\Sigma, K, u, w)=\left(\Sigma_0, 0 , \sqrt{1-\Sigma_0 ^2 C_2}, 0 \right)$,  $\Omega_t=0$. Esta curva de puntos no hiperbólicos contiene a los puntos  $P_{11}$  y $P_{12}$ estudiados en la sección \ref{sect:3.7.1} cuando $u=0$. Los autovalores son $\left\{0,\frac{(3 \gamma -2) \Sigma_0 }{\gamma -1}+4,-\frac{\sqrt{2-2 C_2
   \Sigma_0 ^2}}{\lambda }+\Sigma_0 +2,2\right\}$.    Esta curva es normalmente hiperbólica. En efecto, la curva paramétrica definida por $$\Sigma (\Sigma_0)=\Sigma_0,\quad A(\Sigma_0)=0,\quad u(\Sigma_0)= \sqrt{1-C_2 \Sigma_0 ^2},\quad w(\Sigma_0)=0,$$  tiene como vector tangente en $\Sigma_0$ a $$ \dot{\textbf{r}}(\Sigma_0)=\left(1,0,-\frac{ C_2 \Sigma_0}{\sqrt{1-C_2 \Sigma_0 ^2}},0\right)$$ que  es paralelo al autovector asociado al autovalor nulo dado por $$\textbf{v}(\Sigma_0)=\left(-\frac{\sqrt{1-C_2 \Sigma _0^2}}{C_2 \Sigma _0},0,1,0\right).$$ En este caso particular, de acuerdo a lo comentado a continuación de la definición \ref{normhiper}, sección \ref{seccion1.2}, se puede estudiar la estabilidad de la curva de puntos de equilibrio considerando solo los signos de la parte real de los autovalores no nulos. De esta manera se concluye que:
   \begin{enumerate}
       \item $L$ es una fuente para 
       \begin{enumerate}
           \item $C_2>0,  \Sigma_0=\frac{1}{\sqrt{C_2}},  \lambda >0,  1<\gamma <2$, o  
           \item $C_2>1,  \Sigma_0=-\sqrt{\frac{1}{C_2}},  \lambda >0,  \frac{2 \Sigma_{0}+4}{3 \Sigma_0+4}<\gamma <2$, o  
           \item $C_2>1,  0<\Sigma_0<\sqrt{\frac{1}{C_2}},  \lambda >\sqrt{2} \sqrt{\frac{1-C_2 \Sigma_0^2}{(\Sigma_0+2)^2}}, 1<\gamma <2$, o  
           \item $C_2>0,  0<\Sigma_0<\sqrt{\frac{1}{C_2}},  \lambda >\sqrt{2} \sqrt{\frac{1-C_2 \Sigma_0^2}{(\Sigma_0+2)^2}},  1<\gamma <2$, o  
           \item $C_2>1,  -\sqrt{\frac{1}{C_2}}<\Sigma_0\leq 0,  \lambda >\sqrt{2} \sqrt{\frac{1-C_2 \Sigma_0^2}{(\Sigma_0+2)^2}},  \frac{2 \Sigma_0+4}{3 \Sigma_0+4}<\gamma <2$, o  
           \item $0<C_2\leq 1,  -1<\Sigma_0\leq 0,  \lambda >\sqrt{2} \sqrt{\frac{1-C_2 \Sigma_0^2}{(\Sigma_0+2)^2}},  \frac{2 \Sigma_0+4}{3
   \Sigma_0+4}<\gamma <2$, o  
   \item $C_2>1,  -\sqrt{\frac{1}{C_2}}<\Sigma_0\leq 0,  \lambda >\sqrt{2} \sqrt{\frac{1-C_2 \Sigma_0^2}{(\Sigma_0+2)^2}},  \frac{2
   \Sigma_0+4}{3 \Sigma_0+4}<\gamma <2$, o  
   \item $0<C_2\leq 1,  -1<\Sigma_0\leq 0,  \lambda >\sqrt{2} \sqrt{\frac{1-C_2 \Sigma_0^2}{(\Sigma_0+2)^2}}, 
   \frac{2 \Sigma_0+4}{3 \Sigma_0+4}<\gamma <2$, o  
   \item $C_2>1,  -\sqrt{\frac{1}{C_2}}<\Sigma_0\leq 0,  \lambda >\sqrt{2} \sqrt{\frac{1-C_2 \Sigma_0^2}{(\Sigma_0+2)^2}},  \frac{2 \Sigma_0+4}{3 \Sigma_0+4}<\gamma <2$, o  
   \item $C_2>1,  0<\Sigma_0<\sqrt{\frac{1}{C_2}},  \lambda >\sqrt{2}
   \sqrt{\frac{1-C_2 \Sigma_0^2}{(\Sigma_0+2)^2}},  1<\gamma <2$, o  
   \item $0<C_2\leq 1,  -1<\Sigma_0\leq 0,  \lambda >\sqrt{2} \sqrt{\frac{1-C_2 \Sigma_0^2}{(\Sigma_0+2)^2}},  \frac{2 \Sigma_0+4}{3 \Sigma_0+4}<\gamma <2$, o  
   \item $0<C_2\leq 1,  0<\Sigma_0<\sqrt{\frac{1}{C_2}},  \lambda >\sqrt{2}
   \sqrt{\frac{1-C_2 \Sigma_0^2}{(\Sigma_0+2)^2}},  1<\gamma <2$, o  
   \item $C_2>1,  -\sqrt{\frac{1}{C_2}}<\Sigma_0\leq 0,  \lambda >\sqrt{2} \sqrt{\frac{1-C_2
   \Sigma_0^2}{(\Sigma_0+2)^2}},  \frac{2 \Sigma_0+4}{3 \Sigma_0+4}<\gamma <2$, o  
   \item $C_2>1,  0<\Sigma_0<\sqrt{\frac{1}{C_2}},  \lambda >\sqrt{2}
   \sqrt{\frac{1-C_2 \Sigma_0^2}{(\Sigma_0+2)^2}},  1<\gamma <2$, o  
   \item $0<C_2\leq 1,  -1<\Sigma_0\leq 0,  \lambda >\sqrt{2} \sqrt{\frac{1-C_2 \Sigma_0^2}{(\Sigma_0+2)^2}},  \frac{2 \Sigma_0+4}{3 \Sigma_0+4}<\gamma <2$, o  
   \item $0<C_2\leq 1,  0<\Sigma_0<\sqrt{\frac{1}{C_2}},  \lambda >\sqrt{2}
   \sqrt{\frac{1-C_2 \Sigma_0^2}{(\Sigma_0+2)^2}},  1<\gamma <2$.
       \end{enumerate}
       \item $L$ es una silla para
       \begin{enumerate}
           \item $0<C_2\leq \frac{1}{4},  -2<\Sigma_0\leq -1,  \lambda >\sqrt{2} \sqrt{\frac{1-C_2 \Sigma_0^2}{(\Sigma_0+2)^2}},  1<\gamma <2$, o 
           \item $0<C_2\leq \frac{1}{4}, 
   -1<\Sigma_0<0,  \lambda >\sqrt{2} \sqrt{\frac{1-C_2 \Sigma_0^2}{(\Sigma_0+2)^2}},  1<\gamma <\frac{2 \Sigma_0+4}{3 \Sigma_0+4}$, o 
   \item $\frac{1}{4}<C_2\leq 1,  \Sigma_0=-\frac{1}{\sqrt{C_2}},  \lambda >0,  1<\gamma <2$, o 
   \item $\frac{1}{4}<C_2<1,  -\sqrt{\frac{1}{C_2}}<\Sigma_0\leq -1, 
   \lambda >\sqrt{2} \sqrt{\frac{1-C_2 \Sigma_0^2}{(\Sigma_0+2)^2}},  1<\gamma <2$, o 
   \item $\frac{1}{4}<C_2\leq 1,  -1<\Sigma_0<0,  \lambda >\sqrt{2} \sqrt{\frac{1-C_2
   \Sigma_0^2}{(\Sigma_0+2)^2}},  1<\gamma <\frac{2 \Sigma_0+4}{3 \Sigma_0+4}$, o 
   \item $C_2>1,  \Sigma_0=-\sqrt{\frac{1}{C_2}},  \lambda >0,  1<\gamma
   <\frac{2 \Sigma_0+4}{3 \Sigma_0+4}$, o 
   \item $C_2>1,  -\sqrt{\frac{1}{C_2}}<\Sigma_0<0,  \lambda >\sqrt{2} \sqrt{\frac{1-C_2 \Sigma_0^2}{(\Sigma_0+2)^2}},  1<\gamma <\frac{2 \Sigma_0+4}{3 \Sigma_0+4}$, o 
   \item $0<C_2\leq 1,  0<\Sigma_0<\sqrt{\frac{1}{C_2}},  0<\lambda <\sqrt{2} \sqrt{\frac{1-C_2
   \Sigma_0^2}{(\Sigma_0+2)^2}},  1<\gamma <2$, o 
   \item $0<C_2\leq 1,  -1<\Sigma_0\leq 0,  0<\lambda <\sqrt{2} \sqrt{\frac{1-C_2 \Sigma_0^2}{(\Sigma_0+2)^2}},  \frac{2 \Sigma_0+4}{3 \Sigma_0+4}<\gamma <2$, o 
   \item $C_2>1,  0<\Sigma_0<\sqrt{\frac{1}{C_2}},  0<\lambda <\sqrt{2} \sqrt{\frac{1-C_2 \Sigma_0^2}{(\Sigma_0+2)^2}},  1<\gamma <2$, o 
   \item $C_2>1,  -\sqrt{\frac{1}{C_2}}<\Sigma_0\leq 0,  0<\lambda <\sqrt{2} \sqrt{\frac{1-C_2 \Sigma_0^2}{(\Sigma_0+2)^2}},  \frac{2 \Sigma_0+4}{3 \Sigma_0+4}<\gamma <2$, o 
   \item $0<C_2<\frac{1}{4},  -\sqrt{\frac{1}{C_2}}\leq \Sigma_0\leq -2,  \lambda >0,  1<\gamma <2$, o 
   \item $0<C_2<\frac{1}{4},  -1<\Sigma_0<0,  0<\lambda <\sqrt{2} \sqrt{\frac{1-C_2 \Sigma_0^2}{(\Sigma_0+2)^2}},  1<\gamma <\frac{2 \Sigma_0+4}{3 \Sigma_0+4}$, o 
   \item $0<C_2<\frac{1}{4},  -2<\Sigma_0\leq -1,  0<\lambda <\sqrt{2} \sqrt{\frac{1-C_2 \Sigma_0^2}{(\Sigma_0+2)^2}},  1<\gamma <2$, o 
   \item $\frac{1}{4}\leq C_2<1,  -\sqrt{\frac{1}{C_2}}<\Sigma_0\leq -1,  0<\lambda <\sqrt{2} \sqrt{\frac{1-C_2 \Sigma_0^2}{(\Sigma_0+2)^2}},  1<\gamma <2$, o 
   \item $\frac{1}{4}\leq C_2\leq 1,  -1<\Sigma_0<0,  0<\lambda <\sqrt{2} \sqrt{\frac{1-C_2 \Sigma_0^2}{(\Sigma_0+2)^2}},  1<\gamma <\frac{2 \Sigma_0+4}{3 \Sigma_0+4}$, o 
   \item $C_2>1,  -\sqrt{\frac{1}{C_2}}<\Sigma_0<0,  0<\lambda <\sqrt{2} \sqrt{\frac{1-C_2 \Sigma_0^2}{(\Sigma_0+2)^2}},  1<\gamma <\frac{2 \Sigma_0+4}{3 \Sigma_0+4}$.
       \end{enumerate}
    \item $L$ es no hiperbólico para 
     \begin{enumerate}
         \item $0<C_2<\frac{1}{4}, -1<\Sigma_0<0, \gamma =\frac{2 (\Sigma_0+2)}{3 \Sigma_0+4}, 0<\lambda <\sqrt{2} \sqrt{\frac{1-C_2 \Sigma_0^2}{(\Sigma_0+2)^2}}$, o 
         \item $0<C_2<\frac{1}{4}, -1<\Sigma_0<0, \gamma =\frac{2 (\Sigma_0+2)}{3 \Sigma_0+4}, \lambda >\sqrt{2} \sqrt{\frac{1-C_2 \Sigma_0^2}{(\Sigma_0+2)^2}}$, o 
         \item $0<C_2<\frac{1}{4}, -1<\Sigma_0<0, \sqrt{2} \sqrt{\frac{1-C_2 \Sigma_0^2}{(\Sigma_0+2)^2}}=\lambda , 1<\gamma <2$, o 
         \item $0<C_2<\frac{1}{4}, \sqrt{2} \sqrt{\frac{1-C_2 \Sigma_0^2}{(\Sigma_0+2)^2}}=\lambda , 1<\gamma <2, -2<\Sigma_0\leq -1$, o 
         \item $0<C_2<\frac{1}{4}, 
   \sqrt{2} \sqrt{\frac{1-C_2 \Sigma_0^2}{(\Sigma_0+2)^2}}=\lambda , 1<\gamma <2, 0\leq \Sigma_0<\frac{1}{\sqrt{C_2}}$, o 
   \item $C_2=\frac{1}{4}, -1<\Sigma_0<0, 
   \gamma =\frac{2 (\Sigma_0+2)}{3 \Sigma_0+4}, 0<\lambda <\sqrt{\frac{2}{\Sigma_0+2}-\frac{1}{2}}$, o 
   \item $C_2=\frac{1}{4}, -1<\Sigma_0<0, \gamma =\frac{2 (\Sigma_0+2)}{3 \Sigma_0+4}, 2 \lambda >\sqrt{\frac{8}{\Sigma_0+2}-2}$, o 
   \item $C_2=\frac{1}{4}, -1<\Sigma_0<0, \gamma =\frac{2 (\Sigma_0+2)}{3 \Sigma_0+4}, 
   \sqrt{\frac{8}{\Sigma_0+2}-2}=2 \lambda , 1<\gamma <2$, o 
   \item $C_2=\frac{1}{4}, 1<\gamma <2, \sqrt{\frac{8}{\Sigma_0+2}-2}=2 \lambda , -2<\Sigma_0\leq -1$, o 
   \item $4
   C_2=1, 1<\gamma <2, \sqrt{\frac{8}{\Sigma_0+2}-2}=2 \lambda , 0\leq \Sigma_0<2$, o 
   \item $C_2=\frac{1}{4}, 1<\gamma <2, \Sigma_0=-2, \lambda >0$, o 
   \item $\frac{1}{4}<C_2\leq 1, -1<\Sigma_0<0, \gamma =\frac{2 (\Sigma_0+2)}{3 \Sigma_0+4}, 0<\lambda <\sqrt{2} \sqrt{\frac{1-C_2 \Sigma_0^2}{(\Sigma_0+2)^2}}$, o 
   \item $\frac{1}{4}<C_2\leq 1, -1<\Sigma_0<0, \gamma =\frac{2 (\Sigma_0+2)}{3 \Sigma_0+4}, \lambda >\sqrt{2} \sqrt{\frac{1-C_2 \Sigma_0^2}{(\Sigma_0+2)^2}}$, o 
   \item $\frac{1}{4}<C_2\leq 1, -1<\Sigma_0<0, \sqrt{2} \sqrt{\frac{1-C_2 \Sigma_0^2}{(\Sigma_0+2)^2}}=\lambda , 1<\gamma
   <2$, o \item $\frac{1}{4}<C_2\leq 1, \sqrt{2} \sqrt{\frac{1-C_2 \Sigma_0^2}{(\Sigma_0+2)^2}}=\lambda , 1<\gamma <2, 0\leq \Sigma_0<\frac{1}{\sqrt{C_2}}$, o 
   \item $\frac{1}{4}<C_2\leq 1, \sqrt{2} \sqrt{\frac{1-C_2 \Sigma_0^2}{(\Sigma_0+2)^2}}=\lambda , 1<\gamma <2, -\frac{1}{\sqrt{C_2}}<\Sigma_0\leq -1$, o 
   \item $C_2>1, \gamma =\frac{2 (\Sigma_0+2)}{3 \Sigma_0+4}, \frac{1}{\sqrt{C_2}}+\Sigma_0=0, \lambda >0$, o 
   \item $C_2>1, \gamma =\frac{2 (\Sigma_0+2)}{3 \Sigma_0+4}, -\frac{1}{\sqrt{C_2}}<\Sigma_0<0, 0<\lambda <\sqrt{2} \sqrt{\frac{1-C_2 \Sigma_0^2}{(\Sigma_0+2)^2}}$, o 
   \item $C_2>1, \gamma
   =\frac{2 (\Sigma_0+2)}{3 \Sigma_0+4}, -\frac{1}{\sqrt{C_2}}<\Sigma_0<0, \lambda >\sqrt{2} \sqrt{\frac{1-C_2 \Sigma_0^2}{(\Sigma_0+2)^2}}$, o 
   \item $C_2>1, \sqrt{2} \sqrt{\frac{1-C_2 \Sigma_0^2}{(\Sigma_0+2)^2}}=\lambda , 1<\gamma <2, 0\leq \Sigma_0<\frac{1}{\sqrt{C_2}}$, o 
   \item $C_2>1, \sqrt{2} \sqrt{\frac{1-C_2 \Sigma_0^2}{(\Sigma_0+2)^2}}=\lambda , 1<\gamma <2, -\frac{1}{\sqrt{C_2}}<\Sigma_0<0$.
     \end{enumerate}
   \end{enumerate}
    
    \item $P_{13}: (\Sigma, K, u, w)= \left(\frac{2-3 \gamma }{4 (\gamma -1) C_2} ,  0 ,  0 ,  0\right)$,  $\Omega_t=\frac{1}{\gamma -1}-\frac{(2-3 \gamma )^2}{16 (\gamma -1)^3 C_2}$. Los autovalores son $\left\{\frac{\gamma  (3 \gamma -2)}{8 (\gamma -1)^2 C_2},\frac{(2-3
   \gamma )^2}{8 (\gamma -1)^2 C_2}-2,\frac{(2-3 \gamma )^2}{8 (\gamma
   -1)^2 C_2}-2,\frac{(2-3 \gamma )^2}{4 (\gamma -1)^2 C_2}-2\right\}$.
   
   \item $P_{14}: (\Sigma, K, u, w)=  \left( -\frac{2 (\gamma -1)}{3 \gamma -2} ,  1-\frac{8 (\gamma -1)^2 C_2}{(2-3
   \gamma )^2} ,  0 ,  0\right)$,  $\Omega_t= \frac{4 (\gamma -1) C_2}{(2-3 \gamma )^2}$. Los autovalores son 
   $\left\{-1,\frac{\gamma }{3 \gamma -2},-\frac{1}{2}- \frac{\sqrt{ \left(64 (\gamma -1)^2 C_2-7 (2-3
   \gamma )^2\right)}}{2 (3 \gamma -2)},-\frac{1}{2}+ \frac{\sqrt{ \left(64 (\gamma -1)^2 C_2-7 (2-3
   \gamma )^2\right)}}{2 (3 \gamma -2)}\right\}$. 
   
Los siguientes puntos son nuevos en comparación con los resultados de la sección \ref{sect:3.7.1}.

   \item $P_{13}(\lambda): (\Sigma, K, u, w)= \Bigg(-\frac{2 (\gamma -1) (3 \gamma -2)}{8 C_2 (\gamma -1)^2+\gamma ^2
   \lambda ^2} ,  0 ,  \frac{\gamma  (3 \gamma -2) \lambda }{\sqrt{2}
   \left(8 C_2 (\gamma -1)^2+\gamma ^2 \lambda ^2\right)}$ ,\newline $\frac{
   \sqrt{\gamma } \sqrt{3 \gamma -2} \lambda  \sqrt{4- 16 C_2 (\gamma -1)^2-\gamma  \left(\gamma 
   \left(2 \lambda ^2-9\right)+12\right)}}{\sqrt{2} \left(8 C_2 (\gamma -1)^2+\gamma
   ^2 \lambda ^2\right)}\Bigg)$, \newline $\Omega_t= \frac{\left(\gamma  \lambda ^2-4 (\gamma
   -1) C_2\right) \left(4- 16 C_2 (\gamma -1)^2-\gamma  \left(\gamma 
   \left(2 \lambda ^2-9\right)+12\right)\right)}{\left(8 C_2 (\gamma
   -1)^2+\gamma ^2 \lambda ^2\right){}^2}$.  Se reduce a $P_{13}$  cuando $\lambda\rightarrow 0$.  
 Los autovalores son \newline 
 \\\\
\begin{doublespace}
\noindent\( {\left\{-2+\frac{(2-3 \gamma )^2}{\gamma ^2 \lambda ^2+8 (-1+\gamma )^2 C_2},-2+\frac{2 (2-3 \gamma )^2}{\gamma ^2 \lambda ^2+8 (-1+\gamma
)^2 C_2},\right.}\\
 {-\left((-1+\gamma ) C_2 \left(\gamma ^2 \lambda ^2+8 (-1+\gamma )^2 C_2\right) \left(-4+\gamma  \left(12+\gamma  \left(-9+2 \lambda ^2\right)\right)+16
(-1+\gamma )^2 C_2\right)+\right.}\\
 {\surd \left((-1+\gamma ) C_2 \left(\gamma ^2 \lambda ^2+8 (-1+\gamma )^2 C_2\right){}^2 \left(-4+\gamma  \left(12+\gamma  \left(-9+2 \lambda
^2\right)\right)+16 (-1+\gamma )^2 C_2\right) \right.}\\
 {\left.\left.\left.\left(2 \gamma ^2 (-2+3 \gamma ) \lambda ^2+(-1+\gamma ) C_2 \left(-4+\gamma  \left(28+\gamma  \left(-33+2 \lambda ^2\right)\right)+16
(-1+\gamma )^2 C_2\right)\right)\right)\right)\right/}\\
 {\left(2 (-1+\gamma ) C_2 \left(\gamma ^2 \lambda ^2+8 (-1+\gamma )^2 C_2\right){}^2\right),\left(-(-1+\gamma ) C_2 \left(\gamma ^2 \lambda ^2+8
(-1+\gamma )^2 C_2\right) \right.}\\
 {\left(-4+\gamma  \left(12+\gamma  \left(-9+2 \lambda ^2\right)\right)+16 (-1+\gamma )^2 C_2\right)+}\\
 {\surd \left((-1+\gamma ) C_2 \left(\gamma ^2 \lambda ^2+8 (-1+\gamma )^2 C_2\right){}^2 \left(-4+\gamma  \left(12+\gamma  \left(-9+2 \lambda
^2\right)\right)+16 (-1+\gamma )^2 C_2\right) \right.}\\
 {\left.\left.\left.\left(2 \gamma ^2 (-2+3 \gamma ) \lambda ^2+(-1+\gamma ) C_2 \left(-4+\gamma  \left(28+\gamma  \left(-33+2 \lambda ^2\right)\right)+16
(-1+\gamma )^2 C_2\right)\right)\right)\right)\right/}\\
 {\left.\left(2 (-1+\gamma ) C_2 \left(\gamma ^2 \lambda ^2+8 (-1+\gamma )^2 C_2\right){}^2\right)\right\}}\).
\end{doublespace}
\begin{enumerate}
    \item $P_{13}(\lambda)$ es un pozo para 
    \begin{enumerate}
        \item $1<C_2\leq 2,  \frac{4 C_2-2}{4 C_2-3}<\gamma <2,  \sqrt{\frac{(2-3 \gamma )^2-8 (\gamma -1)^2 C_2}{\gamma ^2}}<\lambda <2 \sqrt{\frac{(\gamma -1) C_2}{\gamma }}$, o  
        \item $C_2>2,  \frac{4 C_2-2}{4 C_2-3}<\gamma <\frac{2 \left(4 C_2-3\right)}{8 C_2-9}+2 \sqrt{2} \sqrt{\frac{C_2}{\left(8 C_2-9\right){}^2}},  \sqrt{\frac{(2-3 \gamma )^2-8 (\gamma -1)^2 C_2}{\gamma ^2}}<\lambda <2 \sqrt{\frac{(\gamma-1)C_2}{\gamma }}$, o
   \item $C_2>2,  \frac{2 \left(4 C_2-3\right)}{8 C_2-9}+2 \sqrt{2} \sqrt{\frac{C_2}{\left(8 C_2-9\right){}^2}}\leq \gamma <2,  0<\lambda <2
   \sqrt{\frac{(\gamma -1)C_2}{\gamma }}$.
    \end{enumerate}
    
    \item Es una fuente para 
     \begin{enumerate}
         \item $0<C_2\leq \frac{1}{2},  1<\gamma <2,  2 \sqrt{\frac{(\gamma-1)C_2}{\gamma }}<\lambda <\frac{1}{2} \sqrt{\frac{2 (2-3 \gamma )^2-32 (\gamma -1)^2 C_2}{\gamma ^2}}$, o
         \item $C_2>\frac{1}{2},  1<\gamma
   <\frac{8 C_2-2}{8 C_2-3},  2 \sqrt{\frac{(\gamma-1)C_2}{\gamma }}<\lambda <\frac{1}{2} \sqrt{\frac{2 (2-3 \gamma )^2-32 (\gamma -1)^2 C_2}{\gamma ^2}}$.
     \end{enumerate}
     \item Es no hiperbólico para 
     \begin{enumerate}
         \item $C_2>\frac{1}{2},  1<\gamma <\frac{8 C_2-2}{8 C_2-3},  \lambda =2 \sqrt{\frac{(\gamma-1)C_2}{\gamma }}$, o
         \item $C_2>\frac{1}{2},  \frac{8 C_2-2}{8 C_2-3}<\gamma <2,  \lambda =2 \sqrt{\frac{\gamma 
   C_2-C_2}{\gamma }}$, o
   \item $C_2>1,  1<\gamma <\frac{2 \left(8 C_2-3\right)}{16 C_2-9}+4 \sqrt{\frac{C_2}{\left(16 C_2-9\right){}^2}},  \lambda =\sqrt{\frac{(2-3 \gamma )^2-8 (\gamma -1)^2 C_2}{\gamma
   ^2}}$, o
   \item $C_2>2,  1<\gamma <\frac{2 \left(4 C_2-3\right)}{8 C_2-9}+2 \sqrt{2} \sqrt{\frac{C_2}{\left(8 C_2-9\right){}^2}},  \lambda =\sqrt{\frac{(2-3 \gamma )^2-8 (\gamma -1)^2 C_2}{\gamma ^2}}$, o
   \item $0<C_2\leq \frac{1}{2},  1<\gamma <2,  \lambda =2 \sqrt{\frac{(\gamma-1)C_2}{\gamma }}$, o
   \item $0<C_2\leq 1,  1<\gamma <2,  \lambda =\sqrt{\frac{(2-3 \gamma )^2-8 (\gamma -1)^2 C_2}{\gamma ^2}}$, o
   \item $0<C_2\leq 2,  1<\gamma <2,  \lambda =\sqrt{\frac{(2-3 \gamma )^2-8 (\gamma -1)^2 C_2}{\gamma ^2}}$.
     \end{enumerate}
    \item Es una silla si 
    \begin{enumerate}
        \item $0<C_2\leq \frac{1}{2},  1<\gamma <2,  0<\lambda <2 \sqrt{\frac{(\gamma-1)C_2}{\gamma }}$, o
        \item $0<C_2\leq \frac{1}{2},  1<\gamma <2,  2 \sqrt{\frac{(\gamma-1)C_2}{\gamma }}<\lambda <\sqrt{\frac{(2-3
   \gamma )^2-8 (\gamma -1)^2 C_2}{\gamma ^2}}$, o
   \item $0<C_2\leq \frac{1}{2},  1<\gamma <2,  \lambda >\sqrt{\frac{(2-3 \gamma )^2-8 (\gamma -1)^2 C_2}{\gamma ^2}}$, o 
   \item $\frac{1}{2}<C_2\leq 1,  1<\gamma \leq \frac{8 C_2-2}{8 C_2-3},  0<\lambda <2 \sqrt{\frac{(\gamma-1)C_2}{\gamma }}$, o 
   \item $\frac{1}{2}<C_2\leq 1,  1<\gamma <\frac{8 C_2-2}{8 C_2-3},  2
   \sqrt{\frac{(\gamma-1)C_2}{\gamma }}<\lambda <\sqrt{\frac{(2-3    \gamma )^2-8 (\gamma -1)^2 C_2}{\gamma ^2}}$, o
   \item $\frac{1}{2}<C_2\leq 1,  1<\gamma <\frac{8 C_2-2}{8
   C_2-3},  \lambda >\sqrt{\frac{(2-3    \gamma )^2-8 (\gamma -1)^2 C_2}{\gamma ^2}}$, o
   \item $\frac{1}{2}<C_2\leq 1,  \frac{8 C_2-2}{8 C_2-3}<\gamma <2,  0<\lambda
   <\sqrt{\frac{(2-3    \gamma )^2-8 (\gamma -1)^2 C_2}{\gamma ^2}}$, o
   \item $\frac{1}{2}<C_2\leq 1,  \frac{8 C_2-2}{8 C_2-3}\leq \gamma <2,  \lambda >2 \sqrt{\frac{(\gamma-1)C_2}{\gamma }}$, o 
   \item $C_2>1,  1<\gamma \leq \frac{8 C_2-2}{8 C_2-3},  0<\lambda <2 \sqrt{\frac{(\gamma-1)C_2}{\gamma }}$, o 
   \item $C_2>1,  1<\gamma <\frac{8 C_2-2}{8 C_2-3},  2 \sqrt{\frac{(\gamma-1)C_2}{\gamma }}<\lambda
   <\sqrt{\frac{(2-3    \gamma )^2-8 (\gamma -1)^2 C_2}{\gamma ^2}}$, o 
   \item $C_2>1,  1<\gamma <\frac{8 C_2-2}{8 C_2-3},  \lambda >\sqrt{\frac{(2-3    \gamma )^2-8 (\gamma -1)^2 C_2}{\gamma ^2}}$, o 
   \item $C_2>1,  \frac{8 C_2-2}{8 C_2-3}<\gamma <\frac{2 \left(8 C_2-3\right)}{16 C_2-9}+4 \sqrt{\frac{C_2}{\left(16 C_2-9\right){}^2}},  0<\lambda
   <\sqrt{\frac{(2-3    \gamma )^2-8 (\gamma -1)^2 C_2}{\gamma ^2}}$, o 
   \item $C_2>1,  \frac{8 C_2-2}{8 C_2-3}\leq \gamma <\frac{2 \left(8 C_2-3\right)}{16 C_2-9}+4
   \sqrt{\frac{C_2}{\left(16 C_2-9\right){}^2}},  \lambda >2 \sqrt{\frac{(\gamma-1)C_2}{\gamma }}$, o 
   \item $C_2>1,  \frac{2 \left(8 C_2-3\right)}{16 C_2-9}+4 \sqrt{\frac{C_2}{\left(16 C_2-9\right){}^2}}\leq \gamma
   <2,  \lambda >2 \sqrt{\frac{(\gamma-1)C_2}{\gamma }}$.
    \end{enumerate}
\end{enumerate}

 \item $P_{14}(\lambda): (\Sigma, K, u, w)= \left(-\frac{2 (\gamma -1)}{3 \gamma -2} ,  \frac{4-8 C_2 (\gamma -1)^2-\gamma
    \left(\gamma  \left(\lambda ^2-9\right)+12\right)}{(2-3 \gamma
   )^2} ,  \frac{\gamma  \lambda }{\sqrt{2} (3 \gamma -2)} , 
   \frac{\lambda }{\sqrt{\frac{4}{\gamma }-6}}\right)$,\newline   $\Omega_t=\frac{4 (\gamma -1)
   C_2-\gamma  \lambda ^2}{(2-3 \gamma )^2}$. Se reduce a $P_{14}$  cuando $\lambda\rightarrow 0$.  
   Las partes reales de los autovaloes $\mu_i$ se representan en la figura \ref{A} para algunos valores de $C_2$, donde se muestra que $P_{14}(\lambda)$ es típicamente una silla para los valores dados del  parámetro $C_2$ (o es no hiperbólico). 
  
\begin{figure*}
       \centering
       \includegraphics[scale=0.37]{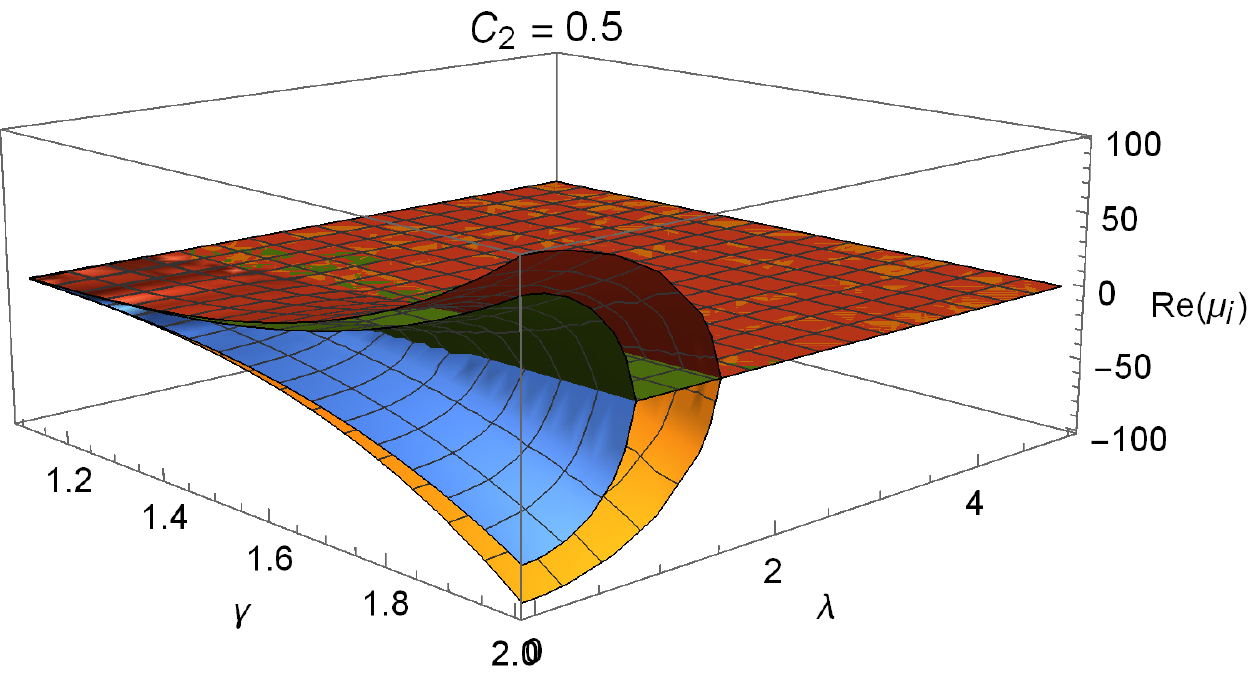} 
       \includegraphics[scale=0.37]{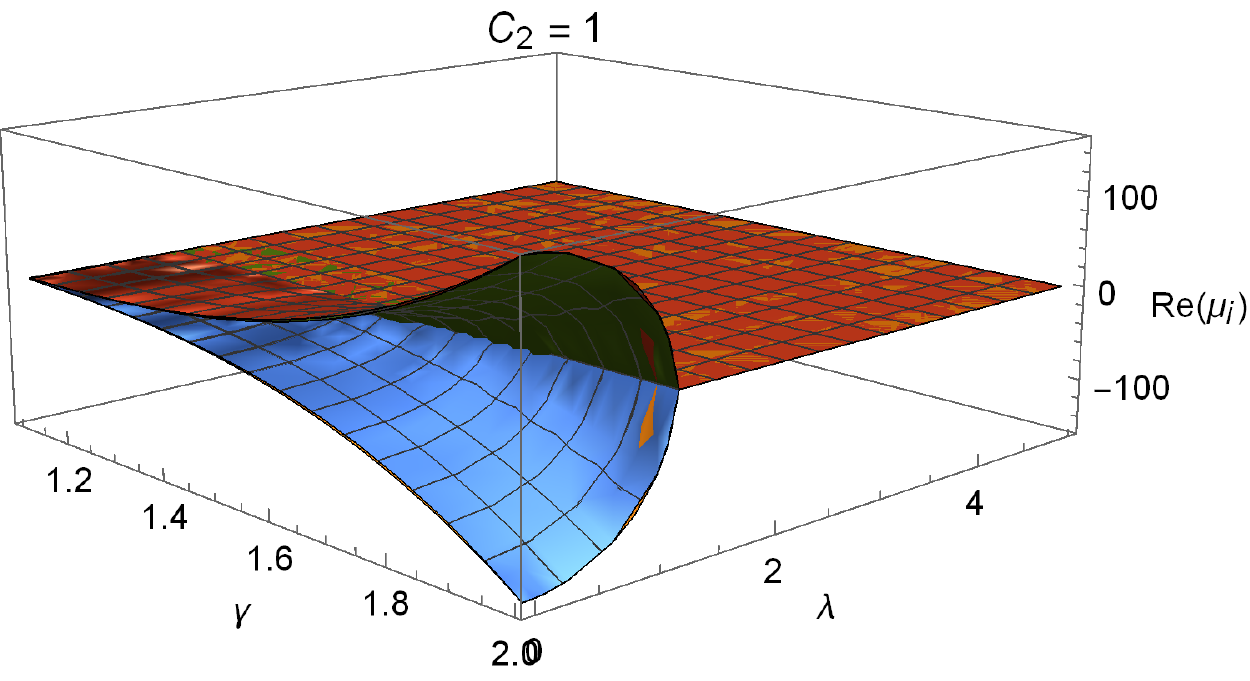} 
       \includegraphics[scale=0.37]{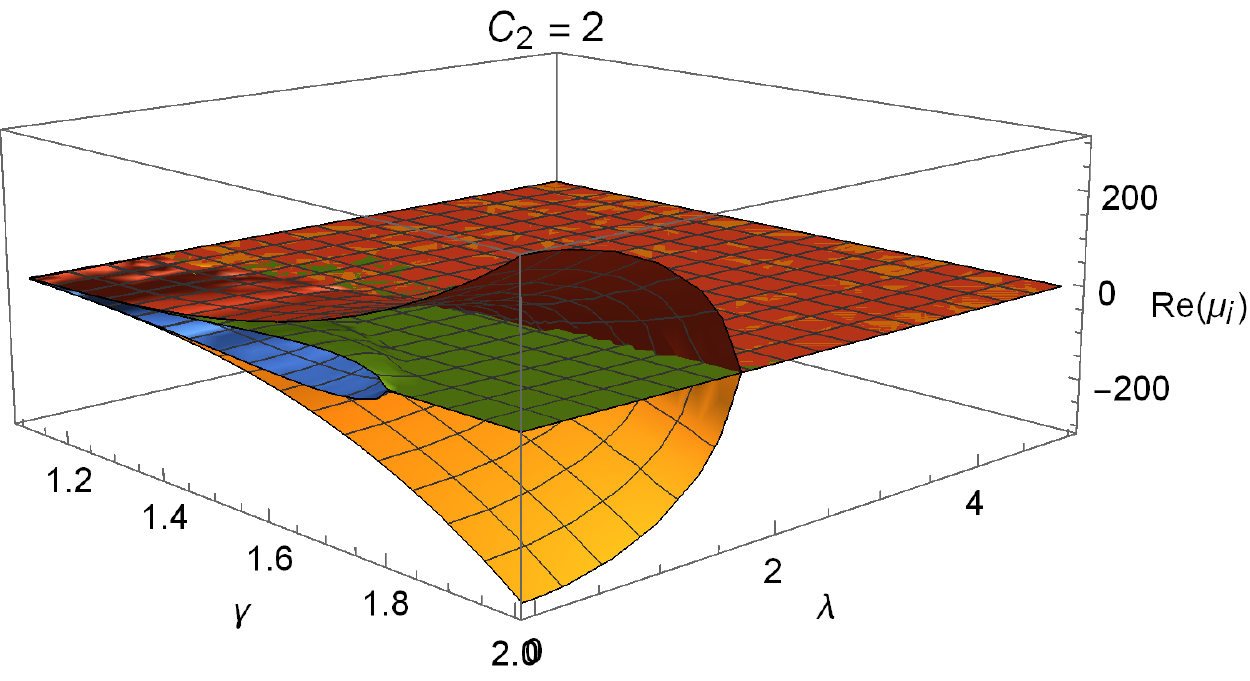}
       \caption{ \label{A} Las partes reales de los autovalores $\mu_i$ correspondientes  a $P_{14}(\lambda)$ para algunos valores de $C_2$. }
         \end{figure*}

   \item $P_{15}(\lambda): (\Sigma, K, u, w)= \left(-\frac{1}{2 C_2} ,  0 ,  \frac{1}{\sqrt{2} \lambda } ,  \frac{\sqrt{\lambda ^2+\left(2-4 \lambda ^2\right) C_2}}{2 \lambda 
   \sqrt{C_2}}\right)$,  $\Omega_t= 0$, con autovalores \newline $\left\{\frac{1}{C_2}+\frac{2}{\lambda ^2}-2,\frac{\gamma }{2 C_2-2
   \gamma  C_2}+\frac{2}{\lambda ^2},\frac{1}{2 C_2}+\frac{1}{\lambda
   ^2}-2,\frac{1}{2 C_2}+\frac{1}{\lambda ^2}-2\right\}$. 
   
   \begin{enumerate}
       \item Es un pozo si 
       \begin{enumerate}
           \item $\frac{1}{2}<C_2\leq 1, 1<\gamma <2, \lambda >\sqrt{2} \sqrt{\frac{C_2}{2 C_2-1}}$, o
        \item $C_2>1, 1<\gamma \leq \frac{4 C_2-2}{4 C_2-3}, \lambda >\sqrt{2} \sqrt{\frac{C_2}{2 C_2-1}}$, o
        \item $C_2>1, \frac{4 C_2-2}{4 C_2-3}<\gamma <2, \lambda >2 \sqrt{\frac{(\gamma-1)C_2}{\gamma }}$
       \end{enumerate}
       \item Es un fuente si 
       \begin{enumerate}
        \item $0<C_2\leq \frac{1}{2}, 1<\gamma <2, 0<\lambda <2 \sqrt{\frac{(\gamma-1)C_2}{\gamma }}$, o
        \item $C_2>\frac{1}{2}, 1<\gamma \leq \frac{8 C_2-2}{8 C_2-3}, 0<\lambda <2 \sqrt{\frac{(\gamma-1)C_2}{\gamma }}$, o 
        \item $C_2>\frac{1}{2}, \frac{8 C_2-2}{8 C_2-3}<\gamma <2, 0<\lambda <\sqrt{2} \sqrt{\frac{C_2}{4 C_2-1}}$.
       \end{enumerate}
       \item No hiperbólico si
       \begin{enumerate}
           \item $0<C_2\leq \frac{1}{4},  1<\gamma <2,  \lambda =2 \sqrt{\frac{(\gamma -1) C_2}{\gamma }}$, o \item $\frac{1}{4}<C_2\leq \frac{1}{2},  1<\gamma <2,  \lambda =2 \sqrt{\frac{(\gamma -1) C_2}{\gamma
   }}$, o 
           \item $\frac{1}{4}<C_2\leq \frac{1}{2},  1<\gamma <2,  \lambda =\sqrt{2} \sqrt{\frac{C_2}{4 C_2-1}}$, o 
           \item $\frac{1}{2}<C_2\leq 1,  \gamma =\frac{8 C_2-2}{8 C_2-3},  \lambda =\sqrt{2}
   \sqrt{\frac{C_2}{4 C_2-1}}$, o 
           \item $\frac{1}{2}<C_2\leq 1,  \gamma =\frac{8 C_2-2}{8 C_2-3},  \lambda =\sqrt{2} \sqrt{\frac{C_2}{2 C_2-1}}$, o 
           \item $\frac{1}{2}<C_2\leq 1,  \frac{8 C_2-2}{8
   C_2-3}<\gamma <2,  \lambda =2 \sqrt{\frac{(\gamma -1) C_2}{\gamma }}$, o 
           \item $\frac{1}{2}<C_2\leq 1,  1<\gamma <\frac{8 C_2-2}{8 C_2-3},  \lambda =2 \sqrt{\frac{(\gamma -1) C_2}{\gamma }}$, o 
           \item $\frac{1}{2}<C_2\leq 1,  \frac{8 C_2-2}{8 C_2-3}<\gamma <2,  \lambda =\sqrt{2} \sqrt{\frac{C_2}{4 C_2-1}}$, o 
           \item $\frac{1}{2}<C_2\leq 1,  1<\gamma <\frac{8 C_2-2}{8 C_2-3},  \lambda =\sqrt{2}
   \sqrt{\frac{C_2}{4 C_2-1}}$, o 
           \item $\frac{1}{2}<C_2\leq 1,  \frac{8 C_2-2}{8 C_2-3}<\gamma <2,  \lambda =\sqrt{2} \sqrt{\frac{C_2}{2 C_2-1}}$, o 
           \item $\frac{1}{2}<C_2\leq 1,  1<\gamma <\frac{8 C_2-2}{8
   C_2-3},  \lambda =\sqrt{2} \sqrt{\frac{C_2}{2 C_2-1}}$, o          \item $C_2>1,  \frac{4 C_2-2}{4 C_2-3}<\gamma <2,  \lambda =2 \sqrt{\frac{(\gamma -1) C_2}{\gamma }}$, o 
            \item $C_2>1,  \frac{8 C_2-2}{8
   C_2-3}<\gamma <\frac{4 C_2-2}{4 C_2-3},  \lambda =2 \sqrt{\frac{(\gamma -1) C_2}{\gamma }}$, o 
            \item $C_2>1,  1<\gamma <\frac{8 C_2-2}{8 C_2-3},  \lambda =2 \sqrt{\frac{(\gamma -1) C_2}{\gamma }}$, o
           \item $C_2>1,  \gamma =\frac{4 C_2-2}{4 C_2-3},  \lambda =\sqrt{2} \sqrt{\frac{C_2}{4 C_2-1}}$, o 
           \item $C_2>1,  \gamma =\frac{8 C_2-2}{8 C_2-3},  \lambda =\sqrt{2} \sqrt{\frac{C_2}{4 C_2-1}}$, o 
           \item $C_2>1,  \frac{4 C_2-2}{4 C_2-3}<\gamma <2,  \lambda =\sqrt{2} \sqrt{\frac{C_2}{4 C_2-1}}$, o 
           \item $C_2>1,  \frac{8 C_2-2}{8 C_2-3}<\gamma <\frac{4 C_2-2}{4 C_2-3},  \lambda =\sqrt{2} \sqrt{\frac{C_2}{4
   C_2-1}}$, o 
           \item $C_2>1,  1<\gamma <\frac{8 C_2-2}{8 C_2-3},  \lambda =\sqrt{2} \sqrt{\frac{C_2}{4 C_2-1}}$, o 
           \item $C_2>1,  \gamma =\frac{4 C_2-2}{4 C_2-3},  \lambda =\sqrt{2} \sqrt{\frac{C_2}{2   C_2-1}}$, o 
           \item $C_2>1,  \gamma =\frac{8 C_2-2}{8 C_2-3},  \lambda =\sqrt{2} \sqrt{\frac{C_2}{2 C_2-1}}$, o 
           \item $C_2>1,  \frac{4 C_2-2}{4 C_2-3}<\gamma <2,  \lambda =\sqrt{2} \sqrt{\frac{C_2}{2   C_2-1}}$, o 
           \item $C_2>1,  \frac{8 C_2-2}{8 C_2-3}<\gamma <\frac{4 C_2-2}{4 C_2-3},  \lambda =\sqrt{2} \sqrt{\frac{C_2}{2 C_2-1}}$, o
           \item $C_2>1,  1<\gamma <\frac{8 C_2-2}{8 C_2-3},  \lambda =\sqrt{2} \sqrt{\frac{C_2}{2 C_2-1}}$.
       \end{enumerate}
       
       Es una silla si 
       \begin{enumerate}
           \item $\frac{1}{2}<C_2\leq 1,  \frac{8 C_2-2}{8 C_2-3}<\gamma <2,  \sqrt{2} \sqrt{\frac{C_2}{4 C_2-1}}<\lambda <2 \sqrt{\frac{(\gamma-1)C_2}{\gamma }}$, or 
           \item $C_2>1,  \frac{4 C_2-2}{4 C_2-3}<\gamma <2, \sqrt{2} \sqrt{\frac{C_2}{4 C_2-1}}<\lambda <\sqrt{2} \sqrt{\frac{C_2}{2 C_2-1}}$, or 
           \item $C_2>1,  \frac{8 C_2-2}{8 C_2-3}<\gamma \leq \frac{4 C_2-2}{4 C_2-3},  \sqrt{2} \sqrt{\frac{C_2}{4 C_2-1}}<\lambda <2\sqrt{\frac{(\gamma-1)C_2}{\gamma }}$, or
           \item $0<C_2\leq \frac{1}{4},  1<\gamma <2,  \lambda >2 \sqrt{\frac{(\gamma-1)C_2}{\gamma }}$, or 
           \item $\frac{1}{4}<C_2\leq \frac{1}{2},  1<\gamma <2,  2 \sqrt{\frac{(\gamma-1)C_2}{\gamma }}<\lambda <\sqrt{2} \sqrt{\frac{C_2}{4 C_2-1}}$, or 
           \item $C_2>\frac{1}{2},  1<\gamma <\frac{8 C_2-2}{8 C_2-3},  2 \sqrt{\frac{(\gamma-1)C_2}{\gamma }}<\lambda <\sqrt{2}    \sqrt{\frac{C_2}{4 C_2-1}}$, or 
           \item $\frac{1}{4}<C_2\leq \frac{1}{2},  1<\gamma <2,  \lambda >\sqrt{2} \sqrt{\frac{C_2}{4 C_2-1}}$, or 
           \item $\frac{1}{2}<C_2\leq 1,  \frac{8 C_2-2}{8 C_2-3}<\gamma <2,  2 \sqrt{\frac{(\gamma-1)C_2}{\gamma }}<\lambda <\sqrt{2} \sqrt{\frac{C_2}{2 C_2-1}}$, or 
           \item $\frac{1}{2}<C_2\leq 1,  1<\gamma \leq \frac{8 C_2-2}{8 C_2-3},  \sqrt{2} \sqrt{\frac{C_2}{4 C_2-1}}<\lambda <\sqrt{2} \sqrt{\frac{C_2}{2 C_2-1}}$, or 
           \item $C_2>1,  \frac{8 C_2-2}{8 C_2-3}<\gamma <\frac{4 C_2-2}{4 C_2-3},  2 \sqrt{\frac{(\gamma-1)C_2}{\gamma }}<\lambda <\sqrt{2} \sqrt{\frac{C_2}{2 C_2-1}}$, or 
           \item $C_2>1,  1<\gamma \leq \frac{8 C_2-2}{8 C_2-3},  \sqrt{2} \sqrt{\frac{C_2}{4 C_2-1}}<\lambda <\sqrt{2} \sqrt{\frac{C_2}{2 C_2-1}}$, or 
           \item $C_2>1,  \frac{4 C_2-2}{4 C_2-3}<\gamma <2,  \sqrt{2} \sqrt{\frac{C_2}{2 C_2-1}}<\lambda <2 \sqrt{\frac{(\gamma-1)C_2}{\gamma }}$.
       \end{enumerate}
   \end{enumerate}

\end{enumerate}

\section{Discusión}
\label{SECT:4-3}
En este capítulo se analizaron cualitativamente algunos sistemas de ecuaciones diferenciales de interés físico, asociados a modelos temporales autosimilares esféricamente simétricos con campo escalar \eqref{modelodos} utilizando las herramientas de sistemas dinámicos presentadas en el capítulo \ref{ch_1}. 

La primera característica notable del presente modelo es que la simetría conformal estática, impone restricciones a los tipos de modelos de campo escalar que se pueden estudiar en el contexto de esta investigación. Es bien conocido que para que el campo escalar no homogéneo $\phi(t,x)$ y su potencial  $V(\phi(t,x))$ satisfagan la simetría homotética, estos tiene que se d ela forma    \cite{Coley:2002je}:
\begin{align*}
& \phi(t,x)=\psi (x)-\lambda t, \quad  V(\phi(t,x))= e^{-2 t} U(\psi(x)), \quad  U(\psi)=U_0 e^{-\frac{2 \psi}{\lambda}}.    
\end{align*}
Se asume $\lambda>0$,tal que para $\psi>0$, $U\rightarrow 0$ cuando $\lambda \rightarrow 0$.
Para que un campo escalar cumpla la simetría conformal estática, $\phi$ and y su potencial  $V(\phi(t,x))$ tienen que tener la forma escrita en  \eqref{phi_respect_symm} \cite{Coley:2002je}. 

Siguiendo argumentos similares a los empleados  en el capítulo \ref{ch_3}, se normalizaron las ecuaciones con la variable $\theta$. Sin embargo, debido a la complejidad computacional del problema resultante, no fue posible obtener y tratar analíticamente todos los puntos de equilibrio del sistema \eqref{reducedsystSF} por lo que solo se consideraron algunos casos especiales de  interés físico. En particular, se estudiaron cuatro casos, estos son \eqref{scalar-field-A} correspondiente a un fluido perfecto en forma de  gas ideal, \eqref{scalar-field-B} correspondiente a las soluciones en el conjunto invariante $\Sigma=0$, \eqref{scalar-field-C} correspondiente al caso de inclinación extrema, y se estudian conjuntos el sistema \eqref{campo-escalar-D} correspondiente a la dinámica en el conjunto invariante $A=v=0$. Los puntos hiperbólicos fueron clasificados en su totalidad de acuerdo a sus condiciones de estabilidad de acuerdo al teorema \ref{hartgrob}.

Cabe destacar que fue posible recuperar los puntos de equilibrio del capítulo \ref{ch_3} para $w=u=0$ y $\lambda\rightarrow 0$, como casos particulares de este modelo.

\section{Trabajo en progreso}
\label{progreso}

A continuación se presenta un modelo que al modelo de la sección \ref{fluidosinpresion} cuando se incluye campo escalar. Debido a la complejidad computacional del problema, los presentamos como trabajo en progreso.  Los resultados del análisis los comunicaremos en un trabajo futuro. Esto es, el caso de gas ideal, $\gamma=1$, donde las ecuaciones  \eqref{YYYYY} y restricciones \eqref{YYYYY-rest} se generalizan a 
\begin{subequations}
\label{ZZZZZ}
\begin{align}
    & \widehat{\theta}= -\sqrt{3} b_2^2-\frac{\sigma \left(2 C_2 \sigma+\theta\right)}{\sqrt{3}}-\sqrt{3} \mu_t v^2-\sqrt{3} \Psi^2,\\
    & \widehat{\sigma}= -\frac{\sqrt{3} \lambda ^2
   b_1^2}{C_2}-\frac{\sqrt{3} \mu_t \left(v^2+1\right)}{2 C_2}+\frac{\sqrt{3} U_0 e^{-\frac{2 \psi}{\lambda }}}{C_2}-\frac{\sigma (2 \theta+\sigma)}{\sqrt{3}},\\
   & \widehat{b_1}=
   \frac{b_1 \sigma}{\sqrt{3}},\\
   & \widehat{b_2}= -\frac{b_2 (\theta+\sigma)}{\sqrt{3}},\\
   & \widehat{\mu_t}= \mu_t \left(\frac{v^2 (\sigma-2 \theta)-\sigma}{\sqrt{3}
   v^2}-2 b_1 v\right),\\
   & \widehat{v}= b_1 \left(v^2-1\right)-\frac{\sigma \left(v^2-1\right)}{\sqrt{3} v},\\
   & \widehat{\Psi}= 2 \lambda  b_1^2-\frac{(2 \theta+\sigma ) \Psi}{\sqrt{3}}-\frac{2 V_0 e^{-\frac{2 \psi}{\lambda }}}{\lambda }, \end{align}
   \begin{align}
 & \widehat{\psi}=\Psi,  
\end{align}
\end{subequations}
con restricciones 
\begin{subequations}
\label{ZZZZZ-rest}
\begin{align}
    & -\frac{2 C_2 b_1 \sigma}{\sqrt{3}}-\lambda  b_1 \Psi+\mu_{t} v=0,\\
    & -3 b_1^2 \left(2 C_2+\lambda ^2\right)-6 b_2^2-2 C_2 \sigma^2+2 \theta^2-6 \mu_t
   v^2+6 U_0 e^{-\frac{2 \psi}{\lambda }}-3 \Psi^2=0. 
\end{align}
\end{subequations}

Se recuerda que para gas ideal, el tensor de energía-momentum de la materia se reduce a \eqref{Tmgas-ideal}. 
Usando, como antes, las variables
\begin{equation}
{\Sigma} =\frac{\sigma}{\theta},\quad  A=\frac{\sqrt{3} b_1}{\theta},\quad K=\frac{3 b_2^2}{\theta^2},\quad {\Omega}_t =\frac{3\mu_t}{\theta^2}, \quad u= \sqrt{\frac{3}{2}} \frac{\Psi}{\theta}, \quad w=\frac{ e^{-\frac{ \psi}{\lambda }}\sqrt{3 U_0}}{\theta},
\end{equation}
junto con la coordenada radial
$$\frac{df}{d \eta} := \frac{\sqrt{3}\widehat{f}}{\theta},$$
y el ``parámetro gradiente de Hubble'' ${r}$, dado por
\begin{equation}
   \widehat{\theta}=-r {\theta}^2,
\end{equation} tal que
\begin{equation}
    r=\frac{2 C_2 \Sigma ^2+K+\Sigma +2 u^2+v^2 \Omega }{\sqrt{3}},
\end{equation}
las ecuaciones \eqref{ZZZZZ} con restricciones \eqref{ZZZZZ-rest}
se transforman en el siguiente sistema (generalizando \eqref{dust-1}): 
\begin{subequations}
\begin{align}
& \Sigma'=-\frac{(A^2 \lambda^2 -w^2)}{C_2}+2 C_2 \Sigma ^3+\Omega_t  \left(\Sigma  v^2-\frac{v^2+1}{2 C_2}\right)+\Sigma  \left(K+2 u^2-2\right),\\
& A'= A \left(2 C_2 \Sigma ^2+K+2 \left(\Sigma +u^2\right)+v^2 \Omega_t \right),\\
& K'= 2 K\left(2 C_2 \Sigma ^2+K+2 u^2+v^2 \Omega_t -1\right),\\
& v'= \frac{\left(v^2-1\right) (A v-\Sigma )}{v},\\
& \Omega_t'=\Omega_t  \left(-2 A v+2 \left(2 C_2 \Sigma ^2+K+\Sigma +2 u^2+v^2 \Omega_t \right)+\Sigma -\frac{\Sigma }{v^2}-2\right),\\
& u'= \frac{\sqrt{2}\left(A^2 \lambda ^2-w^2\right)}{\lambda }+u \left(2 C_2 \Sigma ^2+K+2 u^2+v^2 \Omega_t -2\right),\\
& w'= w \left(2 C_2 \Sigma ^2+K+\Sigma +2 u^2-\frac{\sqrt{2} u}{\lambda }+v^2 \Omega_t \right),
   \end{align}
\end{subequations}
con restricciones 
\begin{subequations}
\label{rest-dust-2}
\begin{align}
& -2 A C_2 \Sigma -\sqrt{2} A \lambda  u+\Omega_t  v=0,\\
& 1-A^2 C_2-\frac{1}{2} A^2 \lambda ^2-C_2 \Sigma ^2-K-u^2-v^2 \Omega_t +w^2=0.
\end{align}
\end{subequations}

Para $v\neq 0$ las restricciones \eqref{rest-dust-2} se pueden resolver globalmente para $\Omega_t$ y $K$.

Luego, se obtiene el sistema reducido 
\begin{subequations}
\label{eq:4.30}
\begin{align}
 &\Sigma'=-\frac{A^2 \left(C_2 \Sigma  \left(2 C_2+\lambda ^2\right)+2 \lambda ^2\right)}{2 C_2}
  -\frac{A \left(v^2+1\right) \left(2 C_2 \Sigma +\sqrt{2} \lambda  u\right)}{2 C_2 v}\nonumber \\
 & +C_2 \Sigma ^3+\frac{w^2}{C_2}+\Sigma 
   \left(u^2+w^2-1\right),\\
&A'= -\frac{1}{2} A^3 \lambda ^2+A C_2 \left(\Sigma ^2-A^2\right)+A \left(2 \Sigma +u^2+w^2+1\right),\\
&v'= \frac{\left(v^2-1\right) (A v-\Sigma )}{v},\\
&u'=C_2 u \left(\Sigma ^2-A^2\right)+u \left(-\frac{A^2 \lambda
   ^2}{2}+w^2-1\right)+\frac{\sqrt{2} \left(A^2 \lambda ^2-w^2\right)}{\lambda }+u^3,\\
& w'= w \left(C_2 \left(\Sigma ^2-A^2\right)-\frac{A^2 \lambda ^2}{2}+\Sigma +u^2-\frac{\sqrt{2} u}{\lambda }+w^2+1\right),   
\end{align}
\end{subequations}
definido en el espacio de estados
\begin{align}
    & \Bigg\{(\Sigma, A, v, u,w)\in\mathbb{R}^5: \frac{A \left(2 C_2 \Sigma +\sqrt{2} \lambda  u\right)}{v}\geq 0,  \nonumber \\ 
    & \quad
    w^2+1\geq C_2 \left(A^2+2 A \Sigma  v+\Sigma ^2\right)+\frac{A^2 \lambda ^2}{2}+\sqrt{2} A \lambda  u v+u^2, \nonumber\\
    & \quad  v\in[-1,0)\cup(0,-1]\Bigg\}
\end{align}
Para $u=w=0$ y $\lambda \rightarrow 0$ se recuperan los puntos de equilibrio de la sección \ref{fluidosinpresion}. 
El análisis de estabilidad de los puntos de equilibrio del sistema más general \eqref{reducedsystSF} y del sistema \eqref{eq:4.30}  se realizará de manera semianalítica  en un trabajo psterior debido a la complejidad computacional de los dos sistema a tratar, los cuáles no puede ser resueltos con nuestras herramientas computacionales actuales. 

\lhead{Capítulo \ref{ch_5}}
\rhead{Conclusiones}
\cfoot{\thepage}
\renewcommand{\headrulewidth}{1pt}
\renewcommand{\footrulewidth}{1pt}
\chapter{Conclusiones}\label{ch_5}
\noindent 

En esta tesis se estudió el espacio de las soluciones de las ecuaciones diferenciales que resultan de considerar fluidos de materia de tipo campo escalar o fluido perfecto en la teoría Einstein-\ae ther que son de interés en cosmología y astrofísica. La teoría de la gravedad de Einstein-\ae ther consiste en la Relatividad General acoplada a un campo vectorial de tipo tiempo unitario, llamado el ``\ae ther''. En esta teoría efectiva, la invarianza de Lorentz es violada, pero la localidad y la covarianza son preservadas en presencia del campo vectorial.

Para la formulación matemática de los modelos se utilizó el formalismo ortonormal 1+3 para escribir las ecuaciones de campo como un sistema de ecuaciones diferenciales parciales en dos variables para métricas esféricamente simétricas. Por otra parte, usando la formulación diagonal homotética, se pudieron escribir las  ecuaciones diferenciales parciales  como ecuaciones diferenciales ordinarias, usando el hecho que la métrica se adapta a la simetría homotética. Las ecuaciones resultantes son muy similares a las de los modelos con hipersuperficies espaciales homogéneas \cite{Goliath:1998mw}, obteniéndose además restricciones algebraicas. Fue  posible entonces usar las técnicas de la teoría cualitativa de los sistemas dinámicos para el análisis de estabilidad de las soluciones de los modelos.  Los resultados analíticos se verificaron mediante integración numérica. 

En el capítulo \ref{ch_1} se hizo una revisión breve de las herramientas de la teoría cualitativa de los sistemas dinámicos que se aplicaron  en la tesis. En el capítulo \ref{ch_2} se discutió el marco teórico.  En la sección \ref{aetheory} se presentó la teoría de la gravedad Einstein-\ae ther, que contiene a la teoría de Relatividad General en un caso límite. En la sección \ref{section2.2} se discutió el formalismo $1+3$ y en la sección \ref{homotetica} se introdujo el formalismo diagonal homotético, que está basado en el formalismo $1+3$. En el capítulo \ref{ch_3} se estudiaron métricas conformalmente estáticas en Teoría de Einstein-aether para modelos de interés físico, como son los fluidos perfectos sin presión y modelos con inclinación extrema. Se obtuvieron y discutieron los criterios de estabilidad de los puntos de equilibrio de los sistemas dinámicos, imponiendo restricciones al espacio de parámetros. También se presentaron retratos de fase para ilustrar el comportamiento cualitativo de las soluciones.

Los puntos de equilibrio  obtenidos por \cite{Goliath:1998mx} se recuperan como casos particulares del presente modelo. 
En la notación $\text{Kernel}^{\text{sgn}(v)}_{\text{sgn}({\Sigma})}$ el kernel indica la interpretación del punto: $M,C$ representan el espacio tiempo de Minkowski; $K$ representa una solución de Kasner; $T$ corresponde a soluciones estáticas; $SL_{\pm}$ corresponde a un espacio plano FLRW y orbitas estáticas dependiendo del parámetro $\gamma$. $H$ se asocia con un cambio de causalidad del campo vectorial homotético. 

\begin{enumerate}

    \item $SL_{\pm}$: Líneas sónicas definidas por ${A}=-\frac{\gamma 
   \varepsilon  (\gamma  (\Sigma
   +2)-2)}{4 (\gamma
   -1)^{3/2}}$,
$v=\varepsilon\sqrt{\gamma -1}$, fueron analizadas en la sección \ref{SL}. A diferencia de relatividad general, para  cuando $1<\gamma<2$ el sistema \eqref{reducedsyst} admite los puntos de equilibrio 
\begin{enumerate}
    \item $SL_1: C_2= \frac{\gamma
   ^2}{4 (\gamma -1)^2},\Sigma =
   \frac{2 (\gamma -1)}{\gamma
   },v=\sqrt{\gamma -1},A=
  - \frac{\gamma  (\gamma  (\Sigma
   +2)-2)}{4 (\gamma
   -1)^{3/2}}$, 
    \item $SL_2: C_2= \frac{\gamma
   ^2}{4 (\gamma -1)^2},\Sigma =
   -\frac{2 (\gamma -1)}{\gamma
   },v= -\sqrt{\gamma -1},A=
   \frac{\gamma  (\gamma  (\Sigma
   +2)-2)}{4 (\gamma
   -1)^{3/2}}$,
\end{enumerate}
los cuáles yacen en la línea sónica.  Si $\gamma=2, C_2=1$ dichos puntos existen, y como $\gamma=2$  el fluido se comporta como materia rígida. Adicionalmente, si $\gamma=2, C_2=1$, estos puntos corresponden a modelos con inclinación extrema ($v=\varepsilon$), $SL_1: \Sigma=1, A=-2, v=1$, y $SL_2: \Sigma=-1, A=0, v=-1$.   $SL_{\pm}$ corresponde a un espacio plano FLRW y orbitas estáticas dependiendo del parámetro $\gamma$

    \item $\widetilde{M}^{\pm}=({\Sigma}, {A}, v)=\left(0,1,\frac{(\gamma -1) \gamma  \pm \left(\sqrt{(\gamma -1) \left((\gamma
   -1) \gamma ^2+(2-\gamma ) (3 \gamma -2)\right)}\right)}{2-\gamma }\right),\quad ({K},\Omega_t)=(0,0)$, existen si $C_2=1$. Representa el espacio tiempo  de Minkowski. 
   
    \item $M^+=({\Sigma}, {A}, v)=(0,1,1),\quad ({K},\Omega_t)=(0,0)$, corresponde a  $P_5$ cuando $C_2=1$.   Representan el espacio tiempo de Minkowski.
    
    \item $M^-=({\Sigma}, {A}, v)=(0,1,-1),\quad ({K},\Omega_t)=(0,0)$, corresponde a $P_6$ cuando $C_2=1$.  Representan el espacio tiempo de Minkowski.
    
    \item $C^0=({\Sigma}, {A}, v)=(0,0,0),\quad ({K},\Omega_t)=(1,0)$, corresponde a $N_1$. Representan el espacio tiempo de Minkowski.
    
    \item $C^{\pm }=({\Sigma}, {A}, v)=(0,0,\pm 1),\quad ({K},\Omega_t)=(1,0)$, corresponden a $N_{2,3}$. Representan el espacio tiempo de Minkowski. 
    
    \item $K^0_-=({\Sigma}, {A}, v)=(-1,0,0),\quad ({K},\Omega_t)=(0,0)$, corresponde a $P_{11}$ cuando $C_2=1$.  Representa una solución de Kasner.
    
    \item $K^0_+=({\Sigma}, {A}, v)=(1,0,0),\quad ({K},\Omega_t)=(0,0)$,  corresponde a $P_{12}$ cuando $C_2=1$.  Representa una solución de Kasner.
    
     \item $K^{\pm}_-=({\Sigma}, {A}, v)=(-1,0,\pm 1),\quad ({K},\Omega_t)=(0,0)$, corresponden a $P_{1,2}$ cuando $C_2=1$.    representa una solución de Kasner.
    
    \item $K^{\pm}_+=({\Sigma}, {A}, v)=(1,0,\pm 1),\quad ({K},\Omega_t)=(0,0)$, corresponden a $P_{3,4}$ cuando $C_2=1$.   Representa una solución de Kasner.
    
    \item $T=({\Sigma}, {A}, v)=\left(-2\frac{\gamma-1}{3\gamma-2},0,0\right),\quad ({K},\Omega_t)=\left(\frac{\gamma^2+4(\gamma-1)}{(3\gamma -2},\frac{4(\gamma-1)}{(3\gamma -2}\right)$,  corresponde a $P_{13}$ cuando $C_2=1$. 
    
    \item $H^-$: La recta definida por $ {A}( {\Sigma})= {\Sigma}  +1$, $v( {\Sigma})=-1$,$\quad (0,-2 {\Sigma} {A})$. Dicha línea  de puntos de equilibrio se asocia con un cambio de causalidad del campo vectorial homotético,
   \end{enumerate}
En el capítulo \ref{ch_4} se estudiaron métricas conformalmente estáticas en Teoría de Einstein-aether para modelos con fluido perfecto inclinado y campo escalar no homogéneo con potential exponencial, por lo que el modelo contiene al modelo estudiado en el capítulo  \ref{ch_3} y, por tanto, contiene al modelo estudiado  en \cite{Goliath:1998mw}. Se estudiaron casos particulares de interés en la Física como son fluido perfecto en forma de gas ideal, soluciones con $\Sigma=0$, modelos con inclinación extrema y el conjunto invariante $A=v=0$. También fue posible obtener condiciones de estabilidad, ya sea numéricamente o analíticamente, para los puntos de equilibrio imponiendo restricciones al espacio parámetros.

Para finalizar, concluimos que   fue posible estudiar un modelo mas general que el estudiado por \cite{Goliath:1998mw}, y se reprodujeron los resultados obtenidos por los autores mediante el uso de  técnicas de la teoría cualitativa de los sistemas dinámicos. También se hizo un análisis cualitativo de algunos cojuntos invariantes para modelos con métricas temporales autosimilares esféricamente simétricas con fluido perfecto y con campo escalar. Obtuvimos puntos de equilibrio para seis sistemas y logramos clasificar todos los puntos en dos grupos, hiperbólicos y no hiperbólicos, luego fueron clasificados como sillas, fuentes o pozos según el caso, imponiendo restricciones al espacio de parámetros del modelo. Para estos se combinaron métodos de solución analíticos y métodos numéricos. Cumpliéndose los objetivos propuestos al inicio de la tesis.

\addcontentsline{toc}{chapter}{\textbf{Bibliografía.}} 
\lhead{Bibliografía}
\lhead{Bibliografía}
\cfoot{\thepage}
\renewcommand{\headrulewidth}{1pt}
\renewcommand{\footrulewidth}{1pt}



\end{document}